\documentclass[11pt, a4paper, oneside]{mythesis}

\usepackage{float}

\usepackage{tabto}

\usepackage{setspace}
\usepackage{dsfont}

\usepackage[english]{babel}

\usepackage{xcolor}

\usepackage{graphicx}

\usepackage[textwidth=12cm,margin=1in]{geometry}

\usepackage[pagebackref]{hyperref}
\hypersetup{colorlinks=true, citecolor=blue, urlcolor=cyan, linkcolor=blue}

\usepackage{amssymb}
\usepackage{amsmath}

\usepackage{lmodern}

\usepackage{mathrsfs}
\usepackage{tikz}
\usepackage{ctable}

\usepackage{physics}

\usepackage{pifont}

\usepackage{comment}

\usepackage{titling}
\usepackage{titlesec}
\usepackage{setspace}

\begin{document}

\frontmatter 

\begin{titlepage}
   \begin{center}
       \vspace{0.5cm}
		\rule{\textwidth}{2pt}
		\vspace{-6.7mm}
		\hrule
            \vspace{.5mm}
		\hrule
		\vspace{2.45mm}
       \textbf{\Large{Multiresolution analysis of quantum theories using Daubechies wavelet basis}}
       \vspace{2.5mm}
       \hrule
       \vspace{.5mm}
       \rule{\textwidth}{2pt}
	   
	  \vspace{1cm}
		
	   {\bf \large THESIS}\\
		\vspace{1.0cm}
		
		submitted in partial fulfillment\\
		of the requirements for the degree of\\
		\vspace{0.5cm}
		{\bf \large DOCTOR OF PHILOSOPHY}\\
		
		\vspace{0.8cm}
		by\\
		\vspace{0.8cm}
		
		\textbf{\large MRINMOY BASAK}\\
		(\large 2018PHXF0411G)\\
		
		\vspace{1.2cm}
		Under the Supervision of\\
		{\bf \large Prof. Raghunath Ratabole}\\
		
		\vspace{1.2cm}     
      \includegraphics[scale=.13]{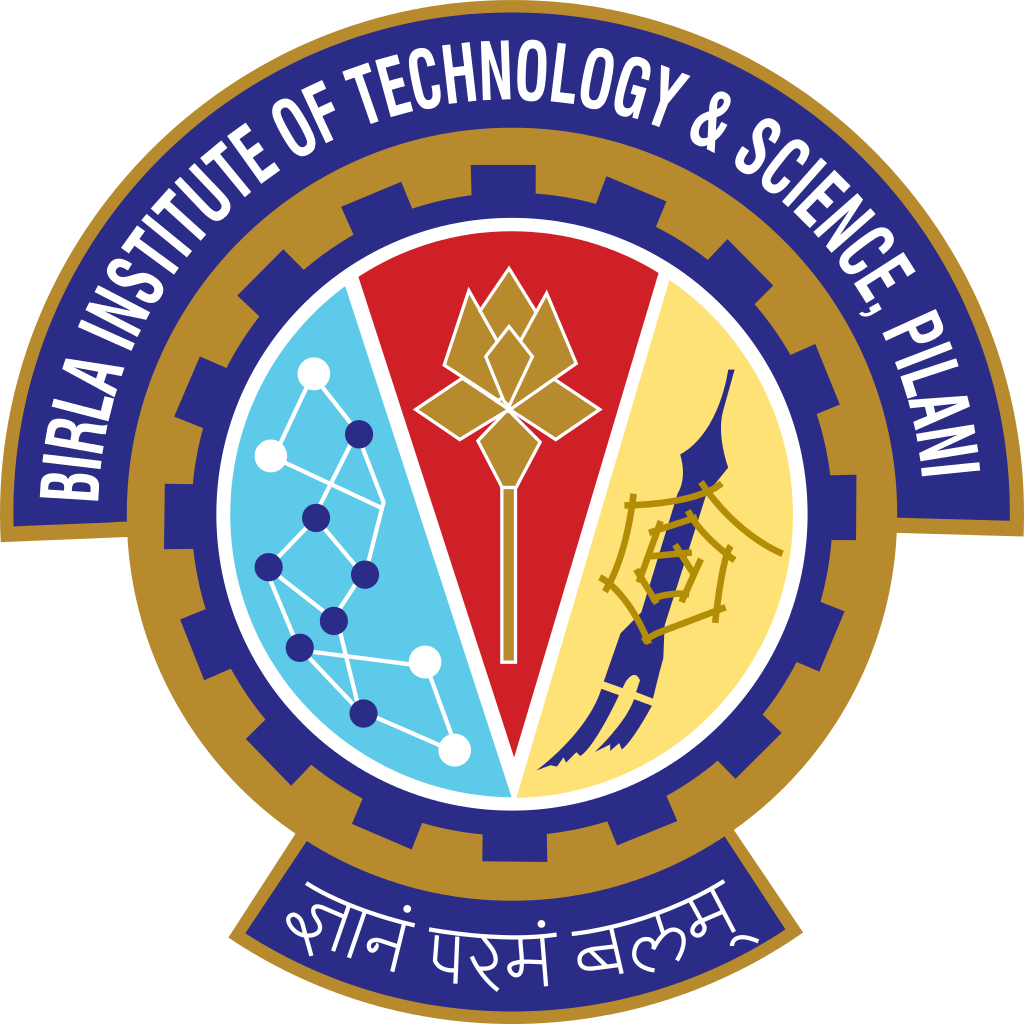}
            
      \vspace{1.0cm}
		
		{\bf  \normalsize  BIRLA INSTITUTE OF TECHNOLOGY \& SCIENCE, PILANI\\
		\today}
            
   \end{center}
\end{titlepage}

\begin{center}
\vspace*{\fill}
This page was purposefully kept blank.
\vspace*{\fill}
\end{center}
\newpage

\addcontentsline{toc}{chapter}{\textbf{Certificate}}
\vspace*{\fill}
\begin{center}
\textbf{BIRLA INSTITUTE OF TECHNOLOGY AND SCIENCE, PILANI}

\vspace{1.1cm}

\textbf{\large CERTIFICATE}

\vspace{1.1cm}
\end{center}
{\onehalfspacing
\noindent
This is to certify that the thesis titled \textbf{``Multiresolution Analysis of Quantum Theories Using Daubechies Wavelet Basis”} submitted by \textbf{Mrinmoy Basak}, ID No \textbf{2018PHXF0411G}, for the award of \textit{Doctor of Philosophy} at this Institute, embodies original research conducted by him under my supervision.

\vspace{2.2cm}
	\begin{flushleft}
       \includegraphics[width=6cm]{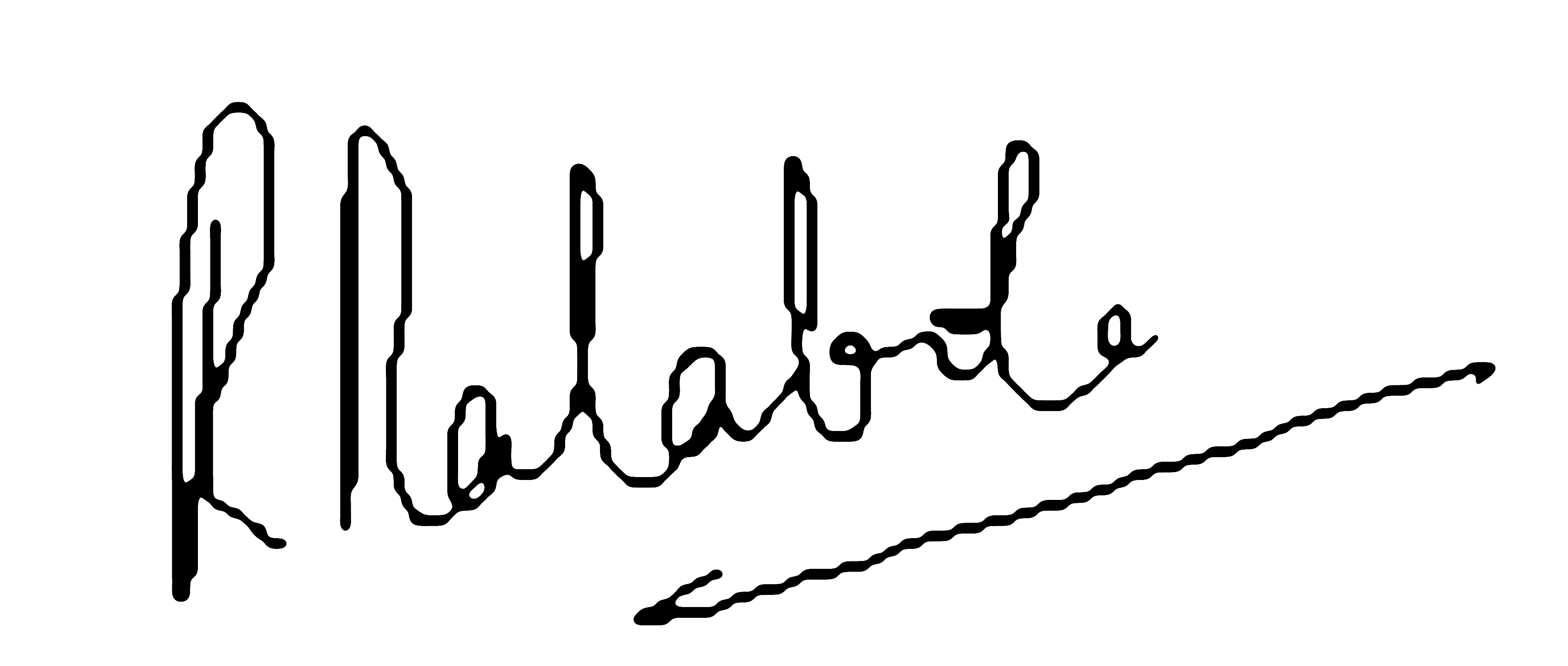}\\
		Supervisor\\
		\vspace{0.25cm}
		Name: \textbf{Prof. Raghunath Ratabole}\\	
		\vspace{0.25cm}
		Designation: \textbf{Professor}\\
		\;\;\;~~~~~~~~~~~~~~ Department of Physics\\
	    \;\;\;~~~~~~~~~~~~~~	BITS-Pilani, K.K. Birla Goa Campus\\
	    \vspace{0.25cm}
		Date: \today
	\end{flushleft}
}
\vspace*{\fill}

\newpage
\addcontentsline{toc}{chapter}{\textbf{Declaration}}
\vspace*{\fill}
\begin{center}
	{\bf \normalsize BIRLA INSTITUTE OF TECHNOLOGY AND SCIENCE, PILANI}\\
	\vspace{1.2cm}
	{\bf \large DECLARATION}
\end{center}
\vspace{1.5cm}
{\onehalfspacing
I hereby declare that the thesis titled \textbf{``Multiresolution Analysis of Quantum Theories Using Daubechies Wavelet Basis”} submitted by me for the degree of \textit{Doctor of Philosophy} at this university is an authentic record of my work, conducted under the supervision of \textbf{Prof. Raghunath Ratabole.}\\
	
\noindent
I also declare that this thesis has not been, and will not be, submitted in part or in full for the award of any other degree or diploma at this institute or any other institute or university.
	
	\vspace{2.2cm}
	\begin{flushleft}
        \includegraphics[width=4.4cm]{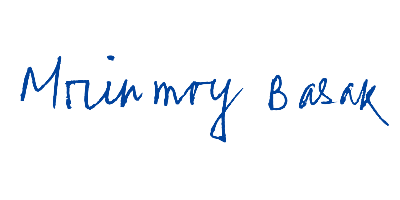}\\
		Name: \textbf{Mrinmoy Basak}\\	
		\vspace{0.25cm}
		ID No.: 2018PHXF0411G\\
		\vspace{0.25cm}
		Date: \today
	\end{flushleft}
}
\vspace*{\fill}

\newpage
\vspace*{\fill}
\begin{center}
\Large
This thesis is dedicated \\
to\\
\LARGE\it
 My Guru Sri Sri Thakur Anukulchandra,\\
 My Reverend Acharyadev,\\
 \Large\rm
and\\
\LARGE\it
My parents.
\end{center}
\vspace*{\fill}

\newpage
\addcontentsline{toc}{chapter}{\textbf{Acknowledgement}}
\begin{center}
	{\bf \large ACKNOWLEDGEMENT}
\end{center}
\vspace{1.2cm}

\noindent
After years of dedicated research, I have a long list of individuals to thank for their invaluable contributions to this thesis. I am deeply grateful to everyone who made this work possible.

\noindent
First and foremost, I extend my heartfelt gratitude to my guide, Prof. Raghunath Ratabole. I am incredibly fortunate to have had his guidance throughout my PhD journey. His unwavering support at every step has profoundly deepened my understanding of physics and broadened my perspective.

\noindent
I am also immensely thankful to my close friends, Monalisa and Kiran. Throughout my PhD, we shared countless thoughts and meals, and I am truly grateful for their integral presence in my life.

\noindent
To my friends Ashoke and Samirul, words cannot express my love and regard for you both. You are true well-wishers, and your lives continue to inspire me.

\noindent
I am grateful to my friend Prashant for our explorations of Goa’s scenic beauty. Your companionship made those experiences unforgettable.

\noindent
My heartfelt thanks go to my brother Sourish and my sister Megha. Our discussions on various topics in physics and life have been invaluable, and I am truly thankful for the lessons I have learned from both of you.

\noindent
I also extend my gratitude to my colleagues and friends, including Manish, Naresh, Harsh, Ashmita, Premchand, Arnab, Saumayen, and others, from whom I have learned so much.

\noindent
To my childhood friends, Suvankar and Himadri, you are two of the most reasonable and intellectual people in my life. My PhD journey likely began when I met you both.

\noindent
I am thankful to my friend Don for always providing real-life advice. Your guidance has never disappointed me, and I am grateful to have you as a friend who always shows me the right path.

\noindent
I also thank my elder brother, Shiladitya, for his constant motivation through both scolding and good advice. Following your guidance has never led me astray.

\noindent
To my friend Olympia, words are insufficient to express my gratitude for your support during the hardest times in my life. Your unconditional love and support have been the driving force propelling me forward.

\noindent
I express my gratitude to Dr. B P Das, who has been a source of inspiration in my life. Despite the challenges, he remains unwavering, and to me, he is a perfect person and one of my heroes.

\noindent
I would also like to extend my heartfelt gratitude to all my family members, especially my uncle, aunt, and my two adorable little sisters, Mili and Misti, for their unwavering support.

\noindent
Lastly, I extend my deepest gratitude to my parents. I am at a loss for words to convey my appreciation for everything you have done for me. My father has been my hero since childhood, always by my side during late-night study sessions. His resilience in overcoming life’s challenges continues to inspire me. As for my mother, her unwavering belief in me and constant encouragement have been my greatest motivation. Thank you for being the best parents I could have asked for.

\vspace{0.5cm}
\begin{flushright}
	{\bf Mrinmoy Basak}
\end{flushright}

\newpage
\addcontentsline{toc}{chapter}{\textbf{Abstract}}
\begin{center}
	{\bf \large ABSTRACT}
\end{center}
\vspace{1.2cm}

\noindent
This thesis studies the application of Daubechies wavelet to quantum theories within the framework of multiresolution analysis (MRA). 

\noindent
We present an overview of the MRA of the space of square-integrable functions $(L^2(\mathbb{R}))$. The explicit construction of Daubechies mother-scaling and wavelet function is described using which the scaling and wavelet functions at various resolutions and locations are constructed. The scaling and wavelet functions together constitute the wavelet basis that spans the space of square-integrable functions. The properties of this wavelet basis and their advantages for analyzing quantum theories are listed.

\noindent
Using the Daubechies wavelets, we study the Hamiltonian eigenvalue problem of the infinite square well potential, the simple harmonic oscillator, the one-dimensional attractive Dirac-delta function potential and the one-dimensional triangular potential as a function of increasing resolution. The nature of multi-scale contributions to eigenvalues and eigenfunctions are highlighted. The solution to the one-dimensional triangular potential is a new and important contribution, as these types of confining potential appear in the Hamiltonian framework of two-dimensional gauge theories such as QCD. 

\noindent
The canonical quantization of a real scalar field theory using the Fourier basis and Daubechies wavelet basis is described. The wavelet-based approach is presented as an alternative to traditional lattice field theory, offering a systematic way to analyze real-time quantum field theories (QFTs) with natural volume and resolution truncations. 

\noindent
The approach to regularization and renormalization in quantum theories is studied within the discrete wavelet framework. For this purpose, we work with the model of the two-dimensional attractive Dirac-delta function potential that is known to demonstrate quintessential features of a typical relativistic quantum field theory. The study showcases the emergence of asymptotic freedom and the renormalization of coupling constants within this framework. 

\noindent
The flow equation methods (more generally Similarity Renormalization Group (SRG) methods) evolved as a means to address multiscale problems (problems in which multiple scales contribute to the observed phenomena). We describe the flow equation method within the wavelet approach and use this to investigate scale (resolution) separation in a two-dimensional scalar field theory. We show that the flow block-diagonalizes the Hamiltonian by resolution at an improved truncation than studied previously.  Using the model of two real scalar fields interacting through an elementary quadratic ‘interaction’, we show how the flow equations effectively filter the low-resolution part from the high-resolution part of the interaction, thereby providing an insight into the construction of effective Hamiltonian using wavelet-based flow equation method.

\tableofcontents

\listoffigures
\addcontentsline{toc}{chapter}{\textbf{List of figures}}

\listoftables
\addcontentsline{toc}{chapter}{\textbf{List of tables}}

\newpage

\mainmatter

\chapter{Introduction}
\label{chap:Introduction}
The discrete wavelet-based formulation of quantum field theory allows one to commit to the Hamiltonian framework while maintaining the discreteness of the lattice approach, yet not compromise the continuum nature of space. The wavelet transform, pioneered by Grossman and Morlet \cite{doi:10.1137/0515056} in the year 1964, is a mathematical tool utilized for decomposing a specific function or continuous-time signal into distinct scale components, is given by,
\begin{eqnarray}
W(a,b)=\int_{-\infty}^{\infty}x(t)\psi^*\left(\frac{t-b}{a}\right)dt,
\end{eqnarray}
where:
\begin{itemize}
\item $W(a,b)$ is the wavelet coefficient at scale $a$ and position $b$,
\item $x(t)$ is the input signal,
\item $\psi(t)$ is the mother wavelet,
\item $a$ is the scale parameter (determining the wavelet's width),
\item $b$ is the translation parameter (shifting the wavelet in time),
\item $\psi^*(t)$ is the complex conjugate of the wavelet function.
\end{itemize}
The scaling $a$ controls how stretched or compressed the wavelet is, allowing the analysis of different frequency components of the signal, while the translation $b$ shifts the wavelet across time to capture temporal features. Initially coined as ``ondelette" in French, which translates to ``small wave," the term was later adapted into English by substituting ``onde" with ``wave," resulting in ``wavelet." In 1988, Ingrid Daubechies, a former student of A Grossman, introduced a family of orthonormal wavelets, known as Daubechies wavelet \cite{doi:10.1137/1.9781611970104}. One-dimensional wavelets ($\phi(x), x \in \mathbb{R}$) are functions that must adhere to the following criteria.
\begin{itemize}
\item The function and the Fourier transform of the function have to be well localized.
\item $\int_{-\infty}^{\infty}\phi(x)dx=1$.
\end{itemize}
Additional requirements are needed for certain applications to simplify the implementation of numerical algorithms. 

There exists a wide variety of wavelet families, including the Daubechies wavelet \cite{doi:10.1137/1.9781611970104,mallat1989multiresolution}, Coiflet wavelet \cite{doi:10.1137/1.9781611970104,beylkin1991fast}, Shannon wavelet \cite{Meyer_1993}, Meyer wavelet \cite{Meyer_1993}, Morlet wavelet \cite{doi:10.1137/1.9781611970104,goupillaud1984cycle}, Mexican hat wavelet, Biorthonormal wavelet \cite{https://doi.org/10.1002/cpa.3160450502,DAHMEN1999132}, and others. When creating a wavelet basis for specific applications, one may benefit from the wavelets having specific attributes. These include their nature and extent of support, smoothness, symmetry, orthogonality etc. The significance of each property can vary depending on the specific application. We compare the properties of different types of wavelet in Table \ref{tab:properties_of_different_wavelets}. 
\begin{table*}[ht]
\begin{center}
\newlength{\digitwidth} \settowidth{\digitwidth}{\rm 0}
\catcode`?=\active \def?{\kern\digitwidth}
\caption{db=Daubechies wavelet, coif=Coiflet wavelet, shan=Shannon wavelet, meyr=Meyer wavelet, mexh=Mexican hat wavelet, morl=Morlet wavelet, bior=Biorthogonal wavelet, batl=Battle-Lemarie wavelet, interp=interpolating wavelet, CWT=Continuous wavelet transformation, DWT=Discrete wavelet transformation.}
\label{tab:properties_of_different_wavelets}
\vspace{2mm}
\begin{tabular}{ c | c | c | c | c | c | c | c | c | c }
\specialrule{.15em}{.0em}{.15em}
\hline
Property & db & coif & shan & meyr & mexh & morl & bior & batl & interp \\
\hline
Compact support 	& $\checkmark$ & $\checkmark$ &  $\times$	&    $\times$  & $\times$ & 	$\times$		& $\checkmark$ &  $\times$	  &  $\checkmark$ \\
\hline
Regularity 			& 		f 	   & 	  f 	  & 	inf 	& 	  inf 	   &	inf 	& 	 inf 	& 	 f 	&       f      &      f\\
\hline
Symmetry 			&   $\times$   &   $\times$   &$\checkmark$ & $\checkmark$ & $\checkmark$ & $\checkmark$ & $\checkmark$ & $\times$ & $\checkmark$ \\
\hline
Orthogonality 		& $\checkmark$ & $\checkmark$ & $\checkmark$ & $\checkmark$ & $\times$ & $\times$ & $\times$ & $\checkmark$ & $\times$ \\
\hline
Explicit expression & $\times$ & $\times$ & $\checkmark$ & $\checkmark$ & $\checkmark$ & $\checkmark$ & $\times$ & $\checkmark$ & $\times$ \\
\hline
CWT 				& $\checkmark$ & $\checkmark$ & $\checkmark$ & $\checkmark$ & $\checkmark$ & $\checkmark$ & $\checkmark$ & $\checkmark$ & $\checkmark$ \\
\hline
DWT 				& $\checkmark$ & $\checkmark$ & $\times$ & $\checkmark$ & $\times$ & $\times$ & $\checkmark$ & $\checkmark$ & $\checkmark$ \\
\hline
\specialrule{.15em}{.15em}{.0em}
\end{tabular}
\end{center}
\end{table*}

The fundamental theories of elementary particles and their interactions are described by local Quantum Field Theories (QFTs) formulated on $(3+1)$-dimensional Minkowski space-time. The quantization of these field theories is usually done using canonical quantization approach or via path integral methods. As a part of the quantization process, it is common to resolve the field using the plane wave basis into its momentum modes. In the plane wave basis, the free field part of the Hamiltonian represents these momentum modes as uncoupled oscillators, while the interaction Hamiltonian represents the couplings between the different momentum modes of the field. When computing the S-matrix elements, it is common to adopt a manifestly covariant approach to perturbation theory to effectively deal with the ultraviolet divergences and reexpress the theory in terms of renormalized masses and couplings \cite{Peskin:1995ev,mandl2010quantum}. However, in doing so, the central role played by the Hamiltonian eigenvalue problem does get compromised.

The lattice approach allows one to analyze QFTs beyond perturbation theory systematically. One defines the field on a Euclidean lattice with a presumed underlying lattice cutoff. The QFT is studied as an equivalent statistical field theory with the partition function defined as a path integral over the Euclidean action. The discrete nature of the lattice makes the field theory computationally tractable. The presence of the explicit cutoff regulates the ultraviolet divergences (beyond perturbation theory), but it explicitly violates Euclidean invariance. There is often a
trade-off between the need for nonperturbative analysis and the desire for full covariance. Within the lattice approach, the continuum limit of QFT is obtained by maintaining criticality in the limit of vanishing cutoff. With the exception of the Hamiltonian lattice approach, Euclidean lattice makes it challenging to work with the Hamiltonian energy eigenvalue problem directly.

Daubechies wavelets and scaling functions constitute an orthonormal basis of compactly supported functions \cite{doi:10.1137/1.9781611970104,f5d23bd3-c0af-3314-bd56-832b7db8db57,https://doi.org/10.1002/cpa.3160410705}. Roughly speaking, each basis function is characterized by its location (translation index) and length scale (resolution). The quantum fields, when expanded in the wavelet basis, lead to its representation as an infinite sequence of operators characterized by a location and resolution index. This approach allows natural volume and resolution truncations of the QFT. The truncated theory is an ordinary quantum mechanical theory with multiple discrete degrees of freedom organized by location and length scale. The maximum resolution plays the role of ultraviolet cutoff.

In this thesis, we utilise Daubechies wavelets as its properties provides distinct advantages in analysis of quantum mechanical and quantum field theoretic (QFT) problems.
\begin{itemize}
\item The Daubechies wavelets form an orthonormal basis for the space of square-integrable functions defined on the real line. One can represent the quantum mechanical wave function and/or the quantum field as a linear combination of wavelet basis functions. 

\item In the quantum field theoretic context the operator valued distributions (quantum fields) is represented in terms of a countable infinite series of discrete operators labelled by discrete resolution (length scale) and location (translation index) indices. The discrete operators associated with distinct resolution and location indices commute with each other. The size of the basis functions' support can be controlled by choosing the order of the wavelets. Additionally, there are infinite number of basis functions with support that can be made as small as desired, making it feasible to utilize these operators to explore the concept of locality and the way in which locality is compromised through truncations. When the quantum fields are represented using the wavelet basis, they can be described as an infinite series of operators, each characterized by indices of position (location) and scale (resolution). This method enables straightforward truncations in volume and resolution within Quantum Field Theory (QFT). The resulting simplified theory becomes a standard quantum mechanical model possessing several discrete variables, categorized by their position and scale. In this framework, the highest resolution acts as the ultraviolet cutoff and the volume act as an infrared cutoff .

\item Daubechies basis functions originate from a single function, referred to as the mother scaling function. This function is the solution of a linear renormalization group equation, known as ``scaling equation''. The basis is natural for formulating renormalization group transformations because it arranges the quantum field's degrees of freedom by length scales (resolutions).

\item The basis elements are created from a scaling function through the use of discrete unitary translations and discrete unitary scale transformations. Despite the fractal nature of the basis functions, integrals that involve multiplying any number of these functions with their low-order derivatives can be precisely calculated by exclusively employing the renormalization group equation and a scaling condition. Additionally, calculations involving the multiplication of basis functions with polynomials are also feasible through the application of the renormalization group equations.
\end{itemize}


Wavelet analysis has emerged as a effective tool across various areas of science and technology. Wavelets have an application in data compression and signal processing \cite{Coifman1994,istepanian2001ecg,agarwal17,vetterli2001wavelets,
khalifa2008compression,860184,4136914,540087,tun2017analysis,801765,doi:10.1080/14639230310001636499,szu1996wavelet}. Wavelets are useful to examine and decompose signals having different frequency component at different time intervals (non-stationary signals). Though short-time Fourier transform (STFT) provide a systematic framework to analyse non-stationary signals, several researchers \cite{4566665,app9071345} have demonstrated the effectiveness of the frequency slice wavelet transform (FSWT) \cite{7726999}, a wavelet framework to analyse non stationary signals, over STFT.  Wavelets' inherent capability for treating multiscale problems makes them a suitable tool for handling problems of turbulence \cite{farge1996wavelets,farge1992wavelet,meneveau1991analysis,
baars2015wavelet,yamada1991orthonormal}. 

An early application of wavelets to quantum mechanical problems was performed by Modisette et al. \cite{MODISETTE1996485}, who used Daubechies wavelets to tackle the quantum harmonic oscillator and the steep double well. They demonstrated that the adaptive nature of the basis functions effectively responds to potential fluctuations in different spatial regions, making this approach particularly useful for problems where such scenarios are present. More recently, several researchers \cite{Behera_Mehra_2015,Goedecker2009WaveletsAT,1997APS..MAR.G2405I} have presented the scheme for systematically solving the Poisson's and the Schrodinger equation. Panja et al. \cite{panja_2016_computing} addresses the problem of an anharmonic oscillator using Daubechies wavelets. Numerous articles \cite{LI1993362,10.1063/1.466951,
Saha2021singh,chawhan2020quantum} authored by different researchers have utilized wavelets to address quantum mechanical problems.

The scale hierarchical nature of wavelets serves as an intrinsic framework for renormalization. Christoph Best was the first to apply the Daubechies wavelet basis in order to deduce the qualitative renormalization flow within the Landau-Ginzburg model. \cite{BEST2000848,best1994variational}.

Smooth wavelets were used by P. Federbush \cite{federbush1995new} to provide regularization for fields in Yang-Mills theories. In contrast to Daubechies wavelets, these wavelets possess smoothness and favorable localization properties, albeit without compact support. The utilization of these wavelets for addressing issues in constructive field theory is thoroughly examined in G. Battle's book \cite{doi:10.1142/3066}.

Evenbly and White, in their work \cite{PhysRevLett.116.140403}, demonstrated a link between entanglement renormalization and discrete wavelet transforms in the context of free particle systems. To approximate the ground state of the critical Ising model, they utilized Daubechies wavelets.

In their work, Halliday and Suranyi \cite{HALLIDAY1995414} introduced the use of Haar wavelet expansions as a new method for simulating $\phi^4$ field theories. This approach leverages the most basic form of Daubechies wavelets, where fields at any given point are represented by averages over adjacent blocks. This technique was then compared to the conventional Metropolis algorithm used in lattice field theory, specifically within the context of a scalar 2D $\phi^4$ field theory.

Brennen, Rohde, Sanders, and Singh \cite{PhysRevA.92.032315} implemented Daubechies wavelets to break down physics into different scales of length or energy, with the goal of understanding field theories better. Their findings revealed that using a wavelet basis allows for the efficient simulation of scalar bosonic quantum field theories on quantum computers.

M. Altaisky, in collaboration with others, has been dedicated to tackling issues in quantum field theory, primarily employing the continuous wavelet transform. His strategy emphasizes the adoption of wavelet techniques for the regularization of local fields, integrating inherent scale cutoffs within quantum field theory. Furthermore, he has suggested employing these techniques in the context of gauge theories \cite{albeverio2010remark,PhysRevD.81.125003,PhysRevD.93.105043,Altaiskii2016,Altaisky_2007,PhysRevD.88.025015,Altaiskii2013,PhysRevD.101.096004}.

In their work, F. Bulut and W. N. Polyzou have advocated for employing Daubechies wavelets to achieve natural truncations in both volume and resolution within field theories \cite{PhysRevD.87.116011,michlinflow,polyzou2020lightfront,kessler2003wavelet,polyzou2018multi, MichlinTracieL2017Uwbt,PhysRevD.95.094501}.

Methods based on discrete wavelets offer techniques for analyzing quantum field theories in a manner akin to the Euclidean lattice method, while they additionally facilitate the study of dynamics in real time. \cite{kessler2003wavelet,polyzou2018multi,MichlinTracieL2017Uwbt,federbush1995new,
BEST2000848,PhysRevLett.116.140403,PhysRevD.106.036025,tomboulis2021wavelet,PhysRevD.108.125008}.The promise of compactly supported wavelets lies in their discrete and multiscale characteristics, offering a systematic framework for both classical and quantum simulations of continuum quantum field theories (QFTs) \cite{best1994variational,HALLIDAY1995414,10.1063/1.1543582,polyzou2023path,PhysRevA.92.032315}. Discrete wavelets have also been used for analyzing statistical field theories \cite{PhysRevD.87.116011,PhysRevD.95.094501,refId0}. Methods developed based
on continuous wavelets provide a complementary perspective \cite{Altaisky_2007,albeverio2010remark,Altaisky2018,PhysRevD.88.025015}. Both approaches, discrete and continuous, have considered regularization, renormalization, and gauge invariance in field theories. Wavelet-based representation of light-front quantum field theories has been formulated \cite{Altaiskii2013,Altaiskii2016,PhysRevD.101.096004} to gain an advantage from the unique properties of being on the light front \cite{polyzou2020lightfront}.

This thesis presents a detailed documentation of a set of studies on quantum mechanical and quantum field theoretic models within the discrete wavelet framework. 

We study the prototypical quantum mechanical one-dimensional models of a particle in an infinite square well potential, a simple harmonic potential, an attractive Dirac-delta function potential and a triangular potential. The Hamiltonian eigenvalue problem is solved using a variational method with a trial function containing linear variational parameters. The construction of the trial function is done using Daubechies scaling functions. The eigenvalues and eigenfunctions are studied as a function of increasing resolution and volume truncation. We demonstrate the convergence of the eigenvalues and eigenfunctions to the exact result with increasing resolution. The contribution of different length scales from different regions of the potential to the eigenvalues is highlighted. The study on the triangular potential using the Daubechies wavelets adds value to the literature. These results will play an important role in analyzing two-dimensional quantum chromodynamics (QCD) on the light front as this confining potential emerges in the Hamiltonian computed within the light front gauge. 

We investigate aspects of renormalization in theories analyzed using wavelet-based methods. We demonstrate the nonperturbative approach of regularization,
renormalization, and the emergence of flowing coupling constant within the context of these methods. This is tested on a model of the particle in an attractive Dirac delta function potential in two spatial dimensions \cite{PhysRevD.107.036015}, which is known to demonstrate quintessential features found in a typical relativistic quantum field theory. 

The discrete wavelet transform provides an exact representation of a field in terms of a countably infinite set of modes labelled by location and resolution indices. The discrete wavelet representation of QFT seeks to analyze the theory in terms of these modes. In an actual computation, one truncates the number of modes by resolution (ultraviolet cutoff) and location (physical space cutoff). Such a truncated version becomes a conventional quantum mechanical model characterized by numerous discrete operator variables characterized by their own commutation relations that are inherited from the commutation relations of the field operators. We also present the formulation of scalar field theory in one spatial dimension within this framework, which is an extension of the work of Bulut and Polyzou \cite{PhysRevD.87.116011}. This includes the construction of the creation and annihilation operators at specific length scales and positions. Additionally, we constructed the number operator by applying the annihilation and creation operators successively to an arbitrary number state in the Fock space within this formalism. In contrast to the Fourier basis, the Hamiltonian will not take the mere diagonal form, but the different scale coupling terms will be present. Diagonalizing the Hamiltonian will yield the energy eigenvalues corresponding to it.

Models of relativistic quantum field theories contain degrees of freedom associated with a full range of energy scales (or length scales), whereas only a small fraction of the lower range of energy scales may be accessible to experiment. This implies that theoretical computations must reliably calculate contributions to low-energy observables from degrees of freedom associated with the full range of energy scales.  Flow equation methods (or, more broadly, Similarity Renormalization Group (SRG)) \cite{wegner1994flow, PhysRevD.48.5863, PhysRevD.49.4214, PERRY1994116, bartlett2003flow, kehrein2007flow, bogner2007similarity, PhysRevC.77.037001, BOGNER201094} achieve this by unitarily evolving the Hamiltonian into a band-diagonal form that has reduced coupling between the low energy scales and the higher energy scales. This approach is useful when systems contain competing energy scales. The implementation of the flow equation method from a wavelet perspective is studied in this thesis.

We present an extension of the work by Michlin and Polyzou \cite{PhysRevD.95.094501} that analyzed the (1+1)-dimensional real scalar field using the wavelet-based flow equation method. We demonstrate that the specifically chosen generator flows the Hamiltonian into a block diagonal form with each diagonal block being associated with a fixed resolution. We also study the low energy dynamics of a system of two coupled real scalar fields in 1+1 dimensions using the same approach. The wavelet basis is known to transform the scalar field theory into a model of coupled localized oscillators, each of which is labelled by location and resolution indices. The chosen interaction in this model represents the coupling between two types of oscillators at the same location and resolution index. There is no coupling between oscillators across locations and resolutions. We show that the wavelet-based flow equation method carries out scale separation while maintaining the interactions between the scalar fields at each resolution, irrespective of the relative size of the scalar masses. The SRG generated effective Hamiltonian is shown to correctly reproduce the exact normal mode frequencies.

The thesis is organized as follows: Chapter \ref{chap:mra_in_a_wavelet_basis}. delves into multiresolution analysis and the Daubechies wavelet basis. The multiresolution analysis is based on the book by M.N Panja \cite{book_panja} and the ``Properties of elements in Daubechies wavelet family" section is based on the work of Polyzou \cite{PhysRevD.87.116011}. Subsequently, Chapter \ref{chap:quantum_mechanics_in_daubechies_wavelet_basis} explores quantum mechanics within the wavelet basis. The quantum mechanical examples discussed here are the original work of us. Chapter \ref{chap:qft_in_a_wavelet_basis} delves into quantum field theory within the same framework. This chapter is based on the work of Polyzou \cite{PhysRevD.87.116011}. We extend their work by calculating the Hamiltonian matrix elements within the Fock basis. Chapter \ref{chap:renormalization_in_a_wavelet_basis} investigates renormalization in a wavelet basis based on the original published work of us \cite{PhysRevD.107.036015}, followed by Chapter \ref{chap:flow_equation_method_in_a_wavelet_basis} which explores the flow equation method within this context. This chapter is based on the our ongoing work that we are planning to submit for publication this December. Finally, Chapter \ref{chap:conclusion_and_outlook} presents the conclusion and outlook of our study.
\chapter{Multiresolution analysis (MRA) of function space ($\mathscr{L}^2(\mathbb{R})$) using Daubechies wavelet basis}
\label{chap:mra_in_a_wavelet_basis}

Approximation theory utilizes a finite set of elements, from a complete basis, which consists of a fixed number of elements (let's say $n$). Within this framework, operators and functions are represented by matrices of dimensions $n\times n$ and $n\times 1$ respectively. The primary challenge arises when the operator acts on the function since this operation typically involves a computation of $n^2$ operations. However, it's worth noting that for diagonal matrices, the number of arithmetic operations can be reduced to just $n$. In approximation theory, the matrix representation of an operator can be simplified if we work in the eigenbasis of that operator. In this case, the matrix becomes diagonal, which greatly simplifies computations. However, it is not always feasible to determine the eigenbasis of a given operator. To overcome this challenge, various approaches have been developed to find alternative frameworks where the matrix representation of the operator is naturally close to being diagonal or band diagonal. These methods aim to reduce the complexity of computations and improve efficiency in approximating functions or operators. During the late 1980s, an extensive mathematical theory knows as multiresolution (MRA) analysis of function space \cite{article01,192463} garnered significant attention in the literature of both physics and mathematics. Belgium mathematician Ingrid Daubechies invented the orthonormal wavelets \cite{https://doi.org/10.1002/cpa.3160410705,doi:10.1137/1.9781611970104} with compact support to address the aforementioned challenge. Wavelets have emerged as a valuable analytical tool in various scientific domains, offering an added advantage compared to the Fourier basis: compact support of the basis function. This additional degree of freedom enables wavelets to efficiently capture localized information, making them highly effective for analysing various problems of different domains of science. Different orthonormal wavelets can be constructed by leveraging the MRA of the function space relevant to the problem at hand. The fundamental concept underlying MRA can be summarized as follows.

\section{MRA of $\mathscr{L}^2(\mathbb{R})$}
MRA is a concept that aims to represent functions in the $\mathscr{L}^2(\mathbb{R})$ space as a sequence of increasingly refined approximations, each of which is smoothed version of $f$, with more and more concentrated smoothing functions \cite{book_panja,https://doi.org/10.1002/cpa.3160410705}. This iterative process allows for a comprehensive analysis of the function's behavior at multiple resolutions, hence the term ``multiresolution analysis". An MRA consists of,

\begin{enumerate}
\item \label{en:1st_point_mra} a family of embedded closed subspaces $V_m\subset \mathscr{L}^2(\mathbb{R})$, $m\in \mathbb{Z}$, and
\begin{eqnarray}
... \subset V_2\subset V_1 \subset V_0 \subset V_{-1}\subset V_{-2} \subset V_{-3} ...\quad ,
\end{eqnarray}
such that,
\item \label{en:2th_point_mra}
\begin{eqnarray}
\bigcap_{m\in \mathbb{Z}} V_m={0}, \quad  \overline{\bigcup_{m\in \mathbb{Z}}V_m}=L^2({\mathbb{R}}),
\end{eqnarray}
and,
\item \label{en:3th_point_mra}
\begin{eqnarray}
f(x)\in V_m \Longleftrightarrow f(2x)\in V_{m+1}\;\; \forall\; m\in\mathbb{Z}.
\end{eqnarray}
\item \label{en:4th_point_mra} There exists a $\phi\in V_0$, such that, for all $m\in\mathbb{Z}$, the $\phi^m_{n}$ constitute an unconditional (where the convergence of the expansion of any vector in the space is not dependent on the order of the basis elements) basis for $V_m$, that is,
\begin{eqnarray}
V_m=\text{linear span} \{\phi^m_n, n\in\mathbb{Z}\}.
\end{eqnarray}
\item \label{en:last_point_mra}The norm of the function is equal to unity,
\begin{eqnarray}
\int \phi(x)dx=1.
\end{eqnarray} 
\end{enumerate}
The sequence $\{V_m,\;m\in\mathbb{Z}\}$ of vector spaces $V_m$, satisfying the properties (\ref{en:1st_point_mra}-\ref{en:last_point_mra}), constitute the MRA of $L^2(\mathbb{R})$.
\subsection{Multiresolution generator}
\label{subsec:Multiresolution_generator}
In the context of multiresolution analysis (MRA), the function $\phi(x)$ mentioned in point (\ref{en:4th_point_mra}) is commonly referred to as the generating function or scale function of the MRA. One of the key properties of MRA, as stated in point (\ref{en:3th_point_mra}), is that the generating function $\phi(x)$ belongs to the space $V_0$ and is also a member of its subspace $V_1$. Based on this property, we can infer that there exists a set of coefficients $h_l$ (where, $l\in \mathbb{Z}$) such that,
\begin{eqnarray}
\label{eq:scaling_equation}
\phi(x)=\sqrt{2}\sum_{l\in \mathbb{Z}}h_l \phi(2x-l).
\end{eqnarray}
This property of scale function is known as a \textit{two-scale relation} or \textit{refinement relation} with \textit{mask} or \textit{low-pass filter} \{$h_l, l\in \mathbb{Z}$\}. From Eq. (\ref{eq:scaling_equation}), we can see that the generating function is formed by the linear combination of scaled and translated copies of itself. Squeezing the function to one half of its original value while maintaining the norm to $1$ is refer to as the increment in resolution to $1$ unit.

The $n$-th translated and $m$-th resolution scaling function is defined as,
\begin{eqnarray}
\phi^m_n(x)=2^{\frac{m}{2}}\phi(2^m x-n).
\end{eqnarray} 
\subsection{Wavelets}
In the context of multiresolution analysis (MRA), which involves a family of spaces $V_m$ and a function $\phi(x)$ satisfying properties (\ref{en:1st_point_mra}) to (\ref{en:last_point_mra}), we can define $W_m$ as the orthogonal complement of $V_m$ within $V_{m+1}$,
\begin{eqnarray}
\label{eq:V_m_and_W_m_relation}
V_{m+1}=V_m\oplus W_m, \quad W_m\perp V_m.
\end{eqnarray}
The nested property of subspaces $V_m$ in (\ref{en:1st_point_mra}) gives
\begin{eqnarray}
W_{j'}\perp W_{j''}\;\; \forall\; j'\neq j''\geq j,
\end{eqnarray}
and,
\begin{eqnarray}
V_J=V_{j_0}\bigoplus_{j=j_0}^{J-1}W_j,\; \text{for}\; j_0<J.
\end{eqnarray}
So, $W_m$ are the scaled versions of $W_0$,
\begin{eqnarray}
\label{eq:relation_W_0_and_W_m}
f\in W_0 \Leftrightarrow f(2^mx)\in W_m.
\end{eqnarray}
Finally, property (\ref{en:2th_point_mra}) of MRA ensures that
\begin{eqnarray}
\mathscr{L}^2(\mathbb{R})=\bigoplus_{j\in \mathbb{Z}}W_j.
\end{eqnarray}
because of properties (\ref{en:1st_point_mra})-(\ref{en:4th_point_mra}) of $V_m$, it turns out that in $W_0$ also there exists a vector $\psi$ such that its integer translates span $W_0$ \cite{mallat1988multiresolution}, i.e.,
\begin{eqnarray}
\overline{\text{span}\{\psi^{0}_{n}\}}=W_0,
\end{eqnarray}
where, as before, $\psi^m_n(x)=2^{\frac{m}{2}}\phi(2^m x-n)$ for $m,n\in \mathbb{Z}$. It follows immediately from Eq.(\ref{eq:relation_W_0_and_W_m}) that then
\begin{eqnarray}
\overline{\text{span}\{\psi^m_n\}}=W_m,\; \text{for all} \; m\in \mathbb{Z}.
\end{eqnarray}
The space denoted as $W_m$ is commonly referred to as the detail space, and within the space $V_{m+1}$, the specific element $\psi(x)$ (belonging to $W_0$) is recognized as the \textit{mother wavelet}. Based on property (\ref{eq:V_m_and_W_m_relation}), when $j=0$, it is evident that $\psi(x)$ can be expressed using the basis of $V_1$ as,
\begin{eqnarray}
\label{eq:wavelet_equation}
\psi(x)=\sqrt{2}\sum_{l\in \mathbb{Z}}g_l \phi(2x-l).
\end{eqnarray}
It is worth noting that the coefficients $g_l$'s appearing in equation (\ref{eq:wavelet_equation}) are commonly referred to as the \textit{high-pass filter} of the Multiresolution Analysis (MRA) produced by the scale function $\phi(x)$. The wavelet and its translated versions at a higher resolution $j$ are defined as, $\psi^j_k(x)=2^{j/2}\psi(2^jx-k)$, where $k$ is the translation parameter. 
\subsection{Basis with compact support}
Many mathematical results in Multiresolution Analysis (MRA) rely on the support of the scale function $\phi(x)$ as well as the non-zero values of the low-pass filters $h_k$'s and high-pass filters $g_k$'s. The process of working with technical formulas involving these filters becomes more manageable when they have a finite number, meaning that the scale functions or wavelets have compact support. Haar basis \cite{Haar1910ZurTD} is the first classical example of compactly supported orthonormal wavelet in $\mathscr{L}^2(\mathbb{R})$. The definition is given by,
\begin{eqnarray}
\phi(x)=\begin{cases}
		& 1,\quad 0\leq x<1\\
		& 0,\quad \text{otherwise}
		\end{cases},\quad 
\psi(x)=\begin{cases}
		& 1,\quad 0\leq x <\frac{1}{2}\\
		& -1,\quad \frac{1}{2}\leq x<1\\
		& 0,\quad	\text{elsewhere}.
		\end{cases}
\end{eqnarray}
In case of the higher-resolution, the set of functions $\left\{\phi^m_n(x)=2^{\frac{m}{2}}\phi(2^mx-n),\, m,n\in \mathbb{Z}\right\}$ forms a orthonormal basis with compact support for the space $V_m$. However, because of the irregularity of this wavelet family, it is desirable to obtain a wavelet family that exhibits higher regularity. 
In the late 20th century, Daubechies \cite{https://doi.org/10.1002/cpa.3160410705,doi:10.1137/1.9781611970104} made significant advancements in the field of wavelets by introducing a generator and wavelets that exhibit increasing regularity as the support of the Multiresolution Analysis (MRA) of $\mathscr{L}^2(\mathbb{R})$ expands. These wavelets are now commonly referred to as the Daubechies wavelet family.
\subsection{Properties of elements in Daubechies wavelet family}
\label{subsec:Properties_of_elements_in_Daubechies_wavelet_family}
Here, we will summarize the key aspects concerning the construction of the Daubechies wavelet basis and its associated properties \cite{https://doi.org/10.1002/cpa.3160410705,doi:10.1137/1.9781611970104,book_panja,PhysRevD.107.036015,kessler2003wavelet,PhysRevD.87.116011}. These elements will be used consistently throughout this thesis. The basis elements comprise both scaling functions and wavelet functions. As explained in Sec. \ref{subsec:Multiresolution_generator}, the scaling functions are derived from a single generating function, denoted as $s(x)$ and often referred to as the mother scaling function. This is defined through the \textit{two-scale relation} or \textit{refinement equation with a mask}  or \textit{low-pass filter or the scaling equation},
\begin{eqnarray}
\label{eq:scaling_equation_1}
s(x)=\sum_{n=0}^{2K-1}h_n \hat{D}\hat{T}^n s(x).
\end{eqnarray}
$\hat{D}$ and $\hat{T}$ are the scaling and translation operations, respectively. These unitary operations are defined by,
\begin{eqnarray}
\label{eq:D_and_T_definition}
\hat{D}s(x)=\sqrt{2}s(2x), \quad\quad \hat{T}s(x)=s(x-1).
\end{eqnarray}
$\hat{T}$ shifts the entire function one unit to the right without altering its form, while $\hat{D}$ compresses the support of the function by a factor of $2$ while preserving its norm.
\begin{eqnarray}
\label{eq:normalization_condition}
\int s(x)dx=1.
\end{eqnarray}
Eq. (\ref{eq:scaling_equation_1}), $K$ denotes a fixed integer that, in turn, determines the level of smoothness and the support of the basis functions, called the \textit{the order of the scaling function}.

\begin{figure}[ht]
\vspace{9pt}
\begin{center}
\includegraphics[scale=.375]{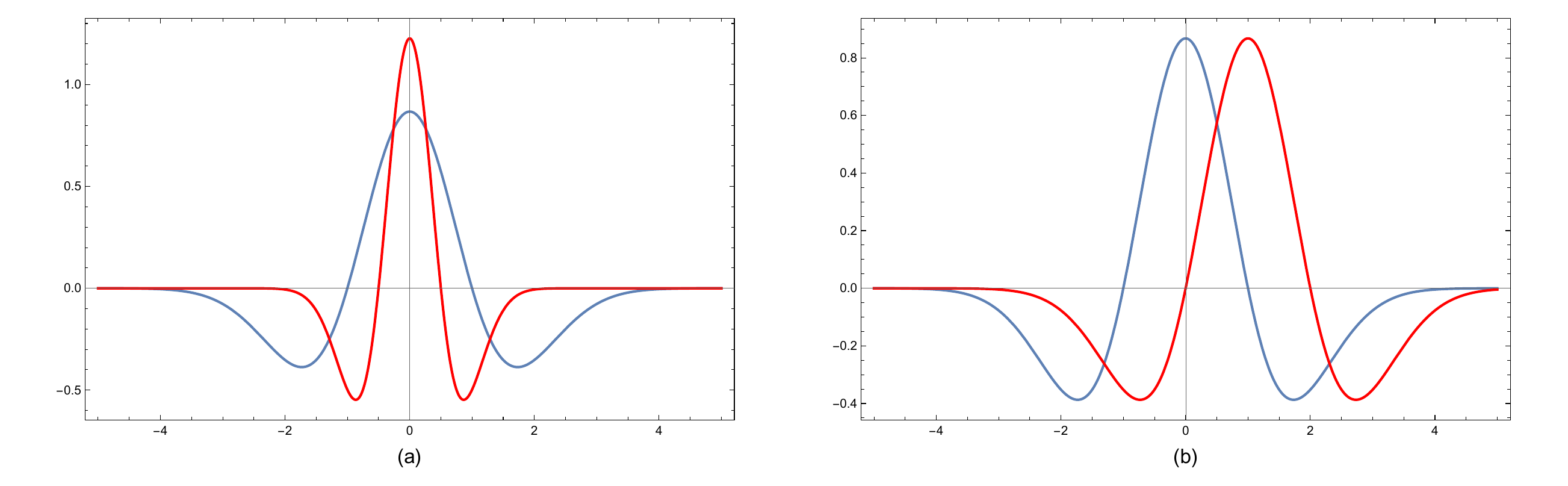}
\end{center}
\caption{Consider any generic function $f(x)$ (depicted in blue). The impact of
(a) the scaling operator $\hat{D}$ and (b) the translation operator $\hat{T}$ on that
function is illustrated in red.}
\label{fig:the_D_and_T_operator}
\end{figure}

The effects of the operators $\hat{D}$ and $\hat{T}$ on a typical function $f(x)$ are illustrated in Fig. \ref{fig:the_D_and_T_operator}.

\begin{figure}[h]
\vspace{9pt}
\begin{center}
\includegraphics[scale=.5]{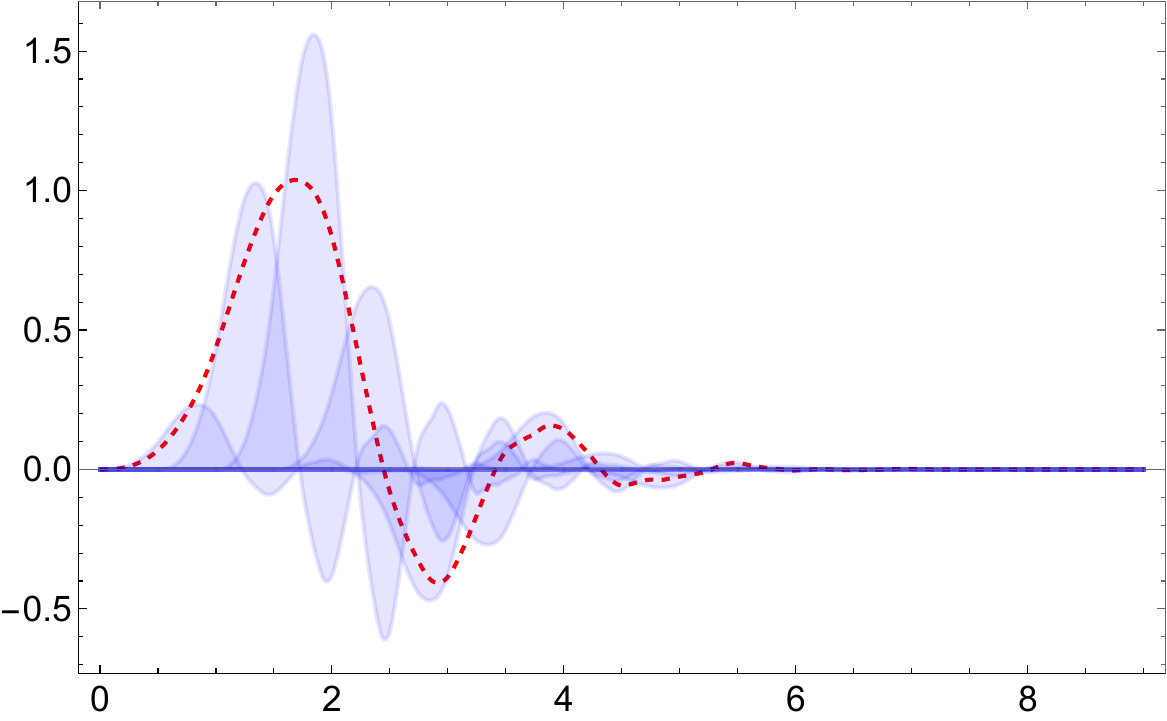}
\end{center}
\caption{
The red dotted line illustrates the mother scaling function $s(x)$ for $K=6$, generated by combining $12$ translated instances of $s(x)$, each scaled to half of its original support.}
\label{fig:scaling_equation}
\end{figure}

Equation (\ref{eq:scaling_equation_1}) expresses $s(x)$ as a particular linear combination involving $2K$ translated and scaled replicas of itself. This relationship is visually depicted in Fig. \ref{fig:scaling_equation} for $K=5$.

Using the solution of Eq. (\ref{eq:scaling_equation_1}), we define the $k$th resolution scaling functions by applying $n$ unit translation followed by $k$ dyadic scale transformation on the mother scaling function,
\begin{eqnarray}
s^k_n(x):=\hat{D}^k\hat{T}^n s(x).
\end{eqnarray}
The scaling functions are orthonormal to each other,
\begin{eqnarray}
\label{eq:normalization_condition_sk_m_sk_n}
\int s^k_m(x)s^k_n(x)dx=\delta_{mn}.
\end{eqnarray}
Any arbitrary linear combination of these functions forms the space $\mathscr{H}^k$ at resolution $k$:
\begin{eqnarray}
\label{eq:arbitrary_linear_combination_H_k}
\mathscr{H}^k=\left\{f(x)|f(x)=\sum_{-\infty}^{\infty}f_n s^k_n(x),\quad \sum_n |f_n|^2<\infty\right\}.
\end{eqnarray}
From scaling equation Eq. (\ref{eq:scaling_equation_1}) and Eq. (\ref{eq:arbitrary_linear_combination_H_k}), it follows that
\begin{eqnarray}
\mathscr{H}^k \subset \mathscr{H}^{k+1},
\end{eqnarray}
more generally, for any $m>0$,
\begin{eqnarray}
\mathscr{H}^k\subset \mathscr{H}^{k+m},
\end{eqnarray}
this means that the $k$th resolution subspace is a linear subspaces of the $(k+m)$th resolution space.

\begin{figure}[ht]
\vspace{9pt}
\begin{center}
\includegraphics[scale=.45]{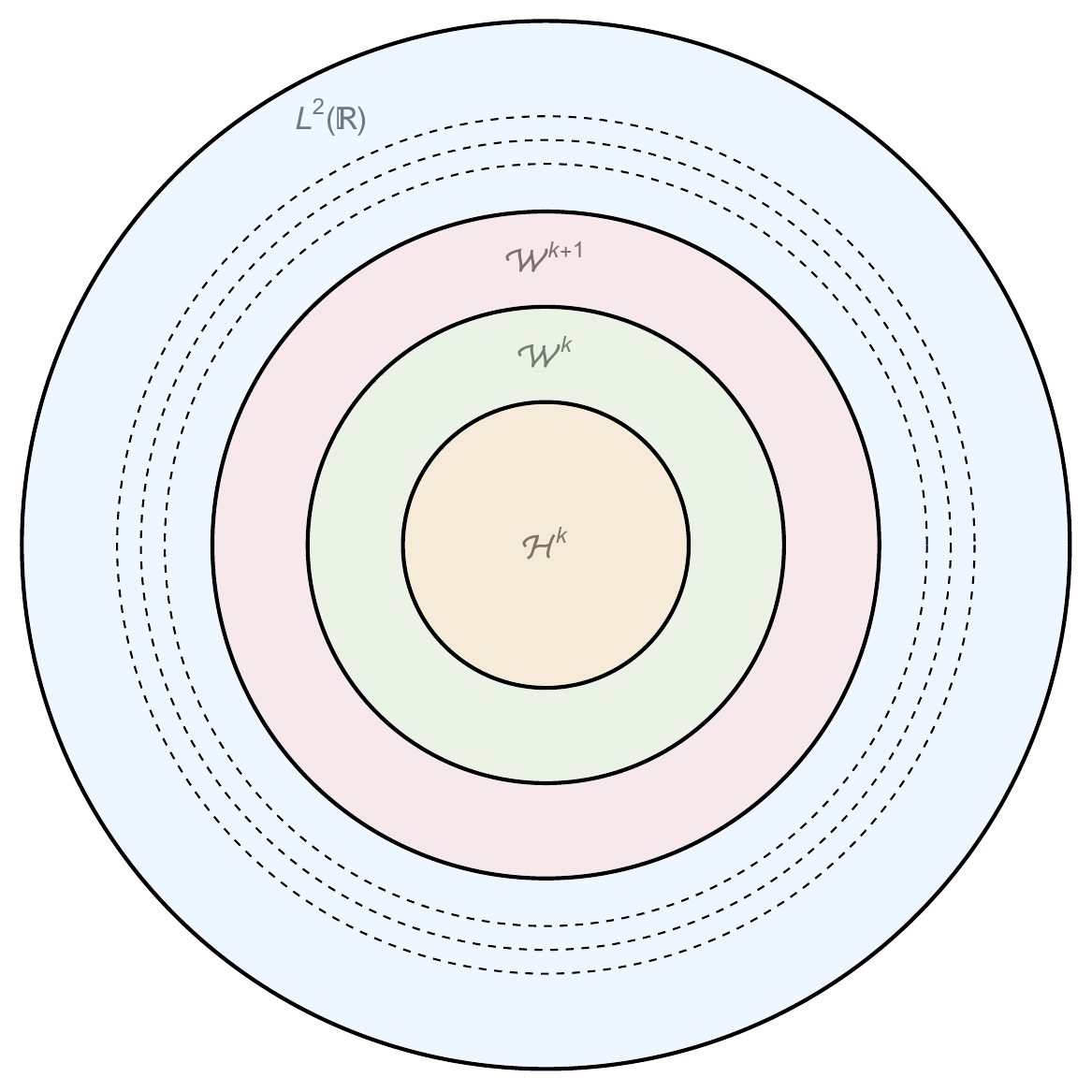}
\end{center}
\caption{Euler diagram for spanning of Hilbert space with wavelet basis.}
\label{fig:Hilbert_Space_Diagram}
\end{figure}

Now we define the mother wavelet function $w(x)$ such that it is orthogonal to the mother scaling function:
\begin{eqnarray}
\label{eq:wavelet_equation_1}
w(x)=\sum_{n=0}^{2K-1}g_n \hat{D}\hat{T}^n s(x),
\end{eqnarray}
where,
\begin{eqnarray}
g_n=(-1)^n h_{2K-1-n}.
\end{eqnarray}
The wavelet  functions are orthonormal to each-other
\begin{eqnarray}
\label{eq:orthonormality_of_wavelet_functions}
\int w^k_m(x)w^k_l(x)dx=\delta_{mn}\delta_{kl},
\end{eqnarray}
and arbitrary linear combination of these generates the space $\mathscr{W}^k$ of resolution $k$:
\begin{eqnarray}
\mathscr{W}^k=\left\{f(x)|f(x)=\sum_{-\infty}^{\infty}f_n w^k_n(x),\quad \sum_n |f_n|^2<\infty\right\}.
\end{eqnarray}
By design,  the scaling functions and wavelet functions are mutually orthonormal,
\begin{eqnarray}
\label{eq:orthogonality_of_scaling_and_wavelet_functions}
\int s^k_m(x)w^{k+l}_n(x)dx=0,\quad l\geq 0,
\end{eqnarray}
and
\begin{eqnarray}
\label{eq:h_k+1_from_h_k_w_k}
\mathscr{H}^{k+1}=\mathscr{H}^k\oplus \mathscr{W}^k.
\end{eqnarray}
The space of square integrable functions, $\mathscr{L}^2(\mathbb{R})$, can be generated be recursive use of Eq. (\ref{eq:h_k+1_from_h_k_w_k}).
\begin{eqnarray}
\mathscr{L}^2(\mathbb{R})=\mathscr{H}^k\oplus\mathscr{W}^k\oplus \mathscr{W}^{k+1}\oplus \mathscr{W}^{k+2}...\quad .
\end{eqnarray}
This is visually represented in Fig. \ref{fig:Hilbert_Space_Diagram}.

There are two potential bases for $\mathscr{H}^k$: one can opt for the scaling function basis at resolution $k$ $\left(\left\{s^k_n(x) \right\}^{\infty}_{n=-\infty} \right)$, or choose a combination of scaling and wavelet functions at resolution $k-1$ $\left(\left\{s^{k-1}_n(x) \right\}^{\infty}_{n=-\infty} \cup \left\{w^{k-1}_n(x) \right\}^{\infty}_{n=-\infty}\right)$. These two bases are interconnected through the following orthogonal transformation given by,
\begin{eqnarray}
&s^{k-1}_n(x)=\sum_{l=0}^{2K-1}h_l s^k_{2n+l}(x),&\\
&w^{k-1}_n(x)=\sum_{l=0}^{2K-1}g_l s^k_{2n+l}(x),&\\
&s^k_n(x)=\sum_{m}h_{n-2m}s^{k-1}_m(x)+\sum_{m} g_{n-2m} w^{k-1}_m(x).&
\end{eqnarray}

By induction, for any given value of $k$, the scaling and wavelet functions,
\begin{eqnarray}
\left\{s^k_n(x)\right\}^{\infty}_{n=-\infty}\cup \left\{w^m_n(x)\right\}^{\infty,\infty}_{n=-\infty,m=k},
\end{eqnarray}
will together constitute the basis of $\mathscr{L}^2(\mathbb{R})$. Any square integrable function,f(x) can be expanded in this basis,
\begin{eqnarray}
\label{eq:square_integrable_function_expansion_in_wavelet_basis}
f(x)=\sum_{n=-\infty}^{\infty}f^s_n s^k_n(x)+\sum_{n=-\infty}^{\infty}\sum_{l=k}^{\infty}f^{w,l}_n w^l_n(x),
\end{eqnarray}
such that,
\begin{eqnarray}
\sum_{n=-\infty}^{\infty}|f^s_n|^2+\sum_{n=-\infty}^{\infty}\sum_{l=k}^{\infty}|f^{w,l}_n|^2<\infty.
\end{eqnarray}
An alternative way to construct $\mathscr{L}^2(\mathbb{R})$ is via an infinite resolution limit of $\mathscr{H}^k$,
\begin{eqnarray}
\mathscr{L}^2(\mathbb{R})=\lim_{k \to \infty}\mathscr{H}^k.
\end{eqnarray}

One another key characteristic of Daubechies wavelet is vanishing moment condition. If the support width of Daubechies scaling $(s(x))$ and wavelet functions $(w(x))$ is $2K-1$, then the associated wavelet $w(x)$ possesses $K$ vanishing moments, namely,
\begin{eqnarray}
\label{eq:vanishing_moment_condition}
\int_{-\infty}^{\infty}x^m w(x)dx=0 \quad \text{for}\quad m=0,1,...,K-1.
\end{eqnarray}
This equation ensures that polynomials of degree $\leq K-1$ can be locally represented by finite linear combinations of scaling functions on a fixed scale due to Strang \cite{strang1989wavelets}. This is a useful property for numerical approximations.

From the scaling equation Eq. (\ref{eq:scaling_equation_1}) and the normalization condition Eq. (\ref{eq:normalization_condition}), we can deduce the following essential condition on the coefficients $h_n$,
\begin{eqnarray}
\label{eq:h_n_normalization_condition}
\sum_{n=0}^{2K-1}h_n=\sqrt{2}.
\end{eqnarray}
This condition must be fulfilled for a scaling equation to have a solution. 

Moreover, the orthonormality among the integer translate of $s^k_n(x)$, $n\in \mathbb{Z}$, introduce an additional condition on $h_n$,
\begin{eqnarray}
\label{eq:h_n_orthonormality_condition}
\sum_{n=0}^{2K-1}h_n h_{n-2m}=\delta_{m0}.
\end{eqnarray}
The scaling equation Eq. (\ref{eq:scaling_equation_1}) and the vanishing moment condition Eq. (\ref{eq:vanishing_moment_condition}) wavelet function gives the additional equations necessary to find Daubechies scaling coefficient, $h_l$:
\begin{eqnarray}
\label{eq:h_n_vanishing_moment_condition}
\sum_{n=0}^{2K-1}n^m g_n=\sum_{n=0}^{2K-1}n^m(-1)^n h_{2K-1-n}=0,\quad m<K.
\end{eqnarray}
The system of equations Eq. (\ref{eq:h_n_normalization_condition}), Eq. (\ref{eq:h_n_orthonormality_condition}) and Eq. (\ref{eq:h_n_vanishing_moment_condition}) can be solved to get the Daubechies scaling coefficients $h_l$. The coefficients $h_n$ for $K=1,2$ and $3$ are given in Table \ref{tab:h_n_coefficients}.
\begin{table*}[ht]
\begin{center}
\setlength{\tabcolsep}{1.5pc}
\catcode`?=\active \def?{\kern\digitwidth}
\caption{The value of $h$ coefficient of Daubechies wavelet basis for different values of $K$.}
\label{tab:h_n_coefficients}
\vspace{2mm}
\begin{tabular}{c c c c}
\specialrule{.15em}{.0em}{.15em}
\hline
$h_n$ & $K=1$ & $K=2$ & $K=3$ \\
\hline
$h_0$ & $1/\sqrt{2}$ & $(1+\sqrt{3})/4\sqrt{2}$ & $(1+\sqrt{10}+\sqrt{5+2\sqrt{10}})/16\sqrt{2}$\\
$h_1$ & $1/\sqrt{2}$ & $(3+\sqrt{3})/4\sqrt{2}$ & $(5+\sqrt{10}+3\sqrt{5+2\sqrt{10}})/16\sqrt{2}$\\
$h_2$ & $0$   		 & $(3-\sqrt{3})/4\sqrt{2}$ & $(10-2\sqrt{10}+2\sqrt{5+2\sqrt{10}})/16\sqrt{2}$\\
$h_3$ & $0$   		 & $(1-\sqrt{3})/4\sqrt{2}$ & $(10-2\sqrt{10}-2\sqrt{5+2\sqrt{10}})/16\sqrt{2}$\\
$h_4$ & $0$   		 & $0$						& $(5+\sqrt{10}-3\sqrt{5+2\sqrt{10}})/16\sqrt{2}$\\
$h_5$ & $0$			 & $0$						& $(1+\sqrt{10}-\sqrt{5+2\sqrt{10}})/16\sqrt{2}$\\
\hline
\specialrule{.15em}{.15em}{.0em}
\end{tabular}
\end{center}
\end{table*}

\begin{figure}[ht]
\vspace{9pt}
\begin{center}
\includegraphics[scale=.375]{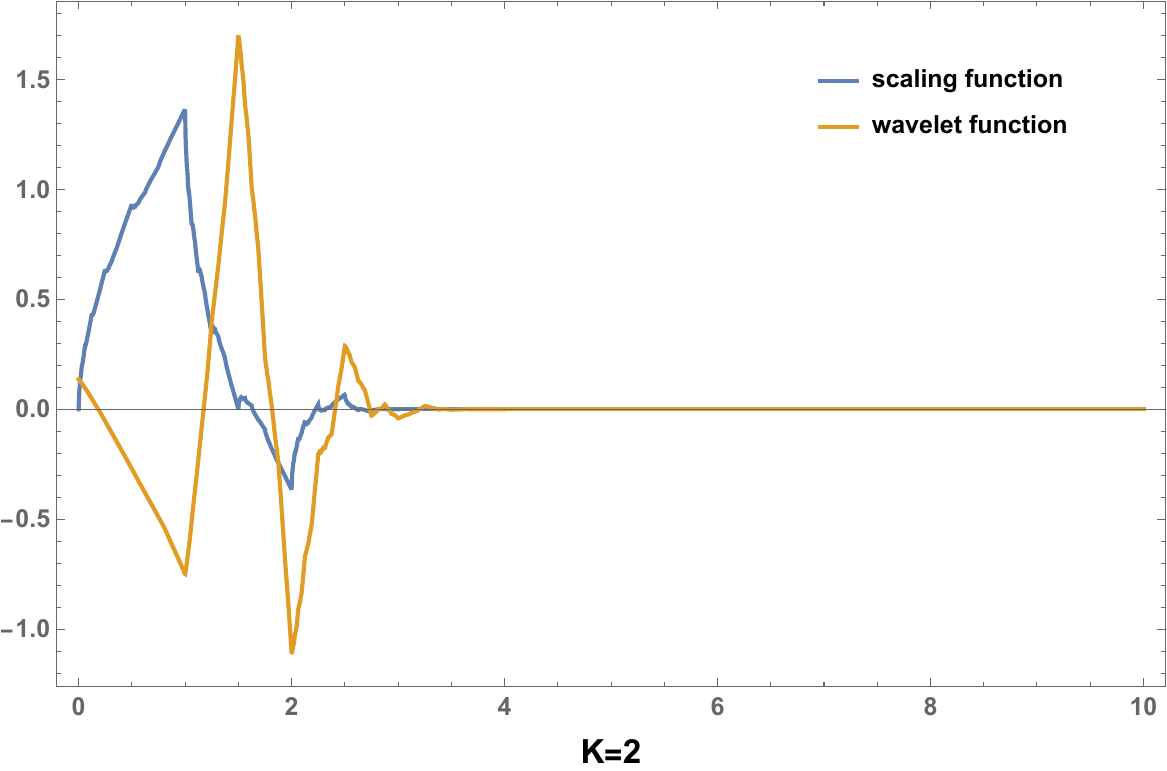}
\includegraphics[scale=.375]{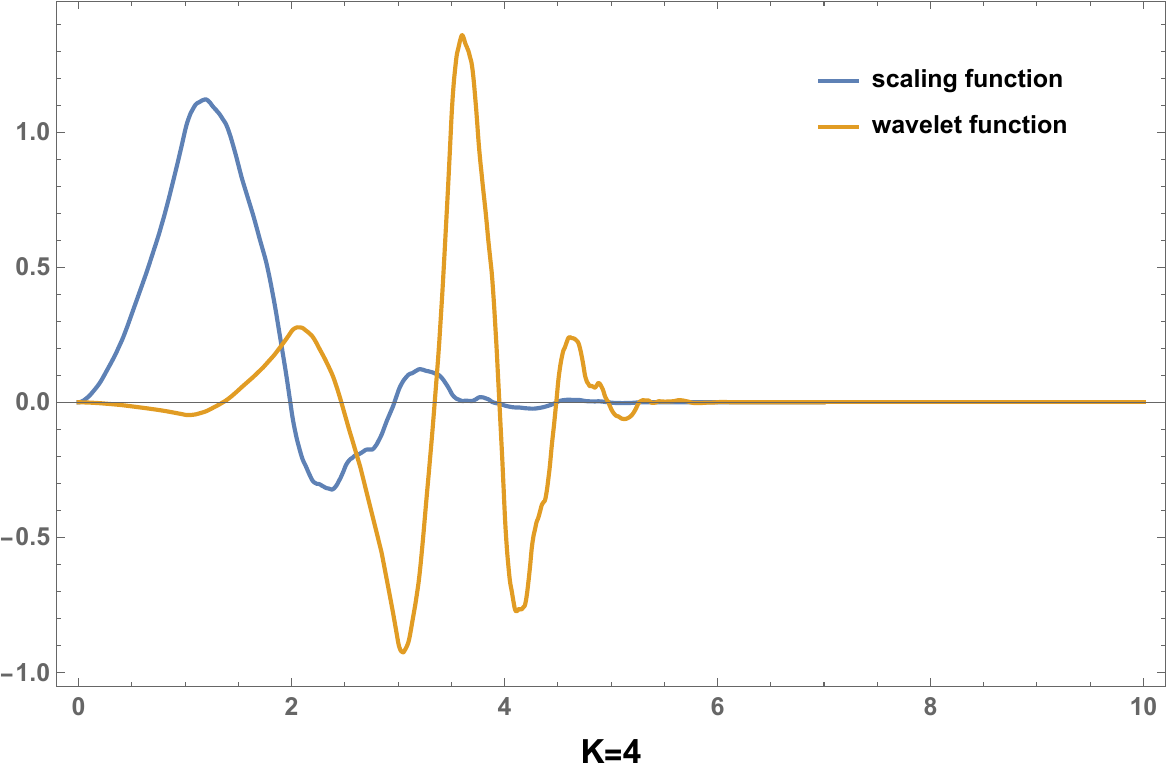}
\includegraphics[scale=.375]{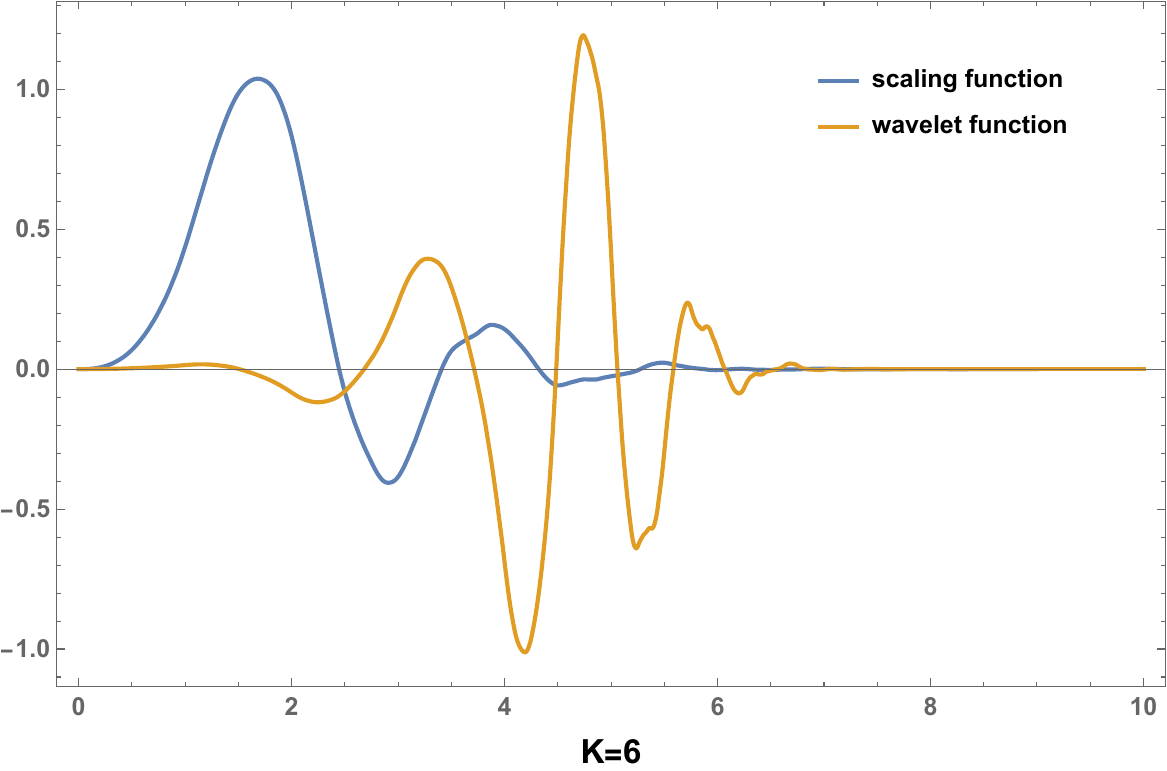}
\includegraphics[scale=.375]{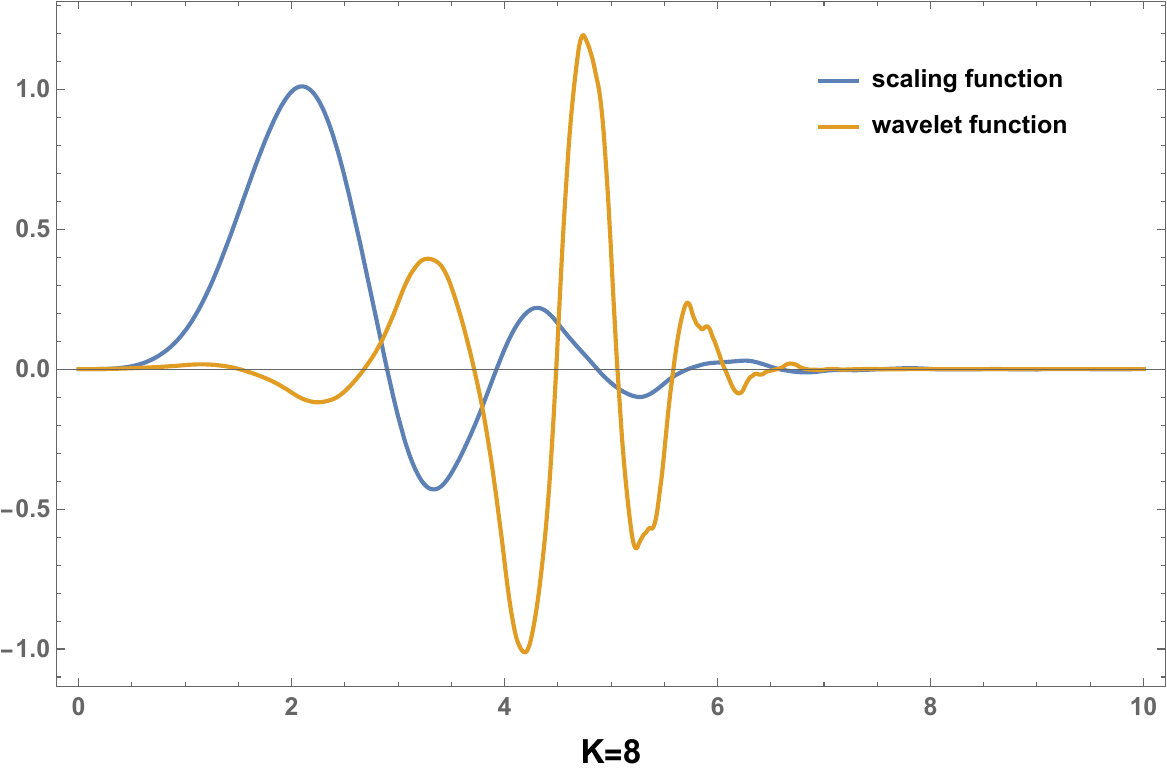}
\end{center}
\caption{Scaling and wavelet functions for different values of $K$}
\label{fig:Scaling_and_wavelet_functions_for_different_K}
\end{figure}

Utilizing the coefficients $h_n$, we can calculate the values of $s(x)$ and $w(x)$ at any given point $x$ using Eqs. (\ref{eq:scaling_equation_1}) and (\ref{eq:wavelet_equation_1}) \cite{Kessler_2003,10.1063/1.168556}. It can be demonstrated that both the mother scaling function $s(x)$ and the mother wavelet function $w(x)$ exhibit compact support within the interval $\left[0,2K-1\right]$. A graphical view of $s(x)$ and $w(x)$ for a sample value of $K=2,4,6$ and $8$ is shown in Fig. \ref{fig:Scaling_and_wavelet_functions_for_different_K}. The basis functions $s^k_n(x)$ and $w^k_n(x)$ have compact support smaller by a factor $2^k$ in comparison with the $s(x)$ and $w(x)$.
\begin{eqnarray}
&&s^k_n(x), w^k_n(x)\neq 0\quad\quad \forall\, x\in \left(\frac{(0-n)}{2^k},\frac{(2K-1-n)}{2^k}\right)\nonumber\\
&&\implies \text{support size}=\frac{(2K-1)}{2^k}.
\end{eqnarray}

The degree of analyticity of the basis functions depends on the value of $K$. For example, the basis functions for $K=1,2$ are not differentiable, $K=3,4$ are slightly differentiable, $K=6$ are double differentiable, and so on.

The extension of the basis function in two and three dimensions can be done by forming direct products of one-dimensional scaling and wavelet functions.

In two-dimension we define:
\begin{eqnarray}
s^k_{\textbf{n}}(\textbf{x}):=s^k_{n_1}(x_1)s^k_{n_2}(x_2),
\end{eqnarray}
and we introduce another notation $w^m_{\textbf{n},\alpha}(\textbf{x})$, which we call generalize wavelets having the following forms:
\begin{eqnarray}
w^m_{\textbf{n},1,k_2}(\textbf{x})&:=&s^k_{n_1}(x_1)w^{k_2}_{n_2}(x_2),\\
w^m_{\textbf{n},2,k_1}(\textbf{x})&:=&w^{k_1}_{n_1}(x_1)s^{k}_{n_2}(x_2),\\
w^m_{\textbf{n},3,k_1,k_2}(\textbf{x})&:=&w^{k_1}_{n_1}(x_1)w^{k_2}_{n_2}(x_2).
\end{eqnarray}
$m$ denotes the smallest wavelet scale present in the product, while $\alpha$ represents the values of $k_1$ and $k_2$, encompassing the four types of products featured in the basis function. Any square-integrable function in two dimensions can be expressed in this basis through the following expansion:
\begin{eqnarray}
f(x_1,x_2)&=&\sum_{n_1,n_2}^{\infty}f^s_{\textbf{n}}s^k_{\textbf{n}}(\textbf{x})+\sum_{n_1,n_2,k_2\geq k}^{\infty}f^{w_1,m}_{\textbf{n}}w^m_{\textbf{n},1,k_2}(\textbf{x})+\sum_{n_1,n_2,k_1\geq k}^{\infty}f^{w_2,m}_{\textbf{n}}w^m_{\textbf{n},2,k_1}(\textbf{x})\nonumber \\
&&+\sum_{n_1,n_2,(k_1,k_2)\geq k}^{\infty}f^{w_3,m}_{\textbf{n}}w^m_{\textbf{n},3,k_1,k_2}(\textbf{x}),
\end{eqnarray}
such that,
\begin{eqnarray}
\sum_{n_1,n_2}^{\infty} |f^s_{\textbf{n}}|^2+\sum_{n_1,n_2,k_2\geq k}^{\infty} |f^{w_1,m}_{\textbf{n}}|^2++\sum_{n_1,n_2,k_1\geq k}^{\infty} |f^{w_2,m}_{\textbf{n}}|^2+\sum_{n_1,n_2,(k_1,k_2)\geq k}^{\infty} |f^{w_3,m}_{\textbf{n}}|^2\leq \infty.
\end{eqnarray}
Here, the summation over $n_1,n_2$ goes from from $-\infty$ to $\infty$ and summation over $k_1$, $k_2$ goes from $k$ to $\infty$.

Now, In three dimension we define \cite{PhysRevD.87.116011}:
\begin{eqnarray}
s^k_{\textbf{n}}(\textbf{x}):=s^k_{n_1}(x_1)s^k_{n_2}(x_2)s^k_{n_3}(x_3),
\end{eqnarray}
and we introduce another notation $w^m_{\textbf{n},\alpha}(\textbf{x})$, which we call generalize wavelets having the following forms:
\begin{eqnarray}
w^m_{\textbf{n},1,k_3}(\textbf{x})&:=&s^k_{n_1}(x_1)s^k_{n_2}(x_2)w^{k_3}_{n_3}(x_3),\\
w^m_{\textbf{n},2,k_2}(\textbf{x})&:=&s^{k}_{n_1}(x_1)w^{k_2}_{n_2}(x_2)s^{k}_{n_3}(x_3),\\
w^m_{\textbf{n},3,k_1}(\textbf{x})&:=&w^{k_1}_{n_1}(x_1)s^{k}_{n_2}(x_2)s^{k}_{n_3}(x_3),\\
w^m_{\textbf{n},4,k_2,k_3}(\textbf{x})&:=&s^{k}_{n_1}(x_1)w^{k_2}_{n_2}(x_2)w^{k_3}_{n_3}(x_3),\\
w^m_{\textbf{n},5,k_1,k_2}(\textbf{x})&:=&w^{k_1}_{n_1}(x_1)w^{k_2}_{n_2}(x_2)s^{k}_{n_3}(x_3),\\
w^m_{\textbf{n},6,k_1,k_3}(\textbf{x})&:=&w^{k_1}_{n_1}(x_1)s^{k}_{n_2}(x_2)w^{k_3}_{n_3}(x_3),\\
w^m_{\textbf{n},7,k_1,k_2,k_3}(\textbf{x})&:=&w^{k_1}_{n_1}(x_1)w^{k_2}_{n_2}(x_2)w^{k_3}_{n_3}(x_3).
\end{eqnarray}
We refer the function, $w^m_{\textbf{n},\alpha}$ as generalized wavelets, where the index, $m$, indicates the minimum scale that appear into the product. The index $\alpha$ represents the values of $k_1,k_2,k_3$ and acts as an identity index to differentiate among the seven products. Any square-integrable function in three dimensions can be expressed in this basis through the following expansion:
\begin{eqnarray}
f(x_1,x_2,x_3)&=&\sum_{n_1,n_2,n_3}^{\infty}f^s_{\textbf{n}}s^k_{\textbf{n}}(\textbf{x})+\sum_{n_1,n_2,n_3,k_3\geq k}^{\infty}f^{w_1,m}_{\textbf{n},k_3}w^m_{\textbf{n},1,k_3}(\textbf{x})+\sum_{n_1,n_2,n_3,k_2\geq k}^{\infty}f^{w_2,m}_{\textbf{n},k_2}w^m_{\textbf{n},2,k_2}(\textbf{x})\nonumber \\
&&+\sum_{n_1,n_2,n_3,k_1\geq k}^{\infty}f^{w_3,m}_{\textbf{n},k_1}w^m_{\textbf{n},3,k_1}(\textbf{x})+\sum_{n_1,n_2,n_3,(k_2,k_3)\geq k}^{\infty}f^{w_4,m}_{\textbf{n},k_2,k_3}w^m_{\textbf{n},4,k_2,k_3}(\textbf{x})\nonumber\\
&&+\sum_{n_1,n_2,n_3,(k_1,k_2)\geq k}^{\infty}f^{w_5,m}_{\textbf{n},k_1,k_2}w^m_{\textbf{n},5,k_1,k_2}(\textbf{x})+\sum_{n_1,n_2,n_3,(k_1,k_3)\geq k}^{\infty}f^{w_6,m}_{\textbf{n},k_1,k_3}w^m_{\textbf{n},6,k_1,k_3}(\textbf{x})\nonumber\\
&&+\sum_{n_1,n_2,n_3,(k_1,k_2,k_3)\geq k}^{\infty}f^{w_7,m}_{\textbf{n},k_1,k_2,k_3}w^m_{\textbf{n},7,k_1,k_2,k_3}(\textbf{x}),
\end{eqnarray}
such that,
\begin{eqnarray}
&&\sum_{n_1,n_2,n_3}^{\infty}|f^s_{\textbf{n}}|^2+\sum_{n_1,n_2,n_3,k_3\geq k}^{\infty}|f^{w_1,m}_{\textbf{n},k_3}|^2+\sum_{n_1,n_2,n_3,k_2\geq k}^{\infty}|f^{w_2,m}_{\textbf{n},k_2}|^2+\sum_{n_1,n_2,n_3,k_1\geq k}^{\infty}|f^{w_3,m}_{\textbf{n},k_1}|^2\nonumber\\
&&+\sum_{n_1,n_2,n_3,(k_2,k_3)\geq k}^{\infty}|f^{w_4,m}_{\textbf{n},k_2,k_3}|^2+\sum_{n_1,n_2,n_3,(k_1,k_2)\geq k}^{\infty}|f^{w_5,m}_{\textbf{n},k_1,k_2}|^2+\sum_{n_1,n_2,n_3,(k_1,k_3)\geq k}^{\infty}|f^{w_6,m}_{\textbf{n},k_1,k_3}|^2\nonumber\\
&&+\sum_{n_1,n_2,n_3,(k_1,k_2,k_3)\geq k}^{\infty}|f^{w_7,m}_{\textbf{n},k_1,k_2,k_3}|^2\leq \infty .
\end{eqnarray}
Here, the summation over $n_1,n_2,n_3$ goes from $-\infty$ to $\infty$ and summation over $k_1,k_2,k_3$ goes from $k$ to $\infty$.

The scaling function, derived as the solution to the two-scale equation, serves as the foundation for other basis functions. These additional functions are formed through linear combinations of translated and scale-transformed versions of the scaling function. Consequently, all the basis functions exhibit fractal support. Representing basis functions in terms of elementary functions with smoothness on a sufficiently small scale proves challenging. Fortunately, applications typically do not require computing the scaling function with extreme accuracy. Instead, it is essential to derive overlap integrals involving products of various basis functions and their derivatives. These integrals adhere to a two-scale relation, and their exact computation is achievable through the utilization of two-scale relation equations and the normalization condition. Remarkably, there's no requirement to ascertain the value of the integrand at specific points during this process. These equations can also be employed for calculating integrals involving products of these functions with polynomials of any degree. 
Given the compact support of the basis functions and the ability to approximate any continuous function with a polynomial within a compact interval, it implies that integrals involving products of these basis functions and continuous functions can be accurately computed to any desired precision. Techniques for computing these quantities are elaborated upon in the Appendix.
\chapter{Quantum mechanics in Daubechies wavelet basis}
\label{chap:quantum_mechanics_in_daubechies_wavelet_basis}
Solving the Hamiltonian eigenvalue problem is central to understanding the physics of any quantum mechanical system. The eigenvalues and eigenfunctions provide the most detailed information about the system. In the first section, we will describe a variational procedure using linear variational parameters by approximately solving the Hamiltonian eigenvalue problem within the wavelet-based framework. We illustrate the procedure to a set of quantum mechanical problems: 
\begin{itemize}
\item The infinite square-well potential,
\item The simple harmonic oscillator,
\item The Dirac-delta function potential,
\item The triangular potential.
\end{itemize} 
A comparison of the wavelet-based analysis with the exact results for eigenvalues and eigenfunction is presented in the following section.

\section{Formalism}
\label{sec:formalism}
In this section, we outline the variational procedure for approximately solving the energy eigenvalue problem,
\begin{eqnarray}
\label{eq:schrodinger_eigenvalue_equation}
H\psi(x)=E\psi(x),
\end{eqnarray}
in wavelet basis. 
Our analysis is confined to a quantum mechanical system consisting of a single particle with mass $m$, moving within one spatial dimension under the influence of potential $V$. In the context of this system, the Hamiltonian $H$ can be expressed as follows:
\begin{eqnarray}
\label{eq:hamiltonian_for_a_particle_confined_in_potential_V}
H=-\frac{\hbar^2}{2m}\frac{\partial^2}{\partial x^2}+V(x).
\end{eqnarray}

The energy eigenfunction $\psi(x)$ associated with the energy eigenvalue $E$, belongs to the Hilbert space (state space) of the system. As described in Sec. \ref{subsec:Properties_of_elements_in_Daubechies_wavelet_family}, any element within the Hilbert space can be approximatively constructed using a wavelet basis with resolution $k$ in the following manner:
\begin{eqnarray}
\label{eq:approximate_psi_H_k}
\psi(x)\approx \sum_{n=-\infty}^{\infty} \psi_n^{s,k}s^k_n(x).
\end{eqnarray}
In other words, the Hilbert space is approximated by the scaling function space $\mathscr{H}^k$. The approximation can be systematically improved by increasing the resolution $k$.

On substituting Eq. (\ref{eq:approximate_psi_H_k}) into Eq. (\ref{eq:schrodinger_eigenvalue_equation}) and leveraging the orthonormality property of the scaling function, we can present the Hamiltonian eigenvalue problem in matrix form,
\begin{eqnarray}
\sum_{n=-\infty}^{\infty} H_{ss,mn}^k \psi^{s,k}_n = E\psi^{s,k}_m,
\end{eqnarray}
where,
\begin{eqnarray}
H_{ss,mn}^{k}=\frac{\hbar^2}{2m}T^k_{ss,mn}+V^k_{ss,mn},
\end{eqnarray}
with,
\begin{eqnarray}
\label{eq:kinetic_energy_term}
T^k_{ss,mn}=-\int_{-\infty}^{\infty}s^k_m(x)\frac{d^2}{dx^2}s^k_n(x)dx,
\end{eqnarray}
and,
\begin{eqnarray}
\label{eq:potential_term_integral}
V^k_{ss,mn}=\int_{-\infty}^{\infty}s^k_m(x)V(x)s^k_n(x)dx.
\end{eqnarray}
Using the property of compact support inherent to the basis functions, the matrix elements of the kinetic energy operator, as given in Eq. (\ref{eq:kinetic_energy_term}), can be reformulated in a manifestly symmetric form,
\begin{eqnarray}
\label{eq:overlap_integrals_derivative_of_scaling_functions}
\label{eq:T_k_ss_mn_1}
T^k_{ss,mn}=\int_{-\infty}^{\infty}\frac{d s^k_m(x)}{dx}\frac{d s^k_n(x)}{dx}dx.
\end{eqnarray}
This constitutes an overlap integral involving the multiplication of the derivatives of the two scaling functions. Such integrals can be evaluated using the method due to Belkin \cite{doi:10.1137/0729097}, as detailed in Appendix \ref{appen:the_kinetic_energy_term}.

The computation of the potential energy matrix elements as given in Eq. (\ref{eq:potential_term_integral}) is conducted in two stages. In the initial stage, we express the potential energy function as an expansion in terms of scaling functions and wavelet functions,
\begin{eqnarray}
\label{eq:potential_term_wavelet_basis}
V(x)=\sum_n V^{s,k}_n s^k_n(x)+\sum_n \sum_{l\geq k} V^{w,l}_n w^l_n(x),
\end{eqnarray}
where,
\begin{eqnarray}
V^{s,k}_n &=& \int_{-\infty}^{\infty} V(x)s^k_n(x)dx,\\
V^{w,l}_n &=& \int_{-\infty}^{\infty} V(x)w^l_n(x)dx.
\end{eqnarray}
The determination of the expansion coefficient $V^{s,k}_n$ and $V^{w,l}_n$ is carried out employing the scheme developed by Sweldens and Piessens \cite{sweldens1994quadrature}. We describe this procedure in the Appendix \ref{subsec:moment_of_scaling_and_wavelet_function}. In the second stage, we substitute Eq. (\ref{eq:potential_term_wavelet_basis}) in Eq. (\ref{eq:potential_term_integral}), we get
\begin{eqnarray}
V^k_{ss,mn}=\sum_{p} V^{s,k}_p I^k_{sss,pmn}+\sum_p \sum_{l\geq k} V^{w,l}_p I^{l,k}_{wss,pmn},
\end{eqnarray}
where,
\begin{eqnarray}
I^k_{sss,pmn}&=&\int_{-\infty}^{\infty} s^k_p(x)s^k_m(x)s^k_n(x)dx,\\
I^{l,k}_{wss,pmn}&=&\int_{-\infty}^{\infty} w^l_p(x)s^k_m(x)s^k_n(x)dx.
\end{eqnarray}
The procedure to evaluate $I^k_{sss,pmn}$ and $I^{l,k}_{wss,pmn}$, is outlined in the Appendix \ref{appen:the_potential_energy_term_1}. 

The Hamiltonian matrix elements can be generated using the procedure outlined above. The Hamiltonian matrix is truncated by selecting an appropriate resolution cut-off $k$ and defining a suitable volume cut-off values by restricting the minimum and the maximum value of the translation indices $m$ and $n$. The eigenvalues and eigenfunctions of the truncated Hamiltonian matrix at resolution $k$ are determined by using one of the standard matrix diagonalization algorithms. The precision of both eigenvalues and eigenfunctions can be systematically enhanced by increasing the truncated volume and resolution $k$.

The quantum mechanical problems in spatial dimensions greater than one can also be investigated with the wavelet framework. This requires extending the basis elements to higher dimensions, a process detailed in Sec. \ref{subsec:Properties_of_elements_in_Daubechies_wavelet_family}. 

\section{The infinite square well potential}
The infinite-square well potential (ISWP) in one spatial dimension is given by \cite{griffiths2017introduction},
\begin{eqnarray}
V(y)=
\begin{cases}
 0\quad &0\leq y\leq a \\
 \infty  \quad &\text{otherwise}.
\end{cases}
\end{eqnarray}
The potential energy within the well is zero; therefore, classically a particle within the boundary of the well is entirely free, except at the two ends where infinite force prevents it from escaping. Because of its simplicity, it's an important example used in quantum mechanics courses to introduce students to the principles of quantum mechanics. Though the infinite square-well is a simplified theoretical construction, it's applications and implications extend to various branches of physics including atomic physics \cite{eisberg1985quantum}, nuclear physics \cite{griffiths2017introduction,krane1987introductory}, and solid-state physics \cite{kittel2004introduction,ashcroft1976solid}. It has practical relevance in understanding and designing systems at quantum level. The infinite square well potential first appears in the textbook by Mott \cite{mott1930outline}. The concept of the infinite well has been utilized as an approximation for physical implementations of atomic mirrors \cite{DOWLING19961,PhysRevLett.71.3083,PhysRevLett.75.629,PhysRevLett.77.1464,PhysRevLett.77.4} and has been employed in discussions regarding the deflection of ultra-cold atoms from mirrors \cite{10.1007/978-1-4419-8907-9_107}. Experiments have been conducted to excite coherent charge oscillations from asymmetric quantum well structures, resembling the asymmetric infinite square well \cite{bonvalet1996femtosecond}. The phenomenon known as the Stark effect \cite{PhysRevLett.53.2173} can be represented by incorporating an electric field into an infinite square well model.

The energy eigenvalue equation of the particle moving in the infinite square-well potential is given by:
\begin{eqnarray}
\label{eq:schrodinger_equation_infinite_square_well_potential}
-\frac{\hbar^2}{2m}\frac{\partial^2 \psi(y)}{\partial y^2}= E \psi (y).
\end{eqnarray}
Because of the infinite extent of the potential, it is impossible for the particle to be outside the well. Consequently, the wavefunction $\psi(y)$ has a non-zero value only inside the well and is zero elsewhere outside the well. 

We convert the Eq. (\ref{eq:schrodinger_equation_infinite_square_well_potential}) into a dimensionless form by defining a new variable $x=\frac{10y}{a}$. Choosing a dimensionless variable in problem-solving offers several advantages. It simplifies the equations, makes them more general, and aids in comparing solutions across different scales. Our choice of the dimensionless variable is governed by our desire to use 3rd order ($K = 3$) Daubechies wavelets for analysing the problem. For $K=3$, the Daubechies wavelet has compact support equal to $5$ which is smaller than the extent of the well. This enables us to accommodate a finite number of resolution $0$ scaling functions within the domain of $x$ ($0 < x < 10$). Changing the variable $y$ to $x$ transforms Eq. (\ref{eq:schrodinger_equation_infinite_square_well_potential}) into the following form:
\begin{eqnarray}
\label{eq:schrodinger_equation_infinite_square_well_dimensionless}
-\frac{1}{2}\frac{\partial^2 \psi(x)}{\partial x^2}= \epsilon \psi (x),
\end{eqnarray}
where, $\epsilon= \frac{2m E a^2}{100\hbar^2}$.\newline
The eigenvalue $\epsilon$ and the eigenfunction $\psi(x)$ can be found using standard analytical method described in \cite{griffiths2017introduction},
\begin{eqnarray}
\epsilon_n = \frac{n^2 \pi^2}{2(10)^2},
\end{eqnarray}
and,
\begin{eqnarray}
\psi_n(x) = \sqrt{\frac{2}{10}} \sin \left(\frac{n\pi x}{10}\right),
\end{eqnarray}
respectively. The index $n$, which labels eigenvalues and eigenfunctions, takes a positive integer value ($n=1,2,3,...$).

\begin{figure}[hbt]
\vspace{9pt}
\begin{center}
\includegraphics[scale=.5]{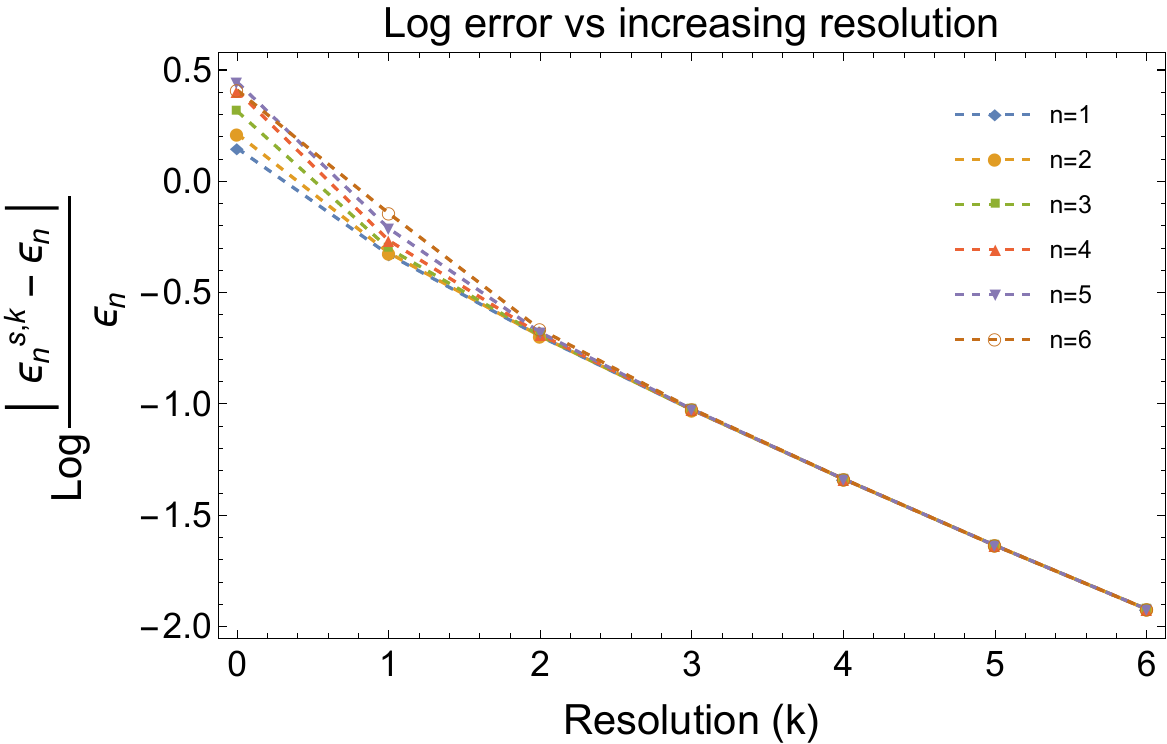}
\caption{\label{fig:log_error_vs_increasing_resolution_ISWP}Log error $\left(\log\frac{|\epsilon_n^{s,k}-\epsilon_n|}{\epsilon_n}\right)$ of different eigenstates with increasing resolution $(k)$ plot for the infinite square well potential.}
\end{center}
\end{figure}

\begin{figure}[hbt]
\begin{center}
\includegraphics[scale=.375]{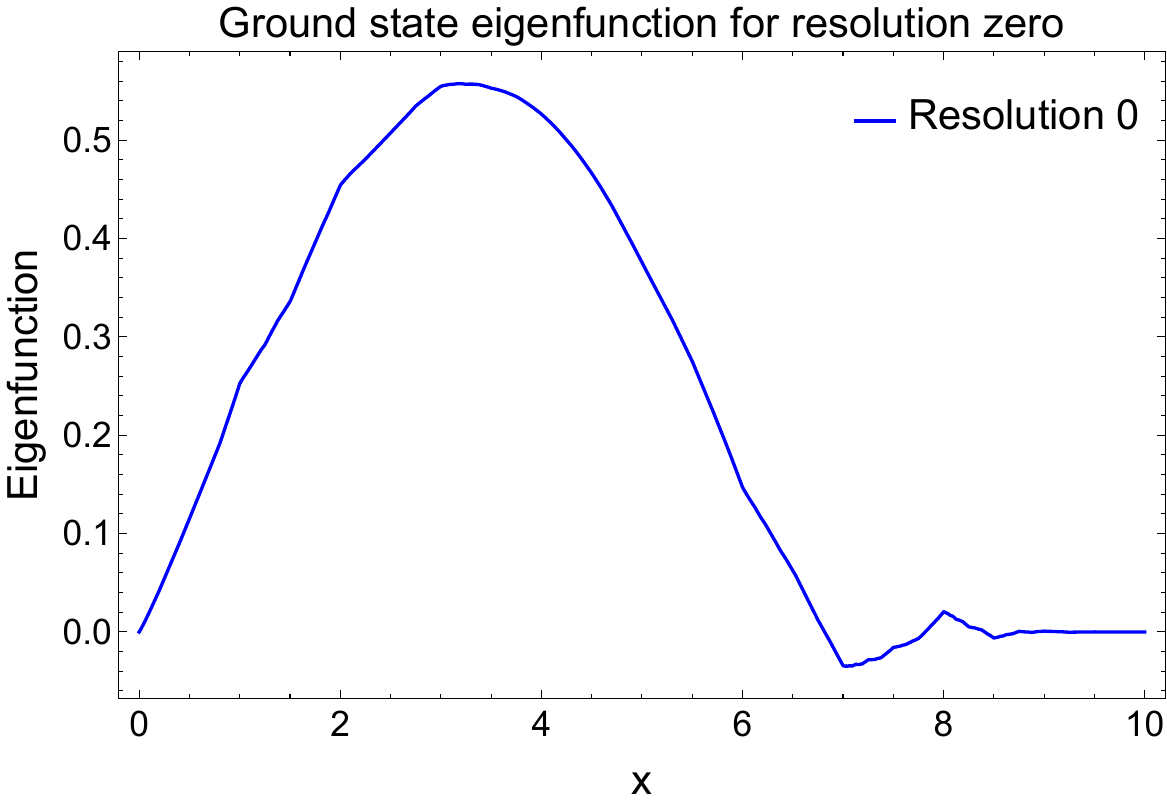}
\includegraphics[scale=.375]{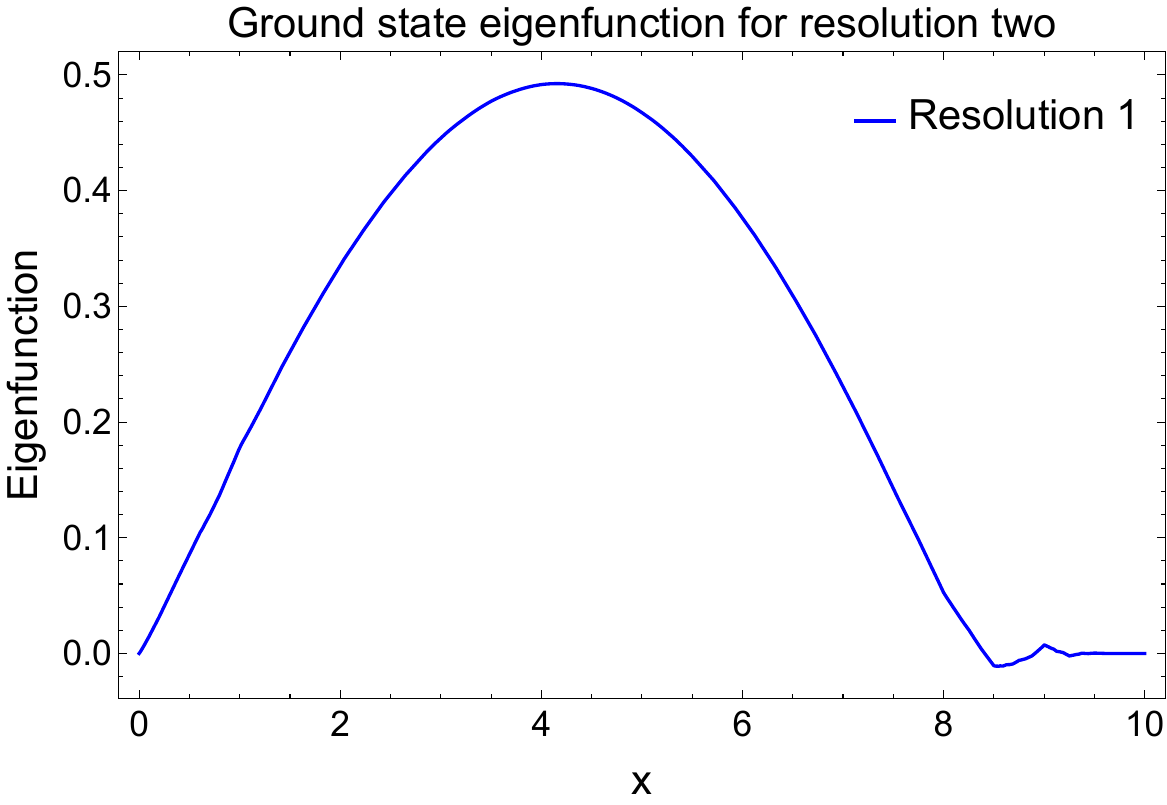}
\includegraphics[scale=.375]{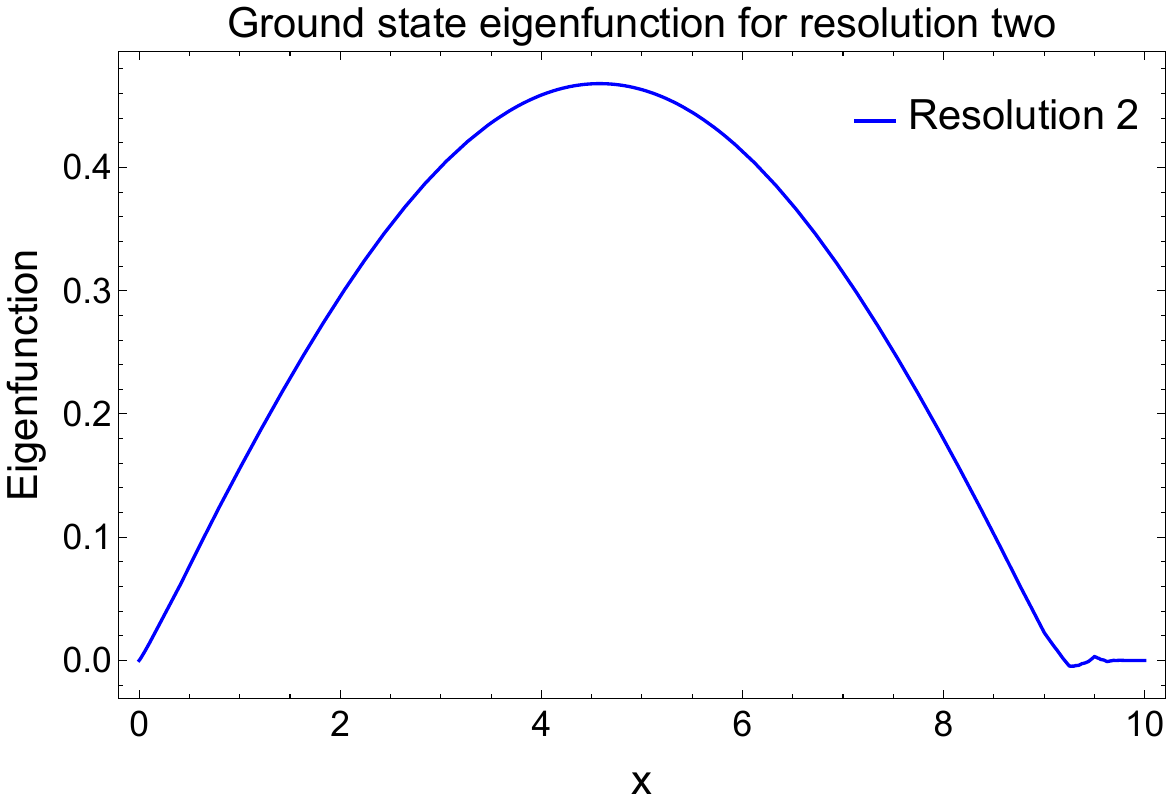}
\includegraphics[scale=.375]{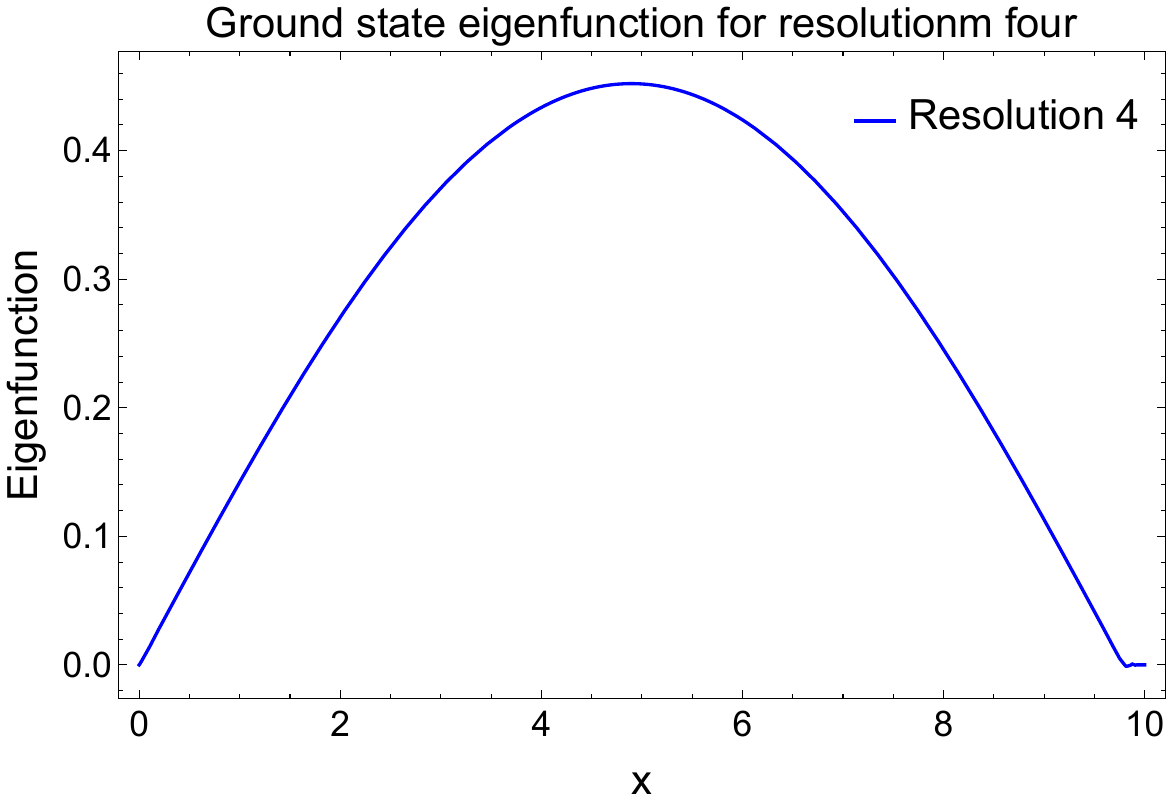}
\includegraphics[scale=.375]{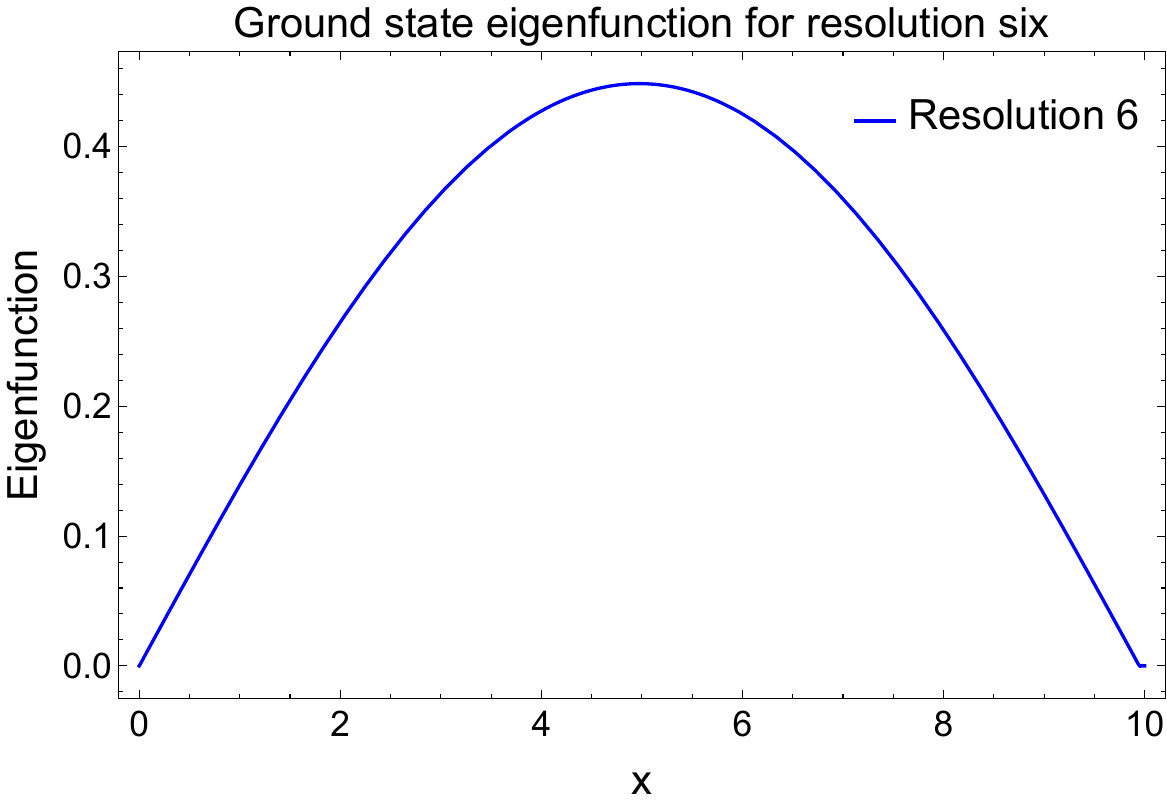}
\includegraphics[scale=.375]{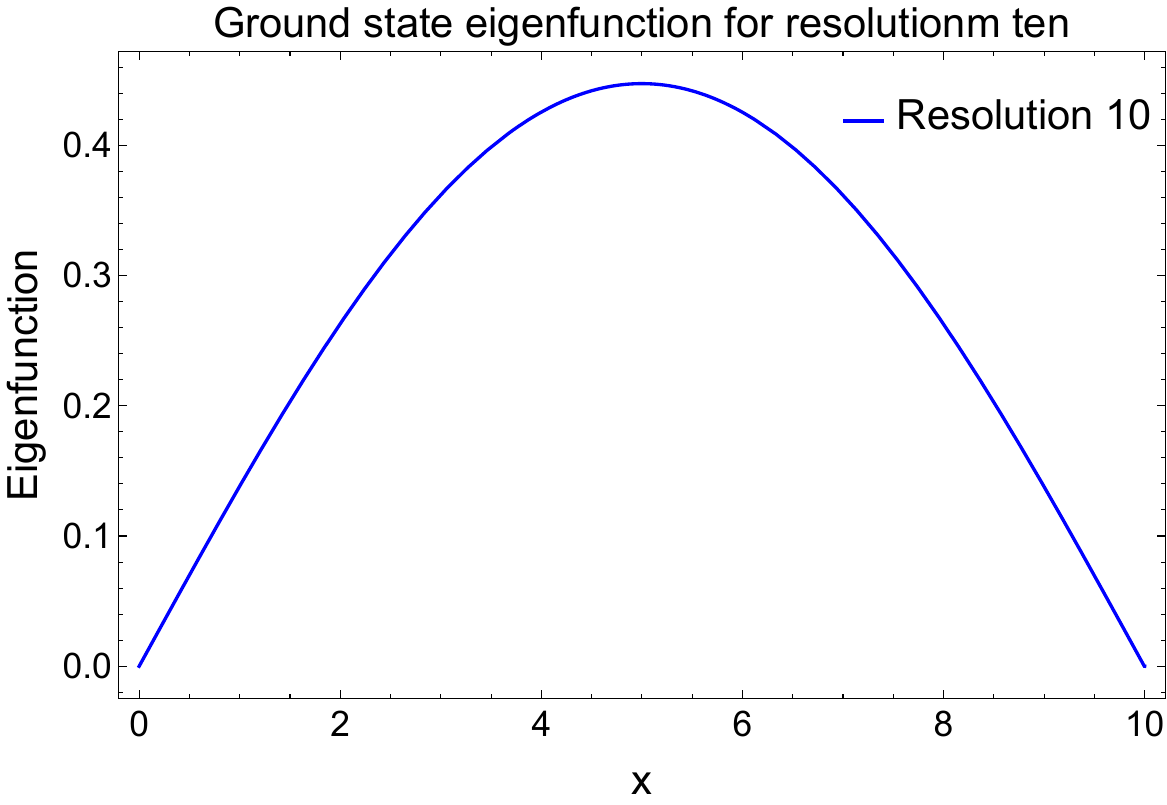}
\caption{\label{fig:Ground_state_for_dif_res_ISWP}The ground state eigenfunction $\psi^{s,k}_0$ of infinite square well potential with increasing resolution.}
\end{center}
\end{figure}

Now, we examine this problem using a wavelet-based approach by approximating the Hilbert space of the problem to resolution space $\mathscr{H}^k$. Within this framework, the energy eigenvalue equation, Eq. (\ref{eq:schrodinger_equation_infinite_square_well_dimensionless}), takes the following form,
\begin{eqnarray}
\label{eq:schrodinger_equation_infinite_square_well_matrix}
\sum_n \frac{1}{2} T^k_{ss,mn} \psi^{s,k}_n=\epsilon^{s,k}\psi^{s,k}_m,
\end{eqnarray}
where, $T^k_{ss,mn}$ is given in Eq. (\ref{eq:T_k_ss_mn_1}). The integral, $T^k_{ss,mn}$, can be computed using the method outlined in Appendix \ref{appen:the_kinetic_energy_term}. $\psi^{s,k}$ and $\epsilon^{s,k}$ represent the resolution $k$ approximations to the exact eigenfunction $\psi(x)$ and the exact eigenvalue $\epsilon$, respectively. $\psi^{s,k}$ encapsulates the physical features of the exact wave function $\psi(x)$ from the length scale, $10$, down to the length scale $\frac{2K-1}{2^k}$. The finite length of the well and the boundary condition on the wave function, which is that wave function should vanish outside the well, naturally impose volume truncation on the problem. For a finite resolution $k$ and order $K$ Daubechies wavelet, one can accommodate $2^k L-2K+2$ number of basis functions within a finite length $L$. With $L = 10$ and $K = 3$, Eq. (\ref{eq:schrodinger_equation_infinite_square_well_matrix}) transforms into a matrix eigenvalue problem with dimensionalities $6, 16, 36, 76,$ and so on, for resolutions $0, 1, 2,$ and $3$ respectively. The eigenvalues and eigenfunctions of these matrices has been determined using Mathematica's library function "$Eigensystem[]$". Other software options, such as Matlab or Python, can also be employed for the same purpose. The obtained eigenvalues are presented in Table \ref{tab:actual_and_approximate_eigenvalues_of_the_infinite_square_well_potential}. The logarithmic error of various energy states with increasing resolution is depicted in Fig. \ref{fig:log_error_vs_increasing_resolution_ISWP}. The ground state eigenfunction for different resolutions is illustrated in Fig. \ref{fig:Ground_state_for_dif_res_ISWP}. 
\begin{table}[ht]
\begin{center}
\setlength{\tabcolsep}{0.85 pc}
\catcode`?=\active \def?{\kern\digitwidth}
\caption{The exact energy eigenvalues along with the approximate energy eigenvalues for different resolution of the infinite square well potential.}
\label{tab:actual_and_approximate_eigenvalues_of_the_infinite_square_well_potential}
\vspace{1mm}
\begin{tabular}{c | c c c c c c }
\specialrule{.15em}{.0em}{.15em}
\hline
$E_n$ & $E_1$ & $E_2$ & $E_3$ & $E_4$ & $E_5$ & $E_6$ \\
\hline
Exact value & $0.049298$ & $0.197192$ & $0.443682$ & $0.788768$ & $1.23245$ & $1.774728$\\

$k=0$ & $0.118741$ & $0.519472$ & $1.360506$ & $2.799268$ & $4.670158$ & $6.335427$ \\

$k=1$ & $0.072412$ & $0.291332$ & $0.665108$ & $1.216576$ & $1.989897$ & $3.05019$ \\

$k=2$ & $0.059202$ & $0.215756$ & $0.485494$ & $0.863235$ & $1.349147$ & $1.943538$ \\

$k=3$ & $0.053937$ & $0.215756$ & $0.485494$ & $0.863235$ & $1.349147$ & $1.943538$ \\

$k=4$ & $0.051566$ & $0.206264$ & $0.464097$ & $0.825073$ & $1.289205$ & $1.856513$ \\

$k=5$ & $0.050439$ & $0.201754$ & $0.453948$ & $0.807019$ & $1.26097$ & $1.815802$ \\

$k=6$ & $0.049889$ & $0.199555$ & $0.448999$ & $0.798221$ & $1.247221$ & $1.795999$ \\

$k=10$ & $0.0493816$ & $0.197526$ & $0.444434$ & $0.790105$ & $1.23454$ & $1.77774$ \\
\hline
\specialrule{.15em}{.15em}{.0em}
\end{tabular}
\end{center}
\end{table}

Table \ref{tab:actual_and_approximate_eigenvalues_of_the_infinite_square_well_potential} reveals that the eigenvalues associated with various energy levels progressively converge towards the exact eigenvalues with increasing resolution.  This is due to the fact that with increasing resolution, the basis will gradually approach the continuum limit. This entails incorporating progressively smaller length-scale contributions to the eigenvalues. Consequently, it is reasonable to anticipate that with infinite resolution, this framework would yield the exact eigenvalues.

In Fig. \ref{fig:log_error_vs_increasing_resolution_ISWP}, we observe that at a given value of lower resolution, the logarithmic error in the eigenvalues is higher for higher excited states. As the resolution increases, the logarithmic error for various states gradually begins to converge. This phenomenon occurs because the infinite square well potential problem is confined within the boundaries of the well and do not have access to the region beyond the boundary. As a result, it exhibits an inherent volume cutoff. Furthermore, the eigenvalues corresponding to higher excited states receive more short-distance contributions compared to the low-lying states. Consequently, at lower resolutions, the accuracy of eigenvalues is lower for higher excited states due to the absence of  contributions from these short-distance degrees of freedom. It is important to note that this analysis provides a quantitative insight into the significant role played by short-distance degrees of freedom when calculating higher-order eigenvalues. We will observe the deviation from the scenario where logarithmic error of eigenvalues do not merge beyond a certain resolution, specifically in the cases where the potential has an infinite extent throughout the space.

\begin{figure}[hbt]
\begin{center}
\includegraphics[scale=.40]{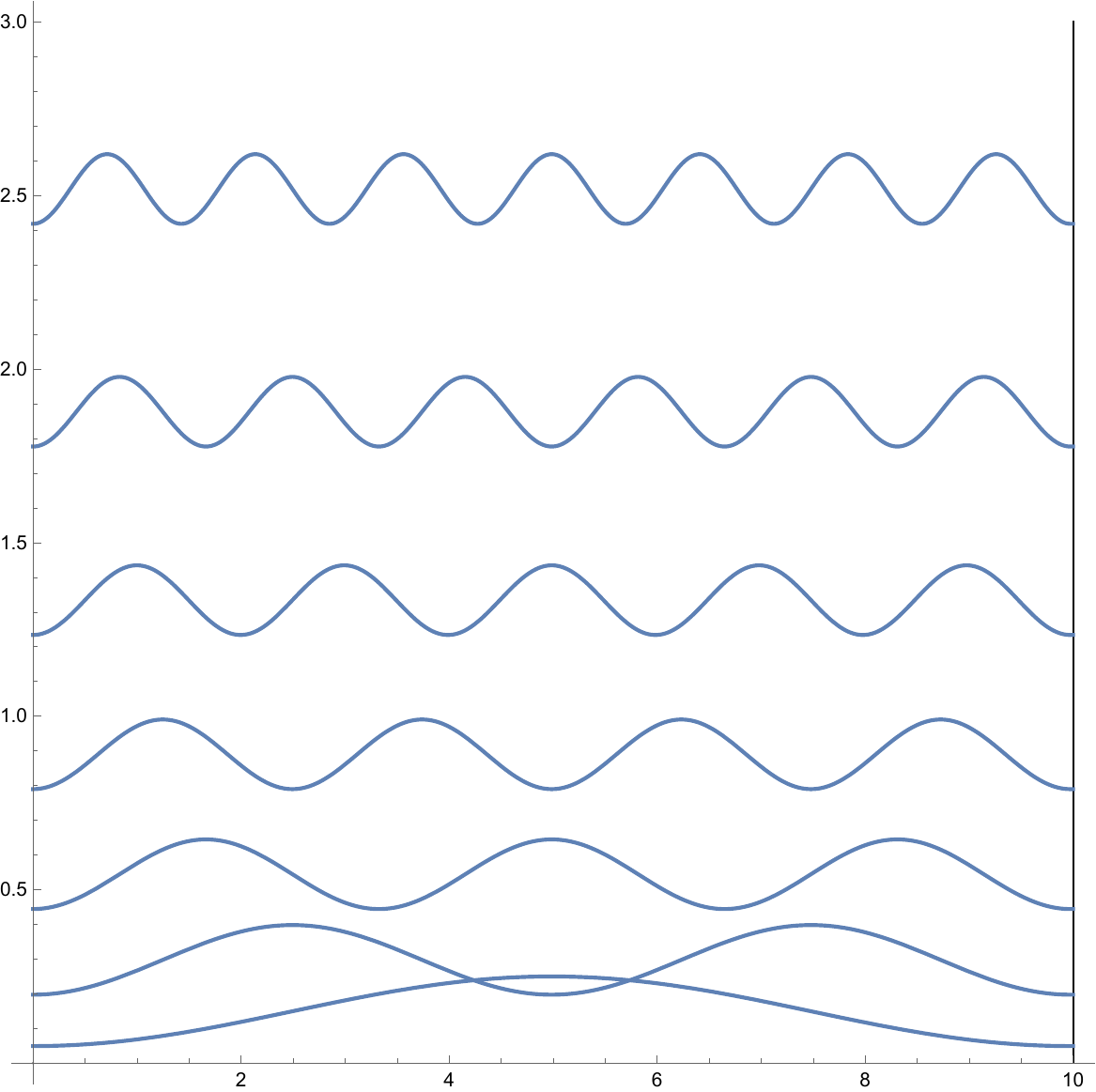}
\caption{\label{fig:the_probability_distribution_of_eigenstates_of_ISWP}The probability distribution for the initial seven eigenstates of the infinite-square well potential was computed utilizing a third-order ($K=3$) Daubechies wavelet basis of resolution $7$ ($k=7$).}
\end{center}
\end{figure}

We fit the ground state energy eigenstate computed using the wavelet-based approach to the curve $a \sin(bx)$ using Mathematica's build-in library \textit{FindFit[]} and compared it to the exact ground state wavefunction of ISWP, expressed as $\sqrt{\frac{2}{10}}\sin\left(\frac{\pi x}{2(10)^2}\right)$. Table \ref{tab:actual_value_and_approximate_value_of_a_b_ISWP} displays the values of $a$ and $b$ corresponding to various resolutions. Notably, as the resolution approaches to $10$, the values of $a$ and $b$ converge towards the exact values of $a=\sqrt{\frac{2}{10}}=0.447214$ and $b=\frac{\pi}{2(10)^2}=0.314159$ associated with the ground state wave function. Figure \ref{fig:the_probability_distribution_of_eigenstates_of_ISWP} illustrates the probability distribution of the first seven eigenstates of the infinite square-well potential for resolution $7$ ($k=7$). This analysis reveals that for the low-lying eigenvalues of a quantum system, this framework can provide reasonably accurate results even with relatively low resolutions, such as $7$.

\begin{table*}[hbt]
\begin{center}
\setlength{\tabcolsep}{0.21pc}
\catcode`?=\active \def?{\kern\digitwidth}
\caption{The actual values of $a$ and $b$ along with the approximate values of $a$ and $b$ with increasing resolution.}
\label{tab:actual_value_and_approximate_value_of_a_b_ISWP}
\vspace{1mm}
\begin{tabular}{c c c c c c c c c c}
\specialrule{.15em}{.0em}{.15em}
\hline
 & Exact value & $k=0$ & $k=1$ & $k=2$ & $k=3$ & $k=4$ & $k=5$ & $k=6$ & $k=10$\\

\hline
$a$ & $0.447214$ & $0.434030$ & $0.466454$ & $0.462151$ & $0.455790$ & $0.451745$ & $0.449536$ & $0.448388$ & $0.447288$\\

$b$ & $0.314159$ & $0.390452$ & $0.357006$ & $0.337663$ & $0.326616$ & $0.320597$ & $0.317435$ & $0.315812$ & $0.314263$\\
\hline
\specialrule{.15em}{.15em}{.0em}
\end{tabular}
\end{center}
\end{table*}

\section{The Simple Harmonic Oscillator}
A particle of mass $m$ moving under the influence of the simple harmonic oscillator potential (SHOP) in one spatial dimension is given by the following Hamiltonian:
\begin{eqnarray}
H=-\frac{\hbar^2}{2m}\frac{\partial^2}{\partial^2 y}+\frac{1}{2}m\omega^2 y.
\end{eqnarray}

The SHOP has a plenty of applications in the construction of many physical systems. In molecular physics and chemistry, the SHOP is used to model vibrational modes of diatomic and polyatomic molecules. The atoms in the molecules are often treated as masses connected by springs (bonds). The harmonic approximation assumes that the displacement of the vibration of the atom is very small, so the potential energy is assumed to be $\frac{1}{2}kx^2$ \cite{Barth2013}. In solid state physics, the SHOP is needed to describe the lattice vibrations in a crystal. Each atoms will vibrate around its equilibrium position, which can be quantized as phonons. The concept of phonons are essential to explain the properties of solids like specific heat \cite{kittel2004introduction}. The harmonic potential serves as a starting point for more complex systems where the actual potential may not be purely harmonic but can be approximated as such near the equilibrium point. This is the basis of perturbation theory and the Taylor expansion, where higher-order terms are added to refine the model.

The energy eigenvalue equation of the particle moving under the influence of the SHOP is given by:
\begin{eqnarray}
\label{eq:Harmonic_oscillator_eigval_eqn}
\left(-\frac{\hbar^2}{2m}\frac{\partial^2 }{\partial y^2}+\frac{1}{2}m\omega^2 y\right)\psi(y)=E \psi(y).
\end{eqnarray}
As the potential extends throughout the space, even the lowest energy of the particle will get contributions from all the energy scales, and it is also possible for the particle to penetrate the boundary of the potential well.

The parameters $m$ and $\omega$ are two dimensionful parameters that can set the scale,  size and energy of the bound states. We employ these parameters to introduce a new set of dimensionless variable, $x=\sqrt{\frac{m\omega}{\hbar}}y$, in which Eq. (\ref{eq:Harmonic_oscillator_eigval_eqn}) can be reformulated as,
\begin{eqnarray}
\label{eq:Harmonic_oscillator_eigenvalue_dimensionless}
-\frac{1}{2}\frac{\partial^2 \psi(x)}{\partial x^2}+\frac{1}{2} x^2\psi(x)=\epsilon \psi(x),
\end{eqnarray}
where, $\epsilon=\frac{E}{\hbar \omega}$.

Equation (\ref{eq:Harmonic_oscillator_eigenvalue_dimensionless}) can be exactly solved analytically resulting the infinite number of bound states with infinitely many eigenvalues \cite{griffiths2017introduction}. The $n$th bound state of the SHOP and its corresponding eigenvalue is given by \cite{griffiths2017introduction},
\begin{eqnarray}
\label{eq:The_exact_eigensystem_of_the_Harmonic_oscillator}
\psi_n(x)=\left(\frac{1}{\pi}\right)^{\frac{1}{4}}\frac{1}{\sqrt{2^n n!}}H_n(x)e^{-\frac{x^2}{2}}\quad;\quad  \epsilon_n=\left(n+\frac{1}{2}\right),
\end{eqnarray}
here, $H_n(x)$ is the Hermite Polynomial, which can be evaluated from the Rodrigues formula,
\begin{eqnarray}
H_n(x)=(-1)^n e^{x^2}\left(\frac{d}{dx}\right)^n e^{-x^2}.
\end{eqnarray}
The ground state eigenfunction and energy of the SHOP is given by,
\begin{eqnarray}
\psi_0(x)=\left(\frac{1}{\pi}\right)^{\frac{1}{4}}e^{-\frac{x^2}{2}}\quad; \quad \epsilon_0=\frac{1}{2}.
\end{eqnarray}

Now, we are going to solve this problem using the wavelet-based approach and compare the accuracy of the results with the exact results given in Eq. (\ref{eq:The_exact_eigensystem_of_the_Harmonic_oscillator}). Following the approach given in Sec. \ref{sec:formalism}, within the wavelet-based framework, the energy eigenvalue equation, Eq. (\ref{eq:Harmonic_oscillator_eigenvalue_dimensionless}), takes the following form,
\begin{eqnarray}
\sum_n \frac{1}{2}\left(T^k_{ss,mn}+V^k_{x^2ss,mn}\right)\psi^{s,k}_n=\epsilon \psi^{s,k}_n.
\end{eqnarray}
$T^k_{ss,mn}$ is the kinetic energy term given in Eq. (\ref{eq:schrodinger_equation_infinite_square_well_matrix}), and can be calculated exactly using the method detailed in the Appendix \ref{appen:the_kinetic_energy_term}. The potential energy term $V^k_{x^2ss,mn}$ is given by,
\begin{eqnarray}
V^k_{x^2ss,mn}=\int x^2 s^k_m(x)s^k_n(x)dx,
\end{eqnarray}
and can be determined using the procedure descried in the Appendix. \ref{appen:The potential of the form x_2n}. After evaluating the kinetic and the potential energy term of the Hamiltonian matrix elements, we construct the truncated Hamiltonian matrix within the resolution subspace $\mathscr{H}^k$, by restricting the volume $V$. The restriction on the volume is achieved by imposing a lower and a upper cutoff on the translation indices $m$ and $n$.

\begin{figure}[hbt]
\begin{center}
\includegraphics[scale=.5]{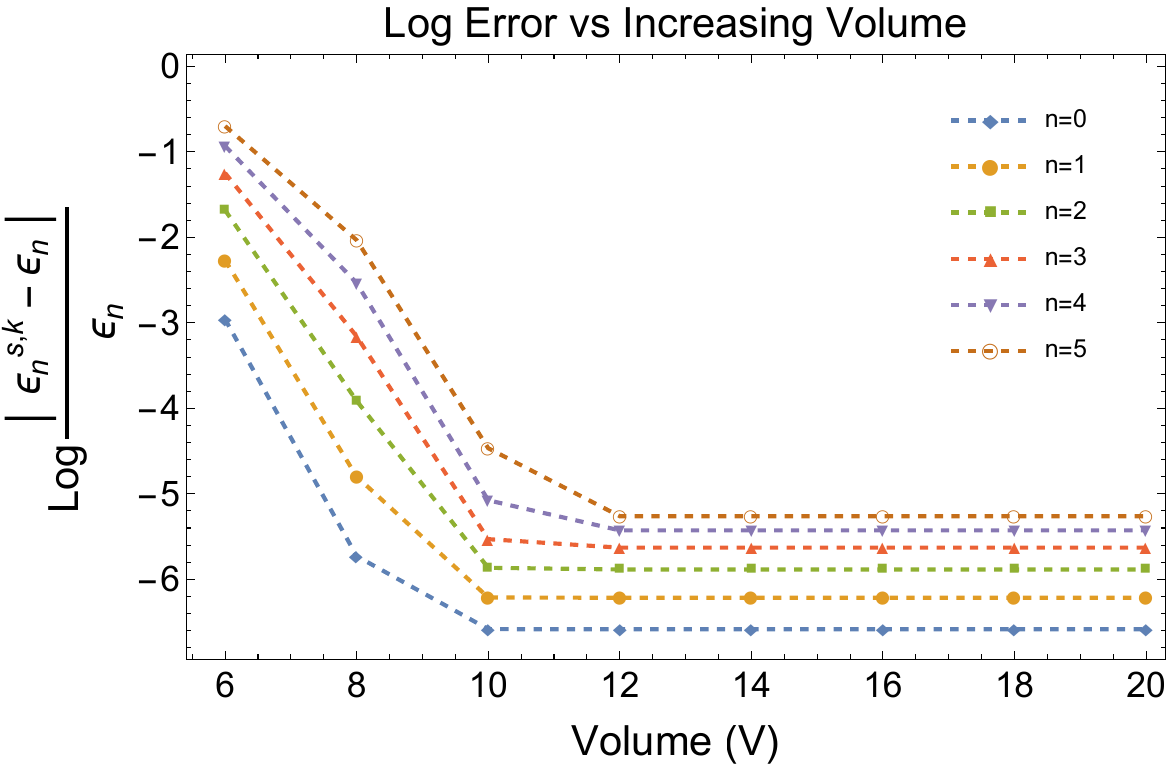}
\caption{\label{fig:Log_error_vs_volume_Plot_SHO}Log error $\left(\log\frac{|\epsilon_{n}^{s,k}-\epsilon_n|}{\epsilon_n}\right)$ of different eigenstates with increasing volume for a fixed resolution ($k=5$) plot for the SHOP.}
\end{center}
\end{figure}

Now, we will present the results obtained using the wavelet-based method. The Log error versus the increasing volume plot for a fixed resolution ($k=5$) is given in the Fig. \ref{fig:Log_error_vs_volume_Plot_SHO}. 
It can be seen that the logarithmic error is not undergoing a significant change beyond a certain volume, $V=14$ $(-7\leq x \leq 7)$. This happens because the low-lying eigenvalues will not get a substantial amount of contributions from the coarser length scale. So, in order to obtain significantly accurate low-lying eigenvalues, we need to choose a volume that is at least $V\geq 14$. We decided to opt for a sufficiently large volume, specifically $V=20$.

Table. \ref{tab:exact_and_approximate_eigenvalues_of_the_SHO}, presents the lowest six eigenvalues of the simple harmonic oscillator potential with increasing resolution for a fixed volume $V=20$ $(-10\leq x\leq 10)$.
\begin{table}[hbt]
\begin{center}
\setlength{\tabcolsep}{0.74 pc}
\catcode`?=\active \def?{\kern\digitwidth}
\caption{The exact energy eigenvalues along with the approximate energy eigenvalues for different resolution of the simple harmonic oscillator potential.}
\label{tab:exact_and_approximate_eigenvalues_of_the_SHO}
\vspace{1mm}
\begin{tabular}{c | c c c c c c }
\specialrule{.15em}{.0em}{.15em}
\hline
$E_n$ & $E_0$ & $E_1$ & $E_2$ & $E_3$ & $E_4$ & $E_5$ \\
\hline
Exact value & $0.5$ & $1.5$ & $2.5$ & $3.5$ & $4.5$ & $5.5$\\

$k=0$ & $0.547051$ & $1.732639$ & $3.066363$ & $4.493113$ & $5.749729$ & $7.439307$ \\

$k=1$ & $0.506325$ & $1.539240$ & $2.621300$ & $3.761093$ & $4.955740$ & $6.198974$ \\

$k=2$ & $0.500500$ & $1.503409$ & $2.511783$ & $3.528567$ & $4.556036$ & $5.595859$ \\

$k=3$ & $0.500033$ & $1.500229$ & $2.500813$ & $3.502032$ & $4.504125$ & $5.507320$ \\

$k=4$ & $0.500002$ & $1.500015$ & $2.500052$ & $3.500131$ & $4.500267$ & $5.500477$ \\

$k=5$ & $0.500000$ & $1.500000$ & $2.500003$ & $3.500008$ & $4.500017$ & $5.500030$ \\

$k=6$ & $0.500000$ & $1.500000$ & $2.500000$ & $3.500001$ & $4.500001$ & $5.500002$ \\

$k=10$ & $0.500000$ & $1.500000$ & $2.500000$ & $3.500000$ & $4.500000$ & $5.500000$ \\
\hline
\specialrule{.15em}{.15em}{.0em}
\end{tabular}
\end{center}
\end{table}
\begin{figure}[hbt]
\begin{center}
\includegraphics[scale=.5]{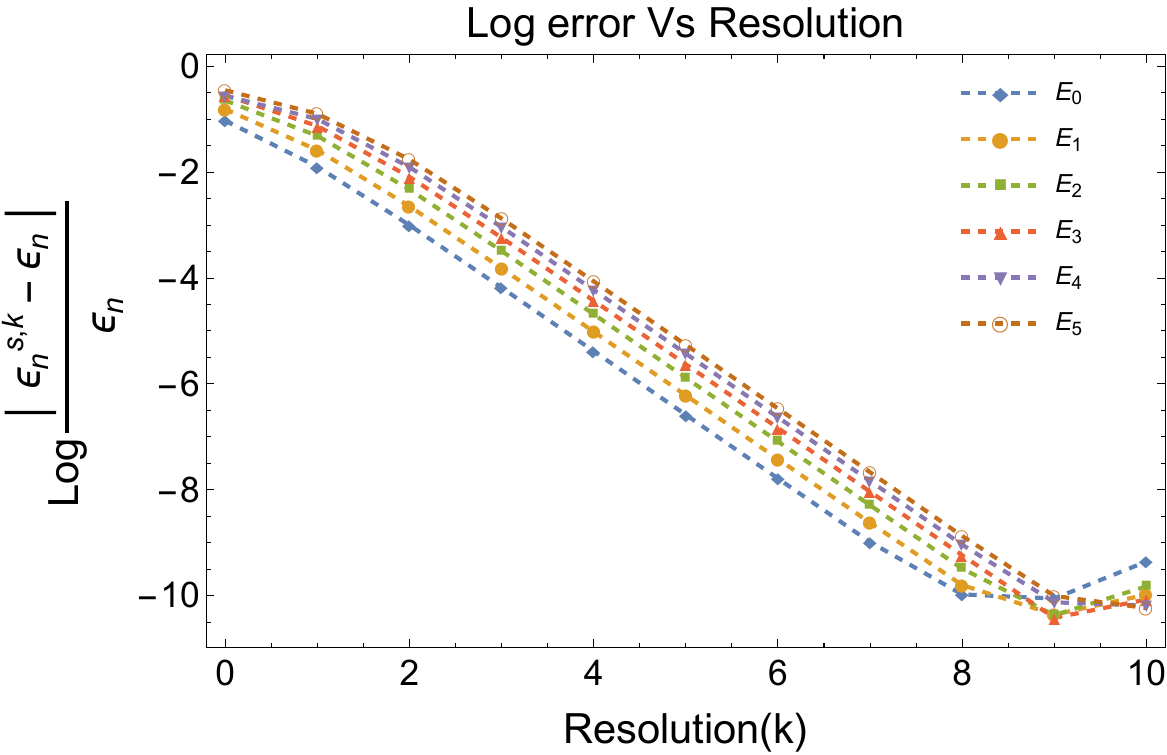}
\caption{\label{fig:Log_error_vs_resolution_Plot_SHO}Log error $\left(\log\frac{|\epsilon^{s,k}_n-\epsilon_n|}{\epsilon_n}\right)$ of different eigenstates with increasing resolution plot for the 1-D SHO potential.}
\end{center}
\end{figure}
Fig. \ref{fig:Log_error_vs_resolution_Plot_SHO}, depicts the log error versus the increasing resolution plot. From this plot, we see that the log error is progressively decreasing with increasing resolution. This occurs because higher resolution incorporates the effect of more short-distance degrees of freedom into the low-lying eigenvalues. Unlike the case of the infinite square well potential the log error curves of different eigenstates are not merging with each-other after a particular resolution, because the problem does not have a rigid boundary. If it had been possible to incorporate all the coarser resolution scales into the problem, we would have observed the same phenomenon as the ISWP.
\begin{figure}[hbt]
\begin{center}
\includegraphics[scale=.374]{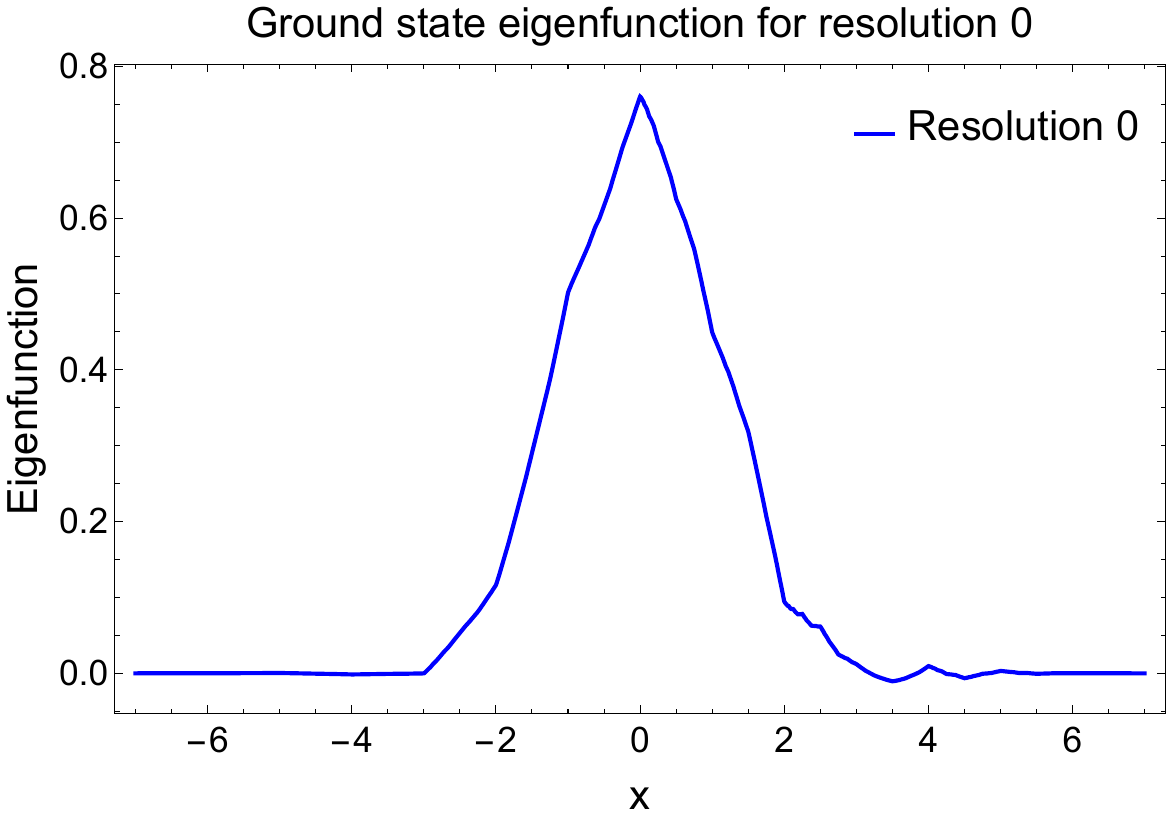}
\includegraphics[scale=.374]{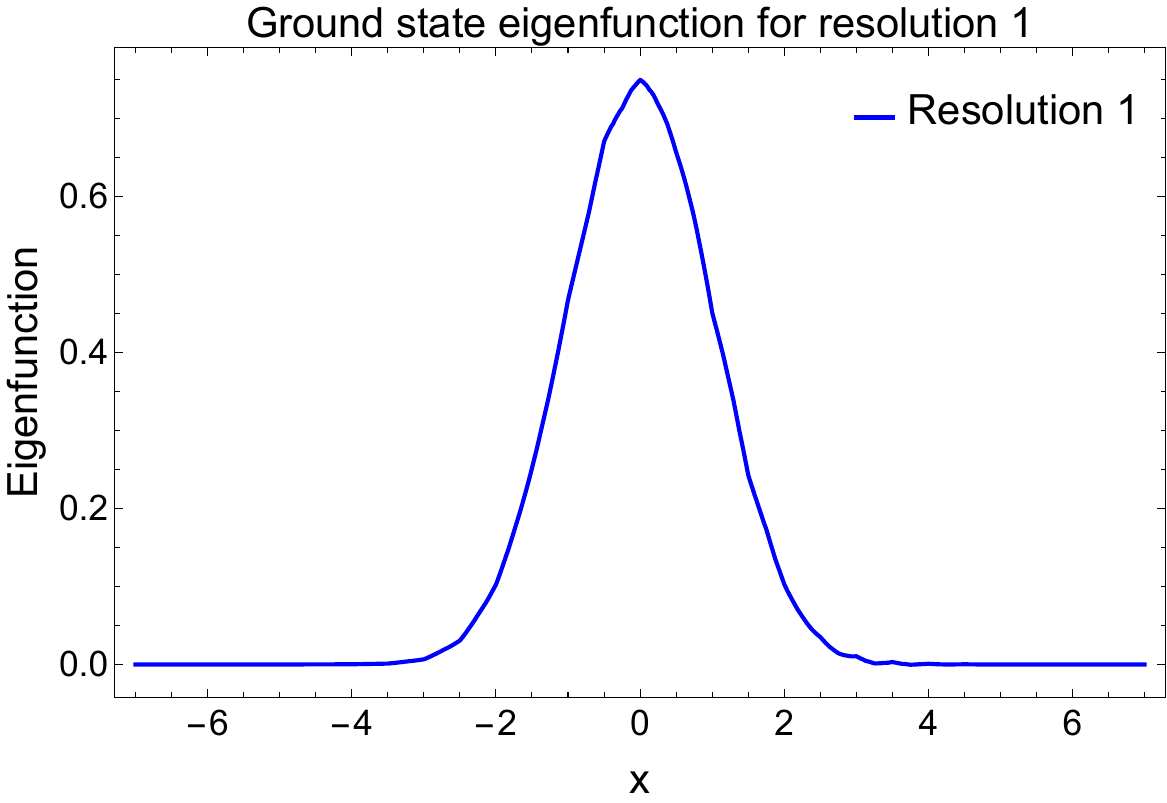}
\includegraphics[scale=.374]{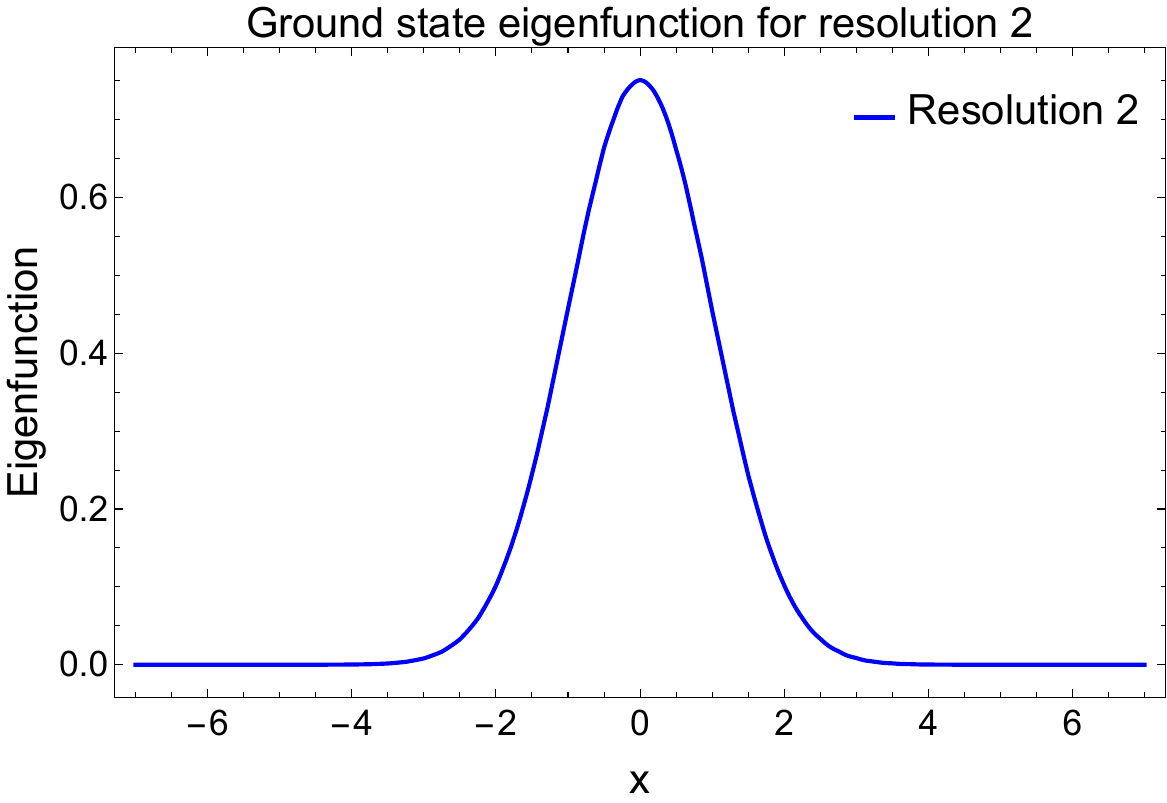}
\includegraphics[scale=.374]{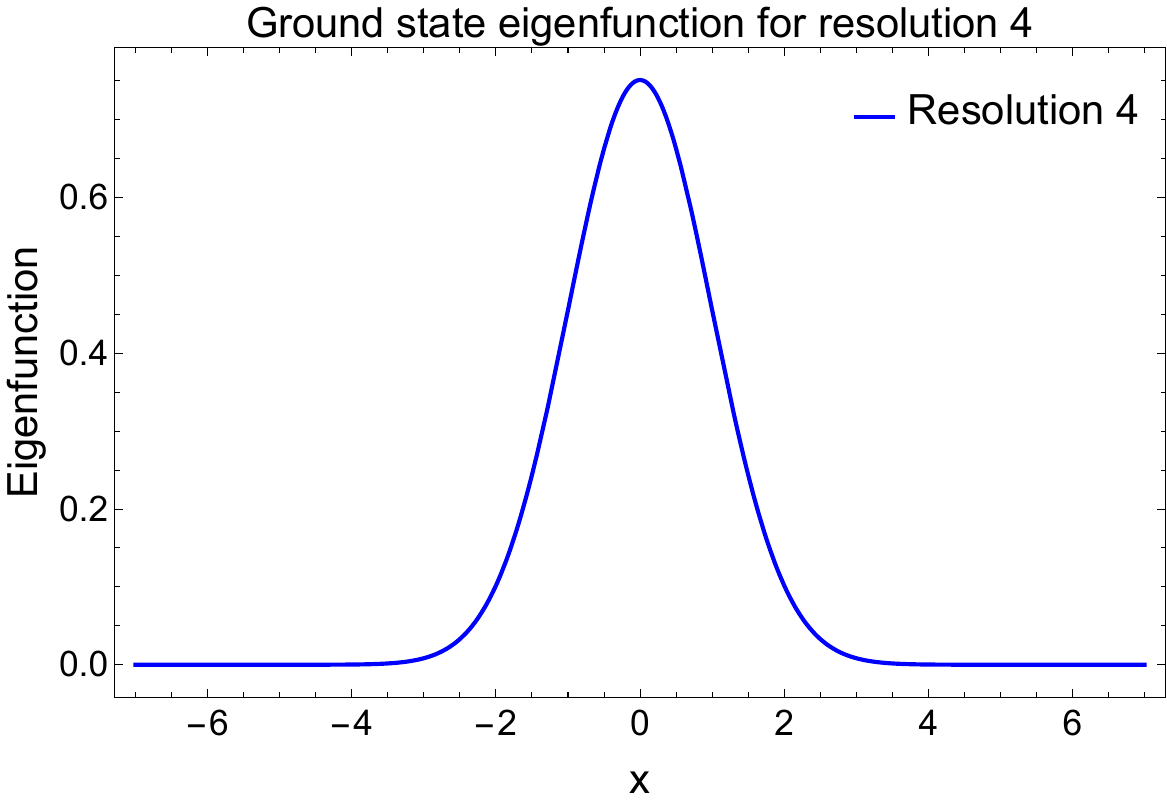}
\includegraphics[scale=.374]{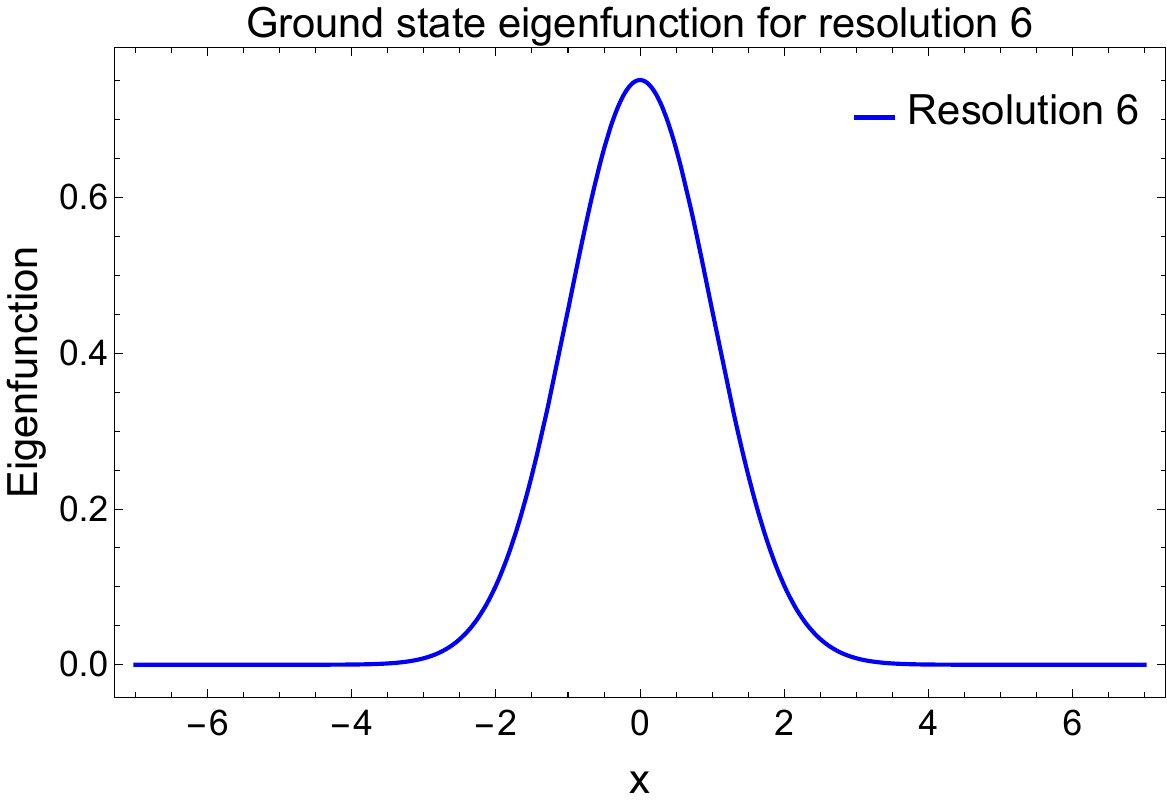}
\includegraphics[scale=.374]{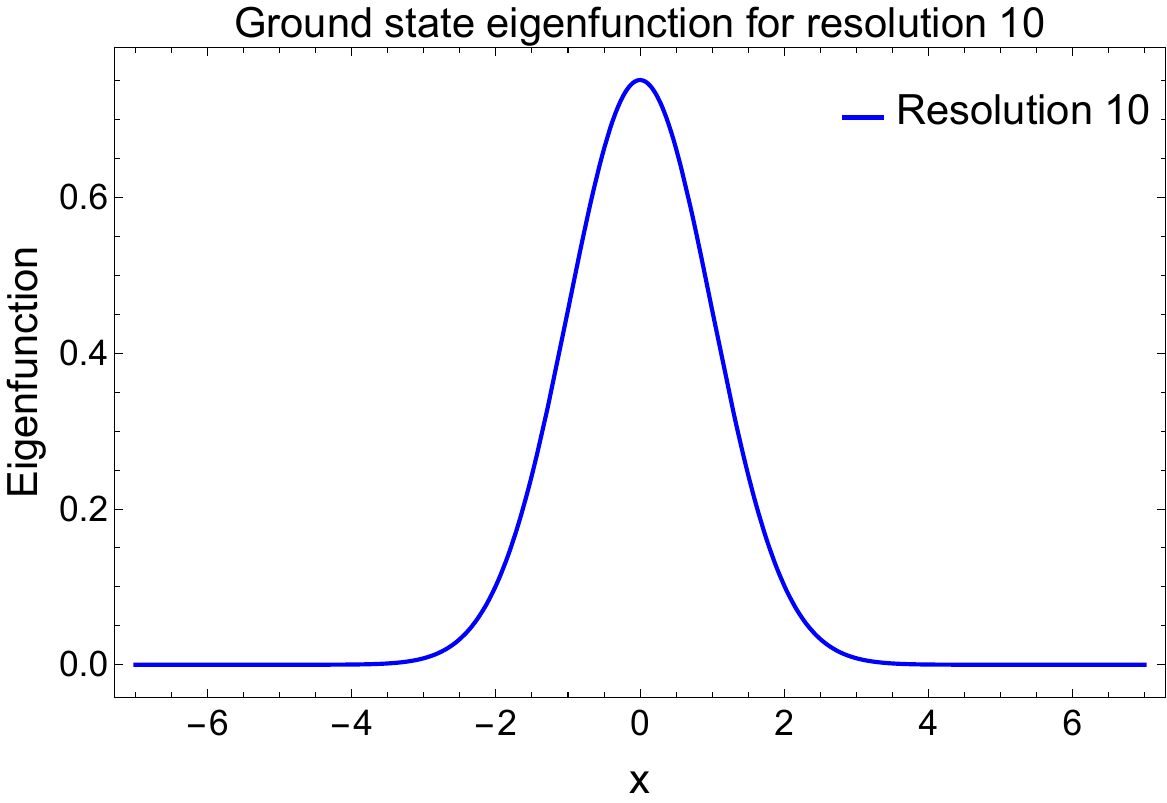}
\caption{\label{fig:Ground_state_for_dif_res_SHO}The ground state eigenfunction $\psi^{s,k}_0$ of 1-D SHO with increasing resolution.}
\end{center}
\end{figure}
\begin{table*}[hbt]
\begin{center}
\setlength{\tabcolsep}{0.21pc}
\catcode`?=\active \def?{\kern\digitwidth}
\caption{The exact values of $a$ and $b$ along with the approximate values of $a$ and $b$ with increasing resolution for 1-D SHO potential.}
\label{tab:actual_value_and_approximate_value_of_a_and_b_SHO}
\vspace{1mm}
\begin{tabular}{c c c c c c c c c c}
\specialrule{.15em}{.0em}{.15em}
\hline
 & Exact value & $k=0$ & $k=1$ & $k=2$ & $k=3$ & $k=4$ & $k=5$ & $k=6$ & $k=10$\\

\hline
$a$ & $0.751126$ & $0.720283$ & $0.745224$ & $0.750594$ & $0.751089$ & $0.751123$ & $0.751125$ & $0.751126$ & $0.751126$\\

$b$ & $0.5$ & $0.420252$ & $0.484000$ & $0.498558$ & $0.499902$ & $0.499994$ & $0.500000$ & $0.500000$ & $0.500000$\\
\hline
\specialrule{.15em}{.15em}{.0em}
\end{tabular}
\end{center}
\end{table*}

In Fig. \ref{fig:Ground_state_for_dif_res_SHO}, we observe that, similar to the case of the ISWP, the ground state eigenvalues of the SHOP approach the form of the exact ground state wave function. We used the built-in function \textit{FindFit[]} in \textit{Mathematica} to fit the curves with the model $ae^{-bx^2}$. This model is inspired by the fact that the exact ground state wave function is already known and has the form $\left(\frac{1}{\pi}\right)^{\frac{1}{4}}e^{-\frac{x^2}{2}}$. Table . \ref{tab:actual_value_and_approximate_value_of_a_and_b_SHO} presents the exact and the approximate values of the constants $a$ and $b$ for increasing resolution. It is observed that the values of $a$ and $b$ approach the exact values of $a=\left(\frac{1}{\pi}\right)^{\frac{1}{4}}=0.751126$ and $b=\frac{1}{2}=0.5$ as the resolution increases.

\begin{figure}[hbt]
\begin{center}
\includegraphics[scale=.40]{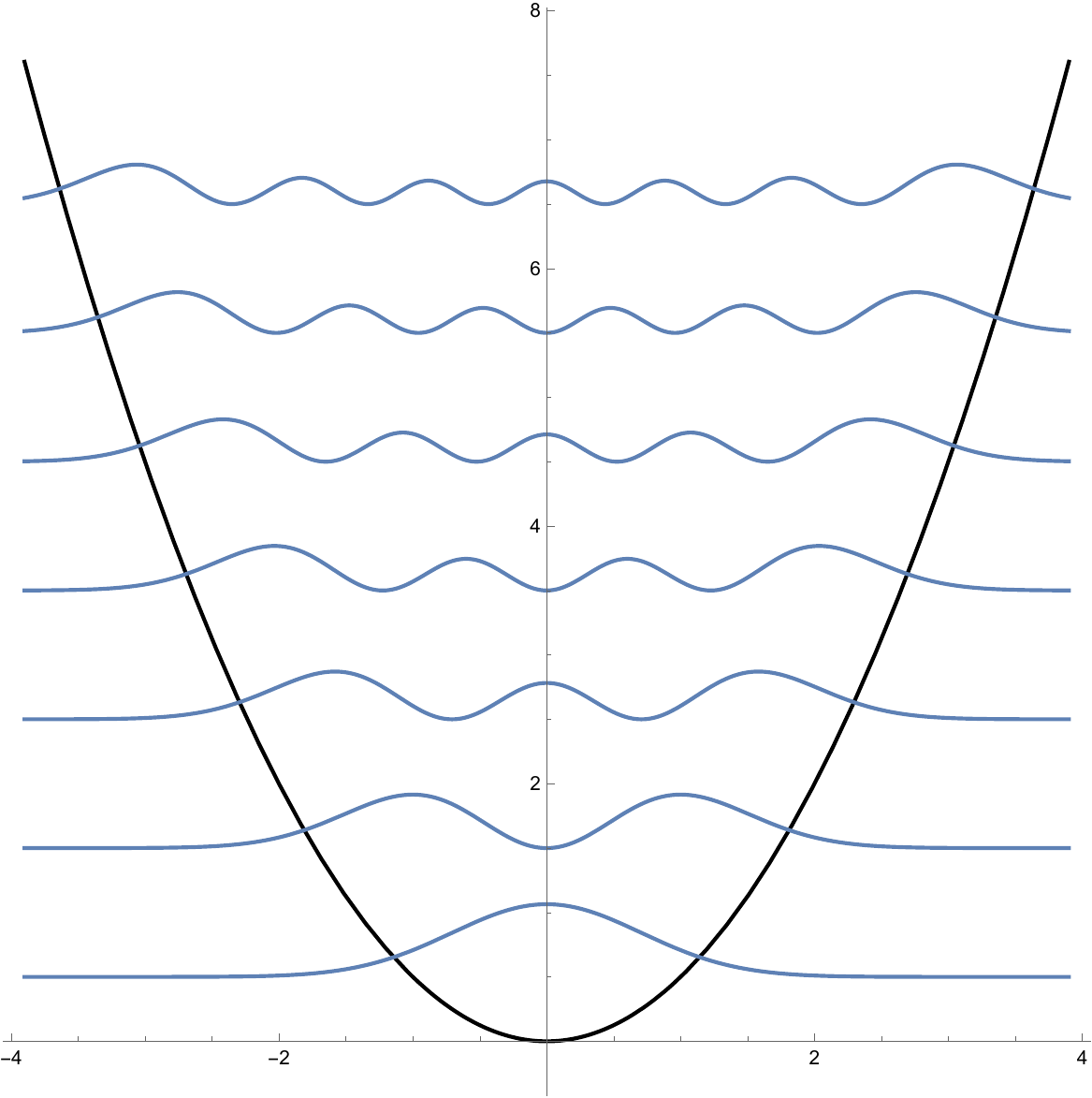}
\caption{\label{fig:Probability_plot_SHO}The probability distribution plot of the SHO potential obtained using wavelet-based formalism at resolution $7$.}
\end{center}
\end{figure}

The results demonstrated here show that the formalism developed here can produce adequately precise results for the low-lying eigenvalues for sufficiently low resolution when there is access to the entire space with the potential is distributed throughout. The probability of finding the particle for the low-lying eigenstates is the highest around the origin compared to the other parts of the space, as the extension of the potential around the origin is less at low energies than at high energies. Therefore, a sufficiently large volume truncation is adequate to obtain an accurate result for the low-lying eigenvalue with the appropriate resolution cutoff. The probability distribution of the lowest $7$ eigenstates of the SHOP obtained using wavelet-based formalism at resolution $7$ is presented in Fig. \ref{fig:Probability_plot_SHO}.
\section{The Dirac-delta function potential}
A particle moving in the presence of an attractive one-dimensional (1-D) Dirac-delta function (DDF) potential is given by the Hamiltonian \cite{griffiths2017introduction,DDF_Balakrishnan_2003},
\begin{eqnarray}
H=-\frac{\hbar^2}{2m}\frac{\partial^2}{\partial y^2} -\alpha \delta(y),
\end{eqnarray}
where, $\delta(y)$ represents the standard one dimensional Dirac-delta function, defined as:
\begin{eqnarray}
\delta(y)=
\begin{cases}
 0\quad &\text{if} \quad y\neq 0 \\
 \infty  \quad &\text{if} \quad y=0
\end{cases}
,\quad \text{with} \quad \int_{-\infty}^{\infty}\delta(y)dy=1.
\end{eqnarray}
The energy eigenvalue equation corresponding to this potential is expressed as,
\begin{eqnarray}
\label{eq:schrodinger_equation_DDF}
-\frac{\hbar^2}{2m}\frac{\partial^2 \psi(y)}{\partial y^2}-\alpha \delta(y)\psi(y)=E\psi(y).
\end{eqnarray}

The introduction of the Dirac delta function potential dates back quite far, with its first appearance in literature credited to Fourier in 1822 \cite{fourier1822théorie}. The first application of the delta function directly to physics \cite{1968max} originated with Helmholtz and Kirchhoff \cite{Kirchhoff2006} around 1882. A decade later, Heaviside \cite{doi:10.1098/rspl.1892.0093} made an explicit connection between the Fourier series and the delta function, further expanding its application to physical problems \cite{doi:10.1098/rsta.2018.0229}. Kirchhoff and Heaviside were likely the first to provide mathematical description of the delta function \cite{https://doi.org/10.1002/zamm.19840640219}. Dirac reintroduced the delta function relatively late, formally reintroducing it in 1926 \cite{dirac1927physical} and later incorporating it into his quantum mechanics text \cite{dirac1981principles}.Dirac delta function potentials have found application in modeling atomic and molecular systems \cite{10.1063/1.1743167,10.1063/1.1740472,10.1063/1.1743168}, such as atomic lattices \cite{doi:10.1098/rspa.1931.0019} and the absorption spectra of organic dyes \cite{https://doi.org/10.1002/hlca.19480310602,10.1063/1.1747011,10.1063/1.1747143}. Moreover, the repulsive Dirac delta function potential can be employed to represent defects in quantum wells. The utilization of Dirac-delta function (DDF) potentials as solvable models for potential barriers or wells has a longstanding history in quantum mechanics, dating back to at least Kronig and Penney \cite{doi:10.1098/rspa.1931.0019}. In their initial work on the \textit{"Quantum mechanics of electrons in crystal lattices"}, they employed a \textit{"series of equidistant rectangular barriers"} and noted that,\newline
\textit{"When the breadth $b$ of these barriers is made infinitely small and their height $V_0$ infinitely large, the results become particularly simple, the influence of the barriers depending then only on the product $bV_0$"}.\newline
Tamm \cite{1932ZPhy...76..849T} utilized such models to predict the existence of surface states, and it took approximately 60 years for these predictions to be experimentally confirmed \cite{PhysRevLett.64.2555}. Applying the standard notation used today, Saxon and Hutner derived \textit{"wave functions and energy levels for monoatomic and diatomic Kronig-Penney models... in which atomic fields are represented by Dirac-delta functions..."} implementing the boundary conditions appropriate for this singular potential. In their renowned work \textit{"Methods of Theoretical Physics"} \cite{morse1954methods}, Morse and Feshbach explicitly examine an attractive Dirac-delta function potential ('potential well'), noting its utility in the study of nuclear forces and discuss its single bound state and explore its scattering solutions. Frost explored both single and multiple attractive DDF potentials as models for "hydrogen-like atoms" \cite{10.1063/1.1743167}, the hydrogen molecule-ion \cite{10.1063/1.1740472}, and more complex systems \cite{frost1956delta}.

Subsequently, the DDF potential has proven to be an essential solvable model for short-range interactions across various applications. It has served as an exemplary case study for exploring novel physics concepts or applying new methods, with the advantage that the resulting mathematical manipulations are often more manageable compared to more realistic systems. Therefore, we have opted for this potential to evaluate the wavelet-based framework, considering it as a second example.

The coupling parameter $\alpha$ in Eq. (\ref{eq:schrodinger_equation_DDF}) is a dimensionful parameter that sets the scale for the size and energy of the bound states. We employ this parameter to introduce a new dimensionless variable, $x=\frac{m \alpha y}{\hbar^2}$, in terms of which Eq. (\ref{eq:schrodinger_equation_DDF}) is reformulated as,
\begin{eqnarray}
\label{eq:schrodinger_equation_DDF_dimensionless}
-\frac{1}{2}\frac{\partial^2 \psi(x)}{\partial x^2}-\delta(x) \psi(x)=\epsilon \psi(x),
\end{eqnarray}
where, $\epsilon=\frac{E\hbar^2}{m \alpha^2}$.

Equation (\ref{eq:schrodinger_equation_DDF_dimensionless}) can be exactly solved, resulting in the presence of a single bound state with the eigenfunction $\psi_0(x)$ and the eigenvalue $\epsilon_0$, given by,
\begin{eqnarray}
\psi_0(x)=e^{-|x|} \quad ;\quad \epsilon_0=\frac{1}{2}.
\end{eqnarray}

Now, we will address this problem utilizing wavelet-based methods. Equation (\ref{eq:schrodinger_equation_DDF_dimensionless}) can be investigated using the wavelet-based approach by approximating the Hilbert space of the problem to the resolution space $\mathscr{H}^k$. Consequently, the energy eigenvalue equation linked to the Dirac-delta function potential will assume the following form.
\begin{eqnarray}
\label{eq:schrodinger_equation_DDF_matrix}
\sum_n \left(\frac{1}{2} T^k_{ss,mn}-V^k_{\delta ss,mn} \right)\psi^{s,k}_n=\epsilon \psi^{s,k}_m,
\end{eqnarray}
where, the kinetic energy matrix element $T^k_{ss,mn}$ is provided by Eq. (\ref{eq:schrodinger_equation_DDF_matrix}), and the potential energy matrix element $V^k_{\delta ss,mn}$ is expressed as,
\begin{eqnarray}
\label{eq:overlap_integral_delta_function_and_scaling_function_1}
V^k_{\delta ss,mn}=\int \delta(x) s^k_m(x) s^k_n(x)dx.
\end{eqnarray}
$V^k_{\delta ss,mn}$ will be non-zero only if $s^k_m(x)$ and $s^k_n(x)$ pass through the origin, as the $\delta$-function potential only has a non-zero value at the origin. Specifically, $V^k_{\delta ss,mn}$ is determined by taking the product of the values of the overlapping scaling functions at the origin,
\begin{eqnarray}
V^k_{\delta ss,mn}=s^k_m(0)s^k_n(0).
\end{eqnarray}
The values of these scaling functions can be determined by solving the scaling equation given in Eq. (\ref{eq:scaling_equation_1}), as outlined in the Appendix \ref{appen:Computation_of_scaling_and_wavelet_function_at_integer_points_and_dyadic_rationals}.
 An alternative approach is to establish the linear equations in the variables $V^k_{\delta ss,mn}$, as detailed in Appendix \ref{appen:The_Dirac_delta_function_potential}. Once the matrix elements of the Hamiltonian are determined, the energy eigenvalue equation of one-dimensional DDF, Eq. (\ref{eq:schrodinger_equation_DDF_matrix}), can be solved to obtain the eigenvalues and the corresponding eigenfunctions.
 
\begin{figure}[hbt]
\begin{center}
\includegraphics[scale=.5]{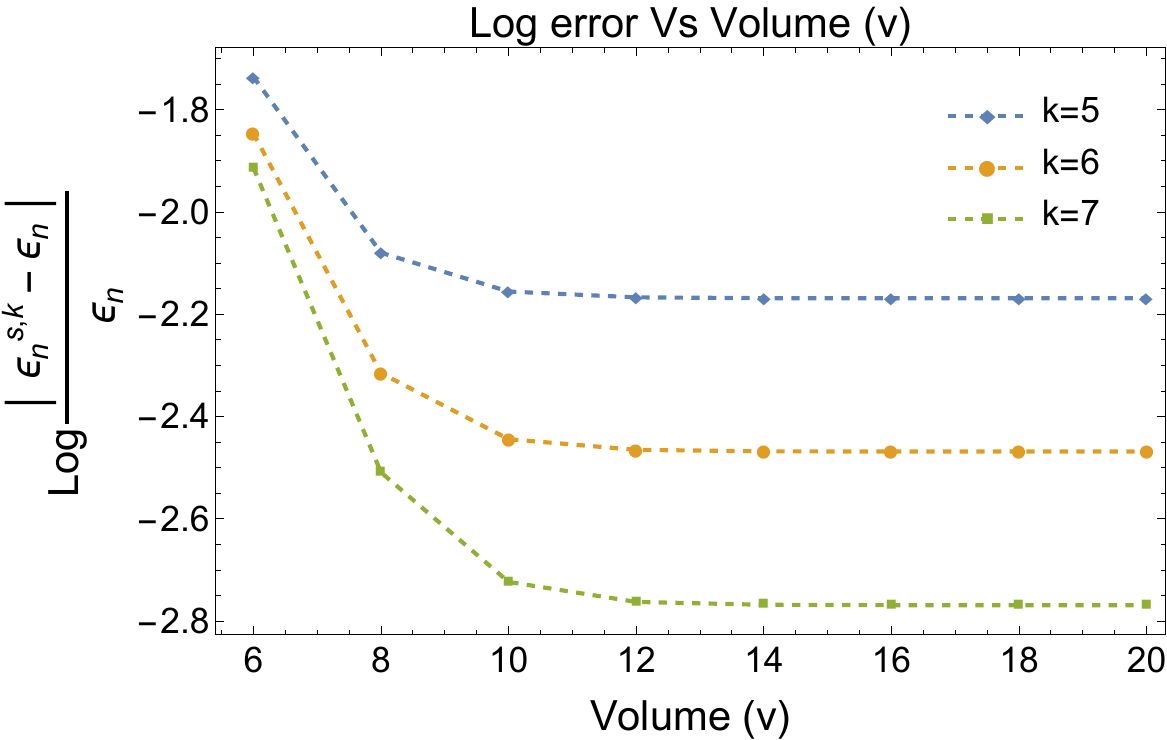}
\caption{\label{fig:Log_error_vs_volume_Plot_DDFP}Log error $\left(\log\frac{|\epsilon^{s,k}_n-\epsilon_n|}{\epsilon_n}\right)$ of different eigenstates with increasing volume plot for the DDF potential.}
\end{center}
\end{figure}

We now present the results of solving this problem using the approach outlined in Sec. \ref{sec:formalism}. The log error of the ground state versus the increasing volume plot for the fixed resolution is given in Fig. \ref{fig:Log_error_vs_volume_Plot_DDFP}. The plot illustrates that the logarithmic error will not undergo significant changes beyond a certain volume, $V=14$ $\left(-7\leq x\leq 7\right)$. This is because matrix elements will have a negligible contribution from scaling functions whose support size exceeds $14$. As we increase the support size of the scaling function, the values of the scaling function at each point in $x$ decrease due to the normalization condition. Consequently, integrating with a potential with such scaling function concentrated at the origin yields lower values. Thus, to achieve the necessary accuracy for the ground state, a volume of $V=14$ would be adequate. However, we opt for a sufficiently large volume of $V=20$ $\left(-10\leq x\leq 10 \right)$ to ensure accurate eigenvalues. 

\begin{figure}[hbt]
\begin{center}
\includegraphics[scale=.5]{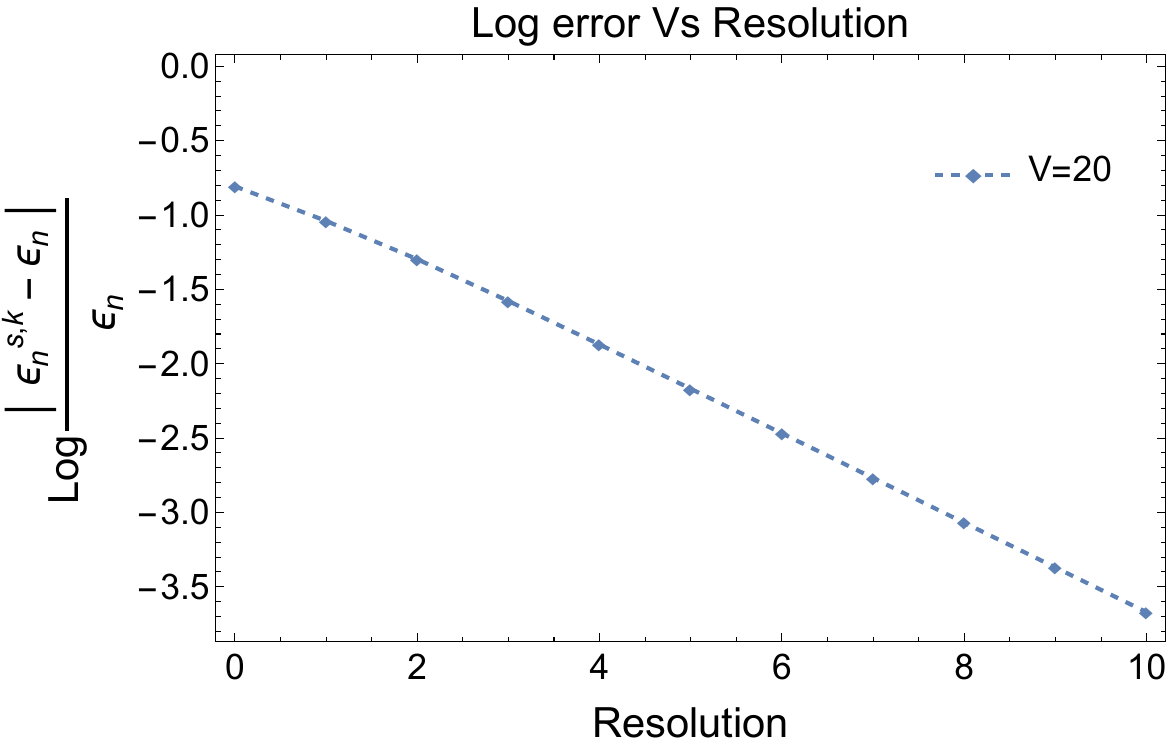}
\caption{\label{fig:Log_error_vs_resolution_Plot_DDFP}Log error $\left(\log\frac{|\epsilon^{s,k}_n-\epsilon_n|}{\epsilon_n}\right)$ of different eigenstates with increasing resolution plot for the 1-D DDF potential.}
\end{center}
\end{figure}

Table. \ref{tab:actual_value_and_approximate_value_of_E_0_1DDDP} presents the ground state energy for various resolutions in this problem, and the plot depicting the logarithmic error with increasing resolution is shown in Fig. \ref{fig:Log_error_vs_resolution_Plot_DDFP}. 
\begin{table}[ht]
\begin{center}
\setlength{\tabcolsep}{1.5 pc}
\catcode`?=\active \def?{\kern\digitwidth}
\caption{The actual values of ground state energy ($\epsilon_0$) along with the approximate values of $\epsilon^{s,k}_0$ with increasing resolution for one dimensional DDF potential.}
\label{tab:actual_value_and_approximate_value_of_E_0_1DDDP}
\vspace{1mm}
\begin{tabular}{c | c}
\specialrule{.15em}{.0em}{.15em}
\hline
$E_n$ & $E_0$ \\
\hline
Exact value & $-0.5$ \\

$k=0$ & $-0.422032$ \\

$k=1$ & $-0.454335$ \\

$k=2$ & $-0.474778$ \\

$k=3$ & $-0.486791$ \\

$k=4$ & $-0.493269$ \\

$k=5$ & $-0.496608$ \\

$k=6$ & $-0.498299$ \\

$k=10$ & $-0.499893$ \\
\hline
\specialrule{.15em}{.15em}{.0em}
\end{tabular}
\end{center}
\end{table}
In Tab. \ref{tab:actual_value_and_approximate_value_of_E_0_1DDDP}, we can see that with increasing resolution, the ground state eigenvalue of the volume-truncated Hamiltonian matrix approaches the actual ground state eigenvalue of the one-dimensional DDF potential. In the Fig. \ref{fig:Log_error_vs_resolution_Plot_DDFP}, it is observable that the logarithmic error of the ground state eigenvalue decreases with increasing resolution. This phenomenon occurs because, similar to the previous case of the infinite square well potential, increasing resolution involves incorporating more contributions from short-distance degrees of freedoms. 

\begin{figure}[hbt]
\begin{center}
\includegraphics[scale=.374]{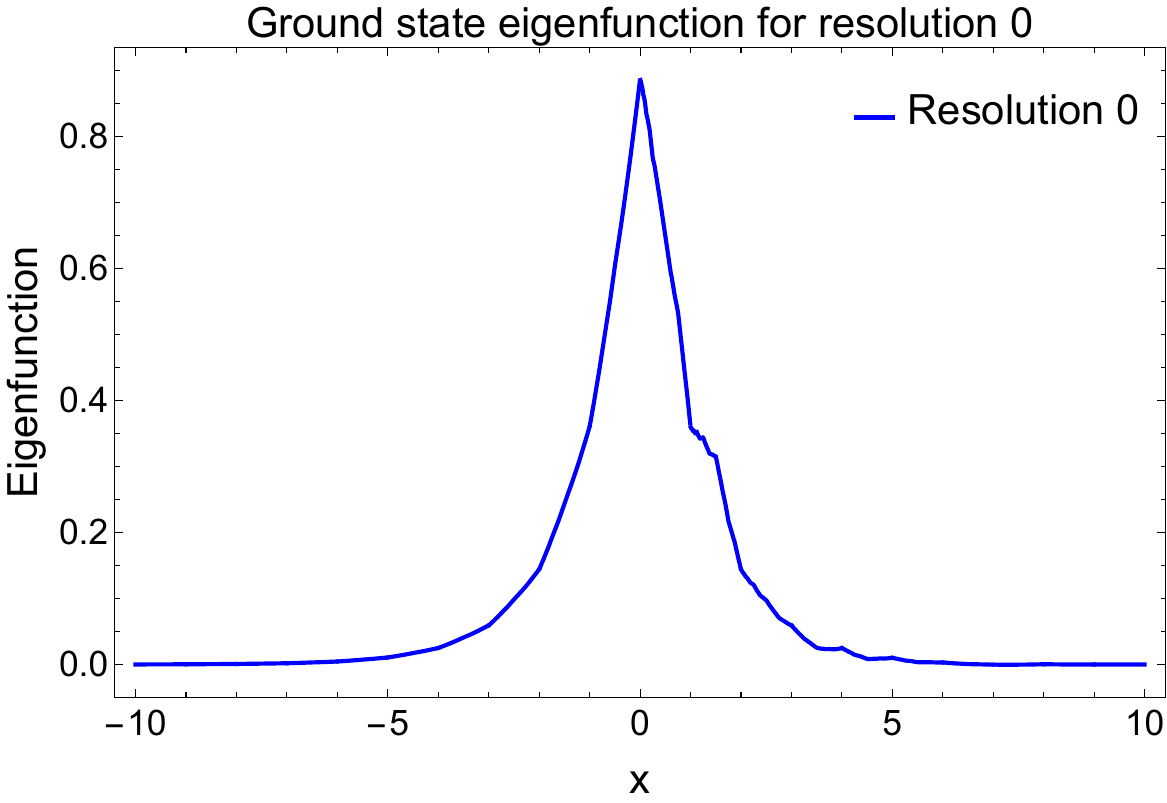}
\includegraphics[scale=.374]{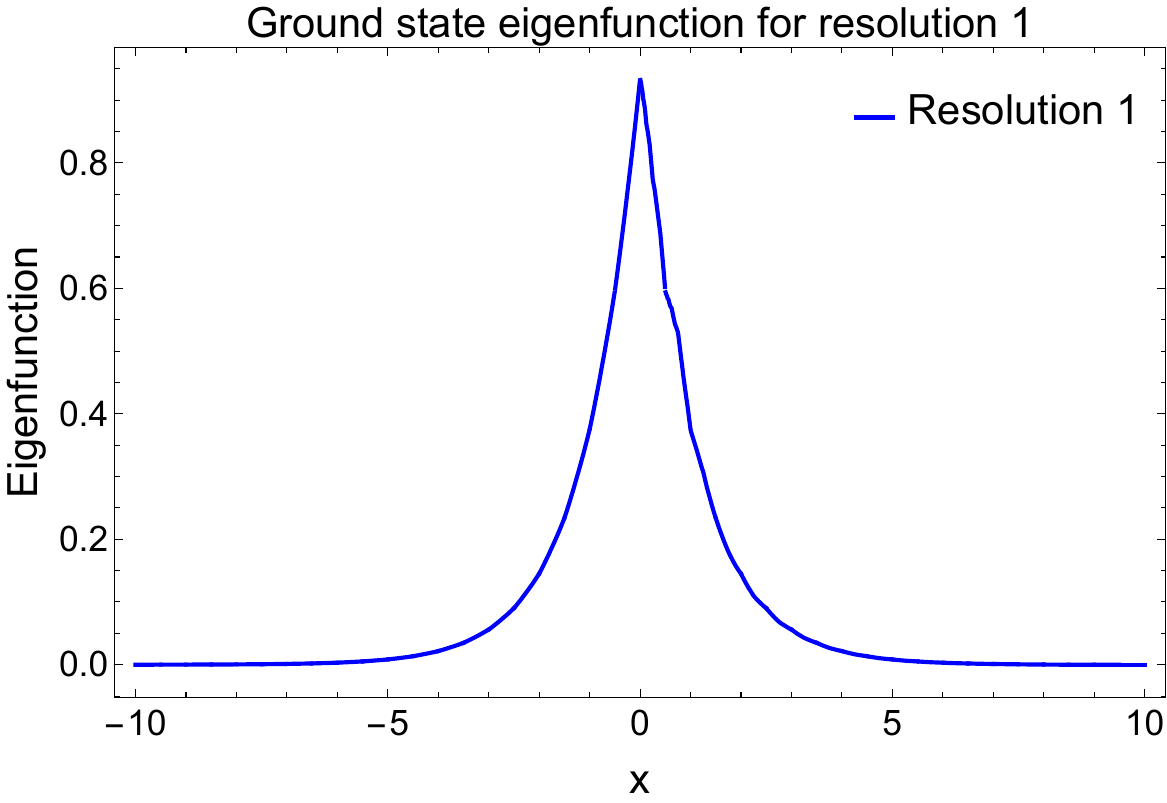}
\includegraphics[scale=.374]{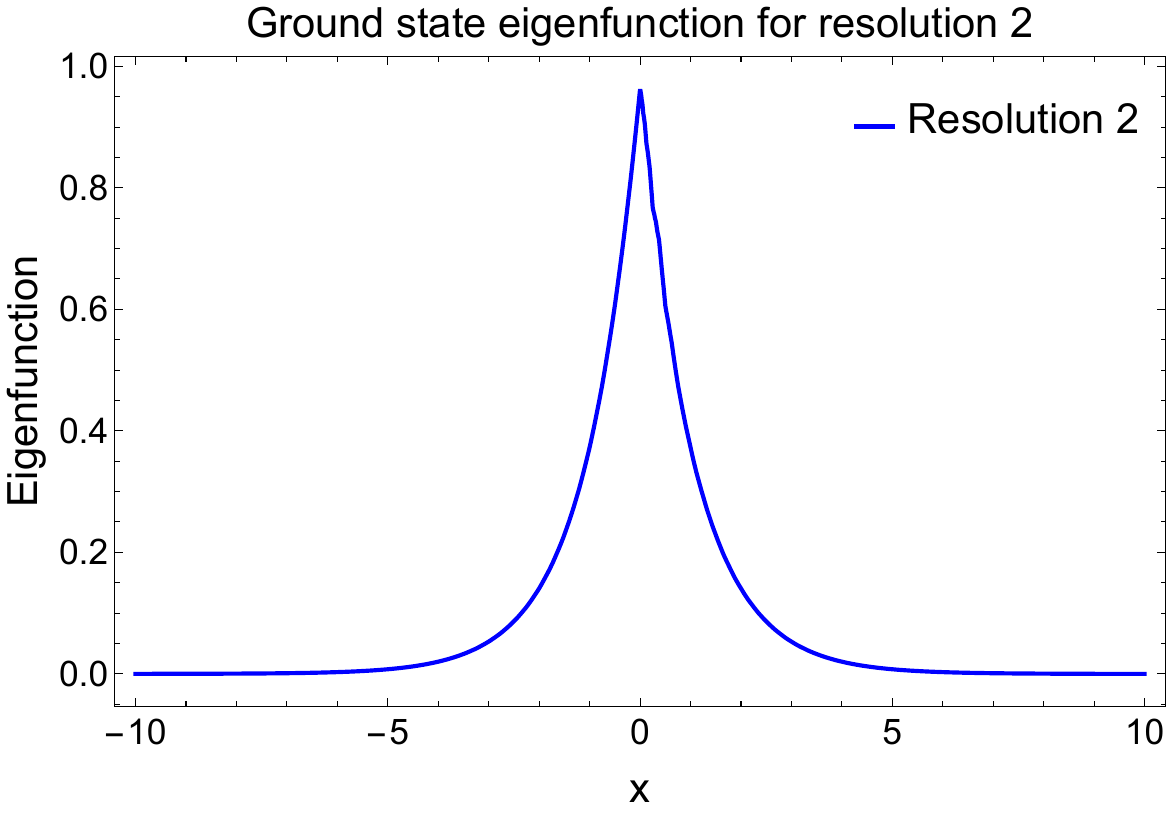}
\includegraphics[scale=.374]{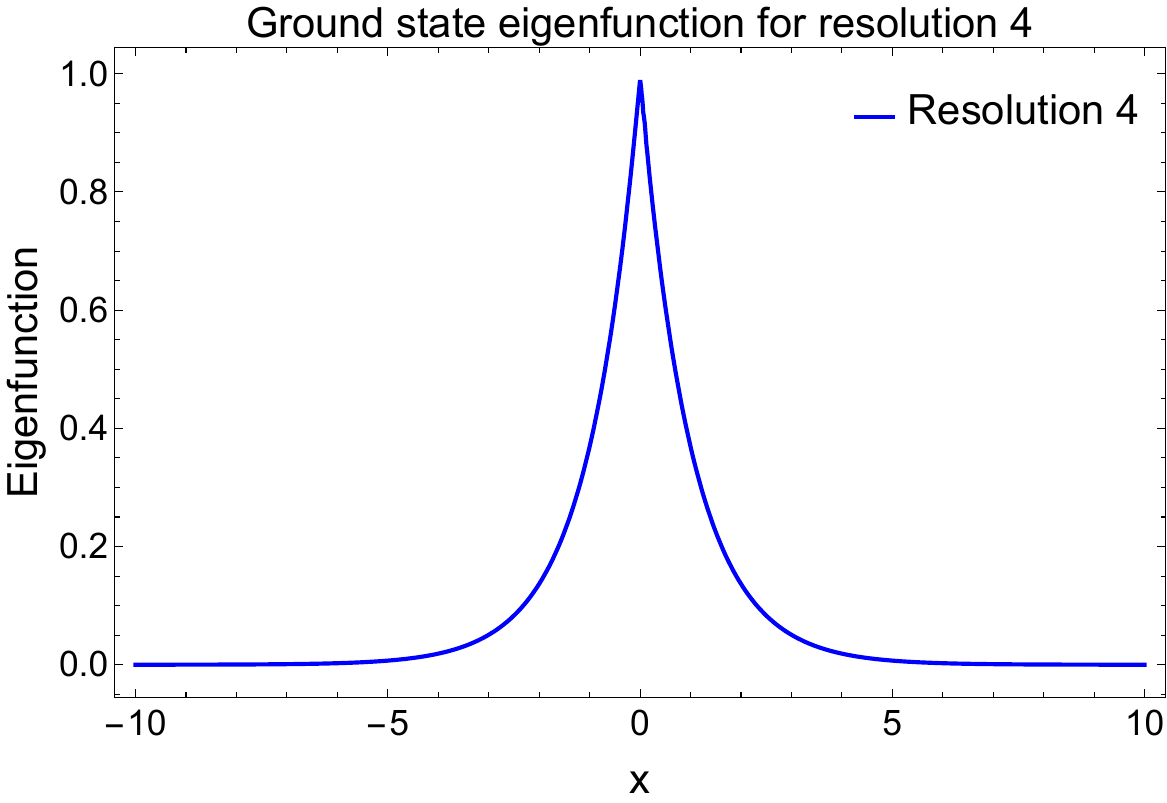}
\includegraphics[scale=.374]{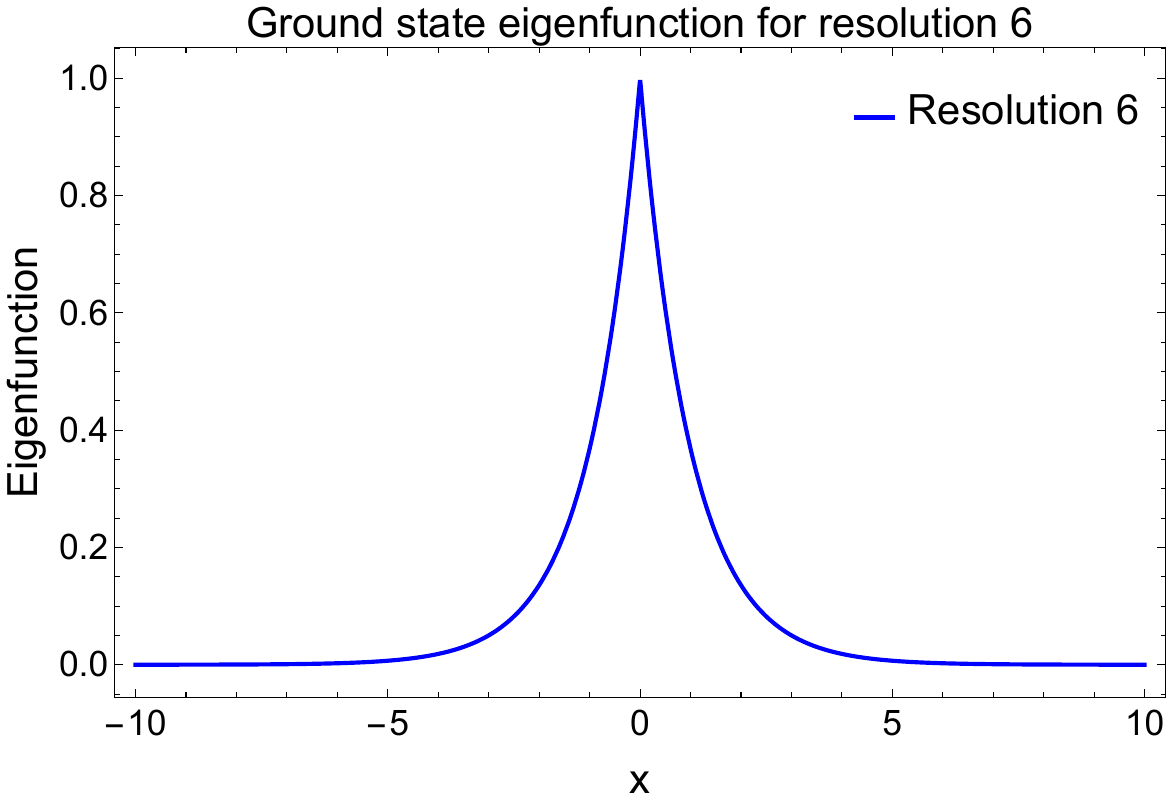}
\includegraphics[scale=.374]{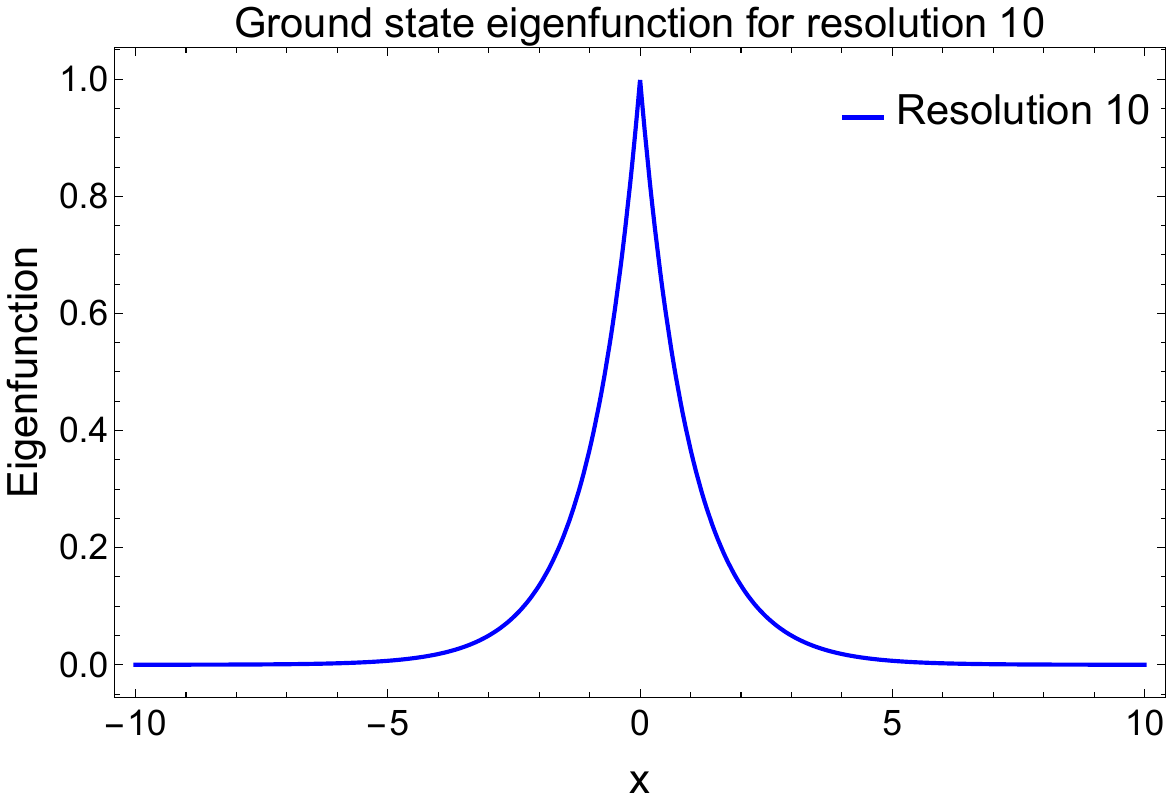}
\caption{\label{fig:Ground_state_for_dif_res_DDFP}The ground state eigenfunction $\psi^{s,k}_0$ of one dimensional DDF with increasing resolution.}
\end{center}
\end{figure}
\begin{table*}[hbt]
\begin{center}
\setlength{\tabcolsep}{0.21pc}
\catcode`?=\active \def?{\kern\digitwidth}
\caption{The exact values of $a$ and $b$ along with the approximate values of $a$ and $b$ with increasing resolution for 1-D DDF potential.}
\label{tab:actual_value_and_approximate_value_of_a_and_b_1DDDP}
\vspace{1mm}
\begin{tabular}{c c c c c c c c c c}
\specialrule{.15em}{.0em}{.15em}
\hline
 & Exact value & $k=0$ & $k=1$ & $k=2$ & $k=3$ & $k=4$ & $k=5$ & $k=6$ & $k=10$\\

\hline
$a$ & $1$ & $0.937583$ & $0.966948$ & $0.984615$ & $0.992851$ & $0.996542$ & $0.998289$ & $0.999143$ & $0.999945$\\

$b$ & $1$ & $0.881903$ & $0.935754$ & $0.969637$ & $0.985783$ & $0.993102$ & $0.996581$ & $0.998291$ & $0.999891$\\
\hline
\specialrule{.15em}{.15em}{.0em}
\end{tabular}
\end{center}
\end{table*}

From Fig. \ref{fig:Ground_state_for_dif_res_DDFP}, we can see that, akin to the infinite square well potential, the resolution-truncated ground state eigenfunction of the DDF potential progressively approaches the form of the actual ground state eigenfunction as the resolution increases. To quantify the accuracy of generating the eigenfunction through this method, we fit these curves with the model $a e^{-b|x|}$. This selection is motivated by the fact that the actual ground state eigenfunction of the one-dimensional Dirac-delta function is given by $e^{-|x|}$. Therefore, the exact values of both $a$ and $b$ are equal to $1$. It can be observed in Table \ref{tab:actual_value_and_approximate_value_of_a_and_b_1DDDP} that the values of $a$ and $b$ approach the exact values as the resolution increases. 

This analysis demonstrates that even with access to the entire space, we can truncate the volume of the space based on the requirements of the problem. For instance, in this case, due to the symmetry of the potential, we can choose a box that is symmetrically spread around the origin. We can also observe that, similar to the previous problem, for a sufficiently low resolution, this method can produce adequately accurate eigenvalues for the ground state. Additionally,  we can anticipate the actual eigenvalue in the scenario where we include the entire space with infinite resolution.

In this problem, the potential is concentrated at the origin. Consequently, truncating the volume at a large length does not make a significant difference. The next question is whether volume truncation will be effective for potentials where the spread of the potential is infinite. To investigate this, we solve the problem of a particle moving in the triangular potential, where the extent of the potential is infinity.
\section{The triangular potential}
A particle moving in the presence of an attractive one-dimensional (1-D) triangular potential (TP) potential is given by the Hamiltonian \cite{griffiths2017introduction,sakurai2011modern,Suzuki2020}, 
\begin{eqnarray}
V(y)=
\begin{cases}
 \kappa y,\quad &  y\geq 0 \\
 -\kappa y, \quad & y\leq 0
\end{cases}
\end{eqnarray}
The energy eigenvalue equation governing the motion of a particle within this potential is,
\begin{eqnarray}
\label{eq:schrodinger_equation_triangular_potential}
\left(-\frac{\hbar^2}{2m}\frac{\partial^2}{\partial y^2}+\kappa |y|\right)\psi(y)=E\psi(y).
\end{eqnarray}

The triangular potential is a theoretical construct used in physics to model certain physical systems, particularly in the context of quantum mechanics. It serves as an idealized representation of a potential energy landscape. The triangular potential is often used to model barriers or wells in quantum mechanics. For instance, it can depict a quantum particle subjected to the influence of a constant electric field, with the potential having the form of a triangular potential \cite{Jain2019}. Another application of the triangular potential is in semiconductor physics, particularly when high electric fields are present across the gate oxide. In such cases, the barrier between the gate metal and the FET (Field-Effect Transistor) channel is no longer a simple rectangular barrier as often depicted in textbooks; instead, it takes on a triangular nature \cite{fowler1928electron}. In the context of ultracold atomic physics, especially noteworthy is the proposition \cite{PhysRevLett.95.127207,PhysRevLett.93.030601,PhysRevLett.95.127205,PhysRevLett.97.190406,PhysRevB.75.174516} that cold atoms loaded into triangular or hexagonal optical lattices manifest exceptionally diverse new phases. These phases include phenomena such as supersolidity \cite{PhysRevLett.95.127207,PhysRevLett.95.127205}, quantum stripe ordered states \cite{PhysRevLett.97.190406}, exotic superconducting states \cite{PhysRevB.75.174516}, or physics reminiscent of graphene \cite{PhysRevA.77.011802,HADDAD20091413}. In semiconductor physics, the triangular potential can be used to approximate the potential energy experienced by an electron in the presence of impurities or defects \cite{wang200910.1007}.

To make this equation scale-invariant, we perform a variable change $y \rightarrow x = \sqrt[3]{\frac{m\kappa}{\hbar^2}}y$, and Eq. (\ref{eq:schrodinger_equation_triangular_potential}) will then assume the following form: 
\begin{eqnarray}
\label{eq:eigen_value_equation_TP}
\left(-\frac{1}{2}\frac{\partial^2}{\partial x^2}+ |x| \right)\psi(x)=\epsilon \psi(x),
\end{eqnarray}
where, $\epsilon=\sqrt[3]{\frac{m}{(\kappa \hbar)^2}}E$.\newline
The general solution to Eq. (\ref{eq:eigen_value_equation_TP}) is expressed as a linear combination of the Airy function $Ai[\sqrt[3]{2}(x-\epsilon)]$ and $Bi[\sqrt[3]{2}(x-\epsilon)]$,
\begin{eqnarray}
\psi(x)=C_1 Ai[\sqrt[3]{2}(x-\epsilon)]+C_2 Bi[\sqrt[3]{2}(x-\epsilon)].
\end{eqnarray}
However, $Bi(x)$ diverges as $x\rightarrow \infty$. This violets the requirement that $\psi(x)\rightarrow 0$ as $x\rightarrow \infty$, which does not make any physical sense. We therefore restrict our solution to the form,
\begin{eqnarray}
\psi_n(x)=C_n Ai[\sqrt[3]{2}(x-\epsilon_n)].
\end{eqnarray}
$C_n$ is the normalization constant for $x>0$. $\epsilon_n$ is the eigenvalue of that particular state.

At this point, we take into account the parity of our solutions. Even and odd eigenstates for $x > 0$ can be extended to provide solutions for $x < 0$ by,
\begin{eqnarray}
\psi^{even}(x)=\psi^{even}(-x),\quad\quad\quad \psi^{odd}(-x)=-\psi^{odd}(x).
\end{eqnarray}
Utilizing these properties, we can deduce the boundary conditions,
\begin{eqnarray}
\frac{\partial \psi^{even}_k(x)}{\partial x}\bigg\vert_{x\rightarrow 0}= Ai'(-\sqrt[3]{2}\epsilon_k)=0,\quad\quad \quad \psi^{odd}_k(0)=Ai(-\sqrt[3]{2}\epsilon_k)=0.
\end{eqnarray}

The exact eigenfunction for the 1-D TP is given by,
\begin{eqnarray}
\label{eq:eigenfunction_of_TP}
\psi_n(x)=C_n Ai\left[\sqrt[3]{2}\left(|x|-\epsilon_n\right)\right].
\end{eqnarray}
These two equations will give the eigenvalues for even and odd parity respectively. The normalization constant $C_n$ can be evaluated numerically as,
\begin{eqnarray}
C_n=\left(\int_{-\infty}^{\infty}Ai^2\left[\sqrt[3]{2}(|x|-\epsilon_n)\right]\right)^{-1/2}.
\end{eqnarray}
The exact  eigenvalues of first six states has been tabulated in Table \ref{tab:actual_and_approximate_eigenvalues_of_the_TP}. Table \ref{tab:values_of_normalization_constant_cn_TP} provides the first six values of the normalization constants $C_n$.
\begin{table}[ht]
\begin{center}
\setlength{\tabcolsep}{1.2 pc}
\catcode`?=\active \def?{\kern\digitwidth}
\caption{The normalization constants, denoted as $C_n$, are presented for the initial six states of the 1-D TP.}
\label{tab:values_of_normalization_constant_cn_TP}
\vspace{1mm}
\begin{tabular}{c c c c c c }
\specialrule{.15em}{.0em}{.15em}
\hline
 $C_0$ & $C_1$ & $C_2$ & $C_3$ & $C_4$ & $C_5$ \\
\hline
 $1.468004$ & $1.131900$ & $1.051006$ & $0.988282$ & $0.950344$ & $0.917356$\\

\hline
\specialrule{.15em}{.15em}{.0em}
\end{tabular}
\end{center}
\end{table}

Next, we will address the problem within the wavelet-based framework. In the truncated scaling function space $\mathscr{H}^k$, the energy eigenvalue equation includes a potential term, $V^k_{\Delta ss,mn}$, in addition to the kinetic energy term, $T^k_{ss,mn}$. The energy eigenvalue equation will assume the following form,
\begin{eqnarray}
\sum_{n}\left(\frac{1}{2} T^k_{ss,mn}+V^k_{\Delta ss,mn} \right)\psi^{s,k}_n=\epsilon \psi^{s,k}_n,
\end{eqnarray}
where, $T^k_{ss,mn}$, the kinetic energy term, can be computed using the method outlined in Sec. \ref{sec:formalism}. The potential energy term is given by,
\begin{eqnarray}
V^k_{\Delta ss,mn}&=&\int x s^k_m(x)s^k_n(x)dx \quad \text{for} \quad x\geq 0,\\
&=& -\int x s^k_m(x)s^k_n(x)dx \quad \text{for} \quad x\leq 0.
\end{eqnarray}
\begin{figure}[hbt]
\begin{center}
\includegraphics[scale=.5]{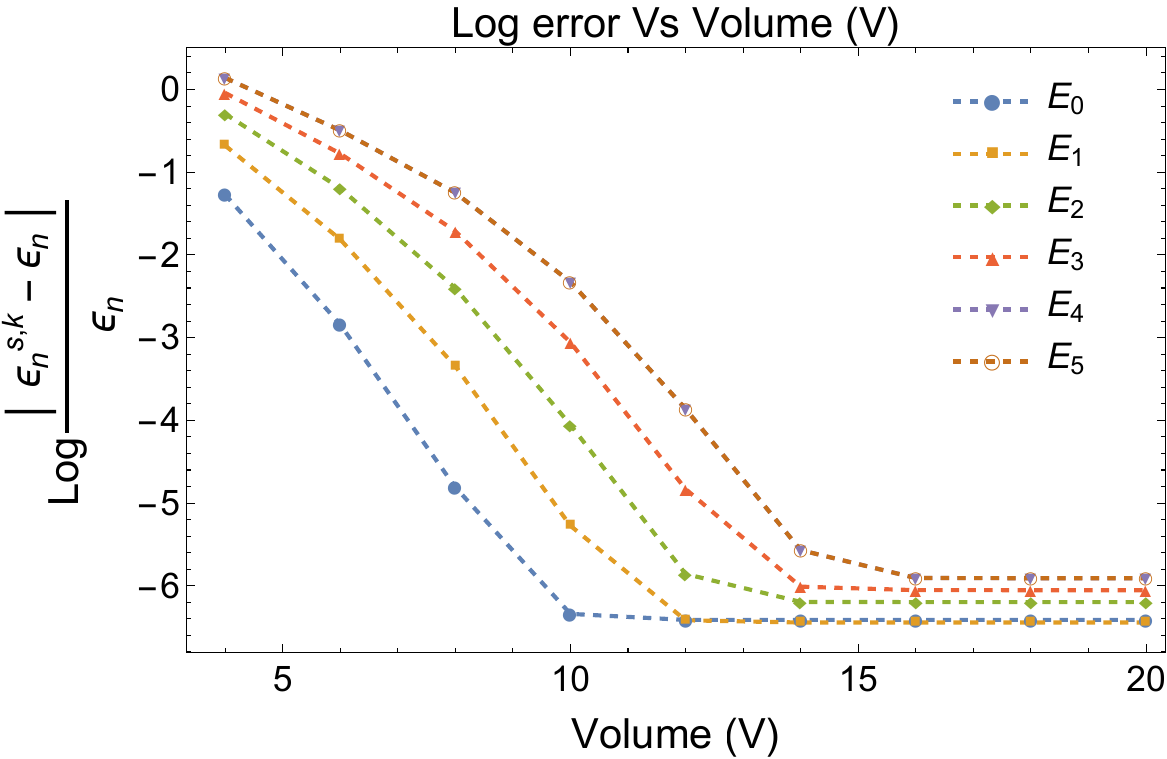}
\caption{\label{fig:Log_error_vs_volume_Plot_TP}Log error $\left(\log\frac{|\epsilon^{s,k}_n-\epsilon_n|}{\epsilon_n}\right)$ of different eigenstates with increasing volume plot for the 1-D TP.}
\end{center}
\end{figure}
We followed the procedure outlined in Sec. \ref{sec:formalism} to compute these overlap integrations, excluding those instances where the scaling functions pass through the origin; in these integrations of this nature, one endpoint of the integral falls within the range covered by the scaling function,  called ``\textbf{partial integral}". So, we have two types of integrals: one where the support of the scaling function is to the left or at the right of the origin and another where the scaling function partially passes through the origin. We evaluated these integrations by the procedure given in the Appendix \ref{appen:The_triangular_potential} \cite{kessler2003scattering}. 

\begin{figure}[hbt]
\begin{center}
\includegraphics[scale=.5]{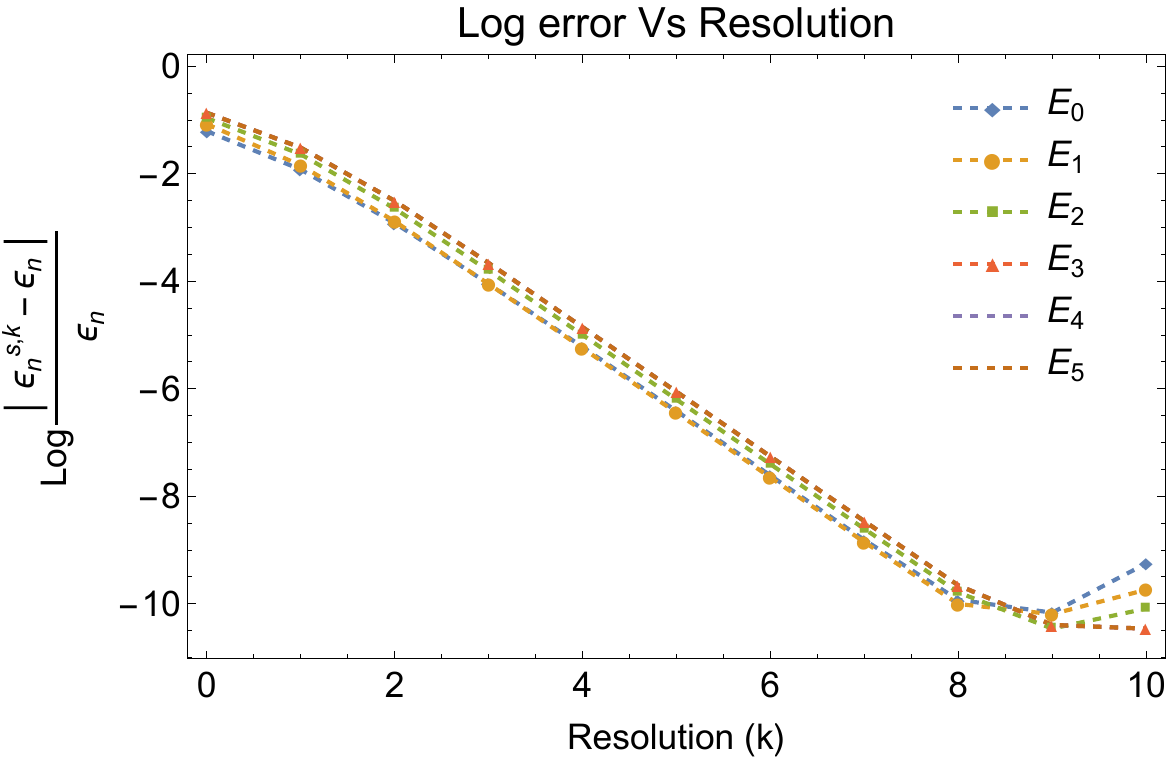}
\caption{\label{fig:Log_error_vs_resolution_Plot_TP}The Log error $\left(\log\frac{|\epsilon^{s,k}_n-\epsilon_n|}{\epsilon_n}\right)$ of different eigenstates with increasing resolution plot for the 1-D TP.}
\end{center}
\end{figure}

Suppose, we want to calculate only a few low-energy eigenvalues for this problem; we must truncate the length to a suitable value so that the low-energy eigenvalues will not change significantly with the increment of size anymore. The log error versus increasing length plot for a few low energy states is given in Fig. \ref{fig:Log_error_vs_volume_Plot_TP}.

\begin{figure}[hbt]
\begin{center}
\includegraphics[scale=.374]{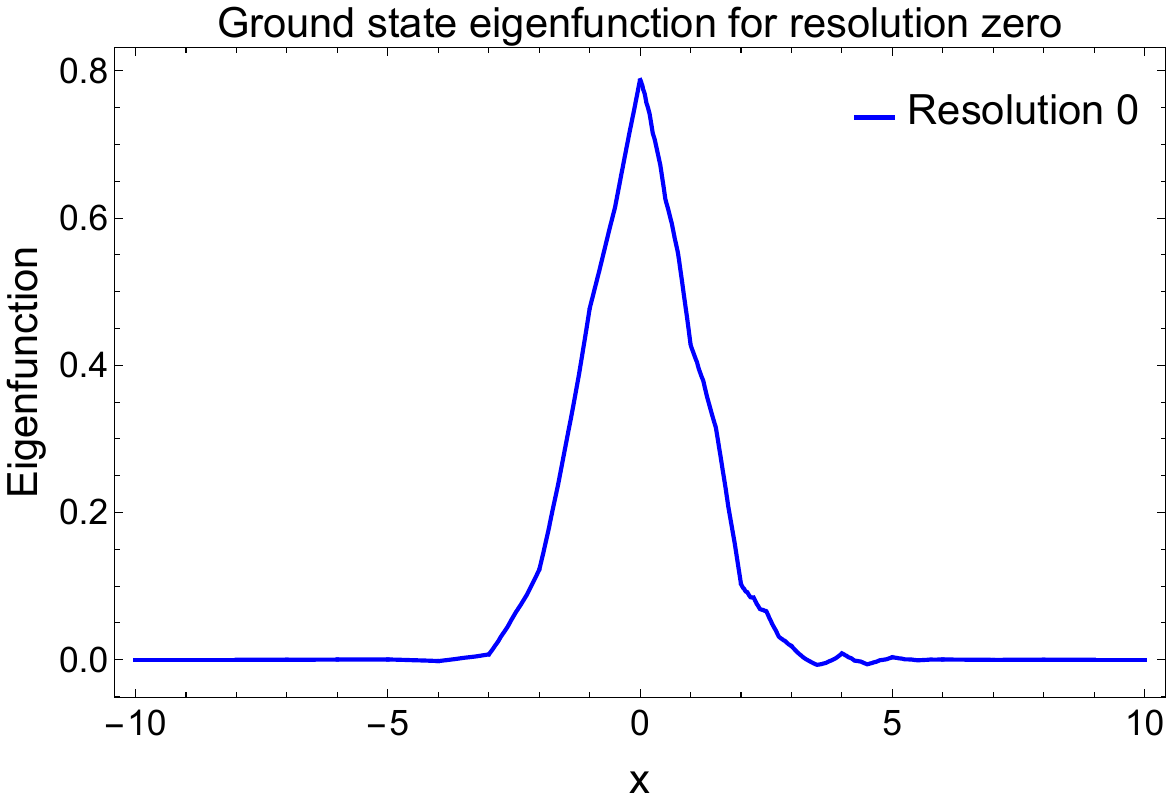}
\includegraphics[scale=.374]{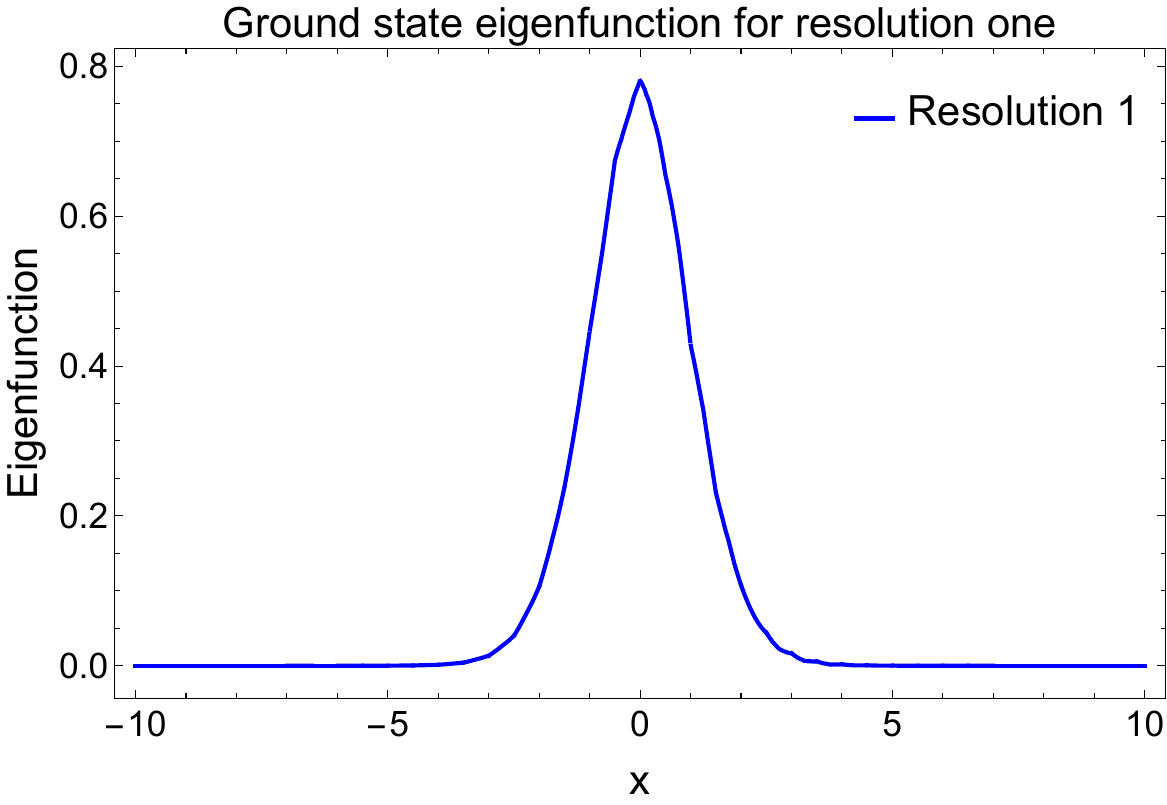}
\includegraphics[scale=.374]{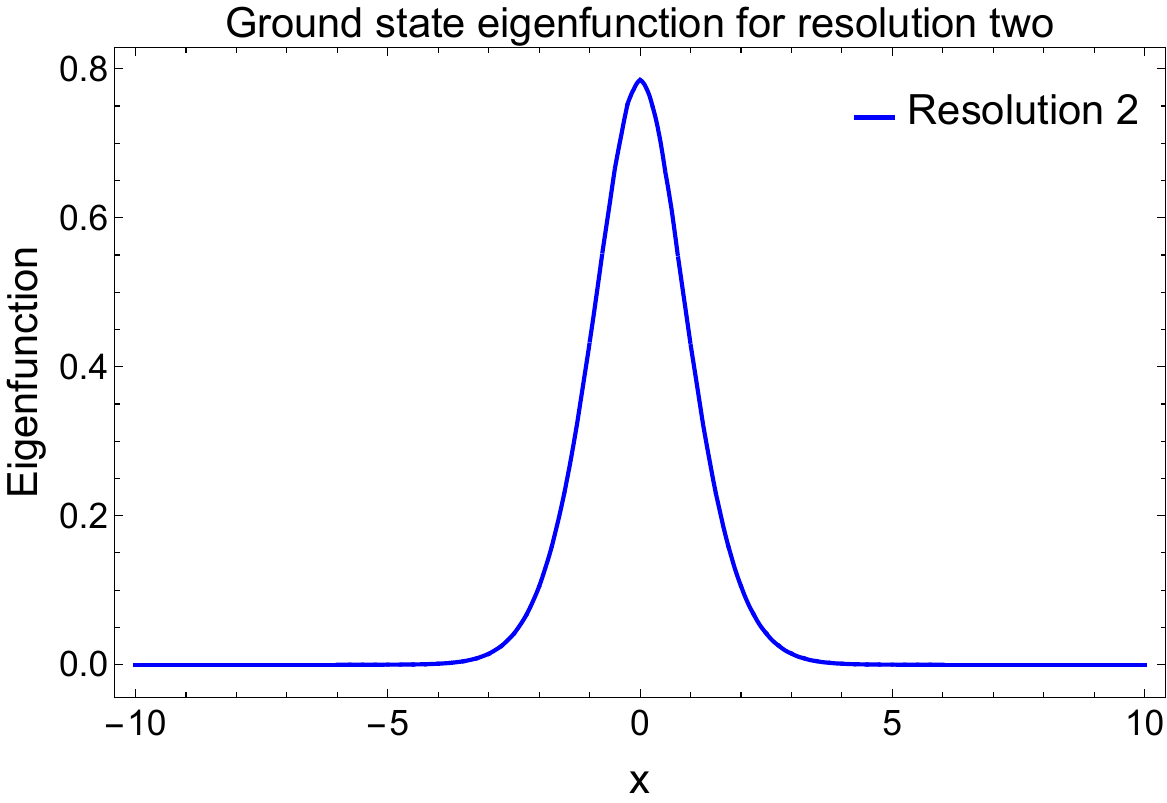}
\includegraphics[scale=.374]{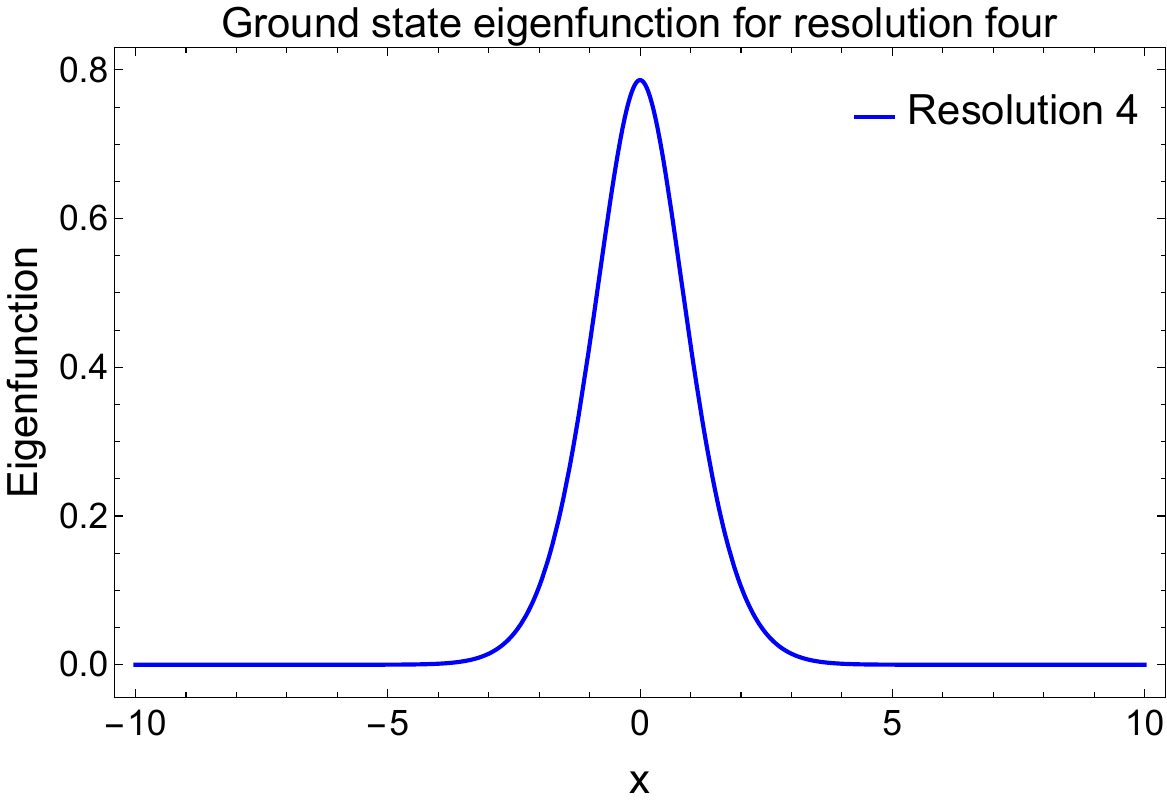}
\includegraphics[scale=.374]{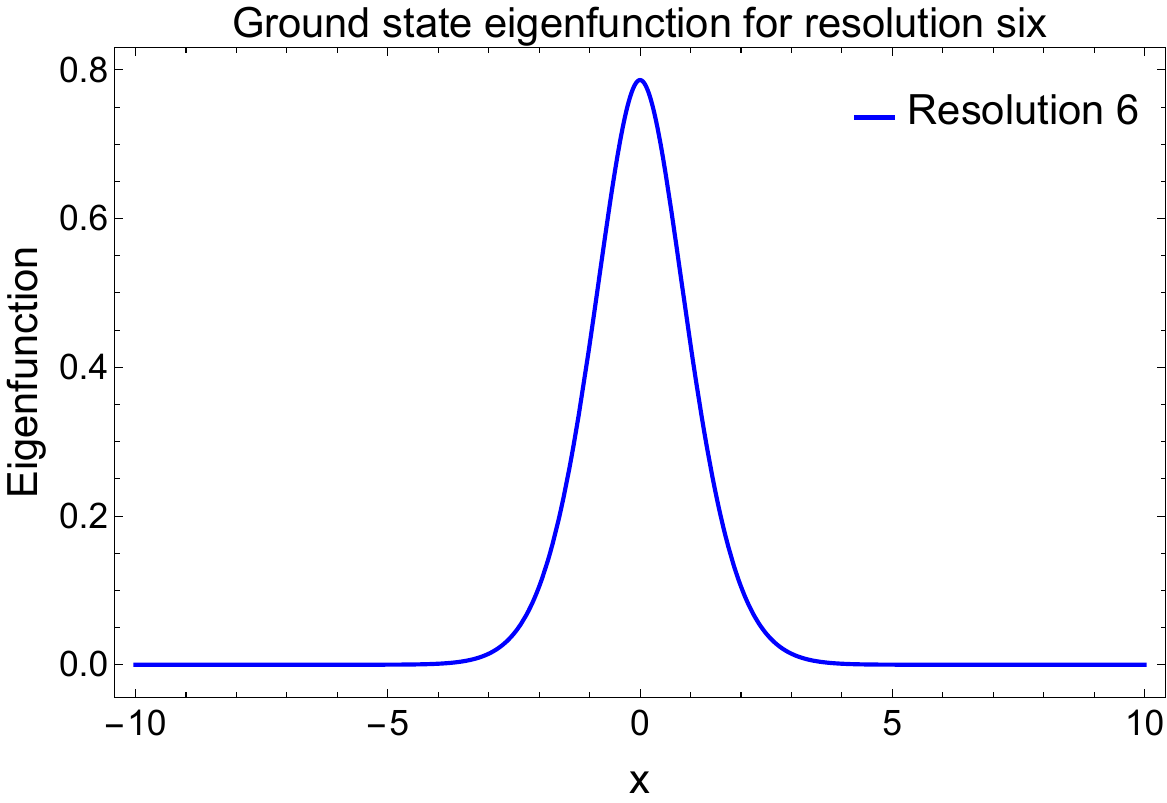}
\includegraphics[scale=.374]{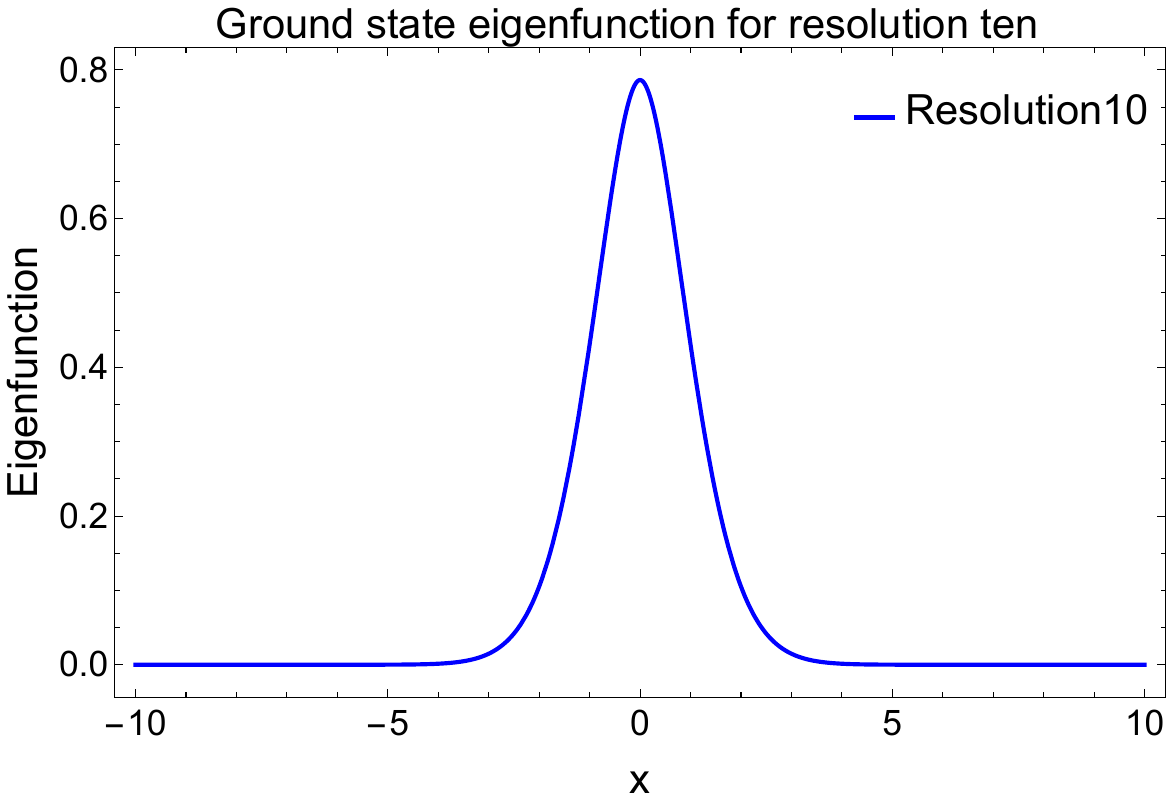}
\caption{\label{fig:Ground_state_for_dif_res_TP}The ground state eigenfunction $\psi^{s,k}_0$ of 1-D TP with increasing resolution.}
\end{center}
\end{figure}

Observing Fig. \ref{fig:Log_error_vs_volume_Plot_TP} reveals that the logarithmic error ceases to exhibit a significant decrease for low-energy eigenvalues beyond $V = 16$ $\left(-8\leq x\leq 8\right)$ as the system volume increases. This is due to the fact that the scaling functions of coarser resolution appear to have minimal contribution on the low-energy eigenvalues. Similar to the situation with the 1-D DDF potential, we opted for a sufficiently large volume, namely $V=20$ $\left(-10\leq x\leq 10 \right)$, to ensure a significant amount of accuracy. 
Table \ref{tab:actual_and_approximate_eigenvalues_of_the_TP} presents the approximate eigenvalues for six low-energy states at a constant volume of $V = 20$ for increasing resolution. The table also includes the corresponding exact eigenvalues for those six states.
\begin{table}[hbt]
\begin{center}
\setlength{\tabcolsep}{0.74 pc}
\catcode`?=\active \def?{\kern\digitwidth}
\caption{The actual energy eigenvalues along with the approximate energy eigenvalues for different resolution of the triangular potential.}
\label{tab:actual_and_approximate_eigenvalues_of_the_TP}
\vspace{1mm}
\begin{tabular}{c | c c c c c c }
\specialrule{.15em}{.0em}{.15em}
\hline
$E_n$ & $E_0$ & $E_1$ & $E_2$ & $E_3$ & $E_4$ & $E_5$ \\
\hline
Exact value & $0.808617$ & $1.855757$ & $2.578096$ & $3.244608$ & $3.825715$ & $4.381671$\\

$k=0$ & $0.858995$ & $2.011962$ & $2.859612$ & $3.688904$ & $4.380506$ & $5.113639$ \\

$k=1$ & $0.818329$ & $1.882558$ & $2.637696$ & $3.344478$ & $3.974205$ & $4.582648$ \\

$k=2$ & $0.809598$ & $1.858190$ & $2.583929$ & $3.254824$ & $3.842099$ & $4.405042$ \\

$k=3$ & $0.808689$ & $1.855924$ & $2.578502$ & $3.245322$ & $3.826884$ & $4.383360$ \\

$k=4$ & $0.808621$ & $1.855768$ & $2.578122$ & $3.244653$ & $3.825790$ & $4.381780$ \\

$k=5$ & $0.808617$ & $1.855758$ & $2.578098$ & $3.244610$ & $3.825720$ & $4.381678$ \\

$k=6$ & $0.808617$ & $1.855757$ & $2.578096$ & $3.244608$ & $3.825716$ & $4.381672$ \\

$k=10$ & $0.808617$ & $1.855757$ & $2.578096$ & $3.244608$ & $3.825715$ & $4.381671$ \\
\hline
\specialrule{.15em}{.15em}{.0em}
\end{tabular}
\end{center}
\end{table}
Figure \ref{fig:Log_error_vs_resolution_Plot_TP} illustrates the log error versus resolution plot for the first six eigenstates of the triangular potential. Upon examining Fig. \ref{fig:Log_error_vs_resolution_Plot_TP}, it can be seen that, similar to the two preceding problems, the logarithmic error associated with the eigenvalues of the initial six eigenstates decreases with the increase in resolution. However, in contrast to the case of an infinite square well potential, the logarithmic error in the eigenvalues does not converge as the resolution increases. This can be attributed to the fact that the 1-D TP features a smoother boundary in comparison to the infinite square well potential.

\begin{figure}[H]
\begin{center}
\includegraphics[scale=0.40]{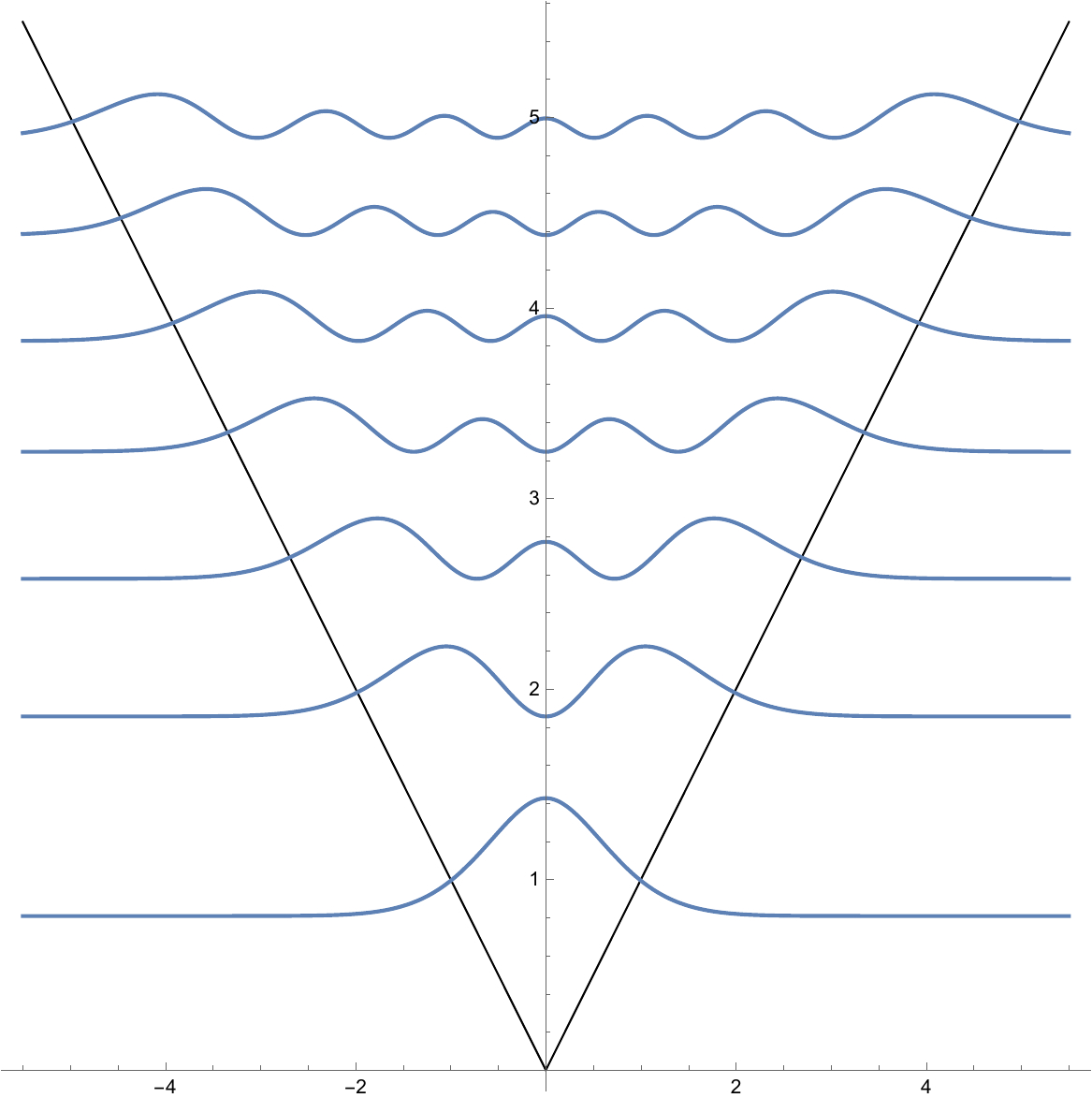}
\caption{\label{fig:Probability_plot_TP}The probability distribution plot of the TP obtained using wavelet-based formalism at resolution $7$ .}
\end{center}
\end{figure}

Similar to the previous two cases, the Fig. \ref{fig:Ground_state_for_dif_res_TP} indicates that as the resolution
increases, the ground state wave function will progressively resemble the actual ground state wave function. To quantitatively analyze the ground state eigenfunction, we fitted those curves with the function $aAi[b(|x|-c)]$, considering that the exact ground state eigenfunction of the triangular potential is given by $1.468004Ai[\sqrt[3]{2}(x-0.808617)]$, according to Eq. (\ref{eq:eigenfunction_of_TP}). The exact values of $a$, $b$, and $c$ are $1.468004$, $1.259921$, and $0.808617$ respectively. Table \ref{tab:actual_value_and_approximate_value_of_a_b_c_TP} demonstrates that the values of $a$, $b$, and $c$ progressively approach the exact values with increasing resolution. 
\begin{table*}[hbt]
\begin{center}
\setlength{\tabcolsep}{0.21pc}
\catcode`?=\active \def?{\kern\digitwidth}
\caption{The exact values of $a$, $b$ and $c$ along with the approximate values of $a$, $b$ and $c$ with increasing resolution.}
\label{tab:actual_value_and_approximate_value_of_a_b_c_TP}
\vspace{1mm}
\begin{tabular}{c c c c c c c c c c}
\specialrule{.15em}{.0em}{.15em}
\hline
 & Exact value & $k=0$ & $k=1$ & $k=2$ & $k=3$ & $k=4$ & $k=5$ & $k=6$ & $k=10$\\

\hline
$a$ & $1.468004$ & $1.399588$ & $1.425686$ & $1.460623$ & $1.467313$ & $1.467953$ & $1.468001$ & $1.468004$ & $1.468004$\\

$b$ & $1.259921$ & $1.120457$ & $1.264444$ & $1.264274$ & $1.260441$ & $1.259963$ & $1.259924$ & $1.259921$ & $1.259921$\\

$c$ & $0.808617$ & $0.887611$ & $0.857299$ & $0.816810$ & $0.809378$ & $0.808672$ & $0.808620$ & $0.808617$ & $0.808617$\\
\hline
\specialrule{.15em}{.15em}{.0em}
\end{tabular}
\end{center}
\end{table*}
Fig. \ref{fig:Probability_plot_TP}, we have plotted the probability distribution of first seven eigenstates of the triangular potential calculated using order $3$ and resolution $7$ Daubechies wavelet functions.

Hence, it can be concluded that the method developed here can be considered as an effective tool for obtaining the eigenvalue and eigenfunction of quantum mechanical eigenvalue problems. Computing the potential energy elements for potential having a singularity at a particular point is given in the reference \cite{Kessler_2003}. However, we did not utilize this technique to tackle any potential of such kind.
\chapter{Quantum Field Theory (QFT) in wavelet basis}
\label{chap:qft_in_a_wavelet_basis}

In this chapter, I will provide a concise historical introduction to Quantum Field Theory (QFT), tracing its origins back to Dirac's initial exploration of Quantum Electrodynamics (QED). What prompted the need for formulating a new theory, and what accomplishments were achieved through its development? Subsequently, we will progress to the evolution of QFT based on wavelets.
\section{An early development, a brief history of QFT}
Its initial accomplishment, specifically the quantization of the electromagnetic field, is recognized as "still the paradigmatic example of a successful quantum field theory" \cite{weinberg1995quantum}. Since photons have rest mass zero, and correspondingly travel in the vacuum at the velocity of light $c$ it is ruled out that a non-relativistic theory such as ordinary QM could give even an approximate description. Indeed, the majority of subjects in the initial stages of quantum theory development (1900–1927) \cite{Helmut1982mehra,doi:10.1126/science.154.3754.1315.a,Arvind2018} revolved around the interaction between radiation and matter, need the application of quantum field theoretical methods. Nevertheless, the quantum mechanical approach articulated by Dirac, Heisenberg, and Schrödinger (1926/27) originated from atomic spectra and did not heavily depend on radiation-related issues. Following the development of QM, a group of researchers including M. Born, W. Heisenberg, and P. Jordan extended the method to electromagnetic fields \cite{10.1119/1.3009634,Dittrich_2015}. P. Jordan, with a background in the literature on light quanta, made significant contributions to QFT. These contributions involved extending the ideas of QM to systems with an infinite number of degrees of freedom. The inception of QFT is usually dated 1927 with Dirac’s famous paper on “The quantum theory of the emission and absorption of radiation” \cite{doi:10.1098/rspa.1927.0039}. Here, Dirac first took the name Quantum Electrodynamics (QED), this is the part of QFT that has been developed first. Using the quantum mechanical framework of the harmonic oscillator, Dirac provided a theoretical account of the quantization of the electromagnetic radiation field and how photons manifest within it. Subsequently, Dirac's method served as a blueprint for extending the quantization to other fields. Over the subsequent three years, initial strides were taken in the advancement of QFT. P. Jordan pioneered the introduction of creation operators for fields obeying the Fermi statistics. Heisenberg and Pauli \cite{Heisenberg1929} introduced the first comprehensive exposition of a general theory of quantum fields, notably detailing the method of canonical quantization, in 1929. Heisenberg and Pauli thus established the basic structure of QFT which can be found in any introduction to QFT up to the present day. 
\section{The Standard Model and the Quantum Field Theory Revolution}
By the early 1950s, QED had evolved into a dependable theory, shedding its preliminary status. It required two decades from the inception of the initial equations until QFT could be systematically applied to address physical problems. Over the subsequent decades, QFT underwent expansion to encompass not only the electromagnetic force but also the weak and strong interactions. This expansion necessitated the discovery of new Lagrangians incorporating novel classes of `particles' or quantum fields. The research aimed to develop a more comprehensive theory of matter and, ultimately, a unified theory of all interactions. Today, there exist reliable theories describing the strong, weak, and electromagnetic interactions among elementary particles, sharing a structure akin to QED. A combined theory associated with the gauge group $SU(3) \otimes SU(2) \otimes U(1)$ is considered as ‘the standard model’ of elementary particle physics which was achieved by Glashow, Weinberg and Salam in 1962 \cite{Salam:1980jd}.
\section{The need for a Non-perturbative Regularization and the Lattice field Theory}
Field theories are characterized by systems possessing an infinite number of degrees of freedom, each associated with a specific point in space. Quantizing field theories presents a subtle challenge as too naive approaches can result in divergent outcomes. To prevent unnecessary divergences, quantum field theories necessitate regularization through the introduction of an ultraviolet cut-off. An approach involves expanding the path integral in powers of the coupling constant. The ensuing Feynman diagrams are then systematically regularized, order by order in the coupling. This perturbative strategy in field theory has yielded remarkable outcomes, particularly in weakly interacting theories. As an illustration, the anomalous magnetic moment of the electron, as derived from QED, stands out as the most comprehensively understood quantity in physics \cite{PhysRev.140.B397,PhysRev.73.416}. Nevertheless, even in cases of weak coupling, the perturbative approach to field theory falls short of complete satisfaction. It is recognized that perturbation theory represents only an asymptotic expansion. The sum of all orders is divergent, meaning it does not provide a definitive definition of the theory beyond perturbation theory. Moreover, for strongly coupled theories, such as Quantum Chromodynamics (QCD) at low energies, the perturbative regularization proves entirely ineffective.

Confinement and the Higgs mechanism \cite{PhysRevLett.13.321,higgs1964broken,PhysRevLett.13.585} are non-perturbative phenomena. To investigate them from fundamental principles, it is imperative to establish the theory beyond the confines of perturbation theory. Lattice regularization offers a systematic approach to address this issue by substituting the continuous space-time continuum with a discrete lattice mesh. It is important not to perceive the lattice as merely an approximation to the continuum theory; instead, it provides a definition for a theory that lacks a direct definition in the continuum. Naturally, to attain the continuum limit, the theory must undergo renormalization by diminishing the lattice spacing to zero and appropriately adjusting the bare coupling constants. This process hinges on the existence of a second-order phase transition in the corresponding 4-dimensional statistical mechanics system. The lattice stands out as an elegant regularization method due to its locality and adherence to local gauge symmetries. The violation of certain space-time symmetries is of lesser consequence, as these symmetries are automatically restored in the continuum limit. 

Preserving chiral symmetry on the lattice poses a nuanced yet crucial challenge. Lattice fermions have several technical problems that have prevented the non-perturbative formulation of the standard model for many years. For example, chiral fermions — like neutrinos — suffer from the lattice fermion doubling problem. Every left-handed neutrino necessarily comes with a right-handed partner. Even the perturbative definition of the standard model has been incomplete beyond one loop, due to ambiguities in treating $\gamma_5$ in dimensional regularization. All these ambiguities are now eliminated, thanks to the new lattice results. It is good to know that the standard model now stands on a firm mathematical basis and that the path integral expressions we write down to define it are completely well-defined even beyond perturbation theory.

\section{Free scalar field theory in plane wave basis}
In this section, we review the canonical quantization of free scalar field of mass-$\mu$ in plane wave basis. 

The Lagrangian density of a free scalar field of mass-$\mu$ is given by,
\begin{eqnarray}
\mathcal{L}(x):=\frac{1}{2}\left(\dot{\Phi}(x)\dot{\Phi}(x)-\nabla\Phi(x).\nabla\Phi(x)-\mu^2 \Phi(x)^2 \right).
\end{eqnarray}
The Euler-Lagrange equations results in Klein-Gordon equation as the equation of motion,
\begin{eqnarray}
\label{eq:the_klein_gordon_equation}
\frac{\partial^2}{\partial t^2}\Phi(x)-\frac{\partial^2}{\partial x^2}\Phi(x)+\mu^2 \Phi(x)=0.
\end{eqnarray}
The generalized momentum is given by,
\begin{eqnarray}
\label{eq:the_momentum_expression}
\Pi(x)=\frac{\partial \mathcal{L}(x)}{\partial \dot{\Phi}},
\end{eqnarray}
and the Hamiltonian is computed through the Lagrange transform,
\begin{eqnarray}
\label{eq:scalar_field_hamiltonian}
H&=&\int \left(\Pi(x) \dot{\Phi}(x)-\mathcal{L}(x)\right)dx \\
&=& \frac{1}{2}\int \left(\Pi(x)\Pi(x)+\nabla\Phi(x).\nabla\Phi(x)+\mu^2\Phi(x)^2 \right)dx.
\end{eqnarray}
The canonical quantization procedure requires that $\Phi(x,t)$ and $\Pi(x,t)$ satisfy the equal time commutation relations,
\begin{eqnarray}
\label{eq:cannonical_commutation_relation_SFT}
\begin{split}
\left[\Phi(x,t),\Phi(y,t)\right]&=& 0,\\
\left[\Pi(x,t),\Pi(y,t)\right]&=& 0,\\
\left[\Phi(x,t),\Pi(y,t)\right]&=&i\delta(x-y).
\end{split}
\end{eqnarray}
The invariance of the action under Poincare group transformations, $x^{\mu}\rightarrow x'^{\mu}=\Lambda^{\mu}_{\nu}x^{\nu}+a^{\mu}$ ; $\phi(x)=\phi'(x')=\phi(\Lambda x+a)=\phi(x)$, leads to the energy-momentum tensor,
\begin{eqnarray}
T^{\mu\nu}(x)&=&-\frac{\partial \mathcal{L}(x)}{\partial (\partial_{\mu}\Phi)}\partial^{\nu}\Phi+\eta^{\mu \nu}\mathcal{L}(x) \nonumber\\
&=& \partial^{\mu}\Phi(x)\partial^{\nu}\Phi(x)-\frac{1}{2}\eta^{\mu \nu}\left(\partial^{\alpha}\Phi(x)\partial_{\alpha}\Phi(x)+\mu^2\Phi(x)^2\right),
\end{eqnarray}
and the angular momentum tensor is,
\begin{eqnarray}
M^{\mu\nu\alpha}=\left(x^{\mu}T^{\nu\alpha}-x^{\nu}T^{\mu\alpha}\right).
\end{eqnarray}
as conserved tensorial densities,
\begin{eqnarray}
\partial_{\mu}T^{\nu\mu}=\partial_{\alpha}M^{\mu\nu\alpha}=0,
\end{eqnarray}
The conserved charges,
\begin{eqnarray}
P^{\mu}=\int_{t=0}dx \mathcal{P}^{\mu}(x),\quad\quad \mathcal{P}^{\mu}(x):=T^{\mu 0}(x)\\
J^{\mu\nu}:=\int_{t=0}dx \mathcal{J}^{\mu\nu}(x),\quad\quad\mathcal{J}^{\mu\nu}(x):=M^{\mu\nu 0}(x),
\end{eqnarray}
are generators of infinitesimal space-time translations and infinitesimal Lorange transformation. The generators correspond to the energy $P^0$,
\begin{eqnarray}
\begin{gathered}
H=P^0=\\
\int_{t=0}:\left(\Pi(x)\dot{\Phi}(x)-\mathcal{L}(x)\right):dx=\frac{1}{2}\int :\left(\Pi(x)\Pi(x)+\nabla\Phi(x).\nabla\Phi(x)+\mu^2 \Phi(x)^2\right):dx,
\end{gathered}
\end{eqnarray}
The linear momentum $P^i$,
\begin{eqnarray}
\begin{gathered}
P^i=-\int_{t=0}:\Pi(x)\nabla_{i}\Phi(x):dx.\\
\end{gathered}
\end{eqnarray}
The angular momentum $J_{ij}$,
\begin{eqnarray}
\begin{gathered}
J^{ij}=-J^{ji}=\epsilon^{ijk}J^k=\nonumber\\
\int_{t=0}dx:\left(\Pi(x)x^j\partial^{i}\Phi(x)-\Pi(x)x^i\partial^{j}\Phi(x)\right):\\
J^{i0}=-J^{0i}=K^{i}=\\
\end{gathered}
\end{eqnarray}
and the boost,
\begin{eqnarray}
\begin{gathered}
\int_{t=0}:\left(\frac{1}{2}\Pi^2(x)x^i+\frac{1}{2}\nabla\Phi(x).\nabla\Phi(x)+\frac{1}{2}x^i\mu^2\Phi(x)^2\right):dx.
\end{gathered}
\end{eqnarray}
where the :: indicate that the operator are normal ordered.

The solution of Klein-Gordon equation, Eq. (\ref{eq:the_klein_gordon_equation}), using the plane wave basis takes the following form,
\begin{eqnarray}
\label{eq:scalar_filed_in_terms_of_a_and_a_dagger}
\Phi(x)=\frac{1}{(2\pi)^{3/2}}\int \frac{dp}{\sqrt{2\omega_{\mu}(p)}}\left(e^{-ip.x}a^{\dagger}(p)+e^{ip.x}a^{\dagger}(p)\right),
\end{eqnarray}
From Eq. (\ref{eq:the_momentum_expression}) and Eq. (\ref{eq:scalar_filed_in_terms_of_a_and_a_dagger}), it follows that,
\begin{eqnarray}
\Pi(x)=\frac{i}{(2\pi)^{3/2}}\int \sqrt{\frac{\omega_{\mu}(p)}{2}}\left(e^{-ip.x}a^{\dagger}(p)-e^{ip.x}a^{\dagger}(p)\right),
\end{eqnarray}
where,
\begin{eqnarray}
\omega_{\mu}(p)=\sqrt{\mu^2+p^2},
\end{eqnarray}
is the single-particle energy. the commutation relations can be computed using Eq. (\ref{eq:cannonical_commutation_relation_SFT}),
\begin{eqnarray}
\begin{aligned}
\left[a(p),a(p)\right]= 0,\\
\left[a^{\dagger}(p),a^{\dagger}(p)\right]= 0,\\
\left[a(p),a^{\dagger}(p')\right]=\delta(p-p').
\end{aligned}
\end{eqnarray}
The vacuum state of the field is the solution of
\begin{eqnarray}
a(p)\ket{0}=0.
\end{eqnarray}
The infinitesimal generators have momentum-space representation as integrals over momentum densities,
\begin{eqnarray}
H&=& \int \mathcal{H}(p)dp=\int dp a^{\dagger}(p)\omega_{m}(p)a(p),\\
P&=& \int \mathcal{P}(p)dp=\int dp a^{\dagger}(p)p a(p),\\
J&=& \int \mathcal{J}(p)dp=\int dp a^{\dagger} (p)\left(i\frac{\partial}{\partial p}\times p\right)a(p),\\
K&=& \int \mathcal{K}(p)dp=\int dp a^{\dagger}(p)\frac{1}{2}\left\{i\frac{\partial}{\partial p},\omega_m(p)\right\}a(p),
\end{eqnarray}
where, we set $x^0=0$ in the last expression.

\section{Quantum Field Theory in wavelet based framework.}
The discrete wavelet-based formulation of quantum field theory allows one to commit to the Hamiltonian framework while maintaining the discreteness of the lattice approach, yet not compromise the continuum nature of space. Daubechies wavelets and scaling functions constitute an orthonormal basis of compactly supported functions \cite{https://doi.org/10.1002/cpa.3160410705,doi:10.1137/1.9781611970104,f5d23bd3-c0af-3314-bd56-832b7db8db57}. Roughly speaking, each basis function is characterized by its location (translation index) and length scale (resolution). The quantum fields, when expanded in the wavelet basis, lead to its representation as an infinite sequence of operators characterized by a
location and resolution index. This approach allows natural volume and resolution truncations of the QFT. The truncated theory is an ordinary quantum mechanical theory
with multiple discrete degrees of freedom organized by location and length scale. The maximum resolution plays the role of ultraviolet cutoff and the volume truncation plays the role of infrared cutoff. In this section, we describe canonical quantization scalar field theory within the framework of discrete wavelets. 

Within the wavelet based canonical quantization, one start by expanding the field operator, $\Phi(\textbf{x},t)$ and its canonical conjugate $\Pi(\textbf{x},t)$ in wavelet basis \cite{PhysRevD.87.116011},
\begin{eqnarray}
\label{eq:phi_x_in_terms_of_phi_k_phi_l}
\Phi(\textbf{x},t)&=&\sum_{\textbf{n}} \Phi^{s,k}(\textbf{n},t)s^k_{\textbf{n}}(\textbf{x})+\sum_{\textbf{n},\alpha,l\geq k} \Phi^{w,l}(\textbf{n},\alpha,t) w^l_{\textbf{n},\alpha}(\textbf{x}),\\
\label{eq:pi_x_in_terms_of_pi_k_pi_l}
\Pi(\textbf{x},t)&=&\sum_{\textbf{n}} \Pi^{s,k}(\textbf{n},t)s^k_{\textbf{n}}(\textbf{x})+\sum_{\textbf{n},\alpha,l\geq k} \Pi^{w,l}(\textbf{n},\alpha,t) w^l_{\textbf{n},\alpha}(\textbf{x}),
\end{eqnarray}
The operator coefficients, $\Phi^{s,k}(\textbf{n},t)$, $\Phi^{w,l}(\textbf{n},\alpha,t)$, $\Pi^{s,k}(\textbf{n},t)$, $\Pi^{w,l}(\textbf{n},\alpha,t)$, correspond to the projections of the field operators onto the orthonormal basis of scaling and wavelets,
\begin{eqnarray}
\label{eq:field_coefficient_phi_sk}
\Phi^{s,k}(\textbf{n},t)&=& \int d\textbf{x} \Phi(\textbf{x},t)s^k_{\textbf{n}}(\textbf{x}),\\
\Phi^{w,l}(\textbf{n},\alpha,t)&=& \int d\textbf{x} \Phi(\textbf{x},t)w^l_{\textbf{n},\alpha}(\textbf{x}) \quad \quad (l\geq k),\\
\Pi^{s,k}(\textbf{n},t)&=& \int d\textbf{x} \Pi(\textbf{x},t)s^k_{\textbf{n}}(\textbf{x}),\\
\label{eq:field_coefficient_pi_wl}
\Pi^{w,l}(\textbf{n},\alpha,t)&=& \int d\textbf{x} \Pi(\textbf{x},t)w^l_{\textbf{n},\alpha}(\textbf{x}) \quad \quad (l\geq k).
\end{eqnarray}

The expansions have been divided into operators distributed over scaling functions and generalized wavelets. This constitutes a separation of scales, where the scale-$1/2^k$ scaling functions capture the coarse-scale structure of the field, while the wavelet components encompass the structure on all finer scales.

When the field is expanded in the wavelet basis, the basis coefficients provide an average of the field's value across a finite area. As the scaling function at scale $1/2^{k+1}$ can be rewritten using scale-$1/2^k$ scaling and wavelet functions, the information contained in fields smeared with scale-$1/2^{k}$ wavelets and scaling functions is equivalent to that in the fields smeared with scale-$1/2^{k+1}$ scaling functions.

When considering all wavelet components, the expansions become exact. However, within any finite region, infinite wavelet basis functions still exist corresponding to the arbitrarily small scales.

Using the equal-time canonical commutation relation, Eq. (\ref{eq:cannonical_commutation_relation_SFT}) and Eq. (\ref{eq:field_coefficient_phi_sk})-Eq. (\ref{eq:field_coefficient_pi_wl}), we find that the operator basis coefficients satisfy,
\begin{eqnarray}
&\left[\Phi^{s,k}(\textbf{n},t),\Phi^{s,k}(\textbf{m},t)\right]=0,\quad \quad \left[\Pi^{s,k}(\textbf{n},t),\Pi^{s,k}(\textbf{m},t)\right]=0,&\\
&\left[\Phi^{s,k}(\textbf{n},t),\Pi^{s,k}(\textbf{m},t)\right]=i\delta_{\textbf{n},\textbf{m}},&\\
&\left[\Phi^{w,r}(\textbf{n},\alpha,t),\Phi^{w,q}(\textbf{m},\beta,t)\right]=0,\quad \quad \left[\Pi^{w,r}(\textbf{n},\alpha,t),\Pi^{w,q}(\textbf{m},\beta,t)\right]=0,&\\
&\left[\Phi^{w,r}(\textbf{n},,\alpha,t),\Pi^{w,q}(\textbf{m},\alpha,t)\right]=i\delta_{\textbf{n},\textbf{m}}\delta_{rq}\delta_{\alpha \beta},&\\
&\left[\Phi^{w,r}(\textbf{n},\alpha,t),\Phi^{s,k}(\textbf{m},t)\right]=0,\quad \quad \left[\Pi^{w,r}(\textbf{n},\alpha,t),\Pi^{s,k}(\textbf{m},t)\right]=0,&\\
&\left[\Phi^{w,r}(\textbf{n},\alpha,t),\Pi^{s,k}(\textbf{m},t)\right]=0,\quad \quad \left[\Pi^{w,r}(\textbf{n},\alpha,t),\Phi^{s,k}(\textbf{m},t)\right]=0,&
\end{eqnarray}
where in all this expressions ${r,q}\geq k$.

The discrete creation and annihilation operators are defined as,
\begin{eqnarray}
\label{eq:a_k_expansion}
a^{s,k}(\textbf{n},t) &:=& \frac{1}{\sqrt{2}}\left(\sqrt{\gamma^{s,k}}\Phi^{s,k}(\textbf{n},t)+i\frac{1}{\sqrt{\gamma^{s,k}}}\Pi^{s,k}(\textbf{n},t)\right),\\
\label{eq:b_k_expansion}
a^{w,r}(\textbf{n},\alpha,t) &:=& \frac{1}{\sqrt{2}}\left(\sqrt{\gamma^{w,\alpha,r}}\Phi^{w,r}(\textbf{n},\alpha,t)+i\frac{1}{\sqrt{\gamma^{w,\alpha,r}}}\Pi^{w,r}(\textbf{n},\alpha,t)\right),
\end{eqnarray}
that satisfy,
\begin{eqnarray}
\label{eq:commutation_relation_a_a_dagger}
\left[a^{s,k}(\textbf{n},t),a^{s,k\dagger}(\textbf{m},t)\right]&=& \delta_{\textbf{n}\textbf{m}},\\
\label{eq:commutaion_relation_b_b_dagger}
\left[a^{w,l}(\textbf{n},\alpha,t),a^{w,j\dagger}(\textbf{m},\beta,t)\right]&=& \delta_{\textbf{n}\textbf{m}}\delta_{jl}\delta_{\alpha \beta},
\end{eqnarray}
in this scenario, all other commutator will vanish, and the expansions involve two real constants $\gamma^{s,k}$ and $\gamma^{w,\alpha, r}$. These variables, along with their complex conjugate, adhere to the specified commutation relations presented in Eq. (\ref{eq:commutation_relation_a_a_dagger}) and Eq. (\ref{eq:commutaion_relation_b_b_dagger}). These rules apply to both creation and annihilation operators irrespective of the value of $\gamma^{s,k}$ and $\gamma^{w,\alpha, r}$. However, to ensure that the annihilation operator effectively annihilate the free field vacuum, choosing an appropriate value for $\gamma^{s,k}$ and $\gamma^{w,\alpha, r}$ are essential. This can be achieved from the following equations,
\begin{eqnarray}
\label{eq:vacuum_expectation_value_a_dagger_a}
\bra{0}a^{s,k\dagger}(\textbf{n},t)a^{s,k}(\textbf{n},t)\ket{0} &=& 0,\\
\label{eq:vacuum_expectation_value_b_dagger_b}
\bra{0}a^{w,r\dagger}(\textbf{n},\alpha,t)a^{w,r}(\textbf{n},\alpha,t)\ket{0} &=& 0,\\
\label{eq:vacuum_expectation_value_a_a_dagger}
\bra{0}a^{s,k}(\textbf{n},t)a^{s,k\dagger}(\textbf{n},t)\ket{0} &=& 1,\\
\label{eq:vacuum_expectation_value_b_b_dagger}
\bra{0}a^{w,r}(\textbf{n},\alpha,t)a^{w,r\dagger}(\textbf{n},\alpha,t)\ket{0} &=& 1.
\end{eqnarray}
This results in quadratic equations for the coefficients $\gamma^{s,k}$ and $\gamma^{w,\alpha,r}$. By adopting this particular choice of $\gamma^{s,k}$ and $\gamma^{w,\alpha,r}$, the annihilation operators effectively annihilate the vacuum state associated with the mass $\mu$ free field.

Utilizing Eq. (\ref{eq:a_k_expansion}), Eq. (\ref{eq:vacuum_expectation_value_a_dagger_a}), Eq. (\ref{eq:vacuum_expectation_value_a_a_dagger}) alongside Eq. (\ref{eq:b_k_expansion}), Eq. (\ref{eq:vacuum_expectation_value_b_dagger_b}), Eq. (\ref{eq:vacuum_expectation_value_b_b_dagger}), we derive the following pair of quadratic equations for $\gamma^{s,k}$ and $\gamma^{w,\alpha,r}$ respectively,
\begin{eqnarray}
&{\gamma^{s,k}}^2 \bra{0}\Phi^{s,k}(\textbf{n},t)\Phi^{s,k}(\textbf{n},t)\ket{0}-\gamma^{s,k} +\bra{0}\Pi^{s,k}(\textbf{n},t)\Pi^{s,k}(\textbf{n},t)\ket{0}=0. &\\
&{\gamma^{w,\alpha , r}}^2 \bra{0}\Phi^{w,r}(\textbf{n},\alpha,t)\Phi^{w,r}(\textbf{n},\alpha,t)\ket{0}-\gamma^{w,\alpha , r} +\bra{0}\Pi^{w,r}(\textbf{n},\alpha,t)\Pi^{w,r}(\textbf{n},\alpha,t)\ket{0}=0. &\quad\quad
\end{eqnarray}

The solution to these equations for the scaling-function fields is,
\begin{eqnarray}
\gamma^{s,k}&=&\frac{1\pm \sqrt{1-4 \bra{0}\Phi^{s,k}(\textbf{m},t)\Phi^{s,k}(\textbf{m},t)\ket{0}\bra{0}\Pi^{s,k}(\textbf{m},t)\Pi^{s,k}(\textbf{m},t)\ket{0}}}{2 \bra{0}\Phi^{s,k}(\textbf{m},t)\Phi^{s,k}(\textbf{m},t)\ket{0}}\nonumber \\
\label{eq:gamma_k_expression}
&=& \frac{1\pm \sqrt{1-4 \bra{0}\Phi^{s,k}(\textbf{0},t)\Phi^{s,k}(\textbf{0},t)\ket{0}\bra{0}\Pi^{s,k}(\textbf{0},t)\Pi^{s,k}(\textbf{0},t)\ket{0}}}{2 \bra{0}\Phi^{s,k}(\textbf{0},t)\Phi^{s,k}(\textbf{0},t)\ket{0}},
\end{eqnarray}
and for the wavelet fields,
\begin{eqnarray}
\gamma^{w,\alpha , r}&=& \frac{1\pm \sqrt{1-4 \bra{0}\Phi^{w,r}(\textbf{m},\alpha,t)\Phi^{w,r}(\textbf{m},\alpha,t)\ket{0}\bra{0}\Pi^{w,r}(\textbf{m},\alpha,t)\Pi^{w,r}(\textbf{m},\alpha,t)\ket{0}}}{2 \bra{0}\Phi^{w,r}(\textbf{m},\alpha,t)\Phi^{w,r}(\textbf{m},\alpha,t)\ket{0}}\nonumber \\
\label{eq:gamma_k_alpha_expression}
&=& \frac{1\pm \sqrt{1-4 \bra{0}\Phi^{w,r}(\textbf{0},\alpha,t)\Phi^{w,r}(\textbf{0},\alpha,t)\ket{0}\bra{0}\Pi^{w,r}(\textbf{0},\alpha,t)\Pi^{w,r}(\textbf{0},\alpha,t)\ket{0}}}{2 \bra{0}\Phi^{w,r}(\textbf{0},\alpha,t)\Phi^{w,r}(\textbf{0},\alpha,t)\ket{0}}.\nonumber\\
\end{eqnarray}
These coefficients depend on the scale $1/2^k$, $\alpha$ and the mass $\mu$; however, they remain independent of $m$ and $t$ due to the space-time translational invariance of the vacuum. With this specific choice, the scaling function fields can be expressed in terms of the creation and annihilation operators as:
\begin{eqnarray}
\label{eq:phi_k_exact_expansion_in_terms_of_a_k}
\Phi^{s,k}(\textbf{m},t)&=&  \frac{1}{\sqrt{2\gamma^{s,k}}}\left(a^{s,k\dagger}(\textbf{m},t)+a^{s,k}(\textbf{m},t)\right),\\
\label{eq:pi_k_exact_expansion_in_terms_of_a_k}
\Pi^{s,k}(\textbf{m},t)&=& i\sqrt{\frac{\gamma^{s,k}}{2}}\left(a^{s,k\dagger}(\textbf{m},t)-a^{s,k}(\textbf{m},t)\right),
\end{eqnarray}
with analogous expressions for the wavelet fields,
\begin{eqnarray}
\label{eq:phi_k_exact_expansion_in_terms_of_b_k}
\Phi^{w,l}(\textbf{m},\alpha,t)&=& \frac{1}{\sqrt{2\gamma^{w,\alpha, l}}}\left(a^{w,l\dagger}(\textbf{m},\alpha,t)+a^{w,l}(\textbf{m},\alpha,t)\right),\\
\label{eq:pi_k_exact_expansion_in_terms_of_b_k}
\Pi^{w,l}(\textbf{m},\alpha,t)&=& i\sqrt{\frac{\gamma^{w,\alpha, l}}{2}}\left(a^{w,l\dagger}(\textbf{m},\alpha,t)-a^{w,l}(\textbf{m},\alpha,t)\right).
\end{eqnarray}
The values of $\gamma^{s,k}$ and $\gamma^{w,\alpha, l}$ depend on the mass term present within the field operators because of the integrals found in Eq. (\ref{eq:gamma_k_expression}) and Eq. (\ref{eq:gamma_k_alpha_expression}).
\begin{eqnarray}
\bra{0}\Phi^{s,k}(\textbf{m},t),\Phi^{s,k}(\textbf{m},t)\ket{0}&=&\frac{1}{(2\pi)^3}\int \frac{s^k_{\textbf{n}}(\textbf{x})s^k_{\textbf{n}}(\textbf{y})}{2\omega_{\mu}(\textbf{p})}e^{i\textbf{p}.(\textbf{x}-\textbf{y})}d^3\textbf{x}d^3\textbf{y}d^3\textbf{p},\\
\bra{0}\Pi^{s,k}(\textbf{m},t),\Pi^{s,k}(\textbf{m},t)\ket{0}&=&\frac{1}{(2\pi)^3}\int \frac{s^k_{\textbf{n}}(\textbf{x})s^k_{\textbf{n}}(\textbf{y})\omega_{\mu}(\textbf{p})}{2}e^{i\textbf{p}.(\textbf{x}-\textbf{y})}d^3\textbf{x}d^3\textbf{y}d^3\textbf{p},\\
\bra{0}\Phi^{w,r}(\textbf{m},\alpha,t),\Phi^{w,r}(\textbf{m},\alpha,t)\ket{0}&=&\frac{1}{(2\pi)^3}\int \frac{w^{r}_{\textbf{n},\alpha}(\textbf{x})w^r_{\textbf{n},\alpha}(\textbf{y})}{2\omega_{\mu}(\textbf{p})}e^{i\textbf{p}.(\textbf{x}-\textbf{y})}d^3\textbf{x}d^3\textbf{y}d^3\textbf{p},\\
\label{eq:vacuum_expectation_value_pi_k_pi_k}
\bra{0}\Pi^{s,k}(\textbf{m},\alpha,t),\Pi^{s,k}(\textbf{m},\alpha,t)\ket{0}&=&\frac{1}{(2\pi)^3}\int \frac{w^r_{\textbf{n},\alpha}(\textbf{x})w^r_{\textbf{n},\alpha}(\textbf{y})\omega_{\mu}(\textbf{p})}{2}e^{i\textbf{p}.(\textbf{x}-\textbf{y})}d^3\textbf{x}d^3\textbf{y}d^3\textbf{p}.\quad\quad\quad
\end{eqnarray}
This entails calculating the integrals of the basis functions across the two-point mass-$\mu$ Wightman functions of the field at a fixed time.

Leveraging Eq. (\ref{eq:phi_k_exact_expansion_in_terms_of_a_k}) through Eq. (\ref{eq:vacuum_expectation_value_pi_k_pi_k}), we can represent Eq. (\ref{eq:phi_x_in_terms_of_phi_k_phi_l}) and Eq. (\ref{eq:pi_x_in_terms_of_pi_k_pi_l}) in terms of discrete creation and annihilation operators as:
\begin{eqnarray}
\Phi(\textbf{x},t)&=&\sum_{\textbf{m}}\frac{s^k_{\textbf{m}}(x)}{\sqrt{2\gamma^{s,k}}}\left(a^{s,k\dagger}+a^{s,k}(\textbf{m},t)\right)\nonumber\\
\label{eq:phi_in_a_and_a_dagger}
&& +\sum_{\textbf{m},l\geq k}\frac{w^l_{\textbf{m},\alpha}(x)}{\sqrt{2\gamma^{w,\alpha, l}}}\left(a^{w,l\dagger}(\textbf{m},\alpha,t)+a^{w,l}(\textbf{m},\alpha,t)\right),\\
\Pi(\textbf{x},t)&=&i\sum_{\textbf{m}}s^k_{\textbf{m}}(x)\sqrt{\frac{\gamma^{s,k}}{2}}\left(a^{s,k\dagger}(\textbf{m},t)-a^{s,k}(\textbf{m},t)\right)\nonumber\\
\label{eq:pi_in_a_and_a_dagger}
&& +i\sum_{\textbf{m},l\geq k}w^l_{\textbf{m},\alpha}\sqrt{\frac{\gamma^{w,\alpha, l}}{2}}\left(a^{w,l\dagger}(\textbf{m},\alpha,t)-a^{w,l}(\textbf{m},\alpha,t)\right).
\end{eqnarray}
In this notation, the operators $a^{s,k}$ and $a^{s,k\dagger}$ are responsible for annihilating and creating a particle of scale $1/2^k$, whereas the operators $a^{w,l}$ and $a^{w,l\dagger}$ responsible for the destruction and creation of particle of scale smaller than $1/2^k$.

To form the Hilbert space of this free field, we consider the limits of finite linear combinations of products of discrete creation operators, $a^{s,k\dagger}(\textbf{m},t)$ and $a^{w,l\dagger}(\textbf{m},\alpha,t)$, when they are applied to the vacuum state at a specific time.

Operators can be decomposed into components that exclusively involve $a^{s,k}$ and $a^{s,k\dagger}$ operators, components that solely involve $a^{w,l}$ and $a^{w,l\dagger}$ operators, and mixed terms encompassing products of at least one operator from each of the aforementioned groups.

The terms involving $a^{s,k}$ and $a^{s,k\dagger}$ operators symbolize the physics at the scale $1/2^{k}$, while the terms exclusively with $a^{w,l}$ and $a^{w,l\dagger}$ operators capture the physics at scales finer than $1/2^{k}$ that do not couple with the scale $1/2^{k}$ operators. Whereas, the mixed term denotes the coupling between degrees of freedom at the scale $1/2^{k}$ and those at finer scales.

The scaler field Hamiltonian Eq. (\ref{eq:scalar_field_hamiltonian}), can be expressed in the following form in this basis as:
\begin{eqnarray}
H= H_{ss}+H_{ww}+H_{sw},
\end{eqnarray}
where,
\begin{eqnarray}
H_{ss}=\frac{1}{2}\left(\sum_{\textbf{n}}{\Pi^{s,k}}^2\left(\textbf{n},t\right)+\sum_{\textbf{m},\textbf{n}}T^k_{ss,\textbf{m}\textbf{n}}\Phi^{s,k}\left(\textbf{m},t\right)\Phi^{s,k}\left(\textbf{n},t\right)+\sum_{\textbf{n}}{\Phi^{s,k}}^2\left(\textbf{n},t\right)\right),
\end{eqnarray}
\begin{eqnarray}
H_{ww}&=&\frac{1}{2}\left(\sum_{\textbf{n}}{\Pi^{w,l}}^2\left(\textbf{n},\alpha, t\right)+\sum_{\textbf{m},\textbf{n},\alpha, \beta, (l,r)\geq k}T^{lr}_{w(\alpha)w(\beta),\textbf{m}\textbf{n}}\Phi^{w,l}\left(\textbf{m},\alpha, t\right)\Phi^{w,r}\left(\textbf{n},\beta, t\right)\right.\nonumber\\
&&\left. +\sum_{\textbf{n}}{\Phi^{w,l}}^2\left(\textbf{n},\alpha, t\right)\right),
\end{eqnarray}
\begin{eqnarray}
H_{sw}&=&\frac{1}{2}\left(\sum_{\textbf{m},\textbf{n},\alpha,l\geq k}T^{lk}_{w(\alpha)s,\textbf{m}\textbf{n}}\Phi^{w,l}(\textbf{m},\alpha, t)\Phi^{s,k}(\textbf{n},t)\right.\nonumber \\
&& \left. +\sum_{\textbf{m},\textbf{n},\beta , r\geq k}T^{kr}_{sw(\beta),\textbf{m}\textbf{n}} \Phi^{s,k}(\textbf{m},t)\Phi^{w,r}(\textbf{n},\beta, t)\right).
\end{eqnarray}

Utilising Eq. (\ref{eq:phi_in_a_and_a_dagger}) and Eq. (\ref{eq:pi_in_a_and_a_dagger}), the Hamiltonian can be rewritten in terms of the creation ($a$) and annihilation ($a^{\dagger}$) operators as,
\begin{eqnarray}
\label{eq:hamiltonian_H_ss}
H_{ss}&:=&\frac{1}{4}\sum_{\textbf{m},\textbf{n}}\left[\left(\frac{1}{\gamma^{s,k}}\left(T^k_{ss,\textbf{m}\textbf{n}}+\mu^2\delta_{\textbf{m}\textbf{n}}\right)-\gamma^{s,k}\delta_{\textbf{m}\textbf{n}}\right)a^{s,k}\left(\textbf{m},t\right)a^{s,k}\left(\textbf{n},t\right)\right.\nonumber\\
&&+\left(\frac{1}{\gamma^{s,k}}\left(T^k_{ss,\textbf{m}\textbf{n}}+\mu^2\delta_{\textbf{m}\textbf{n}}\right)-\gamma^{s,k}\delta_{\textbf{m}\textbf{n}}\right)a^{s,k\dagger}\left(\textbf{m},t\right)a^{s,k\dagger}\left(\textbf{n},t\right)\nonumber\\
&&\left. +\left(\frac{2}{\gamma^{s,k}}\left(T^k_{ss,\textbf{m}\textbf{n}}+\mu^2\delta_{\textbf{m}\textbf{n}}\right)+2\gamma^{s,k}\delta_{\textbf{m}\textbf{n}}\right)a^{s,k\dagger}\left(\textbf{m},t\right)a^{s,k}\left(\textbf{n},t\right)\right],
\end{eqnarray}

\begin{eqnarray}
H_{ww}&:=&\frac{1}{4}\sum_{\textbf{m},\textbf{n},\alpha,\beta,(l,r)\geq k}\left[\left(\frac{1}{\sqrt{\gamma^{w,\alpha, l}\gamma^{w,\beta, r}}}\left(T^{lr}_{w(\alpha)w(\beta),\textbf{m}\textbf{n}}+\mu^2 \delta_{\textbf{m}\textbf{n}}\delta_{lr}\delta_{\alpha\beta}\right)\right.\right.\nonumber\\
&&-\left.\sqrt{\gamma^{w,\alpha , l}\gamma^{w,\beta , r}}\delta_{\textbf{m}\textbf{n}}\delta_{lr}\delta_{\alpha \beta}\right)a^{w,l}(\textbf{m},\alpha ,t)a^{w,r}(\textbf{n},\beta , t)+\left(\frac{1}{\sqrt{\gamma^{w,\alpha , l}\gamma^{w, \beta , r}}}\right.\nonumber\\
&&\left.\left(T^{lr}_{w(\alpha)w(\beta),\textbf{m}\textbf{n}}+\mu^2\delta_{\textbf{m}\textbf{n}}\delta_{lr}\delta_{\alpha\beta}\right)-\sqrt{\gamma^{w,\alpha ,l}\gamma^{w,\beta ,r}}\delta_{\textbf{m}\textbf{n}}\delta_{lr}\delta_{\alpha \beta} \right)\nonumber\\
&& a^{w,l\dagger}(\textbf{m},\alpha , t)a^{w,r\dagger}(\textbf{n},\beta , t)+2\left(\frac{1}{\sqrt{\gamma^{w,\alpha , l}\gamma^{w,\beta , r}}}\left(T^{lr}_{w(\alpha)w(\beta),\textbf{m}\textbf{n}}\right.\right.\nonumber\\
&& +\left. \mu^2 \delta_{\textbf{m}\textbf{n}}\delta_{lr}\delta_{\alpha\beta}\right)+\left.\left.\sqrt{\gamma^{w,\alpha , l}\gamma^{w,\beta , r}}\delta_{\textbf{m}\textbf{n}}\delta_{lr}\delta_{\alpha\beta}\right)a^{w,l\dagger}(\textbf{m},\alpha ,t)a^{w,r}(\textbf{n},\beta ,t)\right],\quad
\end{eqnarray}

\begin{eqnarray}
H_{sw}&:=&\frac{1}{4}\left[\sum_{\textbf{m},\textbf{n},\beta, r\geq k}\frac{T^{kr}_{sw(\beta),\textbf{m}\textbf{n}}}{\sqrt{\gamma^{s,k}\gamma^{w,\beta, r}}}\left(a^{s,k}(\textbf{m},t)a^{w,q}(\textbf{n},\beta, t)+a^{s,k\dagger}(\textbf{m},t)a^{w,r}(\textbf{n},\beta, t) \right.\right.\nonumber\\
&& + \left.a^{w,r\dagger}(\textbf{m},\beta, t)a^{s,k}(\textbf{n},t)+a^{s,k\dagger}(\textbf{m},t)a^{w,r\dagger}(\textbf{n},\beta, t)\right)+\sum_{\textbf{m},\textbf{n},\alpha, l\geq k}\frac{T^{lk}_{w(\alpha)s,\textbf{m}\textbf{n}}}{\sqrt{\gamma^{s,k}\gamma^{w,\alpha,l}}}\nonumber\\
&& \left(a^{w,l}(\textbf{m},\alpha, t)a^{s,k}(\textbf{n},t)+a^{w,l\dagger}(\textbf{m},\alpha, t)a^{s,k}(\textbf{n}, t)+a^{s,k\dagger}(\textbf{n},t)a^{w,l}(\textbf{m},\alpha, t)\right. \nonumber \\
&&\left.\left. +a^{w,l\dagger}(\textbf{m},\alpha,t)a^{s,k\dagger}(\textbf{n},t)\right)\right],
\end{eqnarray}
and the coefficients $T^k_{ss,\textbf{m}\textbf{n}}$, $T^{lr}_{w(\alpha)w(\beta),\textbf{m}\textbf{n}}$ and $T^{lk}_{sw(\alpha),\textbf{m}\textbf{n}}$ are given by,
\begin{eqnarray}
\label{eq:3d_T_k_ss}
T^k_{ss,\textbf{m}\textbf{n}}&=&\int d^3\textbf{x} \nabla s^k_{\textbf{m}}(\textbf{x}).\nabla s^k_{\textbf{n}}(\textbf{x}),\\
\label{eq:3d_T_k_ww}
T^{lr}_{w(\alpha)w(\beta),\textbf{m} \textbf{n}}&=&\int d^3\textbf{x} \nabla w^l_{\textbf{m},\alpha}(\textbf{x}).\nabla w^r_{\textbf{n},\beta}(\textbf{x}),\\
\label{eq:3d_T_k_sw}
T^{kl}_{sw(\alpha),\textbf{m}\textbf{n}}&=&\int d^3\textbf{x} \nabla s^k_{\textbf{m}}(\textbf{x}).\nabla w^l_{\textbf{n},\alpha}(\textbf{x}).
\end{eqnarray}
The integral in equation (\ref{eq:3d_T_k_ss}), can be expressed as the sum of products of integrals, each involving the product of one-dimensional derivatives of scaling functions and delta functions,
\begin{eqnarray}
T^k_{ss,\textbf{m}\textbf{n}}=T^k_{ss,m_1 n_1}\delta_{m_2 n_2}\delta_{m_3 n_3}+\delta_{m_1 n_1}T^k_{ss,m_2 n_2}\delta_{m_3 n_3}+\delta_{m_1 n_1}\delta_{m_2 n_2}T^k_{ss,m_3 n_3},
\end{eqnarray}
where,
\begin{eqnarray}
\label{eq:1d_T_k_ss}
T^k_{ss,mn}=\int dx \frac{\partial s^k_m(x)}{\partial x}\frac{\partial s^k_n(x)}{\partial x}.
\end{eqnarray}
The integral $T^{lr}_{w(\alpha)w(\beta),\textbf{m}\textbf{n}}$ and $T^{kl}_{sw(\alpha),\textbf{m}\textbf{n}}$ can manifest in $64$ and $8$ different forms respectively, each characterized by the summation of the product. The product can either involve integrals with the product of derivatives of scaling functions and delta functions, or integrals with the product of derivatives of wavelet functions and delta functions. This is exemplified in the equations below taking $\alpha=1$ and $\beta=1$,
\begin{eqnarray}
T^{lr}_{w(1)w(1),\textbf{m}\textbf{n}}&=&T^l_{ss,m_1 n_1} \delta_{m_2 n_2} \delta_{l_1 j_1}\delta_{m_3 n_3}+\delta_{m_1 n_1} T^l_{ss,m_2 n_2}  \delta_{l_1 r_1}\delta_{m_3 n_3}+\nonumber\\
&&\delta_{m_1 n_1} \delta_{m_2 n_2}  T^{l_1 j_1}_{ww,m_3 n_3},\quad\quad l_1,r_1\geq l, \\
T^{kl}_{sw(1),\textbf{m}\textbf{n}}&=& \delta_{m_1 n_1} \delta_{m_2 n_2} T^{k l_1}_{sw,m_3 n_3}, \quad l_1\geq l,
\end{eqnarray}
where, $T^k_{ss,mn}$ is given in Eq. (\ref{eq:1d_T_k_ss}) and $T^{lr}_{ww,mn}$ is given in the following equation,
\begin{eqnarray}
\label{eq:1d_T_k_ww}
T^{lr}_{ww,mn}&=&\int dx \frac{\partial w^l_m (x)}{dx} \frac{\partial w^r_n (x)}{dx},\\
\label{eq:1d_T_k_sw}
T^{kl}_{ww,mn}&=&\int dx \frac{\partial w^l_m (x)}{dx} \frac{\partial s^k_n (x)}{dx}.
\end{eqnarray}
The existence of derivatives for the basis functions in Eq. (\ref{eq:1d_T_k_ss}), Eq. (\ref{eq:1d_T_k_ww}), and Eq. (\ref{eq:1d_T_k_sw}) are limited to scaling functions with $K\geq 3$. The computation of these integrals is provided in Appendix \ref{appen:the_kinetic_energy_term}. These integrals exhibit locality, as they vanish when the support of the basis functions does not overlap.

Now, we will present the calculation for evaluating the Hamiltonian matrix element of the free scalar field Hamiltonian in $1+1$ dimensions within this framework. A vacuum state in $1+1$ dimensions can be defined as,
\begin{eqnarray}
\ket{0^{s,k}_0,0^{s,k}_1,0^{s,k}_2,...,0^{s,k}_p,...,0^{s,k}_N:0^{w,k}_0,0^{w,k}_1,0^{w,k}_2,...,0^{w,k}_p,...,0^{w,k}_N},
\end{eqnarray}
here, the subscripts indicate the positions of the particles, and the superscripts denote the scaling and wavelet component particles, respectively.

A scaling and wavelet particle of resolution $k$ can be created at position $p$ by applying the operators $a^{s,k\dagger}(p,t)$ and $a^{w,k\dagger}(p,t)$ on the vacuum state,
\begin{eqnarray}
&&\ket{0^{s,k}_0,0^{s,k}_1,0^{s,k}_2,...,1^{s,k}_p,...,0^{s,k}_N:0^{w,k}_0,0^{w,k}_1,0^{w,k}_2,...,0^{w,k}_p,...,0^{w,k}_N}\nonumber \\
&=& a^{s,k\dagger}(p,t)\ket{0^{s,k}_0,0^{s,k}_1,0^{s,k}_2,...,0^{s,k}_p,...,0^{s,k}_N:0^{w,k}_0,0^{w,k}_1,0^{w,k}_2,...,0^{w,k}_p,...,0^{w,k}_N},\\
&&\ket{0^{s,k}_0,0^{s,k}_1,0^{s,k}_2,...,0^{s,k}_p,...,0^{s,k}_N:0^{w,k}_0,0^{w,k}_1,0^{w,k}_2,...,1^{w,k}_p,...,0^{w,k}_N}\nonumber \\
&=& a^{w,k\dagger}(p,t)\ket{0^{s,k}_0,0^{s,k}_1,0^{s,k}_2,...,0^{s,k}_p,...,0^{s,k}_N:0^{w,k}_0,0^{w,k}_1,0^{w,k}_2,...,0^{w,k}_p,...,0^{w,k}_N}.
\end{eqnarray}
A state having $n0$, $n1$, $n2$,..., $nN$ numbers of scaling particles and $n0$, $n1$, $n2$,..., $nN$ number of wavelet particles of resolution $k$ at position $0$, $1$, $2$,..., $N$ respectively can be defined as follows,
\begin{eqnarray}
\label{eq:the_generic_state_within_wavelet_based_framework}
&&\ket{n0^{s,k}_0,n1^{s,k}_1,n2^{s,k}_2,...,np^{s,k}_p,...,nN^{s,k}_N:n0^{w,k}_0,n1^{w,k}_1,n2^{w,k}_2,...,np^{w,k}_p,...,nN^{w,k}_N}\nonumber\\
&=& \frac{\left(a^{s,k\dagger}(0,t)\right)^{n0}}{\sqrt{n0!}}\frac{\left(a^{s,k\dagger}(1,t)\right)^{n1}}{\sqrt{n1!}}\frac{\left(a^{s,k\dagger}(2,t)\right)^{n2}}{\sqrt{n2!}}...\frac{\left(a^{s,k\dagger}(p,t)\right)^{np}}{\sqrt{np!}}...\frac{\left(a^{s,k\dagger}(N,t)\right)^{nN}}{\sqrt{nN!}}\nonumber\\
&&\frac{\left(a^{w,k\dagger}(0,t)\right)^{n0}}{\sqrt{n0!}}\frac{\left(a^{w,k\dagger}(1,t)\right)^{n1}}{\sqrt{n1!}}\frac{\left(a^{w,k\dagger}(2,t)\right)^{n2}}{\sqrt{n2!}}...\frac{\left(a^{w,k\dagger}(p,t)\right)^{np}}{\sqrt{np!}}...\frac{\left(a^{w,k\dagger}(N,t)\right)^{nN}}{\sqrt{nN!}}\nonumber\\
&&\ket{0^{s,k}_0,0^{s,k}_1,0^{s,k}_2,...,1^{s,k}_p,...,0^{s,k}_N:0^{w,k}_0,0^{w,k}_1,0^{w,k}_2,...,0^{w,k}_p,...,0^{w,k}_N}.
\end{eqnarray}
Applying the operators $a^{s,k\dagger}(p,t)$ and $a^{s,k}(p,t)$ to the above state create and annihilate a scaling particle, while $a^{w,k\dagger}(p,t)$, and $a^{w,k}(p,t)$ will create and annihilate a wavelet particle at position $p$, respectively,
\begin{eqnarray}
&&a^{s,k\dagger}(p,t)\ket{n0^{s,k}_0,n1^{s,k}_1,n2^{s,k}_2,...,np^{s,k}_p,...,nN^{s,k}_N:n0^{w,k}_0,n1^{w,k}_1,n2^{w,k}_2,...,np^{w,k}_p,...,nN^{w,k}_N}\nonumber\\
&=&\sqrt{np^{s,k}+1^{s,k}} \nonumber\\
&&\ket{n0^{s,k}_0,n1^{s,k}_1,n2^{s,k}_2,...,(np^{s,k}+1^{s,k})_p,...,nN^{s,k}_N:n0^{w,k}_0,n1^{w,k}_1,n2^{w,k}_2,...,np^{w,k}_p,...,nN^{w,k}_N},\nonumber\\ \\
&&a^{s,k}(p,t)\ket{n0^{s,k}_0,n1^{s,k}_1,n2^{s,k}_2,...,np^{s,k}_p,...,nN^{s,k}_N:n0^{w,k}_0,n1^{w,k}_1,n2^{w,k}_2,...,np^{w,k}_p,...,nN^{w,k}_N}\nonumber\\
&=&\sqrt{np^{s,k}} \nonumber\\
&&\ket{n0^{s,k}_0,n1^{s,k}_1,n2^{s,k}_2,...,(np^{s,k}-1^{s,k})_p,...,nN^{s,k}_N:n0^{w,k}_0,n1^{w,k}_1,n2^{w,k}_2,...,np^{w,k}_p,...,nN^{w,k}_N},\nonumber\\ 
\\
&&a^{w,k\dagger}(p,t)\ket{n0^{s,k}_0,n1^{s,k}_1,n2^{s,k}_2,...,np^{s,k}_p,...,nN^{s,k}_N:n0^{w,k}_0,n1^{w,k}_1,n2^{w,k}_2,...,np^{w,k}_p,...,nN^{w,k}_N}\nonumber\\
&=&\sqrt{np^{w}+1^{w}} \nonumber\\
&&\ket{n0^{s,k}_0,n1^{s,k}_1,n2^{s,k}_2,...,np^{s,k}_p,...,nN^{s,k}_N:n0^{w,k}_0,n1^{w,k}_1,n2^{w,k}_2,...,(np^{w,k}_p+1^{w,k}),...,nN^{w,k}_N},\nonumber\\ \\
&&a^{w,k}(p,t)\ket{n0^{s,k}_0,n1^{s,k}_1,n2^{s,k}_2,...,np^{s,k}_p,...,nN^{s,k}_N:n0^{w,k}_0,n1^{w,k}_1,n2^{w,k}_2,...,np^{w,k}_p,...,nN^{w,k}_N}\nonumber\\
&=&\sqrt{np^{w,k}}\nonumber\\
&&\ket{n0^{s,k}_0,n1^{s,k}_1,n2^{s,k}_2,...,np^{s,k}_p,...,nN^{s,k}_N:n0^{w,k}_0,n1^{w,k}_1,n2^{w,k}_2,...,(np^{w,k}_p-1^{w,k}),...,nN^{w,k}_N}.\nonumber\\
\end{eqnarray}
The product of the creation and the annihilation operators appearing on the Hamiltonian act on the state defined in Eq. (\ref{eq:the_generic_state_within_wavelet_based_framework}) in the following manner,
\begin{eqnarray}
&& a^{s,k\dagger}(p,t)a^{s,k}(q,t) \nonumber \\ &&\ket{n0^{s,k}_0,n1^{s,k}_1,n2^{s,k}_2,...,np^{s,k}_p,...,nN^{s,k}_N:n0^{w,k}_0,n1^{w,k}_1,n2^{w,k}_2,...,np^{w,k}_p,...,nN^{w,k}_N}\nonumber\\
&=&
\begin{cases}
np^s\ket{n0^{s,k}_0,n1^{s,k}_1,n2^{s,k}_2,...,np^{s,k}_p,...,nN^{s,k}_N:n0^{w,k}_0,n1^{w,k}_1,n2^{w,k}_2,...,np^{w,k}_p,...,nN^{w,k}_N}\\
\text{for}\quad (p=q),\\
\sqrt{(np^s-1)(nq^s-1)}\\
\ket{n0^{s,k}_0,n1^{s,k}_1,...,(np^{s,k}_p+1^{s,k}),...,(nq^{s,k}_q+1^{s,k}),...,nN^{s,k}_N:n0^{w,k}_0,n1^{w,k}_1,...,nN^{w,k}_N}\\
\quad \text{for} \quad (p\neq q),
\end{cases}\nonumber\\
\end{eqnarray}
\begin{eqnarray}
&& a^{s,k\dagger}(p,t)a^{s,k\dagger}(q,t)\nonumber\\
&&\ket{n0^{s,k}_0,n1^{s,k}_1,n2^{s,k}_2,...,np^{s,k}_p,...,nN^{s,k}_N:n0^{w,k}_0,n1^{w,k}_1,n2^{w,k}_2,...,np^{w,k}_p,...,nN^{w,k}_N}\nonumber\\
&=&
\begin{cases}
\sqrt{(np^s+1)(np^s+2)}\\
\ket{n0^{s,k}_0,n1^{s,k}_1,...,(np^{s,k}_p+2^{s,k}),...,nN^{s,k}_N:n0^{w,k}_0,n1^{w,k}_1,n2^{w,k}_2,...,np^{w,k}_p,...,nN^{w,k}_N}\\
\quad\text{for}\quad p=q,\\
\sqrt{(np^s+1)(nq^s-1)}\\
\ket{n0^{s,k}_0,n1^{s,k}_1,...,(np^{s,k}_p+1^{s,k}),...,(nq^{s,k}_q-1^s),...,nN^{s,k}_N:n0^{w,k}_0,n1^{w,k}_1,...,nN^{w,k}_N}\\
\quad \text{for} \quad p\neq q,
\end{cases}
\end{eqnarray}
\begin{eqnarray}
&& a^{s,k}(p,t)a^{s,k}(q,t)\nonumber\\
&&\ket{n0^{s,k}_0,n1^{s,k}_1,n2^{s,k}_2,...,np^{s,k}_p,...,nN^{s,k}_N:n0^{w,k}_0,n1^{w,k}_1,n2^{w,k}_2,...,np^{w,k}_p,...,nN^{w,k}_N}\nonumber\\
&=&
\begin{cases}
\sqrt{np^s(np^s-1)}\\
\ket{n0^{s,k}_0,n1^{s,k}_1,...,(np^{s,k}_p-2^{s,k}),...,nN^{s,k}_N:n0^{w,k}_0,n1^{w,k}_1,n2^{w,k}_2,...,np^{w,k}_p,...,nN^{w,k}_N}\\
\quad\text{for}\quad p=q,\\
\sqrt{(np^s-1)(nq^s-1)}\\
\ket{n0^{s,k}_0,n1^{s,k}_1,...,(np^{s,k}_p-1^s),...,(nq^{s,k}_q-1^{s,k}),...,nN^{s,k}_N:n0^{w,k}_0,n1^{w,k}_1,...,nN^{w,k}_N}\\
\quad \text{for} \quad p\neq q,
\end{cases}
\end{eqnarray}
\begin{eqnarray}
&& a^{w,k\dagger}(p,t)a^{w,k}(q,t)\nonumber\\
&&\ket{n0^{s,k}_0,n1^{s,k}_1,n2^{s,k}_2,...,np^{s,k}_p,...,nN^{s,k}_N:n0^{w,k}_0,n1^{w,k}_1,n2^{w,k}_2,...,np^{w,k}_p,...,nN^{w,k}_N}\nonumber\\
&=&
\begin{cases}
np^{w,k}\\
\ket{n0^{s,k}_0,n1^{s,k}_1,n2^{s,k}_2,...,np^{s,k}_p,...,nN^{s,k}_N:n0^{w,k}_0,n1^{w,k}_1,n2^{w,k}_2,...,np^{w,k}_p,...,nN^{w,k}_N}\\
\quad\text{for}\quad (p=q),\\
\sqrt{(np^{w,k}-1^{w,k})(nq^{w,k}-1^{w,k})}\\
\ket{n0^{s,k}_0,n1^{s,k}_1,...,nN^{s,k}_N:n0^{w,k}_0,n1^{w,k}_1,...,(np^{w,k}_p+1^{w,k}),...,(nq^{w,k}_q+1^{w,k}),...,nN^{w,k}_N}\\
\quad \text{for} \quad (p\neq q),
\end{cases}\nonumber\\
\end{eqnarray}
\begin{eqnarray}
&& a^{w,k\dagger}(p,t)a^{w,k\dagger}(q,t)\nonumber\\
&&\ket{n0^{s,k}_0,n1^{s,k}_1,n2^{s,k}_2,...,np^{s,k}_p,...,nN^{s,k}_N:n0^{w,k}_0,n1^{w,k}_1,n2^{w,k}_2,...,np^{w,k}_p,...,nN^{w,k}_N}\nonumber\\
&=&
\begin{cases}
\sqrt{(np^{w,k}+1^{w,k})(np^{w,k}+2^{w,k})}\\
\ket{n0^{s,k}_0,n1^{s,k}_1,...,nN^{s,k}_N:n0^{w,k}_0,n1^{w,k}_1,n2^{w,k}_2,...,(np^{w,k}_p+2^{w,k}),...,nN^{w,k}_N}\\
\quad\text{for}\quad (p=q),\\
\sqrt{(np^{w,k}+1^{w,k})(nq^{w,k}-1^{w,k})}\\
\ket{n0^{s,k}_0,n1^{s,k}_1,...,nN^{s,k}_N:n0^{w,k}_0,n1^{w,k}_1,...,(np^{w,k}_p+1^{w,k}),...,(nq^{w,k}_q-1^{w,k})...,nN^{w,k}_N}\\
\quad \text{for} \quad (p\neq q),
\end{cases}
\end{eqnarray}
\begin{eqnarray}
&& a^{s,k}(p,t)a^{s,k}(q,t)\nonumber\\
&&\ket{n0^{s,k}_0,n1^{s,k}_1,n2^{s,k}_2,...,np^{s,k}_p,...,nN^{s,k}_N:n0^{w,k}_0,n1^{w,k}_1,n2^{w,k}_2,...,np^{w,k}_p,...,nN^{w,k}_N}\nonumber\\
&=&
\begin{cases}
\sqrt{np^{w,k}(np^{w,k}-1^{w,k})}\\
\ket{n0^{s,k}_0,n1^{s,k}_1,...,nN^{s,k}_N:n0^{w,k}_0,n1^{w,k}_1,...,(np^{w,k}_p-2^{w,k}),...,nN^{w,k}_N}\\
\quad\text{for}\quad (p=q),\\
\sqrt{(np^{w,k}-1^{w,k})(nq^{w,k}-1^{w,k})}\\
\ket{n0^{s,k}_0,n1^{s,k}_1,...,nN^{s,k}_N:n0^{w,k}_0,n1^{w,k}_1,...,(np^{w,k}_p-1^{w,k}),...,(nq^{w,k}_q-1^{w,k}),...,nN^{w,k}_N}\\
\quad \text{for} \quad (p\neq q).
\end{cases}
\end{eqnarray}
So, the matrix element in this basis is given by,
\begin{eqnarray}
&& \bra{m0^s_0,m1^s_1,...,mN^s_N:m0^w_0,m1^w_1,...,mN^w_N}H_{ss}
\ket{n0^s_0,n1^s_1,...,nN^s_N:n0^w_0,n1^w_1,...,nN^w_N}\nonumber \\
\label{eq:hss_in_number_basis}
&&\begin{cases}
=\sum_{p,q}\left(A^{ss}_{p,q}\delta_{m0^{s,k},n0^{s,k}}\delta_{m1^s,n1^s}...\delta_{mp^s,np^s+1^s}...\delta_{mq^{s,k},nq^{s,k}+1^{s,k}}...\delta_{mN^s,nN^s}\delta_{m0^{w},n0^{w}}\delta_{m1^w,n1^w}...\right.\\
\delta_{mN^w,nN^w}+B^{ss}_{p,q}\delta_{m0^{s},n0^{s}}\delta_{m1^s,n1^s}...\delta_{mp^s,np^s+1^s}...\delta_{mq^{s},nq^{s}-1^s}...\delta_{mN^s,nN^s}\delta_{m0^{w},n0^{w}}\delta_{m1^w,n1^w}\\
 ...\delta_{mN^w,nN^w}+C^{ss}_{p,q}\delta_{m0^{s},n0^{s}}\delta_{m1^s,n1^s}...\delta_{mp^s,np^s-1^s}...\delta_{mq^{s},nq^{s}-1^s}...\delta_{mN^s,nN^s}\delta_{m0^{w},n0^{w}}\\
\delta_{m1^{w},n1^{w}}...\delta_{mN^{w},nN^{w}}\quad\text{for}\quad (p\neq q),\\
=\sum_{p,q}\left(A^{ss}_{p,q}\delta_{m0^{s},n0^{s}}\delta_{m1^s,n1^s}...\delta_{mp^s,np^s+2^s}...\delta_{mN^s,nN^s}\delta_{m0^{w},n0^{w}}\delta_{m1^w,n1^w}...\delta_{mN^w,nN^w}\right.\\
+B^{ss}_{p,q}\delta_{m0^{s},n0^{s}}\delta_{m1^s,n1^s}...\delta_{mp^s,np^s}...\delta_{mN^s,nN^s}\delta_{m0^{w},n0^{w}}\delta_{m1^w,n1^w}...\delta_{mN^w,nN^w}\\
+C^{ss}_{p,q}\delta_{m0^{s},n0^{s}}\delta_{m1^s,n1^s}...\delta_{mp^s,np^s-2^s}...\delta_{mN^s,nN^s}\delta_{m0^{w},n0^{w}}\delta_{m1^{w},n1^{w}}...\delta_{mN^{w},nN^{w}}
\,\, \text{for}\,\, (p=q),
\end{cases}
\end{eqnarray}
here,
\begin{eqnarray}
A^{ss}_{p,q}=
\begin{cases}
\frac{\sqrt{(np^s+1^s)(nq^s+1^s)}}{4}\left(\frac{T^k_{ss,pq}}{\gamma^{s,k}}\right)\,\,\text{for}\,\,(p\neq q),\\
\frac{\sqrt{(np^s+1^s)(np^s+2^s)}}{4}\left(\frac{1}{\gamma^{s,k}}\left(T^{k}_{ss,pp}+\mu^2\right)-\gamma^{s,k}\right)\,\,\text{for}\,\,(p=q),
\end{cases}
\end{eqnarray}
\begin{eqnarray}
B^{ss}_{p,q}=
\begin{cases}
\frac{\sqrt{(np^s+1^s)nq^s}}{2}\left(\frac{T^k_{ss,pq}}{\gamma^{s,k}}\right)\,\,\text{for}\,\,(p\neq q),\\
\frac{np^s}{2}\left(\frac{1}{\gamma^{s,k}}\left(T^{k}_{ss,pp}+\mu^2\right)-\gamma^{s,k}\right)\,\,\text{for}\,\,(p=q),
\end{cases}
\end{eqnarray}
\begin{eqnarray}
C^{ss}_{p,q}=
\begin{cases}
\frac{\sqrt{np^s nq^s}}{4}\left(\frac{T^k_{ss,pq}}{\gamma^{s,k}}\right)\,\,\text{for}\,\,(p\neq q),\\
\frac{\sqrt{np^s(np^s-1)}}{4}\left(\frac{1}{\gamma^{s,k}}\left(T^{k}_{ss,pp}+\mu^2\right)-\gamma^{s,k}\right)\,\,\text{for}\,\,(p=q),
\end{cases}
\end{eqnarray}
\begin{eqnarray}
&&\bra{m0^s_0,m1^s_1,...,mN^s_N:m0^w_0,m1^w_1,...,mN^w_N}H_{ww}
\ket{n0^s_0,n1^s_1,...,nN^s_N:n0^w_0,n1^w_1,...,nN^w_N}\nonumber \\
\label{eq:hww_in_number_basis}
&&\begin{cases}
=\sum_{p,q}\left(A^{ww}_{p,q}\delta_{m0^{s},n0^{s}}\delta_{m1^s,n1^s}...\delta_{mN^s,nN^s}\delta_{m0^{w},n0^{w}}\delta_{m1^w,n1^w}...\delta_{mp^{w},np^{w}+1^{w}}...\delta_{mq^{w},nq^{w}+1^w}...\right.\\
\delta_{mN^w,nN^w}+B^{ww}_{p,q}\delta_{m0^{s},n0^{s}}\delta_{m1^s,n1^s}...\delta_{mN^s,nN^s}\delta_{m0^{w},n0^{w}}\delta_{m1^w,n1^w}...\delta_{mp^{w},np^{w}+1^{w}}...\delta_{mq^{w},nq^{w}-1^{w}}\\
...\delta_{mN^w,nN^w}+C^{ww}_{p,q}\delta_{m0^{s},n0^{s}}\delta_{m1^s,n1^s}...\delta_{mN^s,nN^s}\delta_{m0^{w},n0^{w}}\delta_{m1^{w},n1^{w}}...\delta_{mp^{w},np^{w}-1^{w}}...\\
\delta_{mq^{w},nq^{w}-1^{w}}...\delta_{mN^{w},nN^{w}}\quad\text{for}\quad (p\neq q),\\
=\sum_{p,q}\left(A^{ww}_{p,q}\delta_{m0^{s},n0^{s}}\delta_{m1^s,n1^s}...\delta_{mN^s,nN^s}\delta_{m0^{w},n0^{w}}\delta_{m1^w,n1^w}...\delta_{mp^{w},np^{w}+2^{w}}...\delta_{mN^w,nN^w}\right.\\
+B^{ww}_{p,q}\delta_{m0^{s},n0^{s}}\delta_{m1^s,n1^s}...\delta_{mp^s,np^s}...\delta_{mN^s,nN^s}\delta_{m0^{w},n0^{w}}\delta_{m1^w,n1^w}...\delta_{mN^w,nN^w}\\
+C^{ww}_{p,q}\delta_{m0^{s},n0^{s}}\delta_{m1^s,n1^s}...\delta_{mN^s,nN^s}\delta_{m0^{w},n0^{w}}\delta_{m1^{w},n1^{w}}...\delta_{mp^{w},np^{w}-2^{w}}...\delta_{mN^{w},nN^{w}}\,\,\text{for}\,\, (p=q),
\end{cases}
\end{eqnarray}
here,
\begin{eqnarray}
A^{ww}_{p,q}=
\begin{cases}
\frac{\sqrt{(np^w+1^w)(nq^w+1^w)}}{4}\left(\frac{T^{kk}_{ww,pq}}{\gamma^{w,k}}\right)\,\,\text{for}\,\,(p\neq q),\\
\frac{\sqrt{(np^w+1^w)(np^w+2^w)}}{4}\left(\frac{1}{\gamma^{w,k}}\left(T^{kk}_{ww,pp}+\mu^2\right)-\gamma^{w,k}\right)\,\,\text{for}\,\,(p=q),
\end{cases}
\end{eqnarray}
\begin{eqnarray}
B^{ww}_{p,q}=
\begin{cases}
\frac{\sqrt{(np^w+1^w)nq^w}}{2}\left(\frac{T^{kk}_{ww,pq}}{\gamma^{w,k}}\right)\,\,\text{for}\,\,(p\neq q),\\
\frac{np^w}{2}\left(\frac{1}{\gamma^{w,k}}\left(T^{kk}_{ww,pp}+\mu^2\right)-\gamma^{w,k}\right)\,\,\text{for}\,\,(p=q),
\end{cases}
\end{eqnarray}
\begin{eqnarray}
C^{ww}_{p,q}=
\begin{cases}
\frac{\sqrt{np^w nq^w}}{4}\left(\frac{T^{kk}_{ww,pq}}{\gamma^{w,k}}\right)\,\,\text{for}\,\,(p\neq q),\\
\frac{\sqrt{np^w(np^w-1)}}{4}\left(\frac{1}{\gamma^{w,k}}\left(T^{kk}_{ww,pp}+\mu^2\right)-\gamma^{w,k}\right)\,\,\text{for}\,\,(p=q),
\end{cases}
\end{eqnarray}
\begin{eqnarray}
\bra{m0^s_0,m1^s_1,...,mN^s_N:m0^w_0,m1^w_1,...,mN^w_N}H_{sw}
\ket{n0^s_0,n1^s_1,...,nN^s_N:n0^w_0,n1^w_1,...,nN^w_N}\nonumber \\
\label{eq:hsw_in_number_basis}
\begin{cases}
=\sum_{p,q}\frac{T^{kk}_{sw,pq}}{4\sqrt{\gamma^{s,k}\gamma^{w,k}}}\left(A^{sw}_{p,q}\delta_{m0^{s},n0^{s}}\delta_{m1^s,n1^s}...\delta_{mp^s,np^s-1^s}...\delta_{mN^s,nN^s}\delta_{m0^{w},n0^{w}}\delta_{m1^w,n1^w}...\right.\\
\delta_{mq^{w},nq^{w}-1^w}...\delta_{mN^w,nN^w}+B^{sw}_{p,q}\delta_{m0^{s},n0^{s}}\delta_{m1^s,n1^s}...\delta_{mp^s,np^s+1^s}...\delta_{mN^s,nN^s}\delta_{m0^{w},n0^{w}}\\
\delta_{m1^w,n1^w}...\delta_{mq^{w},nq^{w}-1^w}...\delta_{mN^w,nN^w}+C^{sw}_{p,q}\delta_{m0^{s},n0^{s}}\delta_{m1^s,n1^s}...\delta_{mp^s,np^s-1^s}...\delta_{mN^s,nN^s}\\
\delta_{m0^{w},n0^{w}}\delta_{m1^w,n1^w}...\delta_{mq^{w},nq^{w}+1^w}...\delta_{mN^{w},nN^{w}}+D^{sw}_{p,q}\delta_{m0^{s},n0^{s}}\delta_{m1^s,n1^s}...\delta_{mp^s,np^s+1^s}...\\
\left. \delta_{mN^s,nN^s}\delta_{m0^{w},n0^{w}}\delta_{m1^w,n1^w}...\delta_{mq^{w},nq^{w}+1^w}...\delta_{mN^w,nN^w}\right)+\sum_{p,q}\frac{T^{kk}_{ws,pq}}{4\sqrt{\gamma^{w,k}\gamma^{s,k}}}\left(E^{sw}_{pq}\delta_{m0^{s},n0^{s}}\right.\\
\delta_{m1^s,n1^s}...\delta_{mp^s,np^s-1^s}...\delta_{mN^s,nN^s}\delta_{m0^{w},n0^{w}}\delta_{m1^w,n1^w}...\delta_{mq^{w},nq^{w}-1^w}...\delta_{mN^w,nN^w}+F^{sw}_{pq}\\
\delta_{m0^{s},n0^{s}}\delta_{m1^s,n1^s}...\delta_{mp^s,np^s-1^s}...\delta_{mN^s,nN^s}\delta_{m0^{w},n0^{w}}\delta_{m1^w,n1^w}...\delta_{mq^{w},nq^{w}+1^w}...\delta_{mN^w,nN^w}+\\
G^{sw}_{pq}\delta_{m0^{s},n0^{s}}\delta_{m1^s,n1^s}...\delta_{mp^s,np^s+1^s}...\delta_{mN^s,nN^s}\delta_{m0^{w},n0^{w}}\delta_{m1^w,n1^w}...\delta_{mq^{w},nq^{w}-1^w}...\delta_{mN^w,nN^w}\\
+H^{sw}_{pq}\delta_{m0^{s},n0^{s}}\delta_{m1^s,n1^s}...\delta_{mp^s,np^s+1^s}...\delta_{mN^s,nN^s}\delta_{m0^{w},n0^{w}}\delta_{m1^w,n1^w}\\
\left. ...\delta_{mq^{w},nq^{w}+1^w}...\delta_{mN^w,nN^w}\right),
\end{cases}
\end{eqnarray}
where,
\begin{eqnarray}
\begin{cases}
A^{sw}_{p,q}=\sqrt{np^s nq^w},\quad B^{sw}_{p,q}=\sqrt{(np^s+1^s)nq^w},\quad C^{sw}_{p,q}=\sqrt{(np^w+1^w)nq^s},\\
D^{sw}_{p,q}=\sqrt{(np^s+1^s)(nq^w+1^w)},\quad E^{sw}_{p,q}=\sqrt{np^w nq^s},\quad F^{sw}_{p,q}=\sqrt{(np^w+1^w)nq^s},\\
G^{sw}_{p,q}=\sqrt{(np^s+1^s)nq^w},\quad H^{sw}_{p,q}=\sqrt{(np^w+1^w)(nq^s+1^s)}.
\end{cases}
\end{eqnarray}
Adding the three terms given in Eq. (\ref{eq:hss_in_number_basis}), Eq. (\ref{eq:hww_in_number_basis}), and Eq. (\ref{eq:hsw_in_number_basis}), we can get the full Hamiltonian matrix element, which leads to the Hamiltonian eigenvalue equation. We can diagonalize the Hamiltonian to get the energy eigenvalues of the free scalar field theory within a truncated volume. By progressively increasing the resolution we can improve the accuracy of the eigenvalues.

It is instructive to exhibit the structure of resolution $1/2^k$ for a $1+1$ dimensional :$\phi^4(x)$: interaction. It has the following form
\begin{eqnarray}
H^k &=& \frac{1}{2}\sum :\left(\Pi^k_s(n,0)^2+T^k_{ss,mn}\Phi^k_s(m,0)\Phi^k_s(n,0)+\mu^2 \Phi^k_s(n,0)^2 \right.\nonumber\\
&&\left. +\lambda \Gamma^k_{ssss,n_1 n_2 n_3 n_4}\Phi^k_s(n_1,0)\Phi^k_s(n_2,0)\Phi^k_s(n_3,0)\Phi^k_s(n_4,0) \right):
\end{eqnarray}
where the numerical coefficients $T^k_{ss,mn}$ are the overlap integrals given in Eq. (\ref{eq:overlap_integrals_derivative_of_scaling_functions}) and can be evaluated using the method given in the Appendix \ref{appen:the_kinetic_energy_term}. $\Gamma^k_{ssss,n_1 n_2 n_3 n_4}$ are the overlap integrals involving the product of four scaling functions and is given in the following equation:
\begin{eqnarray}
\Gamma^k_{ssss,n_1 n_2 n_3 n_4}=\int s^k_{n_1}(x)s^k_{n_2}(x)s^k_{n_3}(x)s^k_{n_4}(x)dx.
\end{eqnarray}
The evaluation of these integration is given in the Appendix \ref{appen:Overlap_integral_involving_the_product_of_four_scaling_functions}.
\section{Conclusion}
This alternative approach has been first suggested by Wilson \cite{PhysRev.140.B445} in his 1965 work titled ``Model Hamiltonians for Local Quantum Field Theory". One of his aims was to solve the Hamiltonian eigenvalue problem of relativistic QFTs using quantum mechanical techniques that go beyond perturbation theory. In order to establish the validity of these methods, it was important that one could make qualitative order of magnitude estimates of various quantities in the process of computation. This was done using phase space analysis, in which the quantum field was resolved into localized oscillator variables using a complete set of "wavepacket" basis functions. These wavepacket functions were chosen to have specific localization characteristics, and in the absence of an explicit construction, their existence was assumed. The wavepacket functions had properties akin to those of the scaling basis functions of the wavelet theory, which was to take its modern form a couple of decades later. The importance of wavelets for nonperturbative analysis of quantum chromodynamics (QCD) was re-emphasized in an approach pioneered by Wilson et al \cite{PhysRevD.49.6720}. he wavelet analysis presented in this proposal \cite{PhysRevD.49.6720}, while qualitative in nature, depended only on the localized characteristics of wavelets and not on their specific type and form.

The new framework of QFT, described in this chapter, was first advocated by Bulut and Polyzou \cite{PhysRevD.87.116011} by implementing wavelet-based canonical quantization while studying various aspects of Poincare invariance, renormalization group, and gauge invariance from a wavelet perspective. We extended their work by means of the extension of the formalism within the Fock space representation. Further work within this direction of computation of the Fock space basis elements for $\phi^4$ theory is under active development, focusing on improving numerical efficiency, extending to higher truncation levels, and exploring renormalization effects in non-perturbative regimes.

\chapter{Renormalization in a wavelet basis}
\label{chap:renormalization_in_a_wavelet_basis}
The wavelet basis dissects the Hilbert space $L^2(\mathbb{R}^3)$ into an infinite orthogonal direct sum, consisting of successively finer resolution infinite-dimensional subspaces.

We can truncate the Hamiltonian at any desired resolution. The resultant operator resembles a Hamiltonian wherein integrals over fields are substituted by sums of fields, averaged across lattice blocks. The averaging functions in this context are products of scaling functions at a specified scale. In our example of a free field, the truncated Hamiltonian at resolution $1/2^k$ takes the form $H_{ss}$ as specified in Eq. (\ref{eq:hamiltonian_H_ss}).

The inherent scaling properties of the wavelet basis make it natural for the implementation of renormalization group transformations. In the wavelet-based formalism, the procedure starts with a Hamiltonian that is truncated to a specific fixed scale, denoted as $1/2^k$. The truncated Hamiltonian is derived from the formal expansion of the "exact" Hamiltonian by omitting contributions from fields associated with scales smaller than $1/2^k$. The infrared cutoff is achieved by the volume truncation.

The truncated Hamiltonian defines a set of resolution $1/2^k$ Hamiltonians, depending upon the selection of bare parameters $\lambda_{01}, \lambda_{02}, ... \lambda_{0n}$. The initial determination of bare parameters involves computing observables, $O^k_i(\lambda_{01}, ... , \lambda_{0n})$, at $n$ different scales $1/2^k$ and adjusting the bare parameters to align with predetermined ``experimental" values, $O^{k*}_{i}$
\begin{eqnarray}
\label{eq:observable_experimental_and_theoretical}
O^k_i(\lambda_{01},...,\lambda_{0n})=O^{k*}_i\quad\quad 1\leq i\leq n.
\end{eqnarray}
We make the assumption that these equations are solvable, and we can represent the solution of Eq. (\ref{eq:observable_experimental_and_theoretical}) as $\lambda_{01}(k), ..., \lambda_{0n}(k)$, where the factor $k$ denotes the resolution of the truncated Hamiltonian.

The subsequent step in the renormalization process involves examining the category of truncated Hamiltonians at a resolution of $1/2^{k-1}$, meaning a reduction in resolution by a factor of $2$. This Hamiltonians are parametrized by $n$ bare parameters. Again we can solve Eq. (\ref{eq:observable_experimental_and_theoretical}), to get a new set of bare parameters.

We can repeat this process, of solving Eq. (\ref{eq:observable_experimental_and_theoretical}) and getting the new values of bare parameters for different step of renormalization. The new values of bare parameters includes the physics of eliminated degrees of freedom. 

During this procedure, it is necessary to utilize a fixed-size volume cutoff that is sufficiently large to avoid influencing the scale value of $1/2^k$ for the ``experimental" observables, which play a crucial role in determining the sequence of bare coupling constants.

In the next section, we have used an example of the two-dimensional Dirac delta function potential to illustrate aspects of renormalization within the discrete wavelet-based approach. Several authors \cite{10.1119/1.9857,10.1119/1.16691,Jackiw:1991je,10.1119/1.16675,10.1063/1.531271,10.1063/1.532350,cavalcanti2000exact,10.1119/1.19051,CAMBLONG200114,10.1119/1.19485,bound2011lapicki,PhysRevA.65.052123,Erman_2017,ANWONG20182547,Loran_2022,Pazarbası2019} have studied this potential to understand the nuances of renormalization in an elementary setting. Firstly, we will analyze the 2DDF using the Green's function studied by Cavalcanti \cite{cavalcanti2000exact}. Subsequently, we will demonstrate the emergence of asymptotic freedom within the framework of a wavelet-based approach \cite{PhysRevD.107.036015}.
\section{Green's function for 2D Delta-Function Potential, bound state and renormalization}
The 2D DDF has been studied in the literature using various techniques. We opt for the Green's function method due to its close resemblance to the techniques commonly employed in QFT.

The Green's function, $G(E;\textbf{x},\textbf{y})$, associated with the Hamiltonian $H$ is the solution of the differential equation,
\begin{eqnarray}
\label{eq:green's_function_equation_2d_ddf}
\left(E-H\right)G(E;\textbf{x},\textbf{y})=\delta(\textbf{x}-\textbf{y}),
\end{eqnarray}
satisfying the boundary condition
\begin{eqnarray}
\lim_{|\textbf{x}-\textbf{y}|\rightarrow \infty} G(E;\textbf{x},\textbf{y})=0.
\end{eqnarray}
Here, $\textbf{x}$ and $\textbf{y}$ denote points in $D$-dimensional Euclidean space, and $\delta(\textbf{x}-\textbf{y})$ represents a $D$-dimensional Dirac delta function.

We can use the completeness relation of the eigenfunction of $H$ to write the Green's function, $G(E;\textbf{x},\textbf{y})$, in the following form (see Appendix \ref{appen:greens_function} for details),
\begin{eqnarray}
\label{eq:expression_of_greens_function}
G(E;\textbf{x},\textbf{y})=\sum_{\textbf{n}}\frac{\psi_{\textbf{n}}(\textbf{x})\psi^{*}_{\textbf{n}}(\textbf{y})}{E-E_{\textbf{n}}}.
\end{eqnarray}

Let's explore the scenario where the Hamiltonian is written as a sum of two terms,
\begin{eqnarray}
\label{eq:Hamiltonian_free_and_interacting}
H=H_0+\lambda \delta(\textbf{x}).
\end{eqnarray}
If the Green's function corresponding to $H_0$ is known, there exists a straightforward method to determine the Green's function for $H$ \cite{economou2006green}(See Appendix \ref{appen:greens_function}),
\begin{eqnarray}
G(\textbf{x},\textbf{y})&=&G_0(\textbf{x},\textbf{y})+\int d^D z G_0(\textbf{x},\textbf{z})\lambda \delta(\textbf{z}) G(\textbf{z},\textbf{y})\nonumber\\
\label{eq:greens_function_for_delta_function_integral_form}
&=& G_0(\textbf{x},\textbf{y})+\lambda G_0(\textbf{x},\textbf{0})G(\textbf{0},\textbf{y}).
\end{eqnarray}
Now, by substituting $\textbf{x}=\textbf{0}$ into the aforementioned expression and solving for $G(\textbf{0},\textbf{y})$;  we can then insert this result into Eq. (\ref{eq:greens_function_for_delta_function_integral_form}) to derive an explicit expression for the Green's function associated with $H$:
\begin{eqnarray}
\label{eq:Greens_function_for_hamiltonian_having_interaction}
G(\textbf{x},\textbf{y})=G_0(\textbf{x},\textbf{y})+\frac{G_0(\textbf{x},\textbf{0})G_0(\textbf{0},\textbf{y})}{\frac{1}{\lambda}-G_0(\textbf{0},\textbf{0})}.
\end{eqnarray}
It's noteworthy that by successively applying this procedure, one can determine the Green's function for a potential containing an arbitrary number of delta-functions.

Now, let's examine the bound state of the Hamiltonian, Eq. (\ref{eq:Hamiltonian_free_and_interacting}), where $H_0$ represents the Hamiltonian of a free particle in $D$ dimensions. In natural units ($\hbar=2m=1$), the Hamiltonian $H_0$ can be written as,
\begin{eqnarray}
H_0=-\nabla^2=-\sum_{i=1}^{D}\frac{\partial^2}{\partial x_i^2}.
\end{eqnarray}

The energy levels of bound states can be deduced from Eq. (\ref{eq:expression_of_greens_function}), wherein they correspond to the real poles of the Green's function. Given the absence of bound states in the free particle problem, these poles can only manifest as zeros of the denominator in the second term on the right-hand side of Eq. (\ref{eq:Greens_function_for_hamiltonian_having_interaction}). To derive $G_0(E;\textbf{x},\textbf{y})$, we perform a Fourier transform transformation on Eq. (\ref{eq:green's_function_equation_2d_ddf}) with $H$ replaced by $H_0$, yielding,
\begin{eqnarray}
G(E;\textbf{x},\textbf{y})=\int \frac{d^Dk}{(2\pi)^D}\frac{e^{i\textbf{k}.(\textbf{x}-\textbf{y})}}{E-k^2}.
\end{eqnarray}
Hence, to determine the energy of the bound states, we need to solve the equation ($K^2=-E$),
\begin{eqnarray}
\label{eq:bound_state_equation_2dddf}
\frac{1}{\lambda}+\int \frac{d^Dk}{(2\pi)^D}\frac{1}{k^2+K^2}=0.
\end{eqnarray}
In the upcoming discussion, we will analyze Eq. (\ref{eq:bound_state_equation_2dddf}) for $D=2$. In this case, the Green's function, $G_0(E;\textbf{0},\textbf{0})$, is logarithmically divergent. To address this issue, we need to introduce a cut-off in the integral present in Eq. (\ref{eq:bound_state_equation_2dddf}), incorporating the dependence on the cut-off by redefining the parameters of the theory. In QFT this procedure is known as \textit{regularization} and \textit{renormalization}. We will demonstrate those here in the context of this problem, the first step is regularize the integral:
\begin{eqnarray}
\label{eq:divergence_of_bound_state_green's_function}
\int \frac{d^2k}{(2\pi)^2}\frac{1}{k^2+K^2}&=&\frac{1}{2\pi}\int_{0}^{\Lambda}\frac{kdk}{k^2+K^2}\nonumber\\
&=& \frac{1}{4\pi}\ln \left(\frac{\Lambda^2+K^2}{K^2}\right).
\end{eqnarray}
The subsequent step involves absorbing the divergent part of the aforementioned result through a redefinition of the coupling constant:
\begin{eqnarray}
\label{eq:lambda_r_definition}
\frac{1}{\lambda_R}=\frac{1}{\lambda}+\frac{1}{4\pi}\ln \left(\frac{\Lambda^2}{\mu^2}\right).
\end{eqnarray}
The parameter $\mu$ is arbitrary and is introduced to maintain the argument of the logarithm dimensionless. Now, as we approach the limit $\Lambda\rightarrow \infty$, adjusting the bare coupling constant $\lambda$ such that the renormalized coupling constant $\lambda_R$ stays finite, Eq. (\ref{eq:bound_state_equation_2dddf}) transforms into,
\begin{eqnarray}
\frac{1}{\lambda_R}-\frac{1}{4\pi}\ln \left(\frac{K^2}{\mu^2}\right)=0.
\end{eqnarray}
Solving this equation for $K^2$ we find the energy of the bound state:
\begin{eqnarray}
E_B=-K^2=-\mu^2\exp \left(\frac{4\pi}{\lambda_R}\right).
\end{eqnarray}
An interesting observation emerges at this point: even though the Hamiltonian includes only one parameter ($\lambda$), we have obtained an energy ($E_B$) that depends on two parameters ($\lambda_R$ and $\mu$). Nevertheless, the apparent doubling of parameters is illusory. Indeed, it can be demonstrated that the Green’s function depends on a single parameter (besides $E$, \textbf{x} and \textbf{y}). To illustrate this, let's express the denominator of the second term on the right-hand side of Eq. (\ref{eq:Greens_function_for_hamiltonian_having_interaction}) in a regularized form.
\begin{eqnarray}
\frac{1}{\lambda}-G_0(E;\textbf{0},\textbf{0})&=&\frac{1}{\lambda}+\frac{1}{2\pi}\int_{0}^{\Lambda}\frac{kdk}{k^2-E}\nonumber\\
&=& \frac{1}{\lambda}+\frac{1}{4\pi}\ln \left(\frac{\Lambda^2-E}{-E}\right)
\end{eqnarray}
This exemplifies the phenomenon known as \textit{dimensional transmutation} \cite{PhysRevD.7.1888}. We begin with a theory characterized by dimensionless parameters (here, $\lambda$) and conclude with a theory that possesses a dimensionful parameter. This occurs because we had to introduce a dimensionful parameter, $\mu$, thus breaking the length invariance of the theory.
\section{The 2D-DDF potential problem in wavelet basis}
The energy eigenvalue problem for 2D-DDF problem in natural units ($\hbar=1, m=1$) is given by,
\begin{eqnarray}
\label{eq:energy_eigenvalue_equation_2dddf}
\left(-\frac{1}{2}\sum_{i=1}^{2}\frac{\partial^2}{\partial x_i^2}-g \delta(x_1)\delta(x_2)\right)\psi(x_1,x_2)=E\psi(x_1,x_2).
\end{eqnarray}
We approximate the state space of the system to the resolution subspace $\mathscr{H}^k$. Within this approximation, expanding the eigenfunction in scaling function basis,
\begin{eqnarray}
\psi(x_1,x_2)=\sum_{n_1,n_2}\psi^k_{s,n_1,n_2}s^k_{n_1,n_2}(\textbf{x}),
\end{eqnarray}
we can express Eq. (\ref{eq:energy_eigenvalue_equation_2dddf}) as a matrix eigenvalue equation,
\begin{eqnarray}
\sum_{n_3,n_4}H^k_{ss,n_1,n_2;n_3,n_4}\psi_{s,n_3,n_4}^{k}=E \psi^k_{s,n_1,n_2},
\end{eqnarray}
where the Hamiltonian matrix elements are given by,
\begin{eqnarray}
\label{eq:Hamiltonian_matrix_element_scaling_basis}
H^k_{ss,n_1,n_2:n_3,n_4}= &&\int  \left( -\frac{s^k_{n_1,n_2}(x_1,x_2)}{2}\sum_{i=1}^{2}\frac{\partial^2 s^k_{n_3,n_4}(x_1,x_2)}{\partial x_i^2}-g s^k_{n_1,n_2}(x_1,x_2)\delta(x_1)\delta(x_2)\right. \nonumber \\
&& \left.s^k_{n_1,n_2}(x_1,x_2)\right)\times dx_1 dx_2.
\end{eqnarray}
Utilizing integration by parts and the compact support of the scaling functions, we can express Eq. (\ref{eq:Hamiltonian_matrix_element_scaling_basis}) as:
\begin{eqnarray}
H^k_{ss,n_1,n_2:n_3,n_4}=&&\int \left(\frac{1}{2}\sum_{i=1}^{2}\frac{\partial s^k_{n_1,n_2}(x_1,x_2)}{\partial x_i}\frac{\partial s^k_{n_3,n_4}(x_1,x_2)}{\partial x_i}-g s^k_{n_1,n_2}(x_1,x_2)\delta(x_1)\delta(x_2)\right.\nonumber\\
&& \left. s^k_{n_1,n_2}(x_1,x_2)\right)\times dx_1 dx_2.
\end{eqnarray}
Exploiting the separable nature of the two-dimensional scaling basis functions, we can reformulate the Hamiltonian matrix elements that involve distinct overlap integrals as:
\begin{eqnarray}
H^k_{ss,n_1,n_2:n_3,n_4}=\frac{1}{2}\left[T^k_{ss,n_1,n_3}\times \delta_{n_2n_4}+T^k_{ss,n_2,n_4}\times \delta_{n_1n_3}\right]+g V^k_{\delta ss,n_1n_3}\times V^{k}_{\delta ss,n_2n_4},
\end{eqnarray}
where,
\begin{eqnarray}
T^k_{ss,m,n}&=& \int_{-\infty}^{\infty}s^{k'}_{m}(x)s^{k'}_{n}(x)dx,\\
\label{eq:overlap_integral_delta_function_scaling_function_2}
V^k_{\delta ss,m,n}&=& \int_{-\infty}^{\infty} \delta(x) s^k_m(x)s^k_n(x)dx.
\end{eqnarray}
We can evaluate these integrations analytically from the properties of scaling functions using the procedure described in the Appendix (\ref{appen:the_kinetic_energy_term}) and Appendix (\ref{appen:The_Dirac_delta_function_potential}) respectively.

\begin{figure}[h]
\begin{center}
\includegraphics[scale=.45]{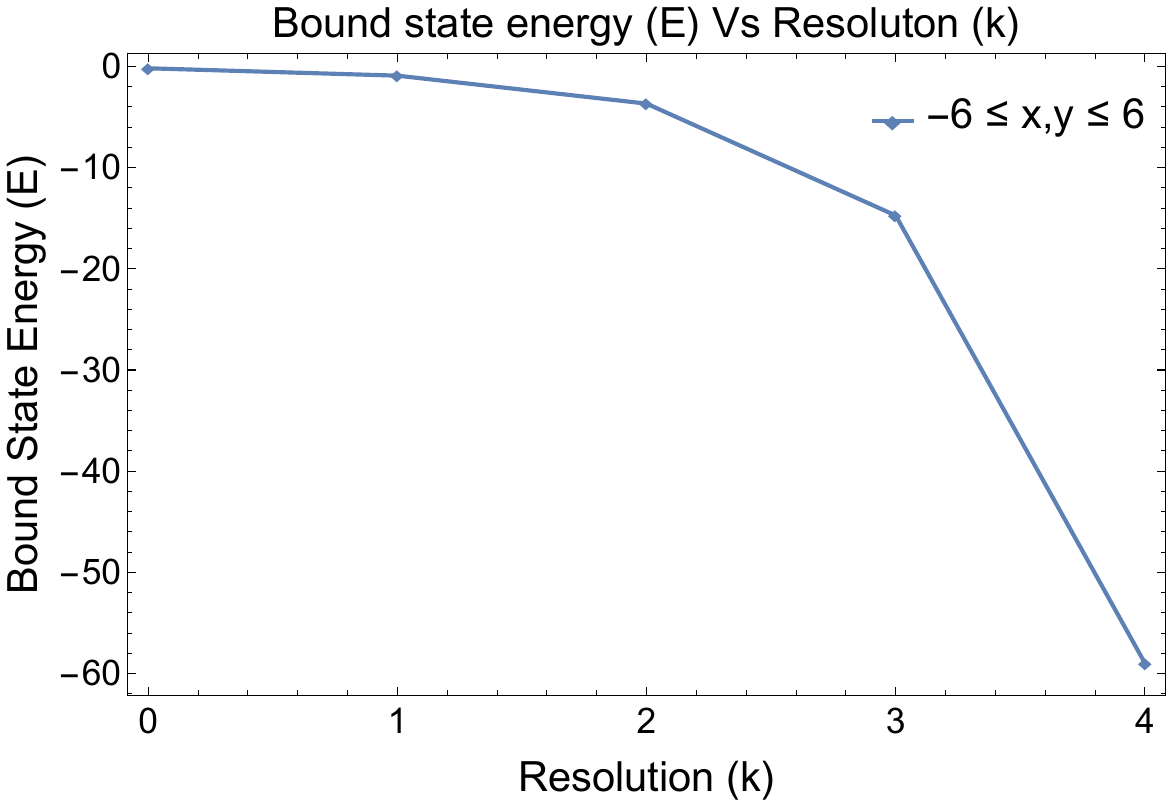}
\caption{\label{fig:negative_divergence_of_eigenvalues_2dddf}Negative divergence of bound state energy for fixed volume ($V=12\times 12$) with fixed coupling constant ($g=1.848694$) versus increasing resolution plot.}
\end{center}
\end{figure}

To perform numerical computations of its eigenvalues, it is necessary to truncate the resulting Hamiltonian matrix. The initial type of truncation is volume truncation, which identifies the region of physical space accessible to the system. In the present problem, we define the volume truncation by $-L\leq x,y \leq L$. involves choosing a value for the resolution $k$. This involves incorporating lengths from scale $2L$ down to the scale $\frac{(2K-1)}{2^k}$ and excluding all length scales finer than this limit. For a fixed
value of bare coupling constant $g$ and resolution, the lowlying eigenvalues remain unchanged with increasing $L$. This saturation of eigenvalues is intuitively anticipated, considering that the majority of the dynamics occur around the origin due to the short range of the potential. Subsequently, all computational results will be presented by selecting a sufficiently large value for $L$. There is a minimum value of the coupling constant $g_{min}$, beyond which precisely one bound state is obtained. As an illustration, when $L=6$ and resolution $k=0$, the minimum value is approximately $g_{min}=1.18$. For any fixed value of the bare coupling constant, the bound state eigenvalue diverges to negative infinity (Fig. \ref{fig:negative_divergence_of_eigenvalues_2dddf}). 
This indicates the occurrence of ultraviolet divergences within the wavelet-based framework. This divergence arises due to the inclusion of increasingly finer length scales into the problem. This reflects the situation described in Eq. (\ref{eq:divergence_of_bound_state_green's_function}). The QFTs defined in the continuum have points in the underlying space-time that can come arbitrarily close to each other. In other words, there does not exist any short-distance (ultraviolet) cutoff. When these quantum field theories are analyzed perturbatively, one encounters divergences at each order of perturbation theory, whose origin can be traced to the lack of underlying short-distance cutoff. The concept of renormalization provides the essential element to derive physical predictions from the perturbative QFT. At each order of perturbation theory, the ultraviolet divergences are regulated by the introduction of an artificial ultraviolet cutoff, following which the dependence of the bare couplings on the ultraviolet cutoff is determined by demanding that it reproduces the experimental values of a finite set of physical observables. For a perturbatively renormalizable theory, this process of renormalization renders all observables of the quantum field theory finite and ultraviolet cutoff independent. In other words, a local limit can be established for a renormalizable theory within the perturbative framework.

Within the wavelet framework, each quantum state of the system can described as an expansion in scaling and wavelet functions. The expansion coefficients of the scaling
functions describe contributions from the length scale scale $2L$ down to $\frac{(2K-1)}{2^k}$, while the expansion coefficients of the wavelet functions represent contribution on all lengths scales finer than $\frac{(2K-1)}{2^k}$. We can impose a short distance cutoff to regulate the theory at a nonperturbative level by truncating the basis function expansion to include only the scaling functions. In other words, the Hilbert space of the system is restricted to $\mathscr{H}^k$. Likewise, all operators (for example, the Hamiltonian) are defined in terms of their action on $\mathscr{H}^k$. The bare
coupling constants of the truncated theory are tuned to reproduce the experimental values of a finite set of physical observables. This step is analogous to solving the set of equations outlined in Eq. (\ref{eq:observable_experimental_and_theoretical}). The process of renormalization consists of constructing the local limit by solving a series of truncated theories with increasing resolution.

\begin{figure}[h]
\begin{center}
\includegraphics[scale=.45]{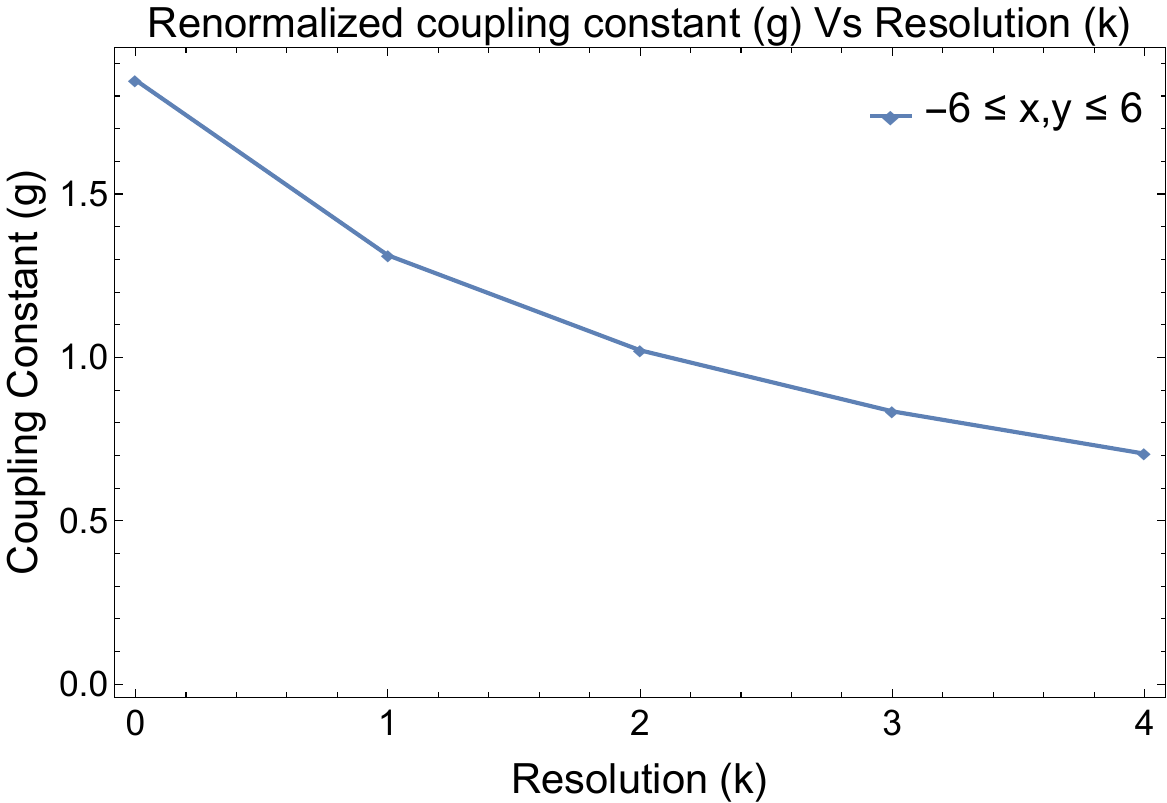}
\caption{\label{fig:renormalized_coupling_constant_2dddf}Renormalized coupling constant for fixed volume ($V=12\times 12$) versus Resolution ($k$) plot.}
\end{center}
\end{figure}

We showcase the application of this wavelet based approach to renormalization in the context of two-dimensional Dirac delta function potential. We have shown in the
previous section that, for a fixed value of bare coupling constant, the ground state energy diverges to negative infinity with increasing resolution. In order to have a
physically meaningful theory containing a bound state, we bring in a renormalization prescription that the theory truncated at resolution $k$ should reproduce the ground state
eigenvalue which we fix at $-1$. We tune the coupling constant value in order for the truncated theory to reproduce this experimental observable. Repeating this process for a
series of truncated theories with increasing resolution, we arrive at the observation that the coupling constant flows with resolution. In particular, it becomes weaker with increasing resolution as is expected in the context of this problem. See Fig. \ref{fig:renormalized_coupling_constant_2dddf} and Table \ref{tab:renormalized_value_of_the_coupling_constant}.
\begin{table}[ht]
\begin{center}
\setlength{\tabcolsep}{1.5pc}
\catcode`?=\active \def?{\kern\digitwidth}
\caption{The values of renormalized coupling constant with a different resolution cutoff.}
\label{tab:renormalized_value_of_the_coupling_constant}
\vspace{1mm}
\begin{tabular}{c c}
\specialrule{.15em}{.0em}{.15em}
\hline
Resolution ($k$) & Coupling constant (g) \\
\hline
$4$ & $0.7053401$ \\

$3$ & $0.8349675$ \\

$2$ & $1.021796$ \\

$1$ & $1.312652$ \\

$0$ & $1.848694$ \\

\hline
\specialrule{.15em}{.15em}{.0em}
\end{tabular}
\end{center}
\end{table}

\section{Summary and Conclusions}
in this chapter, we examine renormaliztion in a discrete wavelet based quantum theory. The attractive two-dimensional Dirac delta function potential was chosen as the model of study as it contains many of the nontrivial features that are observed in a relativistic
quantum field theory such as ultraviolet divergences, asymptotic freedom, and dimensional transmutation. Working with models such as this one will provide insights that will be valuable when working with realistic quantum field theories within the wavelet based framework.

For quantum systems with finite number of dynamical variables, the operator energy eigenvalue problem is converted to matrix eigenvalue problem using the discrete Daubechies wavelet basis, in which the rows and columns of the matrix can be organized by length scales. The off-diagonal Hamiltonian matrix elements have a natural interpretation of coupling between length scales. Specifically, each Hamiltonian matrix element carries a pair “location” and resolution indices. By imposing an upper and lower bound on the “location” index, one can define the region of physical space in which the system
would be studied, which is equivalent to put a infrared cutoff. Setting an upper bound on the resolution index essentially amounts to imposing an ultraviolet cutoff and as such plays the role of the ultraviolet regulator at a nonperturbative level.

We have shown that, if the bare coupling constant is held fixed, then as the resolution is increased the ground state energy diverges as is expected. To make physical sense of
this theory, one demands that the bare coupling constant flows with resolution in such a way as to maintain the physical value of the ground state energy. The coupling
constant becomes weaker as resolution is increased which attests the asymptotically free nature of the two-dimensional Dirac delta function potential.

In the context of QFTs, the field operator can be expanded in terms of scaling and wavelet basis function with operator valued coefficients. This decomposition leads
to the quantum field (the operator valued distribution) being replaced in terms of a countably infinite number of operators with different spatial resolutions. One define a volume truncation by retaining only those basis function terms in the field operator expansion, that have support lying within a specified volume. The resolution truncation
admits only those basis function terms in the field operators expansion that are coarser than a specified resolution. This truncated QFT, which is now a theory with finite number of degrees of freedom, should in principle be solvable. The infinite volume and infinite resolution limit needs to be constructed as a limit of a sequence of truncated theories.
Further investigations in this direction are highly desirable.

\chapter{Flow equation method in a wavelet basis}
\label{chap:flow_equation_method_in_a_wavelet_basis}
In the preceding section, we explored a renormalization approach to address the divergence of eigenvalues in asymptotically free theories, employing a model of 2D-DDF. The current chapter deals with the application of the flow equation method as a means to decouple the interactions between finer and coarser degrees of freedom in a field theory problem. We extended the work of Michilin and Polyzou We study the low energy dynamics of a system of two coupled real scalar fields in 1+1 dimensions using the flow-equation method in a wavelet basis. 

\begin{figure}[hbt]
\caption{Schematic diagram of the generator $K(\lambda)$ with Hermitian matrix, $G(\lambda)$ and $H(\lambda)$ in wavelet basis.}
\label{fig:k_lambda}
\begin{center}
\includegraphics[scale=0.35]{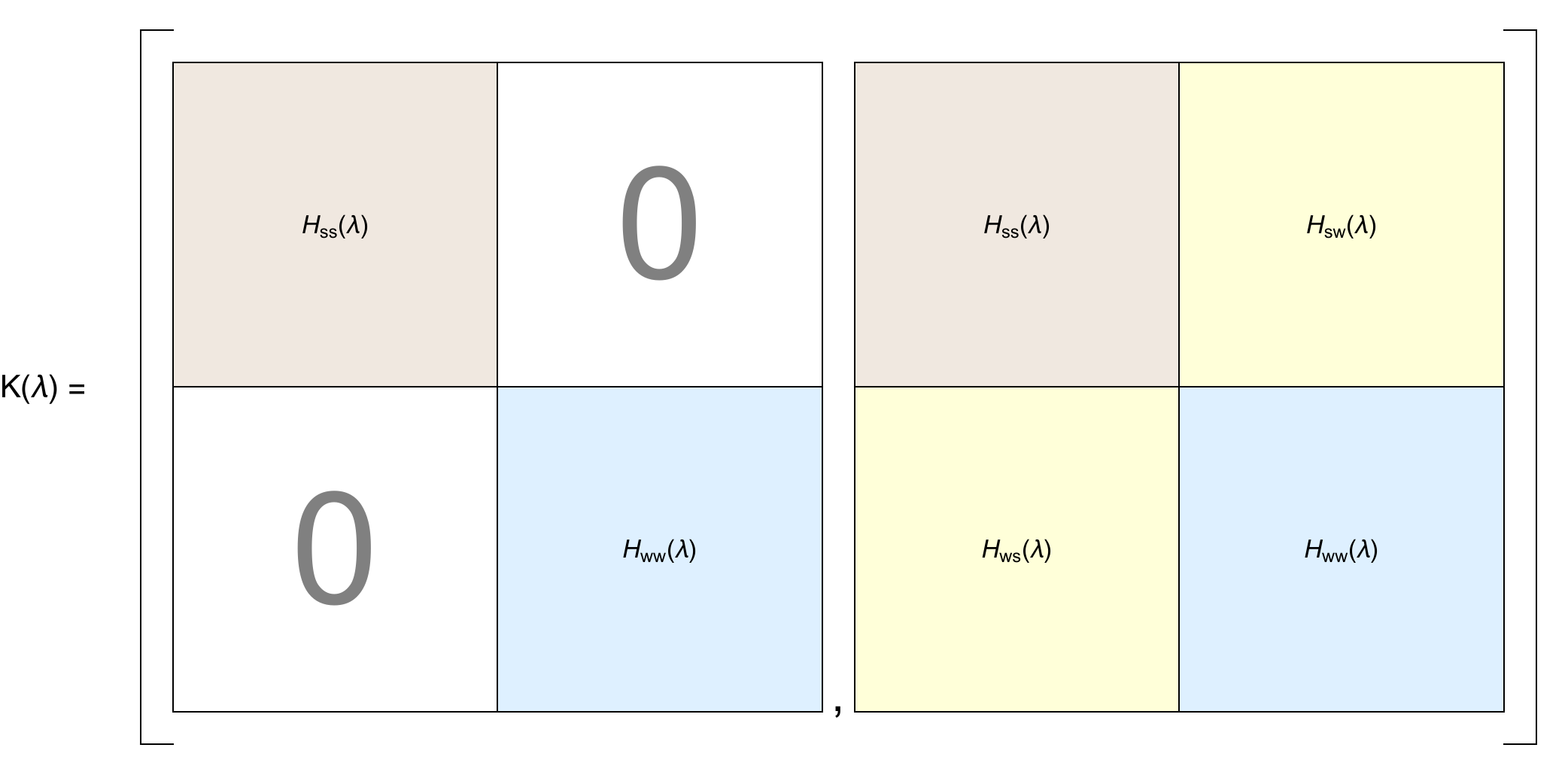}.
\end{center}
\end{figure}

Wegner \cite{wegner1994flow} introduced flow equations as a technique for the gradual evolution of the Hamiltonian towards a unitarily equivalent form. Flow equation methods \cite{PhysRevD.48.5863,PhysRevD.49.4214,PERRY1994116,bartlett2003flow,kehrein2007flow,bogner2007similarity,PhysRevC.77.037001,BOGNER201094} offer an alternative to the direct diagonalization or block diagonalization approaches. Flow equation methods have found application in addressing challenges within quantum field theory and quantum mechanics. An advantageous feature is their simplicity of implementation compared to the integration of short-distance degrees of freedom in a functional integral. Flow equations are specifically designed to achieve diagonalization through a continuously parameterized unitary transformation, denoted as $U(\lambda)$. The transformed Hamiltonian has the following form,
\begin{eqnarray}
H(\lambda)=U(\lambda)H(0) U^{\dagger}(\lambda).
\end{eqnarray}
Here, with $H(0) = H$ representing the original Hamiltonian, the choice of the generator for the flow equation is made to facilitate the continuous evolution of the initial Hamiltonian into the desired form as $\lambda$ increases. Here $\lambda$ is called the flow parameter. As $\lambda$ increases from $0$, the Hamiltonian undergoes evolution toward the desired form. The constructed evolution is designed to exponentially approach the desired form, although there is a possibility for the exponent to become small. However, assessing $H(\lambda)$ at any given value of $\lambda$ consistently results in a Hamiltonian that maintains unitary equivalence to the original Hamiltonian, albeit with weaker scale coupling terms. The problem is to find a generator of the flow that leads to the desired outcome.
The unitarity of $U(\lambda)$ implies that it satisfies the differential equation
\begin{eqnarray}
\frac{dU(\lambda)}{d \lambda}=\frac{dU(\lambda)}{d \lambda}U^{\dagger}(\lambda)U(\lambda)=K(\lambda)U(\lambda),\nonumber
\end{eqnarray}
where,
\begin{eqnarray}
K(\lambda)=\frac{dU(\lambda)}{d \lambda}U^{\dagger}(\lambda)=-K^{\dagger}(\lambda),
\end{eqnarray}
is the anti-Hermitian generator of this unitary transformation. We have the liberty to select a generator that steers the evolution toward the desired outcome. It follows that $H(\lambda)$ satisfies the differential equation,
\begin{eqnarray}
\label{eq:flow_equation_of_Hamiltonian_1}
\frac{dH(\lambda)}{d\lambda}=\left[K(\lambda),H(\lambda)\right].
\end{eqnarray}
In this context, it is beneficial to opt for a generator, denoted as $K(\lambda)$, which is a function of the evolved Hamiltonian:
\begin{eqnarray}
K(\lambda)=\left[G(\lambda),H(\lambda)\right].
\end{eqnarray}
Here, $G(\lambda)$ represents the portion of $H(\lambda)$ where terms involving coupling between different scales are absent. With the choice of $G(\lambda)=G^{\dagger}(\lambda)$ so $K(\lambda)$ is anti-Hermitian. The schematic diagram of $K(\lambda)$, is given in Fig. (\ref{fig:k_lambda}).

It follows that,
\begin{eqnarray}
\label{eq:flow_equation_of_Hamiltonian_2}
\frac{dH(\lambda)}{d\lambda}=\left[K(\lambda),H(\lambda)\right]=\left[\left[G(\lambda),H(\lambda)\right],H(\lambda)\right]=\left[H(\lambda),\left[ H(\lambda),G(\lambda)\right]\right].
\end{eqnarray}
Equation (\ref{eq:flow_equation_of_Hamiltonian_2}) is the desired flow equation for Hamiltonian. A fixed point, $\lambda^{*}$, of this equation occurs when
\begin{eqnarray}
\left[H(\lambda^{*}),\left[H(\lambda^{*}),G(\lambda^{*})\right]\right]=0.
\end{eqnarray}
It can be deduced from the structure of the equation that the generator $K(\lambda)$ exclusively incorporates terms that couple the degrees of freedom associated with wavelet and scaling functions. Here, we will show that this nonlinear equation drives the commutator towards zero, specifically in the context of quantum mechanical problems.

The subsequent considerations are confined to the scenario of quantum mechanical problems and a free field Hamiltonian \cite{PhysRevD.95.094501}. The Hamiltonian, when expressed in the wavelet basis, can be formulated as a sum comprising the same-scale coupling term and the different-scale coupling term as follows,
\begin{eqnarray}
H(\lambda)&=&H_{ss}(\lambda)+H_{ww}(\lambda)+H_{sw}(\lambda)\nonumber\\
&=& G(\lambda)+H_{sw}(\lambda),
\end{eqnarray}
where
\begin{eqnarray}
G(\lambda)=H_{ss}(\lambda)+H_{ww}(\lambda),
\end{eqnarray}
represents the same scale coupling and $H_{sw}(\lambda)$ is the different-scale coupling.
\begin{eqnarray}
\left[G(\lambda),H_{sw}(\lambda)\right]:= H_{sw}(\lambda)'.
\end{eqnarray}
Here, the prime symbol indicates the presence of different scale coupling terms. Commutation of this matrix with $H(\lambda)$ gives,
\begin{eqnarray}
\left[G(\lambda)+H_{sw}(\lambda),H_{sw}'(\lambda)\right]:=H''_{sw}(\lambda)+\left[H_{sw}(\lambda),H'_{sw}(\lambda)\right],
\end{eqnarray}
where
\begin{eqnarray}
H''_{sw}(\lambda)=\left[G(\lambda),H'_{sw}(\lambda)\right]=\left[G(\lambda),\left[G(\lambda),H_{sw}(\lambda)\right]\right].
\end{eqnarray}
The commutation of the scale-coupling terms results in a combination of scaling and wavelet terms,
\begin{eqnarray}
\left[H_{sw}(\lambda),H'_{sw}(\lambda)\right]=H''_{ss}(\lambda)+H''_{ww}(\lambda).
\end{eqnarray}
So, the flow equation, Eq. (\ref{eq:flow_equation_of_Hamiltonian_1}) can be separated into coupled equation for mixed and non-mixed parts. First, we decompose the Hamiltonian into two parts one $H_{sc}(\lambda)$, the same scale coupling term and the second term, $H_{dc}(\lambda)$, the different scale coupling term,
\begin{eqnarray}
H(\lambda)=H_{sc}(\lambda)+H_{dc}(\lambda),
\end{eqnarray}
where
\begin{eqnarray}
H_{sc}(\lambda)=G(\lambda).
\end{eqnarray}
\begin{figure}[hbt]
\caption{The schematic diagram of total Hamiltonian, $H(\lambda)$, divided into two parts $H_b(\lambda)=G(\lambda)$ and $H_c(\lambda)$.}
\label{fig:the_division_of_full_hamiltonian_hb_hc}
\begin{center}
\includegraphics[scale=0.35]{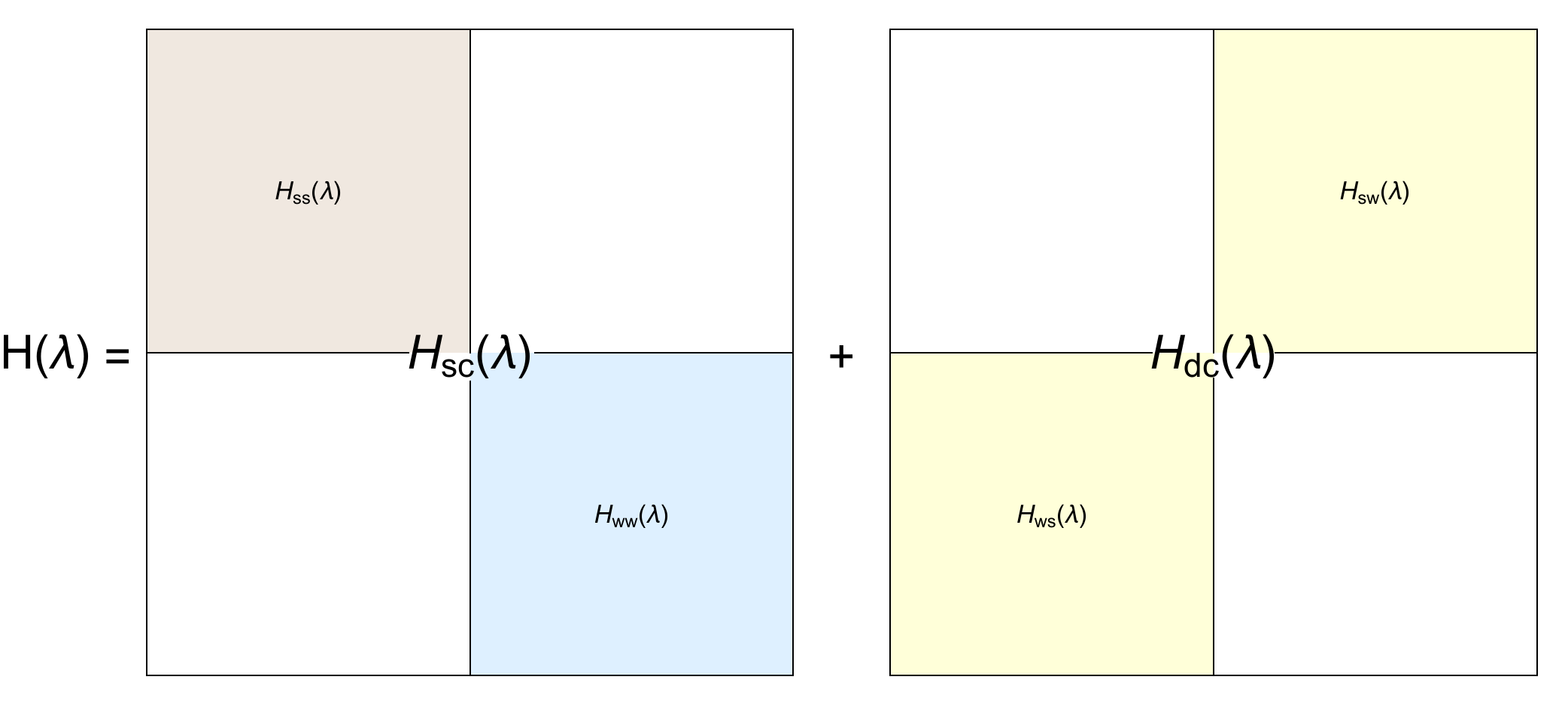}
\end{center}
\end{figure}
The schematic diagram of this decomposition is given in the Fig. (\ref{fig:the_division_of_full_hamiltonian_hb_hc}). Equation (\ref{eq:flow_equation_of_Hamiltonian_1}), can be written as two coupled equation,
\begin{eqnarray}
\label{eq:flow_equation_the_same_coupling_term}
\frac{dH_{sc}(\lambda)}{d\lambda}&=&\left[H_{dc}(\lambda),\left[H_{dc}(\lambda),H_{sc}(\lambda)\right]\right],\\
\label{eq:flow_equation_the_different_coupling_term}
\frac{dH_{dc}(\lambda)}{d\lambda}&=&-\left[H_{sc}(\lambda),\left[H_{sc}(\lambda),H_{dc}(\lambda)\right]\right].
\end{eqnarray}
These equations have a symmetric form under $H_{sc}\leftrightarrow H_{dc}$ except for a sign.

To comprehend the transformation of the Hamiltonian into the desired structure, we represent the equation, Eq. (\ref{eq:flow_equation_the_same_coupling_term}), in a basis constituted by eigenstates of $H_{dc}(\lambda)$ with corresponding eigenvalues $e_{dc}(\lambda)$. Simultaneously, we express the equation, Eq. (\ref{eq:flow_equation_the_different_coupling_term}), in a basis composed of eigenstates of $H_{sc}(\lambda)$ with associated eigenvalues $e_{sc}(\lambda)$. The equations governing the matrix elements within each of these bases take the following form:
\begin{eqnarray}
\frac{d H_{sc,mn}(\lambda)}{d\lambda}&=&\left(e_{dc,m}(\lambda)-e_{dc,n}(\lambda)\right)^2 H_{sc,mn}(\lambda),\\
\frac{d H_{dc,mn}(\lambda)}{d\lambda}&=&-\left(e_{sc,m}(\lambda)-e_{sc,n}(\lambda)\right)^2 H_{dc,mn}(\lambda).
\end{eqnarray}
These equations can be integrated exactly
\begin{eqnarray}
\label{eq:the_evolution_of_same_scale_coupling_part}
H_{sc,mn}(\lambda)&=& e^{\int_0^{\lambda}\left(e_{dc,m}(\lambda')-e_{dc,n}(\lambda')\right)^2 d\lambda'}H_{sc,mn}(0),\\
\label{eq:the_evolution_of_different_scale_coupling_part}
H_{dc,mn}(\lambda)&=& e^{-\int_0^{\lambda}\left(e_{sc,m}(\lambda')-e_{sc,n}(\lambda')\right)^2 d\lambda'}H_{dc,mn}(0).
\end{eqnarray}
These solutions indicate that the matrix elements of $H_{sc}$ exhibit exponential growth, whereas the matrix elements of $H_{dc}$ exhibit exponential decay as the flow parameter $\lambda$ increases. This progression may come to a halt in the presence of eigenvalue degeneracies, approximate degeneracies.

It is seen from these equations that in the high resolution- large volume limit, where the spectrum of the block diagonal Hamiltonian approaches a continuous spectrum, there will be closely spaced eigenvalues, which will lead to slow convergence of some parts of the scale coupling terms of the Hamiltonian.

A system possessing finite energy within a finite volume is anticipated to be dominated by a limited number of degrees of freedom \cite{doi:10.1137/1027082}. These degrees of freedom can be categorized into two groups: those linked to an experimental scale and other pertinent degrees of freedom at smaller scales. Scaling function fields can serve as the degrees of freedom at the experimental scale, while wavelet degrees of freedom represent the smaller scales that remain relevant to the specified volume and energy scale.

Similar considerations hold for Hamiltonians involving interactions; however, it is generally necessary to employ a different flow generator to distinguish the desired degrees of freedom. Additionally, there might be a requirement to initially project the truncated Hamiltonian onto a subspace before addressing the flow equation.

Here, we will illustrate the scale separation using a model of two coupled scalar fields in $1+1$-dimensions. The same methodology can be applied to free scalar field theory, a topic that has already been investigated by Polyzou and his group (refer to the provided reference \cite{PhysRevD.95.094501}). To apply this method to other field theory problems, we might need to modify the flow generator.

\section{The scalar field theory, the normal mode frequencies and the flow-equation in wavelet basis}
\label{sec:The_scalar_field_theory_and_the_normal_mode_frequencies}
The effectiveness of wavelet-based flow equations in the case of a single real free scalar field was studied by Polyzou and Michlin \cite{PhysRevD.95.094501} for resolution $1$. This was substantiated by comparing the normal mode frequencies of the flow equation generated effective Hamiltonian with those of the truncated Hamiltonian. They computed the Hilbert-Schmidt norms to demonstrate the decoupling of the long- and short-distance variables using the flow equation method. In this section, we provide a short review and extension of their work by calculating the square of the normal mode frequency for higher resolution. We showed that with the increasing resolution, the normal mode frequencies approach the exact value of the same. We also demonstrated the efficacy of the flow equation method for resolution $2$ by portraying the schematic diagram of the matrix for increasing values of $\lambda$.

The Lagrangian density of the $1+1$-dimensional real scalar field theory is given by,
\begin{eqnarray}
\label{eq:free_scalar_field_theory_lagrangian}
\mathcal{L}(\phi(\mathrm{x},t),\partial_x \phi(\mathrm{x},t), \dot{\phi}(\mathrm{x},t))=\frac{1}{2}\left[\dot{\phi}^2(\mathrm{x},t)-(\partial_x \phi(\mathrm{x},t))^2-\mu^2 \phi(\mathrm{x},t)^2\right],
\end{eqnarray}
its corresponding Hamiltonian is
\begin{eqnarray}
\label{eq:hamiltonian_scalar_field_theory}
\mathrm{H}(\phi(x,t),\pi(x,t))=\int dx \frac{1}{2}\left[\pi(x,t)^2+\left(\partial_x \phi(x,t)\right)^2+\mu^2 \phi(x,t)^2\right] ,
\end{eqnarray}
where, $\pi(x)$ represents the canonical momentum,
\begin{eqnarray}
\pi(x,t)=\frac{\partial \mathcal{L}}{\partial \dot{\phi}(x,t)}=\dot{\phi}(x,t).
\end{eqnarray}
The instant-form canonical quantization of the scalar field is carried out by demanding that $\phi(\mathrm{x},t)$ and $\pi(\mathrm{x},t)$ satisfy the equal-time canonical commutation relations,
\begin{eqnarray}
\label{eq:cannonical_commutation_relation_phi_pi}
\left[\pi(\mathrm{x},t),\phi(y,t)\right]&=&-\iota \delta(x-y),\\
\label{eq:connonical_commutation_relation_phi_phi_pi_pi}
\left[\pi(x,t),\pi(y,t)\right]&=&\rm\left[\phi(x,t),\phi(y,t)\right]=0.
\end{eqnarray}
On assuming that the scalar field exists only within an interval of length $L$, we write $\phi(\mathrm{x},t)$ and $\pi(\mathrm{x},t)$ in terms of the normal modes as follows,
\begin{eqnarray}
\label{eq:phi_xt_interms_of_normal_modes}
\phi(\mathrm{x},t)&=&\sum_{p=1}^{\infty} \phi_p(t) \sqrt{\frac{2}{L}}\sin(\frac{p\pi\mathrm{x}}{L}),\\
\label{eq:pi_xt_in_terms_of_normal_modes}
\pi(\mathrm{x},t)&=&\sum_{p=1}^{\infty} \pi_p(t) \sqrt{\frac{2}{L}}\sin(\frac{p\pi\mathrm{x}}{L}).
\end{eqnarray}
On substituting Eq. (\ref{eq:phi_xt_interms_of_normal_modes}) and Eq. (\ref{eq:pi_xt_in_terms_of_normal_modes}) into Eq. (\ref{eq:free_scalar_field_theory_lagrangian}), we get,
\begin{eqnarray}
\mathrm{H}=\sum_{p=1}^{\infty}\frac{1}{2} \left(\pi_p^2(t)+\omega_p^2\phi_p^2(t)\right),
\end{eqnarray}
where, $\omega_p$, is given by $\sqrt{\left(\frac{p\pi}{L}\right)^2+\mu^2}$, called the normal mode frequency corresponding to the $p$-mode. This shows that the free scalar field is represented by a set of uncoupled simple harmonic oscillators with normal mode frequencies given by $\omega_p$.

The wavelet based representation of Hamiltonian dynamics is constructed by resolving $\phi(x)$ and $\pi(x)$in the discrete Daubechies wavelet basis:
\begin{eqnarray}
\label{eq:expansion_of_scalar_field_in_wavelet_basis}
\phi(\mathrm{x},t)&=&\sum_{n=-\infty}^{\infty}\phi^{s,k}_n(t)s^k_n(x)+\sum_{l\geq k}^{\infty}\sum_{n=-\infty}^{\infty} \phi^{w,l}_n(t) w^l_n(x),\\
\label{eq:expansion_of_scalar_field_momentum_in_wavelet_basis}
\pi(\mathrm{x},t)&=&\sum_{n=-\infty}^{\infty}\pi^{s,k}_n(t)s^k_n(x)+\sum_{l\geq k}^{\infty}\sum_{n=-\infty}^{\infty} \pi^{w,l}_n(t) w^l_n(x),
\end{eqnarray}
where, the scaling coefficients $\phi^{s,k}_n$ and $\pi^{s,k}_n$, represent variables that describe physics of the scalar field down to the length scale $\frac{2K-1}{2^k}$. The physics of the scalar field on length scales finer than $\frac{2K-1}{2^k}$ is contained in the variables given by the wavelet coefficients $\{\phi^{w,l}_n, \pi^{w,l}_n \,|\, l\ge k \}$. 

Within the wavelet based formulation, the quantum field theory and its quantization are transcribed in terms of the variables given by the scaling and wavelet coefficients. Substituting Eq. (\ref{eq:expansion_of_scalar_field_in_wavelet_basis}) and Eq. (\ref{eq:expansion_of_scalar_field_momentum_in_wavelet_basis}) into Eq. (\ref{eq:hamiltonian_scalar_field_theory}) and using the orthogonality of the wavelet basis, Eq. (\ref{eq:normalization_condition_sk_m_sk_n}), Eq. (\ref{eq:orthonormality_of_wavelet_functions}) and Eq. (\ref{eq:orthogonality_of_scaling_and_wavelet_functions}), gives the Hamiltonian in terms of the scaling and the wavelet basis coefficients.
\begin{eqnarray}
\mathrm{H}=\mathrm{H}_{ss}+\mathrm{H}_{ww}+\mathrm{H}_{sw},
\end{eqnarray}
where, $\mathrm{H}_{ss}$ is the part of the Hamiltonian which includes the physics from the coarsest length scale down to the length scale $\frac{2K-1}{2^k}$.
\begin{eqnarray}
\label{eq:hss_scalar_field_theory}
\mathrm{H}_{ss}:=\frac{1}{2}\left(\sum_n \pi^{s,k}_n(t)\pi^{s,k}_n(t)+\sum_n \phi^{s,k}_m(t)\phi^{s,k}_n(t) \mathcal{D}^k_{ss,mn}+\sum_n \mu^2 \phi^{s,k}_n(t) \phi^{s,k}_n(t)\right).
\end{eqnarray}
 In effect, $\mathrm{H}_{ss}$ represents the Hamiltonian truncated to resolution $k$. Similarly, $\mathrm{H}_{ww}$ is part of the Hamiltonian, which includes the physics on all the length scales finer than $\frac{2K-1}{2^k}$.
\begin{eqnarray}
\label{eq:hww_scalar_field_theory}
\mathrm{H}_{ww}:=\frac{1}{2}\left(\sum_n \pi^{w,l}_n(t)\pi^{w,l}_n(t)+\sum_n \phi^{w,l}_m(t)\phi^{w,q}_n(t) \mathcal{D}^{lq}_{ww,mn}+\sum_n \mu^2 \phi^{w,l}_n(t) \phi^{w,l}_n(t)\right).
\end{eqnarray}
The coupling between the length scales coarser and finer is given by,
\begin{eqnarray}
\label{eq:hsw_scalar_field_theory}
\mathrm{H}_{sw}&:=&\frac{1}{2}\left(\sum_{m,q,n}\phi^{s,k}_m(t)\phi^{w,q}_n(t) \mathcal{D}^{kq}_{sw,mn}\right).
\end{eqnarray}
In Eq. (\ref{eq:hss_scalar_field_theory}), Eq. (\ref{eq:hww_scalar_field_theory}), and Eq. (\ref{eq:hsw_scalar_field_theory}) $\mathcal{D}^k_{ss,mn}$, $\mathcal{D}^{lq}_{ww,mn}$, $\mathcal{D}^{kq}_{sw,mn}$ denote the overlap integrals,
\begin{eqnarray}
\mathcal{D}^{k}_{ss,mn}&=& \int dx \frac{\partial s^k_m(x)}{\partial x}\frac{\partial s^k_n(x)}{\partial x},\\
\mathcal{D}^{lq}_{ww,mn}&=& \int dx \frac{\partial w^l_m(x)}{\partial x}\frac{\partial w^q_n(x)}{\partial x},\\
\mathcal{D}^{kq}_{sw,mn}&=& \int dx \frac{\partial s^k_m(x)}{\partial x}\frac{\partial w^q_n(x)}{\partial x}.
\end{eqnarray}
The procedure to evaluate these overlap integrals analytically is provided in the paper \cite{PhysRevD.87.116011}.

The inverse of Eq. (\ref{eq:expansion_of_scalar_field_in_wavelet_basis}) and Eq. (\ref{eq:expansion_of_scalar_field_momentum_in_wavelet_basis}) (mathematically, the discrete wavelet transform (DWT)), 
\begin{eqnarray}
\phi^{s,k}_n=\int \phi(x)s^k_n(x)dx,\quad \phi^{w,l}_n=\int \phi(x) w^l_n(x)dx,\\
\pi^{s,k}_n=\int \pi(x)s^k_n(x)dx,\quad \pi^{w,l}_n=\int \pi(x) w^l_n(x)dx,
\end{eqnarray}
and the equal time commutation relations Eq. (\ref{eq:cannonical_commutation_relation_phi_pi}) and Eq. (\ref{eq:connonical_commutation_relation_phi_phi_pi_pi}) lead to canonical commutation relations between the scaling coefficients, $\{\phi^{s,k}_{n},\pi^{s,k}_n\}$ and the wavelet coefficients, $\{\phi^{w,l}_{n},\pi^{w,l}_n\}$,
\begin{eqnarray}
\left[\phi^{s,k}_m(t),\pi^{s,k}_n(t)\right]&=&\iota \delta_{mn},\quad \left[\phi^{s,k}_m(t),\phi^{s,k}_n(t)\right]=0, \quad \left[\pi^{s,k}_m(t),\pi^{s,k}_n(t)\right]=0,\\
\left[\phi^{w,l}_m(t),\pi^{w,q}_n(t)\right]&=&\iota \delta_{lq}\delta_{mn},\quad \left[\phi^{w,l}_m(t),\phi^{w,q}_n(t)\right]=0, \quad \left[\pi^{w,l}_m(t),\pi^{w,q}_n(t)\right]=0,\\
&&\left[\phi^{s,k}_m(t),\phi^{w,l}_n(t)\right]=0,\quad \left[\pi^{s,k}_m(t),\pi^{w,l}_n(t)\right]=0,\\
&&\left[\phi^{s,k}_m(t),\pi^{w,l}_n(t)\right]=0, \quad \left[\phi^{w,l}_m(t),\pi^{s,k}_n(t)\right]=0.
\end{eqnarray}

The full Hamiltonian can be rewritten in a compact matrix form as,
\begin{eqnarray}
\label{eq:compact_full_Hamiltonian_scalar_field_theory}
\mathrm{H}&=&\sum_{m,n=-\infty}^{\infty}\frac{1}{2}\left[
\begin{pmatrix}
\pi^{s,k}_m & \pi^{w,l}_m
\end{pmatrix}
\begin{pmatrix}
\delta_{mn} & 0 \\
0 & \delta_{mn}
\end{pmatrix}
\begin{pmatrix}
\pi^{s,k}_n\\
\pi^{w,l}_n
\end{pmatrix}
+\right.\nonumber\\
&&\left.\begin{pmatrix}
\phi^{s,k}_m & \phi^{w,l}_m
\end{pmatrix}
\begin{pmatrix}
\mu^2\delta_{mn}+\mathcal{D}^{k}_{ss,mn} & \mathcal{D}^{kq}_{sw,mn}\\
\mathcal{D}^{qk}_{ws,mn} & \mu^2\delta_{mn}+\mathcal{D}^{lq}_{ww,mn}
\end{pmatrix}
\begin{pmatrix}
\phi^{s,k}_n\\
\phi^{w,l}_n
\end{pmatrix}
\right].
\end{eqnarray}
Eq. (\ref{eq:compact_full_Hamiltonian_scalar_field_theory}) shows that within the wavelet formulation, the free scalar field theory is represented by a set of coupled simple harmonic oscillators.
The matrix 
\begin{eqnarray}
\label{eq:omega_square_matrix_scalar_field_theory}
\Omega^2:= \begin{pmatrix}
\mu^2\delta_{mn}+\mathcal{D}^{k}_{ss,mn} & \mathcal{D}^{kq}_{sw,mn}\\
\mathcal{D}^{qk}_{ws,mn} & \mu^2\delta_{mn}+\mathcal{D}^{lq}_{ww,mn}
\end{pmatrix},
\end{eqnarray}
is a real symmetric matrix that can always be diagonalized using a similarity transformation. There exists an orthogonal matrix $O$ such that
\begin{eqnarray}
O^T \Omega^2 O:=
\begin{pmatrix}
{\omega_s}^2 & 0\\
0 & {\omega_w}^2
\end{pmatrix}.
\end{eqnarray}
where,
${\omega_s}^2$ and ${\omega_w}^2$ are the diagonal matrices containing the eigenvalues of matrix $\Omega^2$. These eigenvalues represent the squares of the normal-mode frequencies. The orthogonal matrix can be used to construct the normal mode coordinates.

We now compare the exact values of the squares of the normal mode frequencies $\omega_p^2 $ with those computed within the wavelet formulation. We choose the spatial interval of length $L=20$, the mass of the scalar field $\mu =1$, and restrict the resolution to $k=10$ for illustration. The results are listed in the following Table \ref{tab:normal_mode_frequencies_exact_and_for_res_10_scalar_fields}.
\begin{table}[hbt]
\begin{center}
\setlength{\tabcolsep}{1.0pc}

\caption{Comparison of normal mode frequencies: exact vs wavelet based computation for $k=10$.}
\label{tab:normal_mode_frequencies_exact_and_for_res_10_scalar_fields}
\vspace{1mm}
\begin{tabular}{c  c | c  c}
\specialrule{.15em}{.0em}{.15em}
\hline
\multicolumn{2}{c|}{Exact $\omega_p^2$} & \multicolumn{2}{c}{Computed $\omega_p^2$ for $k=10$}\\
\hline
$(1:16)$   & $(16:32)$    & $(1:16)$   & $(16:32)$  \\
\hline
$1.024674$ & $8.130789$   & $1.024682$ & $8.133211$ \\
$1.098696$ & $8.994380$   & $1.098730$ & $8.997096$ \\
$1.222066$ & $9.907318$   & $1.222142$ & $9.910344$ \\
$1.394784$ & $10.869604$  & $1.394918$ & $10.872958$\\
$1.616850$ & $11.881239$  & $1.617060$ & $11.884936$\\
$1.888264$ & $12.942221$  & $1.888566$ & $12.946279$\\
$2.209027$ & $14.052552$  & $2.209437$ & $14.056986$\\
$2.579137$ & $15.212230$  & $2.579673$ & $15.217059$\\
$2.998595$ & $16.421257$  & $2.999274$ & $16.426496$\\
$3.467401$ & $17.679631$  & $3.468239$ & $17.685298$\\
$3.985555$ & $18.987354$  & $3.986570$ & $18.993465$\\
$4.553058$ & $20.344425$  & $4.554265$ & $20.350997$\\
$5.169908$ & $21.750843$  & $5.171325$ & $21.757893$\\
$5.836106$ & $23.206610$  & $5.837749$ & $23.214155$\\
$6.551652$ & $24.711725$  & $6.553539$ & $24.719781$\\
$7.316547$ & $26.266187$  & $7.318693$ & $26.274772$\\
\hline
\specialrule{.15em}{.15em}{.0em}
\end{tabular}
\end{center}
\end{table}

\begin{figure}[hbt]
\caption{The evaluation of the matrix $\Omega^2(\lambda)$ for different values of $\lambda$.}
\centering
\label{fig:scalar_field_theory_the_evaluation_of_matrix_for_different_values_of_lambda}
\includegraphics[scale=0.33]{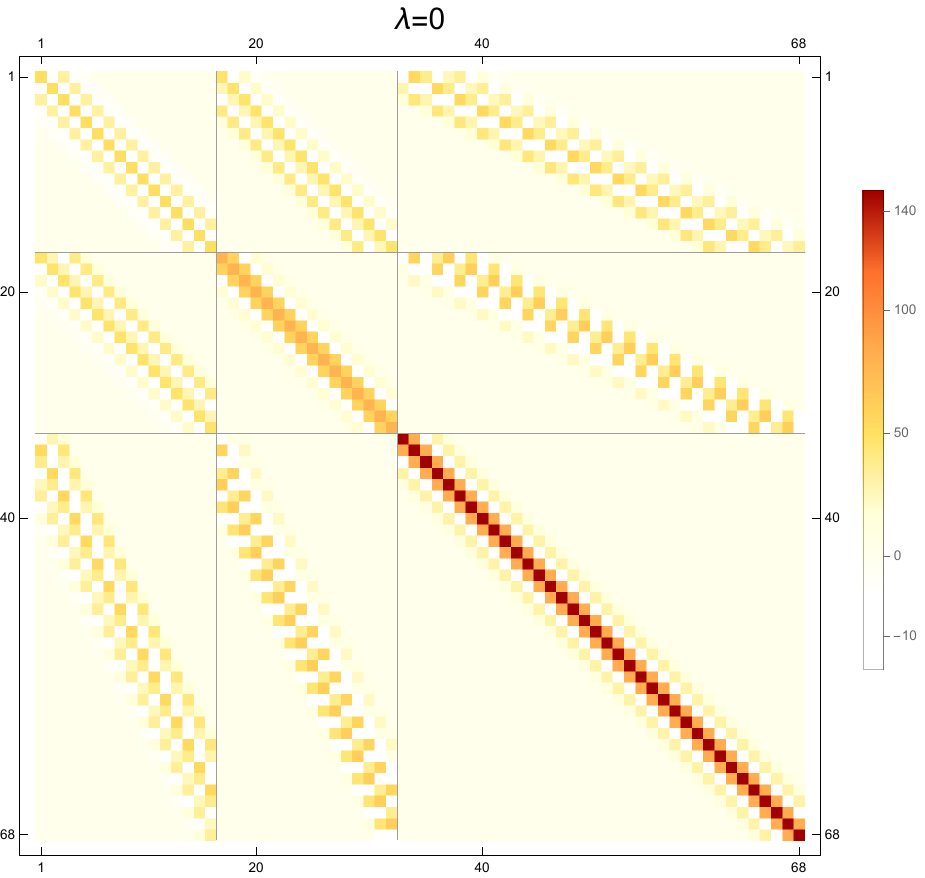}
\includegraphics[scale=0.33]{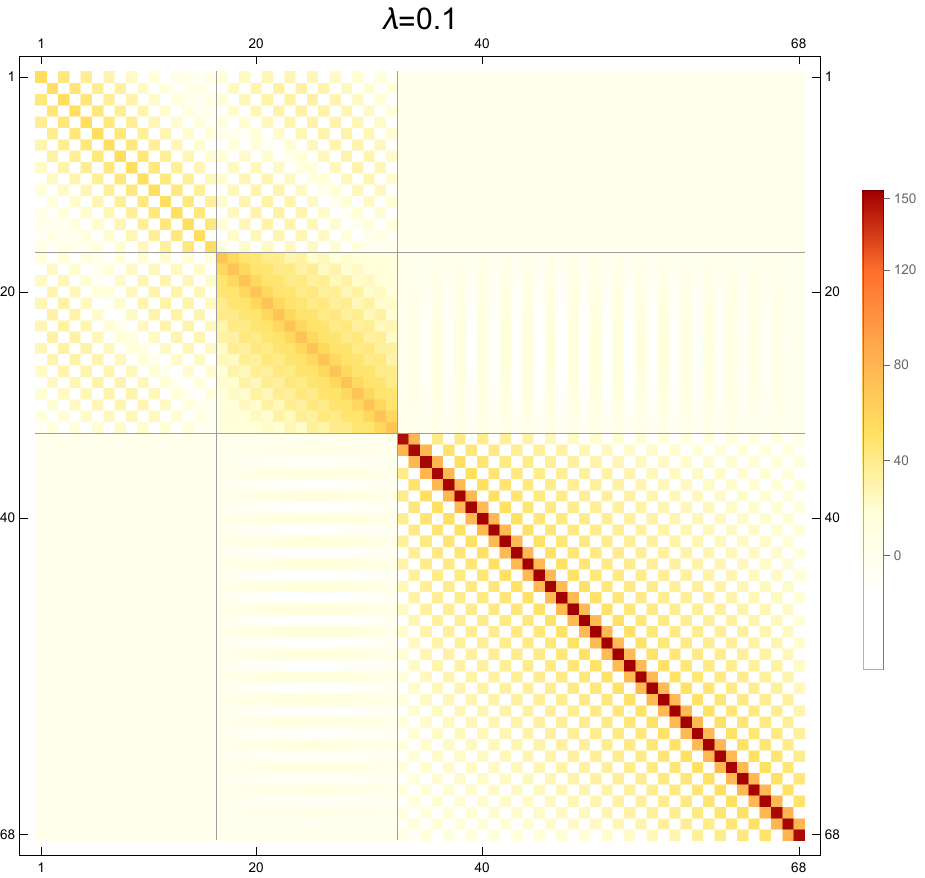}
\includegraphics[scale=0.33]{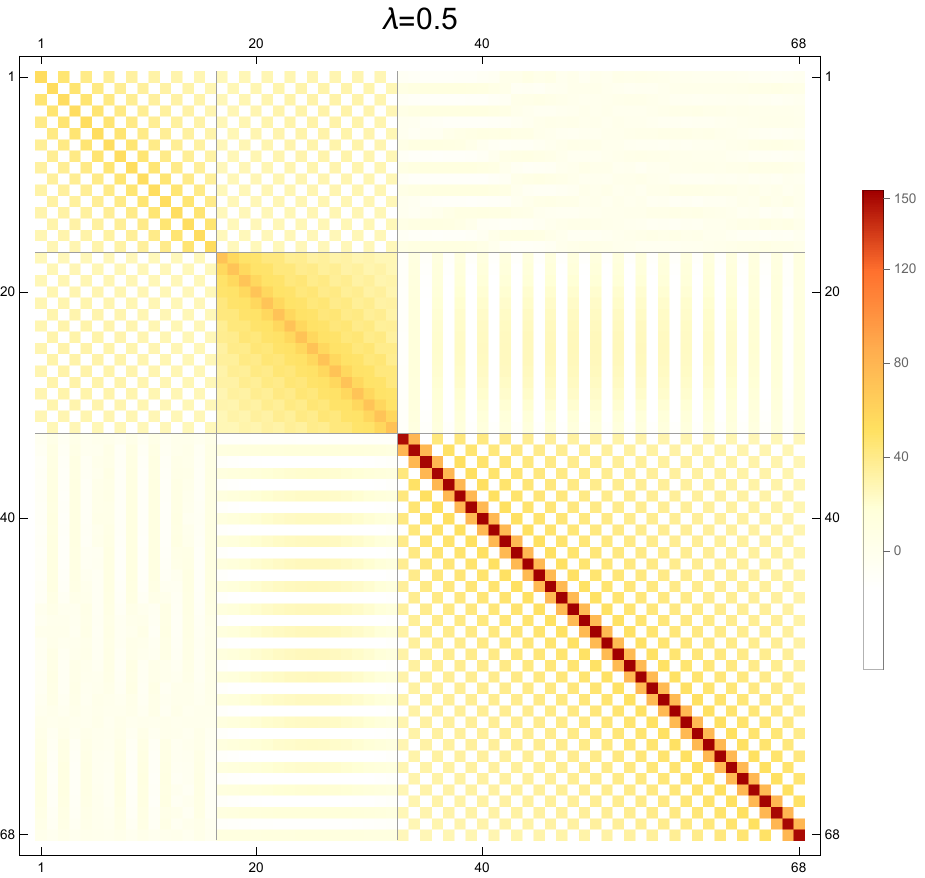}
\includegraphics[scale=0.33]{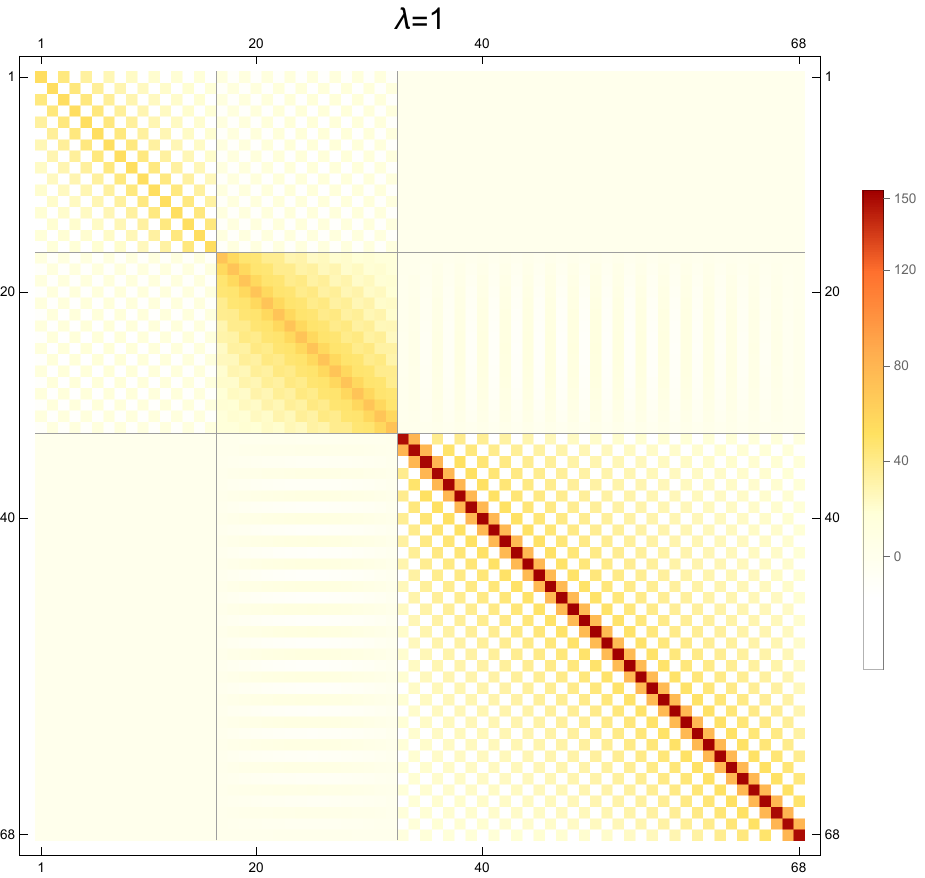}
\includegraphics[scale=0.33]{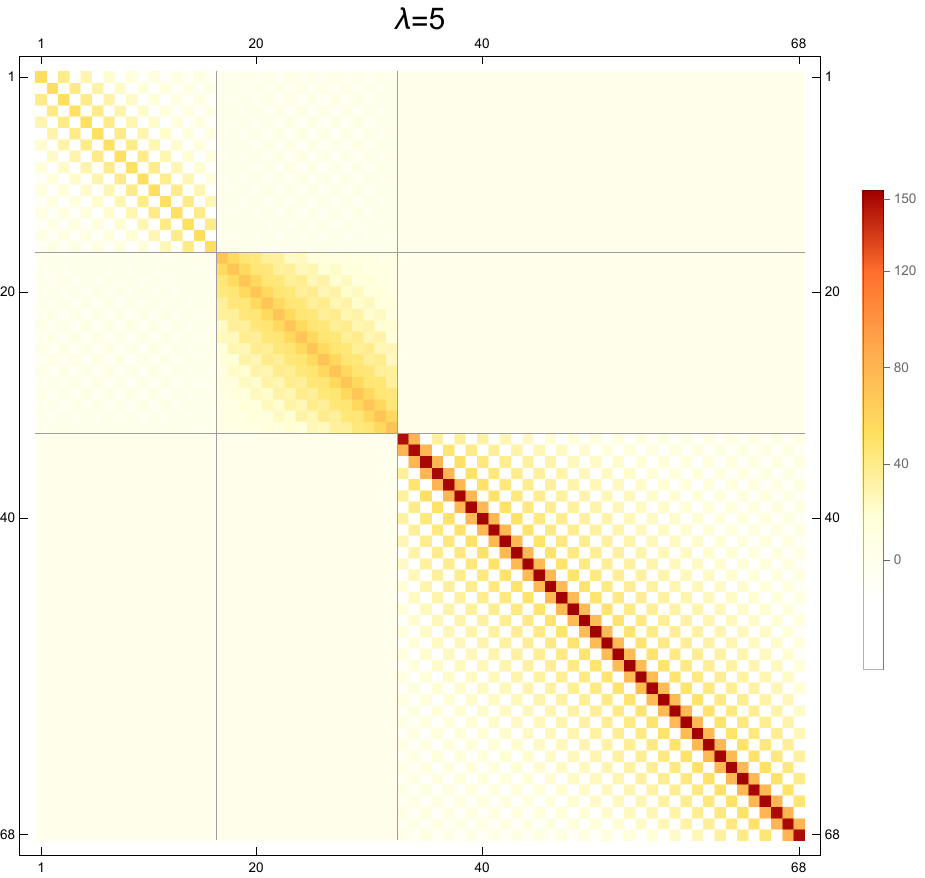}
\includegraphics[scale=0.33]{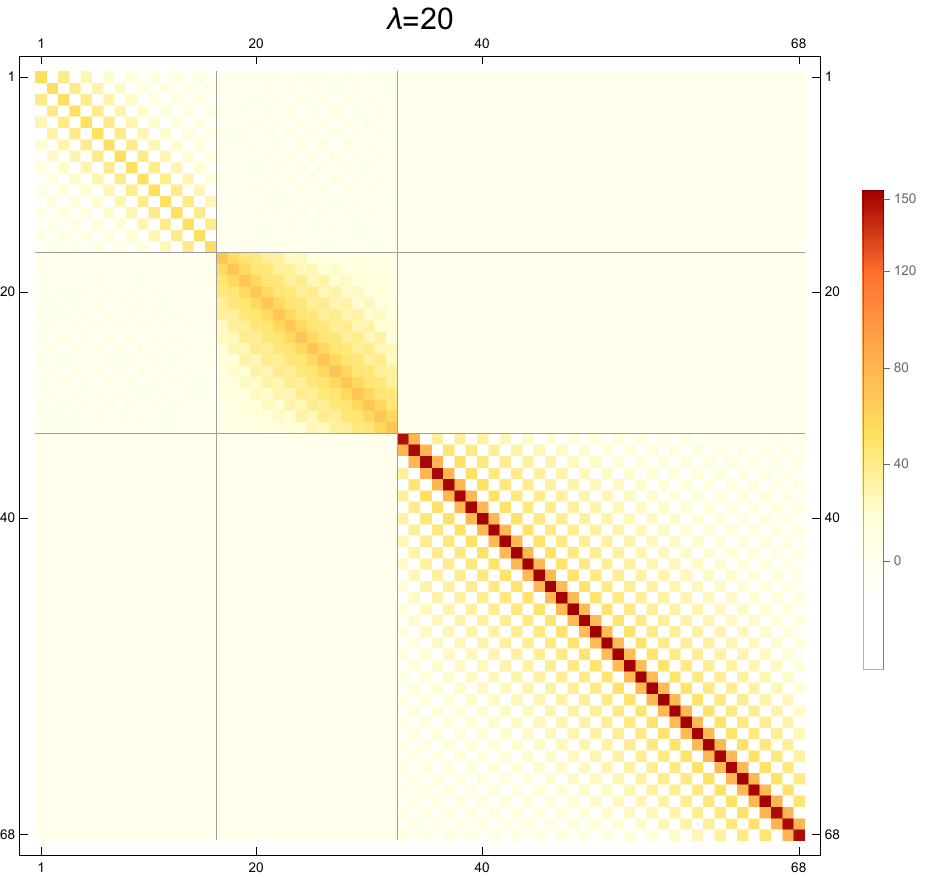}
\end{figure}

Now, we will test the efficacy of the flow-equation, Eq. (\ref{eq:flow_equation_of_Hamiltonian_1}), in block diagonalizing the Hamiltonian for resolution $2$. To do this, first we will rewrite the Hamiltonian, Eq. (\ref{eq:omega_square_matrix_scalar_field_theory}), in the following form using Eq. (\ref{eq:compact_full_Hamiltonian_scalar_field_theory}),
\begin{eqnarray}
\label{eq:compact_full_Hamiltonian_scalar_field_theory_1}
\mathrm{H}=\sum_{m,n=-\infty}^{\infty}\frac{1}{2}\left[
\begin{pmatrix}
\pi^{s,k}_m & \pi^{w,l}_m
\end{pmatrix}
\begin{pmatrix}
\delta_{mn} & 0 \\
0 & \delta_{mn}
\end{pmatrix}
\begin{pmatrix}
\pi^{s,k}_n\\
\pi^{w,l}_n
\end{pmatrix}
+
\begin{pmatrix}
\phi^{s,k}_m & \phi^{w,l}_m
\end{pmatrix}
\Omega^2
\begin{pmatrix}
\phi^{s,k}_n\\
\phi^{w,l}_n
\end{pmatrix}
\right].
\end{eqnarray}
We apply the unitary transformation, parameterized by the continuous parameter $\lambda$,
\begin{eqnarray}
H(\lambda)=U(\lambda) H U^{\dagger}(\lambda)=\sum_{m,n=-\infty}^{\infty}\frac{1}{2}\left[
U(\lambda)
\begin{pmatrix}
\pi^{s,k}_m & \pi^{w,l}_m
\end{pmatrix}
\begin{pmatrix}
\pi^{s,k}_n\\
\pi^{w,l}_n
\end{pmatrix}
U^{\dagger}(\lambda)
+\right.\nonumber\\
\left. U(\lambda)
\begin{pmatrix}
\phi^{s,k}_m & \phi^{w,l}_m
\end{pmatrix}
\Omega^2
\begin{pmatrix}
\phi^{s,k}_n\\
\phi^{w,l}_n
\end{pmatrix}
U^{\dagger}(\lambda)
\right].
\end{eqnarray}
This unitary transformation generates a orthogonal transformation $O$ so the Eq. (\ref{eq:compact_full_Hamiltonian_scalar_field_theory_1}), can be rewritten as,
\begin{eqnarray}
H(\lambda)=\sum_{m,n=-\infty}^{\infty}\frac{1}{2}\left[
\begin{pmatrix}
\pi^{s,k}_m & \pi^{w,l}_m
\end{pmatrix}
\begin{pmatrix}
\pi^{s,k}_n\\
\pi^{w,l}_n
\end{pmatrix}
+
\begin{pmatrix}
\phi^{s,k}_m & \phi^{w,l}_m
\end{pmatrix}
O^{T}(\lambda)
\Omega^2
O(\lambda)
\begin{pmatrix}
\phi^{s,k}_n\\
\phi^{w,l}_n
\end{pmatrix}
\right].
\end{eqnarray}
Differentiating both side with $\lambda$, we get,
\begin{eqnarray}
\frac{dH(\lambda)}{d\lambda}&=&\sum_{m,n=-\infty}^{\infty}\frac{1}{2}\left[
\begin{pmatrix}
\phi^{s,k}_m & \phi^{w,l}_m
\end{pmatrix}\left(
\frac{dO^{T}(\lambda)}{d\lambda}O(\lambda)O^{T}(\lambda)
\Omega^2
O(\lambda)+\right.\right.\nonumber\\
&&\left. \left .O^{T}(\lambda)
\Omega^2 O(\lambda)O^{T}(\lambda)
\frac{dO(\lambda)}{d\lambda}
\right)
\begin{pmatrix}
\phi^{s,k}_n\\
\phi^{w,l}_n
\end{pmatrix}
\right]\nonumber\\
&=& \sum_{m,n=-\infty}^{\infty}\frac{1}{2}\left[
\begin{pmatrix}
\phi^{s,k}_m & \phi^{w,l}_m
\end{pmatrix}\left(
\frac{dO^{T}(\lambda)}{d\lambda}O(\lambda)
\Omega^2(\lambda)-
\Omega^2 (\lambda)
\frac{dO^{T}(\lambda)}{d\lambda}O(\lambda)
\right)
\begin{pmatrix}
\phi^{s,k}_n\\
\phi^{w,l}_n
\end{pmatrix}
\right]\nonumber\\
&=& \sum_{m,n=-\infty}^{\infty}\frac{1}{2}\left[
\begin{pmatrix}
\phi^{s,k}_m & \phi^{w,l}_m
\end{pmatrix}
\left[K(\lambda),\Omega^2(\lambda)\right]
\begin{pmatrix}
\phi^{s,k}_n\\
\phi^{w,l}_n
\end{pmatrix}
\right].
\end{eqnarray}
\normalsize
Where, $K(\lambda)$ is the generator of the transformation and satisfy the following relation,
\begin{eqnarray}
K(\lambda)=\frac{dO^{T}(\lambda)}{d\lambda}O(\lambda)=-O^{T}(\lambda)\frac{dO(\lambda)}{d\lambda}=-K^{\dagger}(\lambda).
\end{eqnarray}
So, the flow equation of the Hamiltonian will become flow equation of the matrix $\Omega^2(\lambda)$ and the corresponding flow equation is,
\begin{eqnarray}
\frac{d\Omega^2(\lambda)}{d\lambda}=\left[K(\lambda),\Omega^2(\lambda)\right].
\end{eqnarray}
To obtain the block diagonal Hamiltonian, we choose the generator in such a way that only the same scale coupling terms are present. The evaluation of matrix elements for different values of $\lambda$ are given in the Fig. \ref{fig:scalar_field_theory_the_evaluation_of_matrix_for_different_values_of_lambda}.

\section{Model}
\label{sec:model_of_two_interacting_scalar_fields}
We consider the system of two scalar fields, $\phi$ and $\psi$, with masses 
$\mu$ and $\nu$, in $1+1$ dimensions, with a mass mixing coupling $g$. Its Lagrangian density is given by,
\begin{eqnarray}
\label{eq:model_lagrangian_density}
&&\mathcal{L}(\phi(\textrm{x},t),\partial_x \phi(\textrm{x},t),\dot{\phi}(\textrm{x},t),\psi(\textrm{x},t),\partial_x \psi(\textrm{x},t),\dot{\psi}(\textrm{x},t)=\frac{1}{2}\left[\dot{\phi}^2(\textrm{x},t)-(\partial_x \phi(\textrm{x},t))^2-\mu^2 \phi(\textrm{x},t)^2\right]\nonumber\\
&&+\frac{1}{2}\left[\dot{\psi}^2(\textrm{x},t)-(\partial_x \psi(\textrm{x},t))^2-\nu^2\psi(\textrm{x},t)^2 \right] -\frac{g^2}{2} \phi(\textrm{x},t) \psi(\textrm{x},t).
\end{eqnarray}
The Hamiltonian has the form, 
\begin{eqnarray}
\mathrm{H}=\int dx\left(\frac{1}{2}\left[\pi_{\phi}^2+\left(\partial_x \phi \right)^2+\mu^2\phi^2\right]+\frac{1}{2}\left[\pi_{\psi}^2+\left(\partial \psi\right)^2+\nu^2\psi^2\right]+\frac{g^2}{2}\phi\psi\right).\quad
\end{eqnarray}
where,
\begin{eqnarray}
\pi_\phi=\frac{\partial \mathcal{L}}{\partial \dot{\phi}}=\dot{\phi}, \,\text{and}\quad \pi_\psi=\frac{\partial \mathcal{L}}{\partial \dot{\psi}}=\dot{\psi},
\end{eqnarray}
represent the canonical momenta corresponding to $\phi$ and $\psi$ respectively.
Expanding the fields in the Fourier series within the interval $L$ yields the following expression of the Hamiltonian,
\begin{eqnarray}
\mathrm{H}=\sum_{p=1}^{\infty}\left(\frac{1}{2}\left[\pi_{\phi,p}(t)^2+(p^2+\mu^2)\phi_p(t)^2+\pi_{\psi,p}(t)^2+(p^2+\nu^2)\psi_p(t)^2\right]+\right.\\
\left. \frac{g^2}{2} \phi_p(t)\psi_p(t)\right),
\end{eqnarray}
where,
\begin{eqnarray}
\phi(\textrm{x},t)&=&\sum_{p=1}^{\infty} \phi_{p}(t)\sqrt{\frac{2}{L}}\sin\left(\frac{p\pi x}{L}\right), \quad \pi_\phi(\textrm{x},t)=\sum_{p=1}^{\infty} \pi_{\phi,p}(t)\sqrt{\frac{2}{L}}\sin\left(\frac{p\pi x}{L}\right),\nonumber \\
\psi(\textrm{x},t)&=&\sum_{p=1}^{\infty} \psi_{p}(t)\sqrt{\frac{2}{L}}\sin\left(\frac{p\pi x}{L}\right), \quad \pi_\psi(\textrm{x},t)=\sum_{p=1}^{\infty} \pi_{\psi,p}(t)\sqrt{\frac{2}{L}}\sin\left(\frac{p\pi x}{L}\right).
\end{eqnarray}
The full Hamiltonian, rewritten in the Fourier variables as,
\begin{eqnarray}
\textrm{H}=\sum_{p=1}^{\infty}\frac{1}{2}\left[\begin{pmatrix}
\pi_{\phi,p}(t)&\pi_{\psi,p}(t)
\end{pmatrix}
\begin{pmatrix}
\mathds{1}&0\\
0&\mathds{1}
\end{pmatrix}
\begin{pmatrix}
\pi_{\phi,p}(t)\\
\pi_{\psi,p}(t)
\end{pmatrix}+\right.\nonumber\\
\left.\begin{pmatrix}
\phi_p(t)&\psi_p(t)
\end{pmatrix}
\begin{pmatrix}
\mu^2+p^2 & \frac{g^2}{2}\\
\frac{g^2}{2}& \nu^2+p^2
\end{pmatrix}
\begin{pmatrix}
\phi_p(t)\\
\psi_p(t)
\end{pmatrix}\right],
\end{eqnarray}
shows that the mass mixing term only couples the same Fourier mode of $\phi_p$ and $\psi_p$. The quadratic nature of the mass mixing term allows one to introduce the normal mode variables, $\Phi_p$ and $\Psi_p$, 
\begin{eqnarray}
\Phi_p(t)&=&\frac{1}{2}\left[\left(\frac{(\nu^2-g^2)-\sqrt{(\nu^2-\mu^2)^2+g^4}}{\sqrt{(\nu^2-\mu^2)^2+g^4}}\right)^{\frac{1}{2}}\phi_p(t)+\right.\nonumber\\
&&\left.\left(\frac{\sqrt{(\nu^2-\mu^2)^2+g^4}-(\nu^2-g^2)}{\sqrt{(\nu^2-\mu^2)^2+g^4}}\right)^{\frac{1}{2}}\psi_p(t)\right],\\
\Psi_p(t)&=&\frac{1}{2}\left[\left(\frac{(\nu^2-g^2)-\sqrt{(\nu^2-\mu^2)^2+g^4}}{\sqrt{(\nu^2-\mu^2)^2+g^4}}\right)^{\frac{1}{2}}\phi_p(t)-\right.\nonumber\\
&&\left.\left(\frac{\sqrt{(\nu^2-\mu^2)^2+g^4}-(\nu^2-g^2)}{\sqrt{(\nu^2-\mu^2)^2+g^4}}\right)^{\frac{1}{2}}\psi_p(t)\right],
\end{eqnarray}
\normalsize
which are not coupled to each other. This can be seen by expressing the Hamiltonian in terms of the normal mode variables $\Phi_p$ and $\Psi_p$,
\begin{eqnarray}
\textrm{H}= \sum_{p=1}^{\infty}\frac{1}{2}\left[\Pi^2_{\Phi,p}(t)+\omega^2_{\Phi,p}\Phi^2_p(t)+\Pi^2_{\Psi,p}(t)+\omega^2_{\Psi,p}\Psi^2_p(t)\right],
\end{eqnarray}
where, 
\begin{eqnarray}
\label{eq:exact_omega_phi_p}
\omega_{\Phi,p}=\sqrt{\left(\frac{p\pi}{L}\right)^2+\frac{\mu^2+\nu^2}{2}+\frac{1}{2}\sqrt{(\mu^2-\nu^2)^2+g^4} },
\end{eqnarray}

\begin{eqnarray}
\label{eq:exact_omega_psi_p}
\omega_{\Psi,p}=\sqrt{\left(\frac{p\pi}{L}\right)^2+\frac{\mu^2+\nu^2}{2}-\frac{1}{2}\sqrt{(\mu^2-\nu^2)^2+g^4}},
\end{eqnarray}
denote the corresponding normal mode frequencies.
\section{Representation of the model in Daubechies wavelet basis using scaling and wavelet functions}
\label{sec:representation_of_Hamiltonian_of_two_coupled_scalr_fields_in_wavelet_basis}
In this section, we outline the canonical quantization of the model defined in section \ref{sec:model_of_two_interacting_scalar_fields}. We begin by resolving both the scalar fields, $\phi$ and $\psi$, and their canonical conjugates, $\pi_\phi$ and $\pi_\psi$, in Daubechies wavelet basis,
\begin{eqnarray}
\label{eq:discretized_phi_field}
\phi(x,t)&=&\sum_{n}\phi^{s,k}_{n}(t)s^{k}_n(x)+\sum_{n,l\ge k}\phi^{w,l}_{n}(t)w^l_n(x),\\
\label{eq:discretized_pi_phi_field}
\pi_{\phi}(x,t)&=&\sum_{n}\pi^{s,k}_{\phi ,n}(t)s^k_n(x)+\sum_{n,l\ge k}\pi^{w,l}_{\phi, n}(t)w^l_n(x),
\end{eqnarray}
and,
\begin{eqnarray}
\label{eq:discretized_psi_field}
\psi(x,t)&=&\sum_{n}\psi^{s,k}_{n}(t)s^k_n(x)+\sum_{n,l\ge k}\psi^{w,l}_{n}(t)w^l_n(x),\\
\label{eq:discretized_pi_psi_field}
\pi_{\psi}(x,t)&=&\sum_{n}\pi^{s,k}_{\psi, n}(t)s^k_n(x)+\sum_{n,l\ge k}\pi^{w,l}_{\psi, n}(t)w^l_n(x).
\end{eqnarray}

The discrete scaling and wavelet basis coefficients can be extracted from the fields and their canonical conjugates using,
\begin{eqnarray}
\begin{aligned}
\label{eq:definition_of_phi_coefficients}
\phi^{s,k}_{n}(t)&=&\int dx \phi(x,t)s^k_n(x),\;\;\;
\phi^{w,l}_{n}(t)&=&\int dx \phi(x,t)w^l_n(x),\\
\pi^{s,k}_{\phi, n}(t)&=&\int dx \pi_\phi(x,t)s^k_n(x),\;\;\;
\pi^{w,k}_{\phi, n}(t)&=&\int dx \pi_\phi(x,t)w^l_n(x),
\end{aligned}
\end{eqnarray}
and
\begin{eqnarray}
\begin{aligned}
\label{eq:definition_of_psi_coefficients}
\psi^{s,k}_{n}(t)&=&\int dx \psi(x,t)s^k_n(x),\;\;\;
\psi^{w,l}_{n}(t)&=&\int dx \psi(x,t)w^l_n(x),\\
\pi^{s,k}_{\psi n}(t)&=&\int dx \pi_\psi(x,t)s^k_n(x),\;\;\;
\pi^{w,l}_{\psi, n}(t)&=&\int dx \pi_\psi(x,t)w^l_n(x).
\end{aligned}
\end{eqnarray}
The equal time canonical commutation relation of the fields and their conjugates,
\begin{eqnarray}
\begin{aligned}
&&\left[\phi(x,t),\pi_\phi(y,t)\right]=-i \delta(x-y),\quad
\left[\phi(x,t),\phi(y,t)\right]=0,\\
&&\left[\psi(x,t),\pi_\psi(y,t)\right]=-i \delta(x-y),\quad
\left[\psi(x,t),\psi(y,t)\right]=0,
\end{aligned}
\end{eqnarray}
written in terms of the scaling and wavelet basis coefficients, using Eq. (\ref{eq:definition_of_phi_coefficients}) and Eq. (\ref{eq:definition_of_psi_coefficients}), reveal as mutually independent canonical pairs of Hamilton's variables. The nontrivial commutation relations are given by,
\begin{eqnarray}
\begin{aligned}
\left[\phi^{s,k}_n(t),\pi^{s,k}_{\phi,m}(t)\right]=i \delta_{mn},\quad
\left[\phi^{w,l}_{n}(t),\pi^{w,r}_{\phi,m}(t)\right]=i \delta_{mn}\delta_{lr},\\
\end{aligned}
\end{eqnarray}
and,
\begin{eqnarray}
\begin{aligned}
\left[\psi^{s,k}_n(t),\pi^{s,k}_{\psi,m}(t)\right]=i \delta_{mn},\quad
\left[\psi^{w,l}_{n}(t),\pi^{w,r}_{\psi,n}(t)\right]=i \delta_{mn}\delta_{lr},\\
\end{aligned}
\end{eqnarray}

We express the Hamiltonian in terms of the scaling and wavelet canonical variables by substituting Eq. (\ref{eq:discretized_phi_field})-Eq. (\ref{eq:discretized_pi_psi_field}) into Eq. (\ref{eq:model_lagrangian_density}). The result shows that the Hamiltonian is a sum of three terms, 
\begin{eqnarray}
\label{eq:phipsi_hamiltonian}
\mathrm{H}=\mathrm{H}_{ss}+\mathrm{H}_{ww}+\mathrm{H}_{sw},
\end{eqnarray}
where $\mathrm{H}_{ss}$ describes the physics of the model on coarse length scales down to resolution $k$,
\begin{eqnarray}
\label{eq:phipsi_h_ss}
\mathrm{H}_{ss}:=&&\frac{1}{2}\left(\sum_n \pi^{s,k}_{\phi,n}\pi^{s,k}_{\phi,n}+\sum_n \pi^{s,k}_{\phi,n}\pi^{s,k}_{\phi,n}
+\sum_{n,m}\phi^{s,k}_{n}\phi^{s,k}_{m}\mathcal{D}^k_{ss,mn}+\sum_{n,m}\psi^{s,k}_{ss,n}\psi^{s,k}_{m}\mathcal{D}^k_{s,mn}\right.\nonumber\\
&&\left.+\sum_n \mu^2 \phi^{s,k}_{n}\phi^{s,k}_{n}+\sum_n \nu^2 \psi^{s,k}_{n}\psi^{s,k}_{n}
+\sum_n g^2 \phi^{s,k}_{n}\psi^{s,k}_{n}\right),
\end{eqnarray}
$\mathrm{H}_{ww}$ describes the physics of the model on length scales finer than resolution $k$, 
\begin{eqnarray}
\label{eq:phipsi_h_ww}
\mathrm{H}_{ww}:=&&\frac{1}{2}\left(\sum_{\substack{n \\ l\ge k}} \pi^{w,l}_{\phi,n}\pi^{w,l}_{\phi,n}+\sum_{\substack{n \\l\ge k}} \pi^{w,l}_{\psi,n}\pi^{w,l}_{\psi,n}
+\sum_{\substack{n,m \\(l,q)\ge k}}\phi^{w,l}_{n}\phi^{w,q}_{m}\mathcal{D}^{lq}_{ww,mn}+\sum_{\substack{n,m \\(l,q)\ge k}}\psi^{w,l}_{n}\psi^{w,q}_{m}\mathcal{D}^{lq}_{ww,mn}\right.\nonumber\\
&&\left.+\sum_{\substack{n \\ {l}\ge k}} \mu^2 \phi^{w,l}_{n}\phi^{w,l}_{n}+\sum_{\substack{n \\ l\ge k}} \nu^2 \psi^{w,l}_{n}\psi^{w,l}_{n}
+\sum_{\substack{n \\ l\ge k}} g^2 \phi^{w,l}_{n}\psi^{w,l}_{n}\right),
\end{eqnarray}
and, $\mathrm{H}_{sw}$ represents interactions between lengths scales coarser and finer than resolution $k$,
\begin{eqnarray}
\label{eq:phipsi_h_sw}
\mathrm{H}_{sw}:=&&\frac{1}{2}\left(\sum_{\substack{n,m \\ q\ge k}} \phi^k_{n}\phi^{w,q}_{m}\mathcal{D}^{kq}_{sw,nm}+\sum_{\substack{n,m \\ l\ge k}} \phi^{w,l}_{m}\phi^{s,k}_{n}\mathcal{D}^{kl}_{sw,mn} 
+\sum_{\substack{n,m \\ q\ge k}} \psi^{s,k}_{n}\psi^{w,q}_{m}\mathcal{D}^{kq}_{sw,nm}+\right.\nonumber\\
&&\left.\sum_{\substack{n,m \\ l\ge k}} \psi^{w,l}_{m}\psi^k_{n}\mathcal{D}^{kl}_{sw,mn}\right).
\end{eqnarray}
The form of $\mathrm{H}_{sw}$ shows that the nature of the coupling between $\phi$ and $\psi$ is such that they don't couple across different length scales.  The coefficients $\mathcal{D}^k_{ss,nm}$, $\mathcal{D}^{lq}_{ww,nm}$ and $\mathcal{D}^{kq}_{sw,nm}$ are the constant matrices given by,
\begin{eqnarray}
\mathcal{D}^k_{ss,nm}&=&\int \frac{d}{dx}s^k_n(x)\frac{d}{dx}s^k_m(x)dx,\\
\mathcal{D}^{lq}_{ww,nm}&=&\int \frac{d}{dx}w^l_n(x)\frac{d}{dx}w^q_m(x)dx,\\
\mathcal{D}^{kq}_{sw,nm}&=&\int \frac{d}{dx}s^k_n(x)\frac{d}{dx}w^q_m(x)dx \;\;\; q\ge k .
\end{eqnarray}
These matrices can be evaluated using the procedure due to Beylkin \cite{doi:10.1137/0729097}. This procedure is outlined in detail by Polyzou \cite{PhysRevD.87.116011}. and the rest of the pair combinations commute with each other. The procedure to evaluate this integrations are given in the Appendix 

\section{Analysis}
\label{sec:analysis}
In this section, we demonstrate the flow equation method in the context of the elementary model of interacting field theory introduced in Sec \ref{sec:representation_of_Hamiltonian_of_two_coupled_scalr_fields_in_wavelet_basis}. This constitutes an extension of the analysis of the free scalar field theory due to Michlin and Polyzou \cite{PhysRevD.95.094501}.

The Hamiltonian of this quantum field theory, Eq. (\ref{eq:phipsi_h_ss})-Eq. (\ref{eq:phipsi_h_ww}), can be rewritten in a compact form,
\begin{eqnarray}
&&\mathrm{H}=\sum_{m,n=-\infty}^{\infty}\frac{1}{2}\left[\begin{pmatrix}
\pi^{s,k}_{\phi,m}\;&\pi^{s,k}_{\psi,m}\;&\pi^{w,l}_{\phi,m}\;&\pi^{w,l}_{\psi,m}
\end{pmatrix}
\begin{pmatrix}
\delta_{mn}&0&0&0\\
0&\delta_{mn}&0&0\\
0&0&\delta_{mn}&0\\
0&0&0&\delta_{mn}\\
\end{pmatrix}
\begin{pmatrix}
\pi^{s,k}_{\phi,n}\\
\pi^{s,k}_{\psi,n}\\
\pi^{w,q}_{\phi,n}\\
\pi^{w,q}_{\psi,n}
\end{pmatrix}+ \right.\nonumber \\
\label{eq:psiphi_hamiltonian_matrix_form}
&&\begin{pmatrix}
\phi^{s,k}_{m} & \psi^{s,k}_{m} & \phi^{w,l}_{m} & \psi^{w,l}_{m}
\end{pmatrix}\nonumber\\
&&\left.\underbrace{\left(
\begin{array}{@{}c|c@{}}
\begin{matrix}
\mu^2\delta_{mn}+\mathcal{D}^k_{ss,mn} & \frac{g^2}{2}\delta_{mn} \\
\frac{g^2}{2}\delta_{mn} & \nu^2\delta_{mn}+\mathcal{D}^k_{ss,mn}
\end{matrix}
&
\begin{matrix}
\mathcal{D}^{kq}_{sw,mn} & 0 \\
0 & \mathcal{D}^{kq}_{sw,mn}
\end{matrix}\\
\cmidrule[0.4pt]{1-2}\\
\begin{matrix}
\mathcal{D}^{qk}_{ws,mn} & 0 \\
0 & \mathcal{D}^{qk}_{ws,mn}
\end{matrix}
&
\begin{matrix}
\mu^2\delta_{mn}+\mathcal{D}^{lq}_{ww,mn} & \frac{g^2}{2}\delta_{mn}\\
\frac{g^2}{2}\delta_{mn} & \nu^2\delta_{mn}+\mathcal{D}^{lq}_{ww,mn}
\end{matrix}
\end{array}
\right)}_{\Omega^2}
\begin{pmatrix}
\phi^{s,k}_{n} \\ \psi^{s,k}_{n} \\ \phi^{w,q}_{n} \\ \psi^{w,q}_{n}
\end{pmatrix}
\right]\quad,\nonumber\\
\end{eqnarray}

\normalsize
where, $\Omega^2$ denotes a matrix that represents the coupling of each scalar field across length scales. It is important to note that in this model, the two scalar fields couple to each other only on the same scale. They do not have any coupling across scales.

The symmetric nature of the coupling matrix $\Omega^2$ implies the existence of an orthogonal matrix $O$, using which we can compute the normal mode frequencies and construct the corresponding normal mode variables. The Stone-Von Neumann theorem guarantees the existence of a unitary operator $U$ which generates this orthogonal transformation $O$,
\begin{eqnarray}
\label{eq:the_transformation_of_filed_operators}
U
\begin{pmatrix}
\phi^{s,k} \\
\psi^{s,k} \\
\phi^{w,l} \\
\psi^{w,l}
\end{pmatrix}
U^{\dagger}
=
\begin{pmatrix}
\Phi^{s,k} \\
\Psi^{s,k} \\
\Phi^{w,l} \\
\Psi^{w,l}
\end{pmatrix}
=O
\begin{pmatrix}
\phi^{s,k} \\
\psi^{s,k} \\
\phi^{w,l} \\
\psi^{w,l}
\end{pmatrix}
\quad \text{and}\quad
U
\begin{pmatrix}
\pi^{s,k} \\
\pi^{s,k} \\
\pi^{w,l} \\
\pi^{w,l}
\end{pmatrix}
U^{\dagger}
=
\begin{pmatrix}
\Pi^{s,k} \\
\Pi^{s,k} \\
\Pi^{w,l} \\
\Pi^{w,l}
\end{pmatrix}
=O
\begin{pmatrix}
\pi^{s,k} \\
\pi^{s,k} \\
\pi^{w,l} \\
\pi^{w,l}
\end{pmatrix}.
\end{eqnarray}
Here, the superscripts on the normal mode variables $\Phi^{s,k}$, $\Phi^{w,l}$, $\Psi^{s,k}$ and $\Psi^{w,l}$, serve as a mere reminder that they arose from $\phi^{s,k}$, $\phi^{w,l}$, $\psi^{s,k}$ and $\psi^{w,l}$ through an orthogonal rotation. The rotation $O$ itself mixes resolutions. The unitary transform of the Hamiltonian gives the normal mode representation of the model,
\begin{eqnarray}
H'&=& UHU^{\dagger}=\nonumber\\
&&\frac{1}{2}
\left[
\begin{pmatrix}
\Pi^{s,k} & \Pi^{s,k} & \Pi^{w,l} & \Pi^{w,l}
\end{pmatrix}
\begin{pmatrix}
\mathds{1}&0&0&0\\
0&\mathds{1}&0&0\\
0&0&\mathds{1}&0\\
0&0&0&\mathds{1}
\end{pmatrix}
\begin{pmatrix}
\Pi^{s,k} \\
\Pi^{s,k} \\
\Pi^{w,l} \\
\Pi^{w,l}
\end{pmatrix}\right.\nonumber\\
&&+
\left.\begin{pmatrix}
\Phi^{s,k} & \Psi^{s,k} & \Phi^{w,l} & \Psi^{w,l}
\end{pmatrix}
\begin{pmatrix}
\omega^2_{s,\Phi} & 0 & 0 & 0\\
0 & \omega^2_{s,\Psi} & 0 & 0\\
0 & 0 & \omega^2_{w,\Phi} & 0\\
0 & 0 & 0 & \omega^2_{w,\Psi}
\end{pmatrix}
\begin{pmatrix}
\Phi^{s,k} \\
\Psi^{s,k} \\
\Phi^{w,l} \\
\Psi^{w,l}
\end{pmatrix}
\right],
\end{eqnarray}
\normalsize
where, $\omega^2_{s,\Phi}$, $\omega^2_{s,\Phi}$, $\omega^2_{w,\Psi}$, and $\omega^2_{w,\Psi}$ are the diagonal matrices containing the squares of the normal mode frequencies associated with $\Phi$ and $\Psi$ respectively. The correspondence between the normal mode variables and normal mode frequencies in the Fourier based analysis given in Sec. \ref{sec:model_of_two_interacting_scalar_fields} and the present wavelet based analysis is given by,
\begin{eqnarray}
\begin{pmatrix}
\Phi^{s,k} \\
\Phi^{w,l}
\end{pmatrix}
\rightarrow
\Phi_p,
\quad
\begin{pmatrix}
\Pi^{s,k}_{\Phi} \\
\Pi^{w,l}_{\Phi}
\end{pmatrix}
\rightarrow
\Pi_{\Phi,p},
\quad
\begin{pmatrix}
\omega^2_{s,\Phi} \\
\omega^2_{w,\Phi}
\end{pmatrix}
\rightarrow
\omega^2_{\Phi,p},\\
\begin{pmatrix}
\Psi^{s,k} \\
\Psi^{w,l}
\end{pmatrix}
\rightarrow
\Psi_p,
\quad
\begin{pmatrix}
\Pi^{s,k}_{\Psi} \\
\Pi^{w,l}_{\Psi}
\end{pmatrix}
\rightarrow
\Pi_{\Psi,p},
\quad
\begin{pmatrix}
\omega^2_{s,\Psi} \\
\omega^2_{w,\Psi}
\end{pmatrix}
\rightarrow
\omega^2_{\Psi,p},
\end{eqnarray}

As an illustration, in Table \ref{tab:compare_omega_and_omega_for_resolution_10}, we compare the first sixteen exact values of the normal mode frequencies for $\Phi$ and $\Psi$ with the corresponding wavelet based estimates for resolution $k=1$ and $k=6$, within a spatial interval of length $L=20$.
\begin{table}[hbt]
\begin{center}
\caption{Comparison between the exact and wavelet based estimates for normal mode frequencies}
\label{tab:compare_omega_and_omega_for_resolution_10}
\setlength{\tabcolsep}{0.5pc}
\vspace{1mm}
\begin{tabular}{c  c | c  c | c  c}
\specialrule{.15em}{.0em}{.15em}
\hline
\multicolumn{2}{c|}{Exact $\omega_p^2$} & \multicolumn{2}{c|}{$\omega_p^2$ for $k=1$} & \multicolumn{2}{c}{$\omega_p^2$ for $k=6$} \\
\hline
$\Phi$ & $\Psi$ & $\Phi$ & $\Psi$ & $\Phi$ & $\Psi$ \\
\hline
$0.52467$ & $1.52467$ & $0.53595$ & $1.53595$ & $0.53587$ & $1.53587$ \\
$0.59870$ & $1.59870$ & $0.64408$ & $1.64408$ & $0.64365$ & $1.64364$ \\
$0.72207$ & $1.72207$ & $0.82537$ & $1.82537$ & $0.82387$ & $1.82387$ \\
$0.89478$ & $1.89478$ & $0.97526$ & $1.97526$ & $1.07745$ & $2.07745$ \\
$1.11685$ & $2.11685$ & $1.08189$ & $2.08189$ & $1.40536$ & $2.40536$ \\
$1.38826$ & $2.38826$ & $1.41694$ & $2.41694$ & $1.80809$ & $2.80809$ \\
$1.70903$ & $2.70903$ & $1.83517$ & $2.83517$ & $1.70944$ & $2.70944$ \\
$2.07914$ & $3.07914$ & $2.34243$ & $3.34243$ & $2.28520$ & $3.28520$ \\
$2.49859$ & $3.49859$ & $2.94602$ & $3.94602$ & $2.83516$ & $3.83516$ \\
$2.96740$ & $3.96740$ & $3.65564$ & $4.65564$ & $3.45602$ & $4.45602$ \\
$3.48556$ & $4.48556$ & $4.48633$ & $5.48633$ & $4.14729$ & $5.14729$ \\
$4.05306$ & $5.05306$ & $5.46196$ & $6.46196$ & $4.91207$ & $5.91207$ \\
$4.66991$ & $5.66991$ & $6.61455$ & $7.61455$ & $5.75707$ & $6.75707$ \\
$5.33611$ & $6.33611$ & $7.97792$ & $8.97792$ & $6.68965$ & $7.68965$ \\
$6.05165$ & $7.05165$ & $9.58140$ & $10.5814$ & $7.71480$ & $8.71480$ \\
$6.81655$ & $7.81655$ & $11.4468$ & $12.4468$ & $8.83446$ & $9.83446$ \\
\hline
\specialrule{.15em}{.15em}{.0em}
\end{tabular}
\end{center}
\end{table}

Table \ref{tab:compare_omega_and_omega_for_resolution_10} presents a comparison of the exact values of the lowest sixteen normal mode frequencies associated with the normal variables, $\Phi$ and $\Psi$ with those computed from the model truncated at resolution $k=0$, $k=1$ and $k=10$. The number of degrees of freedom increases from $32$ to $72$ to $40952$. We can see that the accuracy of the lowest $\omega_p^2$, obtained through the numerical diagonalization of the coupling matrix $\Omega^2$, increases from $97.80\% (k=0)$ to $99.06\% (k=1)$ to $99.99\% (k=10)$. The number of degrees of freedom scales exponentially with resolution $k$ and linearly with the length $L$ of the interval.   In what follows, we study the efficacy of the flow equation method in constructing an effective Hamiltonian for this model. Starting from the QFT truncated at $k=1$, we use the flow equation method to construct an effective Hamiltonian within the $k=0$ sector.

To test the flow equation, Eq. (\ref{eq:flow_equation_of_Hamiltonian_1}), we apply the unitary transformation, parameterized by the continuous flow parameter $\lambda$, to the Hamiltonian described in Eq. (\ref{eq:psiphi_hamiltonian_matrix_form}). The transformed Hamiltonian will take the following form,
\begin{eqnarray}
&&\mathrm{H}(\lambda)=U(\lambda) \mathrm{H} U^{\dagger}(\lambda)\nonumber\\
&&=\sum_{m,n=-\infty}^{\infty}\frac{1}{2}\left[U(\lambda)\begin{pmatrix}
\pi^{s,k}_{\phi,m}\;&\pi^{s,k}_{\psi,m}\;&\pi^{w,l}_{\phi,m}\;&\pi^{w,l}_{\psi,m}
\end{pmatrix}
\begin{pmatrix}
\pi^{s,k}_{\phi,n}\\
\pi^{s,k}_{\psi,n}\\
\pi^{w,q}_{\phi,n}\\
\pi^{w,q}_{\psi,n}
\end{pmatrix}U^{\dagger}(\lambda)+\right. \nonumber\\
&&\left. U(\lambda)\begin{pmatrix}
\phi^{s,k}_{m} & \psi^{s,k}_{m} & \phi^{w,l}_{m} & \psi^{w,l}_{m}
\end{pmatrix}
\Omega^2
\begin{pmatrix}
\phi^{s,k}_{n} \\ \psi^{s,k}_{n} \\ \phi^{w,q}_{n} \\ \psi^{w,q}_{n}
\end{pmatrix}U^{\dagger}(\lambda)
\right]\nonumber\\
&&=\sum_{m,n=-\infty}^{\infty}\frac{1}{2}\left[\begin{pmatrix}
\Pi^{s,k}_{\phi,m}\;&\Pi^{s,k}_{\psi,m}\;&\Pi^{w,l}_{\phi,m}\;&\Pi^{w,l}_{\psi,m}
\end{pmatrix}
\begin{pmatrix}
\Pi^{s,k}_{\phi,n}\\
\Pi^{s,k}_{\psi,n}\\
\Pi^{w,q}_{\phi,n}\\
\Pi^{w,q}_{\psi,n}
\end{pmatrix}+ \right.\nonumber\\
&&\left.\begin{pmatrix}
\Phi^{s,k}_{m} & \Psi^{s,k}_{m} & \Phi^{w,l}_{m} & \Psi^{w,l}_{m}
\end{pmatrix}
\Omega^2
\begin{pmatrix}
\Phi^{s,k}_{n} \\ \Psi^{s,k}_{n} \\ \Phi^{w,q}_{n} \\ \Psi^{w,q}_{n}
\end{pmatrix}
\right].
\end{eqnarray}
Using Eq. (\ref{eq:the_transformation_of_filed_operators}), the above equation can be rewritten as follows:
\begin{eqnarray}
\label{eq:effective_Hamiltonian_at_scale_lambda}
 \mathrm{H}(\lambda)&=&\sum_{m,n=-\infty}^{\infty}\frac{1}{2}\left[\begin{pmatrix}
\pi^{s,k}_{\phi,m}\;&\pi^{s,k}_{\psi,m}\;&\pi^{w,l}_{\phi,m}\;&\pi^{w,l}_{\psi,m}
\end{pmatrix}
\begin{pmatrix}
\pi^{s,k}_{\phi,n}\\
\pi^{s,k}_{\psi,n}\\
\pi^{w,q}_{\phi,n}\\
\pi^{w,q}_{\psi,n}
\end{pmatrix}+\right. \nonumber\\
&&\left.\begin{pmatrix}
\phi^{s,k}_{m} & \psi^{s,k}_{m} & \phi^{w,l}_{m} & \psi^{w,l}_{m}
\end{pmatrix}
O^{T}(\lambda)\Omega^2 O(\lambda)
\begin{pmatrix}
\phi^{s,k}_{n} \\ \psi^{s,k}_{n} \\ \phi^{w,q}_{n} \\ \psi^{w,q}_{n}
\end{pmatrix}
\right].
\end{eqnarray}
\normalsize
By differentiating Eq. (\ref{eq:effective_Hamiltonian_at_scale_lambda}), we obtain the following flow equation:
\begin{eqnarray}
\frac{d\mathrm{H}(\lambda)}{d\lambda}=\sum_{m,n=-\infty}^{\infty}\frac{1}{2}
\left[
\begin{pmatrix}
\phi^{s,k}_{m} & \psi^{s,k}_{m} & \phi^{w,l}_{m} & \psi^{w,l}_{m}
\end{pmatrix}
\left[K(\lambda),\Omega^2(\lambda)\right]
\begin{pmatrix}
\phi^{s,k}_{n} \\ \psi^{s,k}_{n} \\ \phi^{w,q}_{n} \\ \psi^{w,q}_{n}
\end{pmatrix}
\right].
\end{eqnarray}
Here, $\Omega^2(\lambda)=O^T(\lambda) \Omega^2 O(\lambda)$ and $K(\lambda)=\left[G(\lambda),\Omega^2(\lambda)\right]$. $G(\lambda)$ is the part of the matrix $\Omega^2(\lambda)$, where different scale coupling terms are not present.

We only keep resolution $0$ scaling and resolution $0$, wavelet functions. The resolution has been kept at $k=1$ to reduce the complexity of solving a large number of differential equations. The flow equation is designed in such a way that at any value of $\lambda$, we will still get a Hamiltonian which is unitary equivalent to the initial Hamiltonian. The truncated fields are given by an expansion in a finite number of basis functions of two resolutions:
\begin{eqnarray}
\label{eq:expansion_of_phi_field_in_scaling_wavelet_basis}
\phi(x)&=& \sum_{n=0}^{15} \phi^{s,0}_{n}(t)s_n(x)+\sum_{n=0}^{15} \phi^{w,0}_{n}(t)w_{n}(x),\\
\label{eq:expansion_of_piphi_field_in_scaling_wavelet_basis}
\pi_{\phi}(x)&=& \sum_{n=0}^{15} \pi^{s,0}_{\phi,n}(t)s_n(x)+\sum_{n=0}^{15} \pi^{w,0}_{\phi,n}(t)w_{n}(x),
\end{eqnarray}
and
\begin{eqnarray}
\label{eq:expansion_of_psi_field_in_scaling_wavelet_basis}
\psi(x)&=& \sum_{n=0}^{15} \psi^{s,0}_{n}(t)s_n(x)+\sum_{n=0}^{15} \psi^{w,0}_{n}(t)w_{n}(x),\\
\label{eq:expansion_of_pipsi_field_in_scaling_wavelet_basis}
\pi_{\psi}(x)&=& \sum_{n=0}^{15} \pi^{s,0}_{\psi,n}(t)s_n(x)+\sum_{n=0}^{15} \pi^{w,0}_{\psi,n}(t)w_{n}(x).
\end{eqnarray}
The truncated fields from Eq. (\ref{eq:expansion_of_phi_field_in_scaling_wavelet_basis}) to Eq. (\ref{eq:expansion_of_pipsi_field_in_scaling_wavelet_basis}) will vanish smoothly at the boundary $x=0$ and $x=20$. We have used this boundary condition to avoid complications.

As shown in Eq. (\ref{eq:psiphi_hamiltonian_matrix_form}), the $\pi$ coefficient of the Hamiltonian matrix is already diagonalized and it will remain diagonal after defining the new fields; we applied the flow equation to the $\phi$ coefficient of the Hamiltonian to see the evaluation with the flow parameter $\lambda$. The $\phi$ coefficient matrix is given by,
\begin{eqnarray}
\Omega^2:=\begin{pmatrix}
\mu^2\delta_{mn}+\mathcal{D}^k_{ss,mn} & \frac{g^2}{2}\delta_{mn} & \mathcal{D}^{kq}_{sw,mn} & 0 \\
\frac{g^2}{2}\delta_{mn} & \nu^2\delta_{mn}+\mathcal{D}^k_{ss,mn} & 0 & \mathcal{D}^{kq}_{sw,mn} \\
\mathcal{D}^{lk}_{sw,mn} & 0 & \mu^2\delta_{mn}+\mathcal{D}^{lq}_{ww,mn} & \frac{g^2}{2}\delta_{mn} \\
0 & \mathcal{D}^{lk}_{sw,mn} & \frac{g^2}{2}\delta_{mn} & \nu^2\delta_{mn}+\mathcal{D}^{lq}_{ww,mn}
\end{pmatrix}.
\end{eqnarray}
The upper two rows are the scaling function field, and the lower two rows are the wavelet function fields, respectively. We can always diagonalise the matrix to get the uncoupled masses of two new fields. This is a real symmetric matrix, so there exists an orthogonal transformation which can diagonalize this, 
\begin{eqnarray}
O^T \Omega^2 O =\begin{pmatrix}
\tilde{\mu}^s & 0 & 0 & 0\\
0 & \tilde{\nu}^s & 0 & 0\\
0 & 0 & \tilde{\mu}^w & 0\\
0 & 0 & 0 & \tilde{\nu}^w
\end{pmatrix},
\end{eqnarray}
$\tilde{\mu}^s$, $\tilde{\nu}^s$, $\tilde{\mu}^w$, and $\tilde{\nu}^w$ are the diagonal matrices consists of the eigenvalues of the matrix $M$.

The exact square of normal mode frequencies calculated by diagonalizing the truncated matrix, along with the eigenvalues of $\textrm{H}_{ss}$ block for the flowing parameter $s=0$ and $s=20$, has been tabulated in the Table. \ref{tab:normal_mode_frequencies}.
\begin{table}[hbt]
\begin{center}
\caption{Comparison among the $\omega_p^2$ governed by the exact diagonalization of the matrix $\Omega^2$ and the $\textrm{H}_{ss}$ part of the matrix for different values of the flow parameter $\lambda=0$ and $\lambda=20$.}
\label{tab:normal_mode_frequencies}
\setlength{\tabcolsep}{0.5pc}
\vspace{1mm}
\begin{tabular}{c  c | c  c | c | c | c | c }
\specialrule{.15em}{.0em}{.15em}
\hline
\multicolumn{4}{c|}{Exact eigenvalues} & \multicolumn{2}{c|}{$\lambda$=0} & \multicolumn{2}{c}{$\lambda$=20}\\
\hline
$\mu(1:16)$ & $\mu(16:32)$ & $\nu(1:16)$ & $\nu(16:32)$ & $\mu(1:16)$ & $\nu(1:16)$ & $\mu(1:16)$ & $\nu(1:16)$\\
\hline
$0.53595$ & $16.0036$ & $1.53595$ & $17.0036$ & $0.53621$ & $1.53621$ & $0.53595$ & $1.53595$\\
$0.64408$ & $18.6863$ & $1.64408$ & $19.6863$ & $0.64567$ & $1.64567$ & $0.64408$ & $1.64408$\\
$0.82537$ & $21.6110$ & $1.82537$ & $22.6110$ & $0.83255$ & $1.83255$ & $0.82537$ & $1.82537$\\
$1.08189$ & $24.7402$ & $2.08189$ & $25.7402$ & $1.10829$ & $2.10829$ & $1.08189$ & $2.08189$\\
$1.41694$ & $28.0249$ & $2.41694$ & $29.0249$ & $1.49495$ & $2.49495$ & $1.41694$ & $2.41694$\\
$1.83517$ & $31.4082$ & $2.83517$ & $32.4082$ & $2.02510$ & $3.02510$ & $1.83517$ & $2.83517$\\
$2.34243$ & $34.8285$ & $3.34243$ & $35.8285$ & $2.73639$ & $3.73639$ & $2.34243$ & $3.34243$\\
$2.94602$ & $38.2224$ & $3.94602$ & $39.2224$ & $3.66170$ & $4.66170$ & $2.94602$ & $3.94602$\\
$3.65564$ & $41.5241$ & $4.65564$ & $42.5241$ & $4.81698$ & $5.81698$ & $3.65564$ & $4.65564$\\
$4.48633$ & $44.6643$ & $5.48633$ & $45.6643$ & $6.18994$ & $7.18994$ & $4.48633$ & $5.48633$\\
$5.46196$ & $47.5703$ & $6.46196$ & $48.5703$ & $7.73254$ & $8.73254$ & $5.46196$ & $6.46196$\\
$6.61455$ & $50.1696$ & $7.61455$ & $51.1696$ & $9.35997$ & $10.35997$ & $6.61455$ & $7.61455$\\
$7.97792$ & $52.3950$ & $8.97792$ & $53.3950$ & $10.95703$ & $11.95703$ & $7.97792$ & $8.97792$\\
$9.58140$ & $54.1883$ & $10.5814$ & $55.1883$ & $12.39138$ & $13.39138$ & $9.58140$ & $10.5814$\\
$11.4468$ & $55.5034$ & $12.4468$ & $56.5034$ & $13.53144$ & $14.53144$ & $11.4468$ & $12.4468$\\
$13.5869$ & $56.5034$ & $12.4468$ & $56.5034$ & $14.53144$ & $15.53144$ & $13.5869$ & $14.5869$\\
\hline
\specialrule{.15em}{.15em}{.0em}
\end{tabular}
\end{center}
\end{table}
It follows that the unitary operator $U$ does a complete diagonalization that separates different scale degrees of freedom. The transformation matrix $O$ is not unique because permutations of its columns also permute the eigenvalues.

The advantage of this representation is that we can understand the effect of changes in resolution and volume on the truncated Hamiltonian. The number of oscillators is proportional to the cutoff volume, and the oscillator frequencies are the square roots of the eigenvalues of the matrix $\Omega^2$. Increasing the resolution will increase the accuracy of the eigenvalues, and the number of modes within a given volume will also increase. Considering the resolution to be infinite, there will be an infinite number of states for a given volume. The increase in volume will introduce additional modes between the two states. In the case of free field theory, the energy spectrum will be continuous and infinite. To achieve the continuum limit within the wavelet-based framework, we have to increase the volume to the infinite limit and to enhance precision, it is essential to increase the resolution.

The precision can also be improved by increasing the order, $K$, of the wavelet basis. For a given scale (resolution), the basis functions can locally represent polynomials of a higher degree with an increasing value of $K$ \cite{dahmen1997multiscale}. The cost is that, the basis functions have larger support for a given level of resolution. The increment of the accuracy by increasing the value $K$ is demonstrated in \cite{singh2016holographic}.

Now, we will understand whether the flow equation can eliminate the coupling between two scales in the matrix $\Omega^2$ with the increasing value of parameter $\lambda$. The mass parameter $\mu$ and $\nu$ are set to be $1$. The equations were solved using \textit{Mathematica}.

\begin{figure}[hbt]
\caption{Hilbert-Schmidt norm of all four types of non-zero quadratic expressions as a function of the flow parameter.}
\label{fig:HS_norm_of_four_types_of_quadretic expressions}
\includegraphics[scale=0.40]{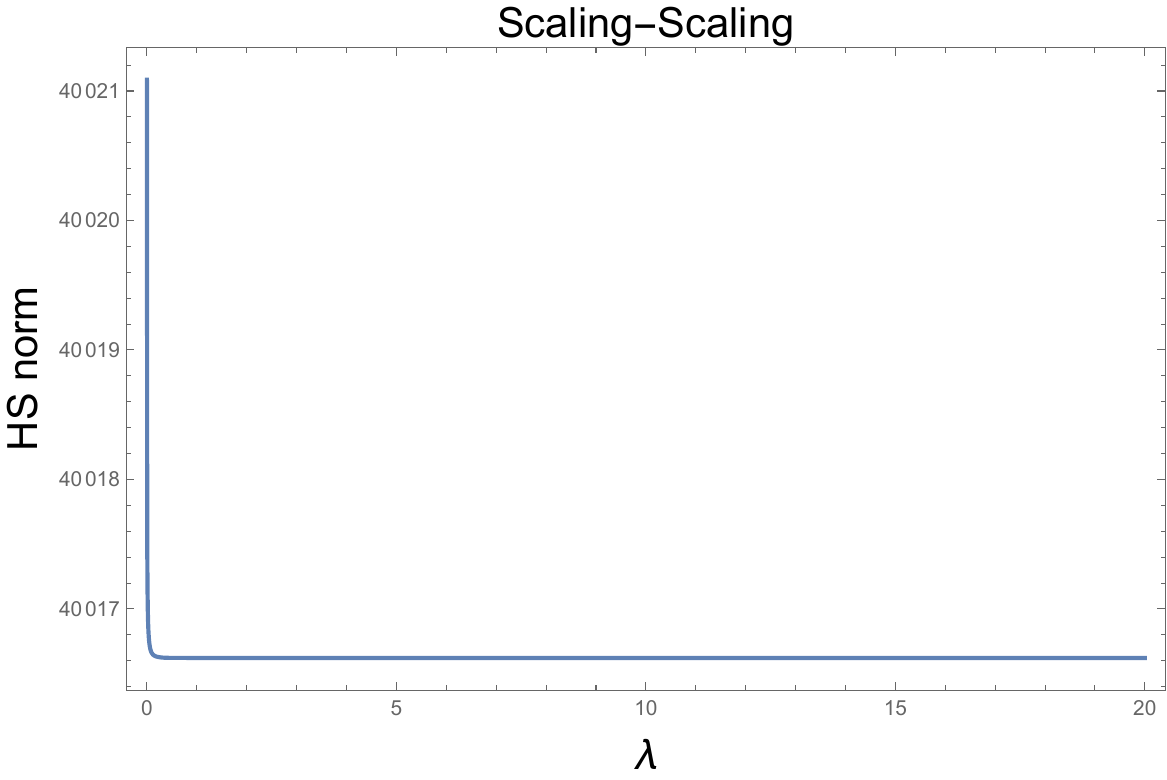}
\includegraphics[scale=0.388]{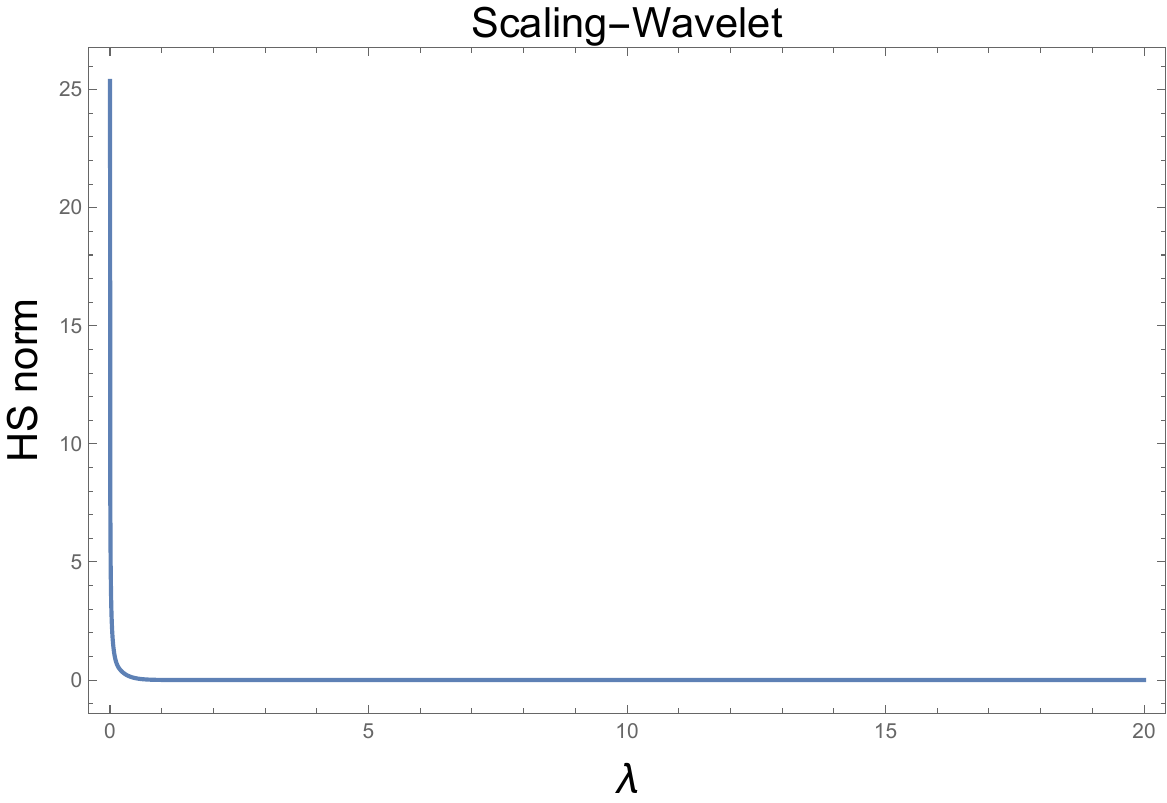}
\includegraphics[scale=0.388]{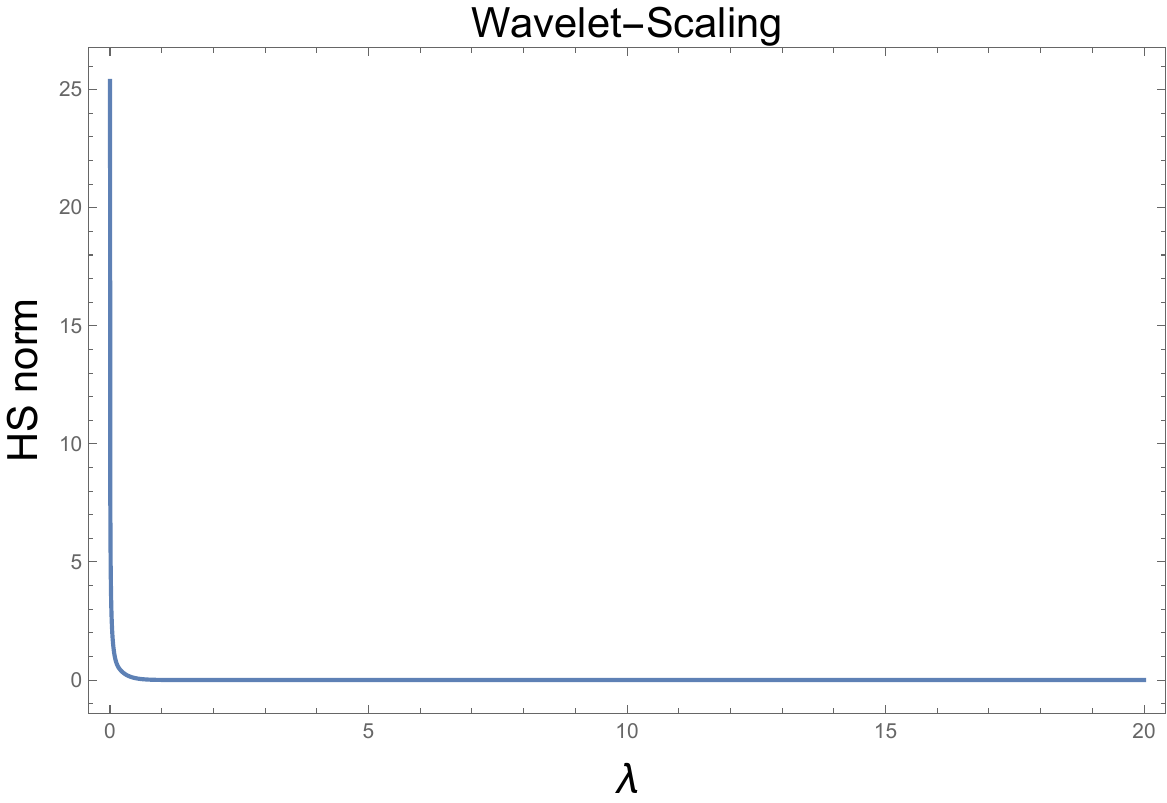}
\includegraphics[scale=0.40]{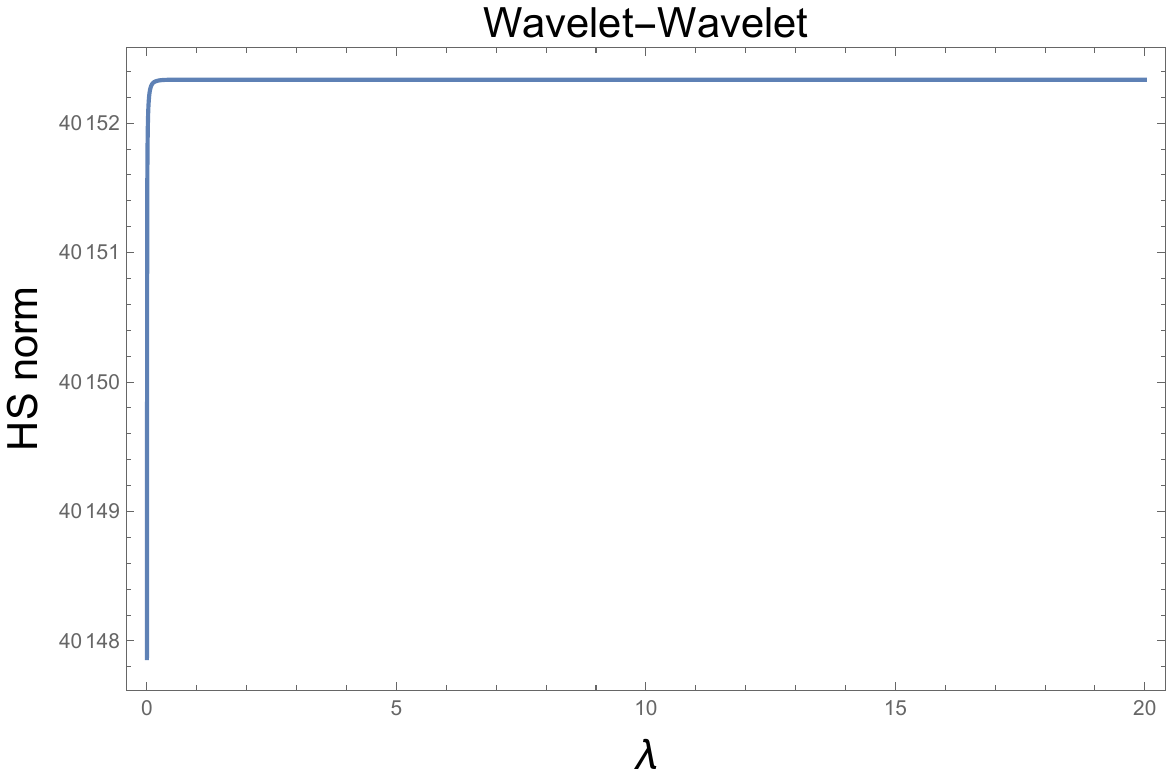}
\end{figure}

To illustrate the evaluation of the matrix elements in each block, we calculated the Hilbert-Schmidt norm for each block of the matrix as a function of $\lambda$. The norms are defined as follows,
\begin{eqnarray}
&&\sqrt{\sum_{ij}\Omega^{2*}_{ss,ij}(\lambda)\Omega^2_{ss,ij}}(\lambda),\quad \sqrt{\sum_{ij}\Omega^{2*}_{sw,ij}(\lambda)\Omega^2_{sw,ij}}(\lambda),\nonumber\\ &&\sqrt{\sum_{ij}\Omega^{2*}_{ws,ij}(\lambda)\Omega^2_{ws,ij}}(\lambda),\quad \text{and} \quad\sqrt{\sum_{ij}\Omega^{2*}_{ww,ij}(\lambda)\Omega^2_{ww,ij}}(\lambda).\,\quad\quad
\end{eqnarray}
\normalsize
Fig. \ref{fig:HS_norm_of_four_types_of_quadretic expressions}, depict the Hilbert-Schmidt norms of the coefficients of each blocks as a function of the flow-parameter.
All scaling or wavelet field coefficients will become non-zero values with the increasing value of $\lambda$, while the coupling matrices will all become zero values. Initially, the coupling coefficients will decrease very quickly, but with the increment of $\lambda$, the rate of decrease will slow down significantly. At $\lambda=20$, the norm will reduce by $90\%$ of its original value.

\begin{figure}[hbt]
\label{fig:matrix_plot_for_different_values_of_lambda}
\caption{The evaluation of the matrix elements with the increasing value of $\lambda$.}
\includegraphics[scale=0.33]{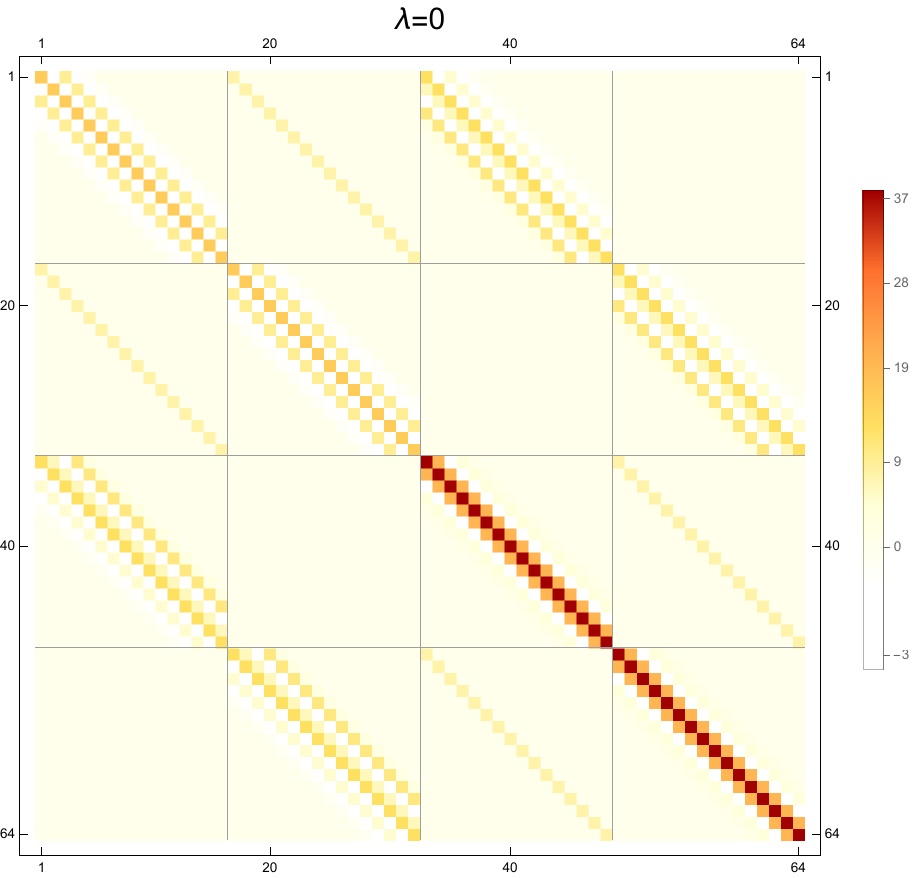}
\includegraphics[scale=0.33]{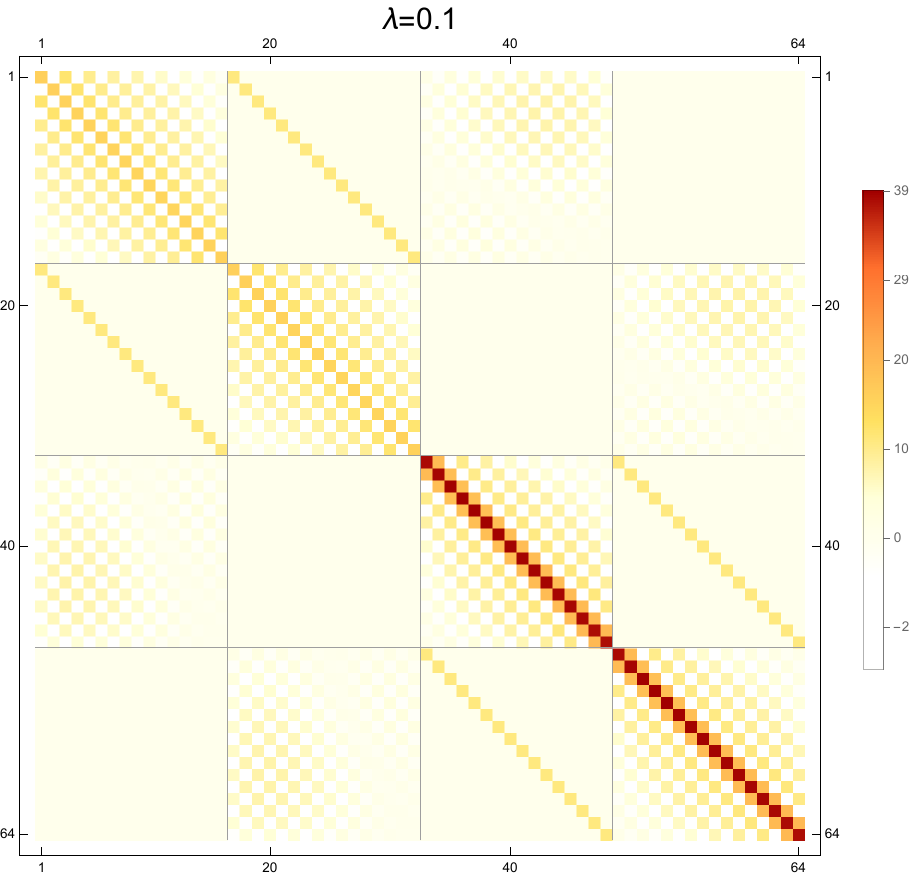}
\includegraphics[scale=0.33]{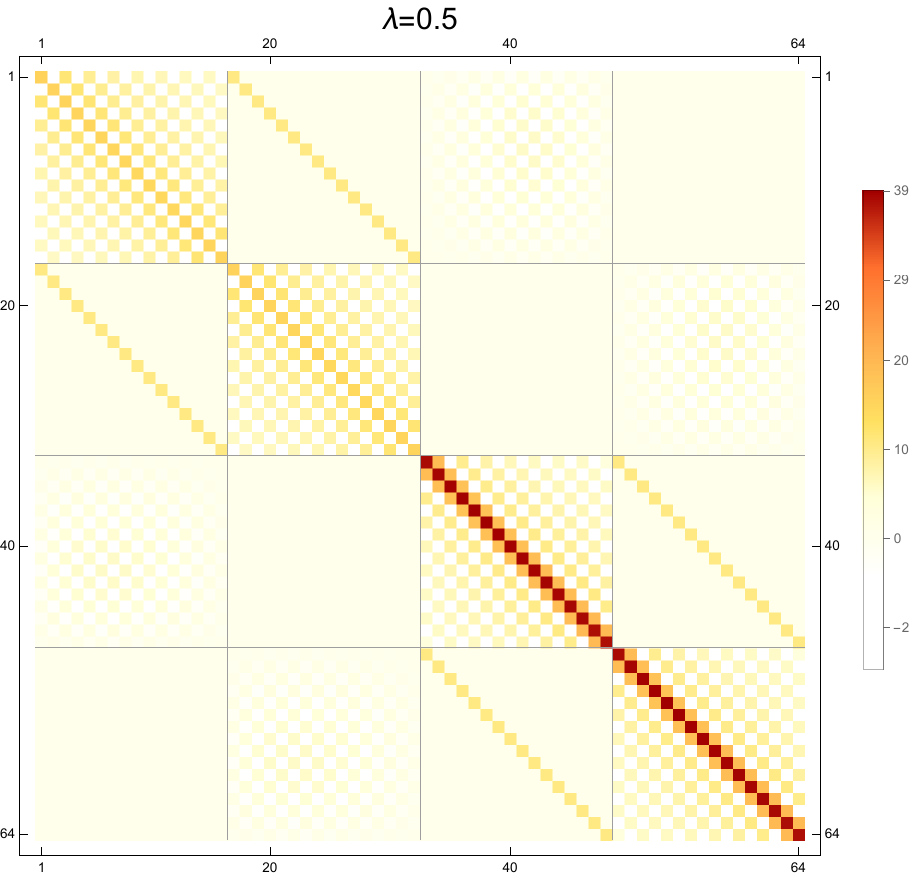}
\includegraphics[scale=0.33]{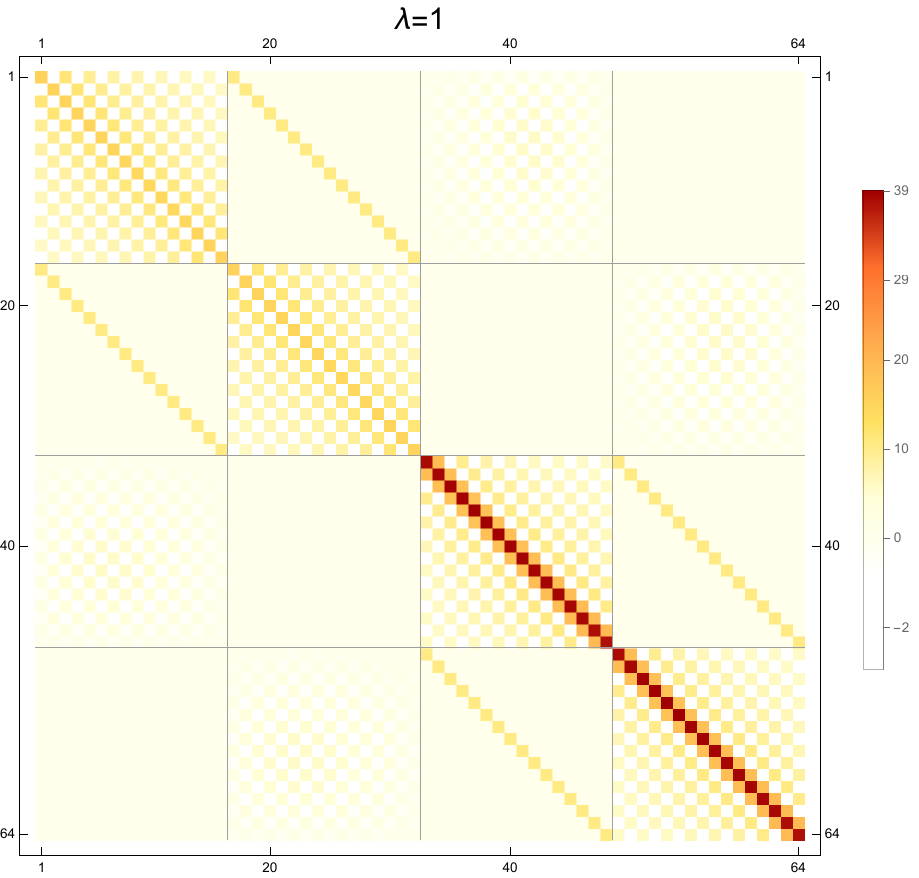}
\includegraphics[scale=0.33]{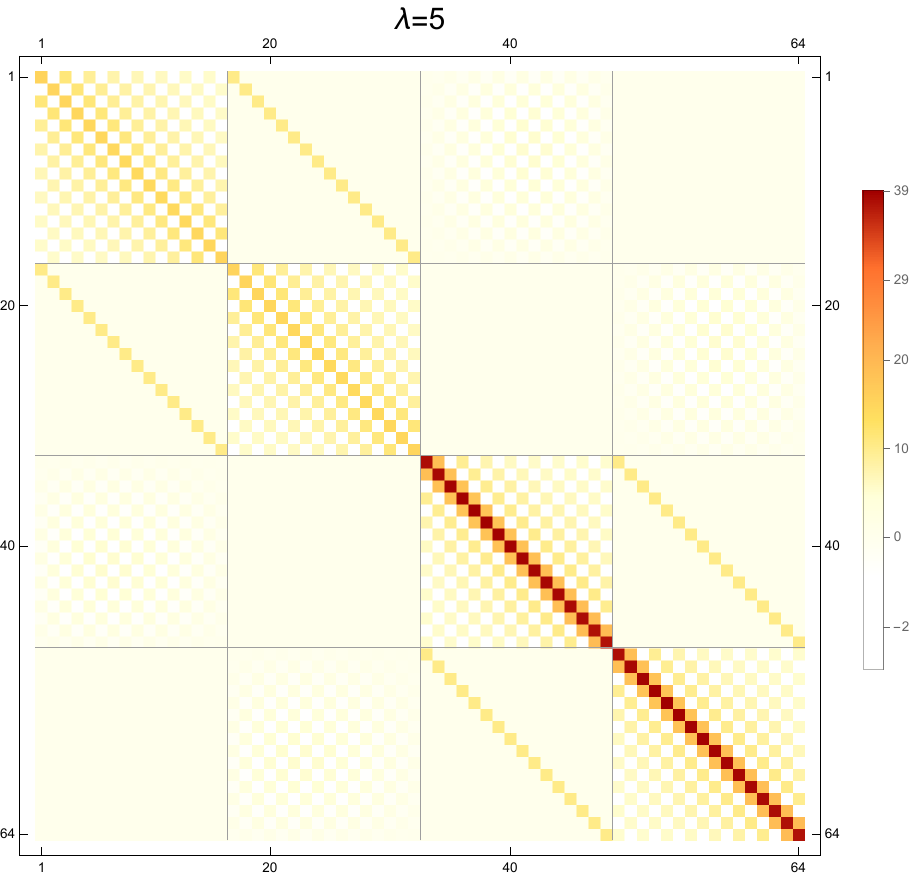}
\includegraphics[scale=0.33]{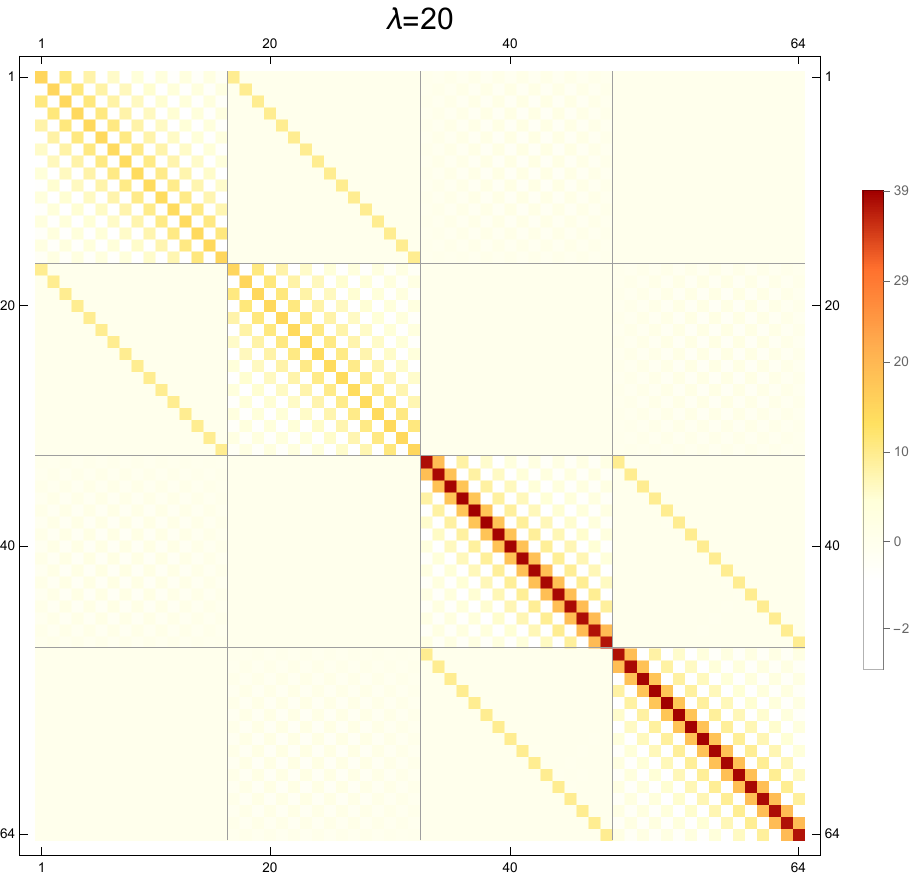}
\end{figure}

As the Hilbert-Schmidt norm is dominated by the largest matrix elements. It is also useful to understand how the individual matrix elements will evolve with the increasing value of $\lambda$. In Fig. \ref{fig:matrix_plot_for_different_values_of_lambda}, we have listed the graphical representation of the individual matrices for the different values of $\lambda$. 

In each plot, there are 16 blocks of matrices, with each block being a 16x16 matrix. The first two blocks of the first and second rows make up a larger block containing only scaling-scaling coupling terms. Within these four blocks, the first block of the first row contains terms corresponding to $\phi$ field, the second block of the first row contains terms corresponding to the coupling between $\phi$ and $\psi$ fields, the first block of the second row also contains the coupling terms between $\phi$ and $\psi$ filed, the second block of the second row will contain terms corresponding to the field $\psi$. The last two blocks within the first and second rows form a larger block that includes the scaling wavelet coupling terms. Of these four blocks, the third block of the first row consists of terms related to the $\phi$ $\phi$ fields, whereas the fourth block of the first row contains terms relating to the coupling between $\phi$ and $\psi$ fields on a larger scale. However, in this representation, it's seen from the matrix plot that the field coupling terms in higher and coarser scale coupling elements are absent. Akin to this arrangement, we will have similar arrangements within the first two blocks of the third and fourth rows, containing the scaling wavelet coupling terms. The third and fourth blocks of the third and fourth rows contain the wavelet coupling terms. The third block of the third row contains elements corresponding to finer scale $\phi$ fields, and the fourth block of the third row contains the coupling terms between two fields on a finer scale. Similarly, the third block of the fourth column also contains the coupling terms between two fields, and lastly, the fourth block of the last row contains the finer scale coupling terms corresponding to the $\psi$ field. In Fig. \ref{fig:matrix_plot_for_different_values_of_lambda}, five figures correspond to different values of the flow parameter $\lambda$: $0$, $0.1$, $0.5$, $5$, and $20$.

The first picture in Fig. (\ref{fig:matrix_plot_for_different_values_of_lambda}) shows the initial values. In the first block of the first row and the second block of the second row, the elements are nearly diagonal due to the existence of a few overlap integrals, $\mathcal{D}^k_{ss,mn}$, resulting from the compact support of the scaling function. Similarly, in the third block of the third row and the fourth block of the fourth row, only a few diagonal terms and some adjacent diagonal terms exist due to the existence of limited overlap integrals, $\mathcal{D}^k_{ww,mn}$, which results from the compact support of wavelet functions. The off-diagonal block matrices, which include the scaling-wavelet coupling terms, are almost diagonal because there are only a limited number of overlap integrals ($\mathcal{D}^k_{sw,mn}$) due to the compact support of both scaling and wavelet functions.

In Fig. (\ref{fig:matrix_plot_for_different_values_of_lambda}), it is seen that at $\lambda=1$, the off-diagonal terms become almost zero. The value of the off-diagonal elements are not changing significantly with the increasing values of $\lambda$ beyond $\lambda=1$. At $\lambda=20$, the matrix will be almost diagonal, that means it will preserve the local nature of the truncated theory.

In the case of two fields coupled via pseudo interaction, the flow equation can decouple the coarser and the finer scale degrees of freedom. At each value of the flow parameter the flow equation will produce a new equivalent Hamiltonian. Ignoring the coupling terms at each value of $\lambda$, the Hamiltonian is a sum of two operators associated with coarser and finer scale degrees of freedom. But those degrees of freedom will include the effects of the eliminated coupled degrees of freedom.

\chapter{Conclusion and outlook}
\label{chap:conclusion_and_outlook}
In this thesis, we have advocated employing the Daubechies wavelet basis for the analysis of problems in quantum mechanics and quantum field theory. Because of the multi-resolution characteristics and the compact support of the basis functions, it's possible to allocate high-resolution basis functions where necessary without requiring equivalent resolutions elsewhere. This multi-resolution advantage can be leveraged without compromising the orthogonality of the basis functions. Additionally, owing to the discrete nature of the basis elements, it becomes feasible to construct a discrete Hamiltonian matrix for quantitatively determining the eigenvalues and eigenfunctions of the Hamiltonian eigenvalue problem. 

The wavelet-based formulation is similar in spirit to lattice field theory. It has the potential to provide a robust framework for the non-perturbative examination of quantum field theories. Discrete wavelet-based techniques offer a means to examine these quantum field theories (QFTs) in a manner reminiscent of the Euclidean lattice approach while also allowing the investigation of real-time dynamics. The hope is that the discrete and multi-scale characteristics of compactly-supported wavelets will provide a conducive environment for carrying out classical and quantum simulations of continuum QFTs. 

As the wavelet basis functions can be derived from a renormalization group equation, this is a natural basis to formulate renormalization. We showcase this feature by working with the 2-dimensional attractive Dirac delta function potential as an illustrative example to showcase aspects of renormalization within the discrete wavelet-based approach. This well-studied potential was chosen as a model because it demonstrates several features commonly encountered in relativistic quantum field theories, such as ultraviolet divergences, asymptotic freedom, and dimensional transmutation. Our study shows that when the bare coupling constant remains constant, the ground state energy diverges as the resolution increases which is indicative of ultraviolet divergences in this problem. We also show how the concept of renormalization results in a resolution-dependent coupling constant.

The Similarity Renormalization Group (SRG) emerged from the independent works of Glazek-Wilson and Wegner. Both researcher(s) were driven by the aim of addressing strongly correlated systems within a perturbative approach. Wavelet-based SRG seeks to transform the in the block diagonalized form. The Hamiltonian is expressed using a wavelet basis, which incorporates both physically relevant scales and chosen minimal scales. The Hamiltonian will be arranged into different blocks: the diagonal blocks will have the same scale coupling and the off diagonal blocks feature the different scale coupling. By applying the SRG technique, the various scale coupling terms are progressively eliminated, resulting in an effective Hamiltonian that contains only the physically relevant degrees of freedom. It does so while assuring no change in the low-energy physics. By choosing different forms of the generator for the similarity flow, one may be able to control the growth of the Hamiltonian elements which effectively contain the scale-dependent coupling constant. Keeping our general goal in sight, we developed SRG within the wavelet framework by choosing an appropriate generator of the flow. In this thesis, we extended the work of Polyzou and Michlin \cite{PhysRevD.95.094501} by applying the formalism to a theory where two scalar fields are coupled together in $1+1$ dimensions. In contrast, they utilised the free scalar field theory to illustrate this formalism.

The next logical step is to extend multi-resolution methods to realistic theories in $3+1$                                                                                                                                                                                                     dimensions. This presents greater computational challenges compared to lower dimensions. Several important problems to consider include the bound state problem, scattering problem, the analysis of correlation functions, and the adaptation of these methods to gauge theories. The next class of problems that may be interesting is the $1+1$ dimensional solvable field theory like free Dirac field theory, free electromagnetic theory. Another area where wavelet representation is advantageous is in quantum computing.

\appendix

\chapter{Some properties of scaling and wavelet function}
\section{Computation of scaling and wavelet function at integer points and dyadic rationals}
\label{appen:Computation_of_scaling_and_wavelet_function_at_integer_points_and_dyadic_rationals}
In this section of the appendix, we will discuss the approach to solving the scaling equation Eq. (\ref{eq:scaling_equation_1}) and wavelet equation Eq. (\ref{eq:wavelet_equation_1}) to compute the values of those functions at different points in space. To get the unique values of these functions at different points in space, we need the normalization conditions of the scaling function Eq. (\ref{eq:normalization_condition}). 

\subsection{Computing scaling function at integer points}
Using the properties of scaling and translation operation Eq. (\ref{eq:scaling_equation_1}) can be rewritten in the following form,
\begin{eqnarray}
\label{eq:scaling_equation_2}
s(x)=\sqrt{2}\sum_{n=0}^{2K-1}h_n s(2x-n).
\end{eqnarray}
The scaling function $s(x)$ has the support in the interval $\left[0,2K-1\right]$ with $s(0)=0$ and $s(2K-1)=0$ for $K \geq 2$ \cite{doi:10.1137/1.9781611970104,Mehra2017,Kessler_2003}. For $K=3$ substituting $x=n=0,1,2,...5,$ we get the a set of homogeneous equations,
\begin{eqnarray}
\label{eq:homogeneous_equation_of_scaling_function}
s(n)=\sqrt{2}\sum_{i=0}^{5}h_i s(2n-i).
\end{eqnarray}
To uniquely determine the solution, we need an inhomogeneous equation which we can get following the normalization condition of the scaling function Eq. (\ref{eq:normalization_condition}) by putting $x=0,1,2,...5$ and converting the integration into summation as,
\begin{eqnarray}
\label{eq:inhomogeneous_equation_of_scaling_function}
\sum_{n=0}^{5} s(n)=1.
\end{eqnarray}
Solving Eq. (\ref{eq:homogeneous_equation_of_scaling_function}) and Eq. (\ref{eq:inhomogeneous_equation_of_scaling_function}), we get the values of the scaling function at all the integer points. See Table \ref{tab:scaling_function_integer_pointsand_dyad_points}.

\subsection{Computation of scaling functions at dyadic rationals}
To get the values of x at different dyadic rational points, substituting x in Eq. (\ref{eq:scaling_equation_2}) with $\frac{1}{2},\frac{3}{2},...\frac{9}{2}$ we get the following set of equations which can be written in the matrix form as,
\begin{eqnarray}
\begin{pmatrix}
s(\frac{1}{2})\\
s(\frac{3}{2})\\
s(\frac{5}{2})\\
s(\frac{7}{2})\\
s(\frac{9}{2})\\
\end{pmatrix}
=\sqrt{2}
\begin{pmatrix}
h_1 & h_0 & 0 & 0 & 0\\
h_3 & h_2 & h_1 & h_0 & 0\\
h_5 & h_4 & h_3 & h_2 & h_1\\
0 & 0 & h_5 & h_4 & h_3\\
0 & 0 & 0 & 0 & h_5\\
\end{pmatrix}
\times
\begin{pmatrix}
s(0)\\
s(1)\\
s(2)\\
s(3)\\
s(4)\\
\end{pmatrix}.
\end{eqnarray}
Knowing the values of $s(x)$ at all integer points within the support of the scaling function, we can get the values of the scaling function at different dyadic rational
points, which are listed in Table. \ref{tab:scaling_function_integer_pointsand_dyad_points}. Recursively using the scaling equation Eq. (\ref{eq:scaling_equation_2}), we will get the values of the scaling function at all dyadic rationals. Similarly, we can get the values of wavelet function at all integers and dyadic rationals using Eq. (\ref{eq:wavelet_equation_1}).
\begin{table}[ht]
\begin{center}
\setlength{\tabcolsep}{1.5pc}
\catcode`?=\active \def?{\kern\digitwidth}
\caption{ The values of the scaling function for $K=3$ at all integer points and dyadic rationals.}
\label{tab:scaling_function_integer_pointsand_dyad_points}
\vspace{1mm}
\begin{tabular}{c c c c}
\specialrule{.15em}{.0em}{.15em}
\hline
$s(x)$ at int. pt.s & Values & $s(x)$ at dyad. $\mathbb{R}$ pt.s & Values \\
\hline
$s(0)$ & $0$ & $s(\frac{1}{2})$ & $0.605178$\\

$s(1)$ & $1.28634$ & $s(\frac{3}{2})$ & $0.441122$\\

$s(2)$ & $-0.385837$ & $s(\frac{5}{2})$ & $-0.0149706$\\

$s(3)$ & $0.0952675$ & $s(\frac{7}{2})$ & $-0.0315413$\\

$s(4)$ & $0.00423435$ & $s(\frac{9}{2})$ & $0.000210945$\\

$s(5)$ & $0$\\
\hline
\specialrule{.15em}{.15em}{.0em}
\end{tabular}
\end{center}
\end{table}

In this section, first, we will deduce certain properties that will be useful in determining the overlap integrals.
\section{The moment of scaling and wavelet function:} 
\label{subsec:moment_of_scaling_and_wavelet_function}
The moment of $k$th resolution scaling and wavelet functions are defined as,
\begin{eqnarray}
\expval{x^m}_s^k = \int x^m s^k(x)dx,\quad\quad \expval{x^m}_w^k = \int x^m w^k(x)dx.
\end{eqnarray}
By changing the variable, we obtain the following relations:
\begin{eqnarray}
\expval{x^m}^k_s &=& 2^{k(m+\frac{1}{2})}\expval{x^m}_s,\\
\expval{x^m}^k_w &=& 2^{k(m+\frac{1}{2})}\expval{x^m}_w,
\end{eqnarray}
where,
\begin{eqnarray}
\label{eq:moment_of_the_scaling_function}
\expval{x^m}_s &=& \int x^m s(x)dx,\\
\expval{x^m}_w &=& \int x^m w(x)dx,
\end{eqnarray}
is the "moment of the $0$th resolution scaling function", which can be constructed recursively starting from the normalization condition of scaling function.
\begin{eqnarray}
\expval{x^0}_s = \int s(x)dx =1.
\end{eqnarray}
Using the unitary property of the dilation operator, we can express Eq. (\ref{eq:moment_of_the_scaling_function}) as,
\begin{eqnarray}
\expval{x^m}_s &=& \int D^{-1}x^m D^{-1} s(x) dx\nonumber\\
&=& \frac{1}{\sqrt{2}}\frac{1}{2^m}\sum_l h_l \int x^m s(x-l)dx\nonumber\\
&=& \frac{1}{\sqrt{2}}\frac{1}{2^m}\sum_l h_l \int (x+l)^m s(x)dx\nonumber\\
\label{eq:moment_of_scaling_function_1}
\expval{x^m}_s &=& \frac{1}{\sqrt{2}}\frac{1}{2^m}\sum_l h_l \sum_{k=0}^{m} \frac{m!}{k!(m-k)!} \int x^m s(x)dx.
\end{eqnarray}
Utilizing $\sum_l h_l = \sqrt{2}$ and relocating the term with $k=m$ to the left side of Eq. (\ref{eq:moment_of_scaling_function_1}), we derive the following recursion relation:
\begin{eqnarray}
\label{eq:moment_of_scaling_function_2}
\expval{x^m}_s=\int x^m s(x)dx=\frac{1}{2^m-1}\sum_{k=0}^{m-1} \frac{m!}{k!(m-k)!}\left(\sum_{l=1}^{2K-1}h_l l^{m-k} \right)\int x^k s(x)dx. 
\end{eqnarray}
Moments of the wavelets are obtained by replacing $h_l$ in Eq. (\ref{eq:moment_of_scaling_function_2}) by $g_l$:
\begin{eqnarray}
\label{eq:moment_of_wavelet_function_1}
\expval{x^m}_w=\int x^m s(x)dx=\frac{1}{2^m-1}\sum_{k=0}^{m-1} \frac{m!}{k!(m-k)!}\left(\sum_{l=1}^{2K-1}g_l l^{m-k} \right)\int x^k s(x)dx.
\end{eqnarray}
\section{The partial moment of scaling and wavelet function}
\label{appen:The_partial_moment_of_scaling_and_wavelet_function}
When solving integral equations on a finite or semi-infinite interval, it becomes essential to compute integrals where an endpoint of the interval lies within the support of a scaling basis function \cite{shann_quadreture93,shann_quadreture98}. Through the utilization of scaling and dilation operators, these integrals can be related to the following integrals:
\begin{eqnarray}
I_{x^ks,m+}:= \int_{0}^{\infty} \hat{T}^m s(x) x^k dx, \quad\quad I_{x^ks,m-}:=\int_{-\infty}^{0} \hat{T}^m s(x) x^k,
\end{eqnarray}
and
\begin{eqnarray}
I_{ss,mn+}:=\int_{0}^\infty \hat{T}^m s(x) \hat{T}^n s(x)dx,\quad\quad I_{ss,mn-}:=\int_{-\infty}^0 \hat{T}^m s(x) \hat{T}^n s(x)dx.
\end{eqnarray}
Utilizing the unitary property of the dilation operator, and the scaling equation, Eq. (\ref{eq:scaling_equation_1}), we get the following linear equation,
\begin{eqnarray}
\label{eq:partial_moment_involving_x_k_and_scaling_function}
I_{x^q s,m+}&=& 2^{-(q+\frac{1}{2})}\sum_{l=0}^{2K-1}h_l I_{x^q s,2m+l,+},\\
\label{eq:partial_moment_involving_scaling_and_scaling_function}
I_{ss,mn+}&=& \sum_{r=0}^{2K-1} \sum_{s=0}^{2K-1} h_r h_s I_{ss,2m+r,2m+s,+}.
\end{eqnarray}
For $m \geq 0$, $I^k_{x^qs,m+}$ represents the complete moment, as elaborated in the preceding section, and its evaluation proceeds as follows:
\begin{eqnarray}
I_{x^q s,m+} &=& \int_0^{\infty} x^q s^k_m(x)dx\nonumber\\
&=& \int_0^{\infty} x^q s^k(x-m)dx\nonumber\\
&=& \int_0^{\infty} (x+m)^q s^k(x)dx \nonumber\\
&=& \int_0^{\infty} \left(x^q+\binom n 1 m x^{q-1} +\binom n 2 m x^{q-2}+...+m^q \right) s^k(x)dx \nonumber\\
&=& \expval{x^q}^k_s+\binom n1 m \expval{x^{q-1}}^k_s+\binom n 2 m \expval{x^{q-2}}^k_s+...+\frac{ m^q}{2^{\frac{k}{2}}}.
\end{eqnarray}
Here, $\expval{x^q}^k_s$, is the moment of the scaling function, can be evaluated using the procedure of the preceding section. So, Eq. (\ref{eq:partial_moment_involving_x_k_and_scaling_function}) becomes a linear system for the unknown partial moment in terms of the full moments. These equations can
be solved for the non-trivial $I_{x^q s,m+}$. The $I_{x^q s,m-}$ are obtained using $I_{x^q s,m-}$ using $I_{x^q s,m-}=I_{x^q s,m}-I_{x^q s,m+}$.

For $m \geq 0$ and $n \geq 0$, the value of $I_{ss,mn+}$ is obtained from the orthonormality relation of the scaling function:
\begin{eqnarray}
I_{ss,mn+}=\delta_{mn}.
\end{eqnarray}
This leads to a small linear system that relates the unknown integrals $I_{ss,mn+}$ to the known $I_{ss,mn}=\delta_{mn}$. The $I_{ss,mn-}$ can again be obtained by subtraction. 
This approach can be employed to calculate the overlap integral for the potential energy term associated with the triangular potential.
\section{Partition of unity}
We can expand $1$ and $x$ in scaling function basis. Expanding $1$ in this basis will give us the first "partition of unity".
\begin{eqnarray}
\label{eq:expansion_1_and_x}
1= \sum_n a_n s_n(x),\quad \text{and}\quad x=\sum_n b_n s_n(x),
\end{eqnarray}
where
\begin{eqnarray}
a_n= \int s_n(x)dx \quad \text{and} \quad b_n=\int x s_n(x)dx.
\end{eqnarray}
$a_n$ can be determined using the normalization condition in Eq. (\ref{eq:normalization_condition})
\begin{eqnarray}
\label{eq:expansion_a_n}
a_n = 1.
\end{eqnarray}
And following the subsequent procedure, we can determine $b_n$.
\begin{eqnarray}
b_n &=& \int x s_n(x)dx \nonumber\\
&=& \int (x+n)s(x)dx\nonumber \\
&=& \left(n+\expval{x}_s\right),
\end{eqnarray}
$\expval{x}$ is the moment of the scaling function at resolution $0$, which can be determined by putting the value of $m$ to be equal to $1$ in Eq. (\ref{eq:moment_of_scaling_function_1}).
\begin{eqnarray}
\label{eq:expansion_b_n}
\expval{x}_s =  \frac{1}{\sqrt{2}}\sum_l lh_l.
\end{eqnarray}
So, the first "partition of unity" equation is given by,
\begin{eqnarray}
\label{eq:partition_of_unity_1}
1= \sum_n s_n(x).
\end{eqnarray}
The second equation can be obtain from the expansion of $x$ in scaling function basis. From Eq. (\ref{eq:expansion_b_n}) and Eq. (\ref{eq:expansion_1_and_x}) it follows that,
\begin{eqnarray}
\label{eq:expansion_of_x}
x= \sum_n ns_n(x)+\frac{1}{\sqrt{2}}\sum_l lh_l.
\end{eqnarray}
Upon differentiating both sides of Eq. (\ref{eq:expansion_of_x}), we obtain another "partition of unity" for the derivative of the Scaling function:
\begin{eqnarray}
\label{eq:partition_of_unity_2}
1= \sum_n n s'_n(x).
\end{eqnarray}
\chapter{The overlap integrals}
\label{appen:the_overlap_integrals}
In the majority of applications within the wavelet framework, it is generally unnecessary to compute the scaling functions at every point in space. Nevertheless, it is crucial to determine the overlap integrals that involve the product of scaling functions and wavelet functions, the product of the derivative of the scaling and wavelet functions, or the product of polynomial scaling and wavelet functions.

In this section of the appendix, we will outline the general procedure for evaluating these integrations without requiring the knowledge of the scaling function and wavelet functions at each point in space. This process will be illustrated through specific examples of overlap integrals.

\section{The procedure for determining the overlap integrals}
\label{appen:the_procedure_for_determining_the_overlap_integrals}
In this section, initially, we will deduce certain properties that will be useful in determining the overlap integrals. 

The generic integral can be expressed in the following manner,
\begin{eqnarray}
\Gamma_n=\int f_1(x)f_2(x)f_3(x)...f_n(x)dx,
\end{eqnarray}
here, $f_i(x)$ represents the scaling function, wavelet function, the first derivative of the scaling or the wavelet function, or the power of $x$.

Computation of these quantities can be simplified to solving a set of linear equations by adhering to the procedure outlined below. The computation involves utilizing the following relations of scaling functions,
\begin{eqnarray}
&\int s^k_n(x)dx = \frac{1}{\sqrt{2^k}},&\\
&DT^{2k} = T^k D,&\\
\label{eq:commutator_d_dx_D}
&\frac{d}{dx} D = 2D \frac{d}{dx},&\\
&Dx = 2x D,&\\
&Tx = (x-1)T.&
\end{eqnarray}
In addition to these identities, we require the scaling equation, the wavelet equation, and their derivatives in the following form:
\begin{eqnarray}
\label{eq:scaling_equation_3}
s^k_m(x) &=& \sum_{n} H_{mn} s^{k+1}_n(x),\\
\label{eq:wavelet_equation_3}
w^k_m(x) &=& \sum_{n} G_{mn} s^{k+1}_n(x),\\
\label{eq:derivative_scaling_equation}
s^{k'}_m(x) &=& 2\sum_{n} H_{mn} s^{(k+1)'}_n(x),\\
\label{eq:derivative_wavelet_equation}
w^{k'}_m(x) &=& 2\sum_{n} G_{mn} s^{(k+1)'}_n(x),
\end{eqnarray}
where,
\begin{eqnarray}
H_{mn} = h_{n-2m}, \quad \text{and}\quad G_{mn}= g_{n-2m}.
\end{eqnarray}
These equations establish a connection between the scale $2^{-k}$ and scale $2^{-(k+1)}$ for scaling functions, wavelet functions, and their derivatives. Iterative application of these equations allows for the increment of $k$ in each of the functions by any desired amount.

Moreover, the scale factor $k$ of all these functions in the integral can be adjusted by the same amount, whether increased or decreased, using the following relation,
\begin{eqnarray}
\label{eq:relation_of_k_and_lower_0}
\int D^k f_1(x) D^k f_2(x) D^k f_3(x)... D^k f_n(x)dx= 2^{k(\frac{n}{2}-1)}\int f_1(x)f_2(x) f_3(x)... f_n(x)dx.
\end{eqnarray}
This relation can be derived by employing the definition of the operator $\hat{D}$ as given in Eq. (\ref{eq:D_and_T_definition}), in conjunction with Eq. (\ref{eq:commutator_d_dx_D}).

To calculate the integral $\Gamma_n$ the following steps are employed:
\begin{enumerate}
\item[Step 1:] Utilize Eq. (\ref{eq:relation_of_k_and_lower_0}) to establish a connection between the integral and another integral in which the finest scale appearing in the integrand is $\frac{1}{2^0}$. Subsequently, the scale of each function in the integrand becomes $\frac{1}{2^0}$ or coarser (negative $k$).

\item[Step 2:] Employ repeated application of Eq. (\ref{eq:scaling_equation_3})- Eq. (\ref{eq:derivative_wavelet_equation}) to substitute all coarse-scale functions with linear combinations of scale $1/2^0$ scaling functions. Following these steps, the original integral can be represented as a finite sum of coefficients, where all functions in the integrand are scale $1/2^0$ scaling functions or their derivatives. It is noteworthy that the wavelet contribution can always be expressed in terms of scaling functions using Eq. (\ref{eq:wavelet_equation_3}) or Eq. (\ref{eq:derivative_wavelet_equation}).

\item[Step 3:] Leveraging the integer translation invariance property of the scaling function to shift the support of the leftmost function such that it starts at $0$. Given that each function exhibits compact support within an interval of width $2K-1$, excluding the leftmost function, other functions may vary within the range, $[-2K+2,2K-2]$. Consequently, for $n$ functions, the count of non-zero coefficient is bounded by $(4K-3)^{n-1}$.

\item[Step 4:] Apply the scaling equation to the integrand of the resulting integral and utilize translational invariance to shift the leftmost index to zero. This leads to a set of $(4K-3)^{n-1}$ homogeneous linear equations for the coefficient functions with the leftmost index set to zero.

\item[Step 5:] 
Utilize the partition of unity property to derive additional inhomogeneous equations for these coefficients.

\item[Step 6:] Solving this set of homogeneous and inhomogeneous equations yields the unique set of solutions for the non-zero coefficients.
\end{enumerate}
\section{The kinetic energy term}
\label{appen:the_kinetic_energy_term}
Here, we will illustrate the process outlined in the preceding section for assessing overlap integrals by calculating the overlap integrals that entails the product of the derivative of the scaling function and the wavelet function, as specified in Eq. (\ref{eq:overlap_integrals_derivative_of_scaling_functions}), Eq. (\ref{eq:1d_T_k_sw}) and Eq. (\ref{eq:1d_T_k_ww}).
\begin{eqnarray}
\label{eq:overlap_integral_derivative_of_scaling_function_1}
T^{k}_{ss,mn} &=& \int s^{k'}_{m}(x)s^{k'}_{n}(x)dx,\\
\label{eq:overlap_integral_derivative_of_scaling_and_wavelet_function_1}
T^{kl}_{sw,mn} &=& \int s^{k'}_{m}(x)w^{l'}_{n}(x)dx,\quad l\geq k,\\
\label{eq:overlap_integral_derivative_of_wavelet_and_wavelet_function_1}
T^{lp}_{ww,mn} &=& \int w^{l'}_{m}(x)w^{p'}_{n}(x)dx,\quad p\geq l.
\end{eqnarray}

As depicted in the first step, it is necessary to adjust the scale of each function present in the integrand to a scale of $\frac{1}{2^0}$ or a coarser level. Leveraging Eq. (\ref{eq:commutator_d_dx_D}) and the unitary property of the operator $\hat{D}$,  Eq. (\ref{eq:overlap_integral_derivative_of_scaling_function_1}) can be reformulated in the following manner,
\begin{eqnarray}
T^{k}_{mn} &=& \int s^{k'}_{m}(x)s^{k'}_{n}(x)dx\nonumber \\
&=& \int \frac{d}{dx} \hat{D}^k s_m(x) \frac{d}{dx} \hat{D}^k s_n(x) dx\nonumber \\
&=& 2^{2k}\int  \hat{D}^k \frac{d}{dx} s_m(x) \hat{D}^k \frac{d}{dx} s_n(x) dx \nonumber\\
\label{eq:overlap_integral_derivative_scaling_function_2}
&=& 2^{2k} T^0_{ss,mn}.
\end{eqnarray}
For the case of Eq. (\ref{eq:overlap_integral_derivative_of_scaling_and_wavelet_function_1}) and Eq. (\ref{eq:overlap_integral_derivative_of_wavelet_and_wavelet_function_1}), the functions within the integral exhibit distinct scales. Hence, it is imperative to standardize the scales by representing both functions in relation to the same scale scaling functions. This objective can be accomplished by employing Eq. (\ref{eq:commutator_d_dx_D}), Eq. (\ref{eq:derivative_scaling_equation}), Eq. (\ref{eq:derivative_wavelet_equation}), and the unitary property of the operator $\hat{D}$ in the subsequent manner,
\begin{eqnarray}
T^{kl}_{sw,mn} &=& \int s^{k'}_m(x) w^{l'}_{n}(x)dx,\quad l\geq k\nonumber\\
&=& \int \frac{d}{dx} \left( H^{l+1-k}_{mm'} D^{l+1} s_{m'}(x)\right) \frac{d}{dx} \left( G_{nn'} D^{l+1} s_{n'}(x)\right)dx\nonumber \\
&=& \int  H^{l+1-k}_{mm'} \frac{d}{dx} \left( D^{l+1} s_{m'}(x)\right) G_{nn'} \frac{d}{dx} \left(D^{l+1} s_{n'}(x)\right)dx\nonumber\\
&=& 2^{2(l+1)}H^{l+1-k}_{mm'} G_{nn'}\int  D^{l+1} \frac{d}{dx} s_{m'}(x) D^{l+1} \frac{d}{dx} s_{m'}(x)dx\nonumber\\
&=& 2^{2(l+1)}H^{l+1-k}_{mm'} G_{nn'}\int \frac{d}{dx} s_{m'}(x) \frac{d}{dx} s_{n'}(x)dx\nonumber\\
\label{eq:overlap_integral_derivative_scaling_and_wavelet_function_2}
&=& 2^{2(l+1)}H^{l+1-k}_{mm'} G_{nn'} T^{0}_{ss,m'n'},
\end{eqnarray}
and
\begin{eqnarray}
T^{lp}_{sw,mn} &=& \int w^{l'}_m(x) w^{p'}_{n}(x)dx,\quad p\geq l\nonumber\\
&=& \int \frac{d}{dx} \left( G_{mm'} D^{l+1} s_{m'}(x)\right) \frac{d}{dx} \left( G_{nn'} D^{p+1} s_{n'}(x)\right)dx\nonumber \\
&=& \int \frac{d}{dx} \left( (GH^{p-l})_{mm'} D^{p+1} s_{m'}(x)\right) \frac{d}{dx} \left( G_{nn'} D^{p+1} s_{n'}(x)\right)dx\nonumber \\
&=& \int (GH^{p-l})_{mm'} \frac{d}{dx} \left( D^{p+1} s_{m'}(x)\right) G_{nn'} \frac{d}{dx} \left(D^{p+1} s_{n'}(x)\right)dx\nonumber\\
&=& 2^{2(p+1)} (GH^{p-l})_{mm'} G_{nn'}\int  D^{l+1} \frac{d}{dx} s_{m'}(x) D^{l+1} \frac{d}{dx} s_{n'}(x)dx\nonumber\\
&=& 2^{2(p+1)} (GH^{p-l})_{mm'} G_{nn'}\int \frac{d}{dx} s_{m'}(x) \frac{d}{dx} s_{n'}(x)dx\nonumber\\
\label{eq:overlap_integral_derivative_wavelet_and_wavelet_function_2}
&=& 2^{2(p+1)}(GH^{p-l})_{mm'} G_{nn'} T^{0}_{ss,m'n'}.
\end{eqnarray}
From Eq. (\ref{eq:overlap_integral_derivative_scaling_function_2}), Eq. (\ref{eq:overlap_integral_derivative_scaling_and_wavelet_function_2}), and Eq. (\ref{eq:overlap_integral_derivative_wavelet_and_wavelet_function_2}), it is seen that the integrals can be expressed as linear combinations of matrices $T^0_{ss,mn}$. Consequently, determining the values of $T^0_{ss,mn}$, we can subsequently determine the integrals. In the integral $T^0_{ss,mn}$, all functions in the integrand are derivatives of the scale $\frac{1}{2^0}$ scaling functions. Therefore, we can advance to step three by shifting the leftmost function such that it starts from the origin by employing the translation invariance property of the scaling function as follows,
\begin{eqnarray}
T^0_{ss,mn} = \int s(x)s_{n-m}(x)dx=T^0_{ss,0(n-m)}=T^0_{ss,0l},
\end{eqnarray}
where, $l=(n-m)$.

Proceeding with the derivation of homogeneous equations for these integrals, which corresponds to step four, we will reduce the resolution of the scaling function by a factor of $-1$ in the integral $T^0_{ss,0l}$ by utilizing the unitarity of operator $\hat{D}$,
\begin{eqnarray}
T^0_{ss,0l}= T^{-1}_{0l}=\int s^{-1'}(x)s^{-1'}_{l}(x)dx.
\end{eqnarray}
Subsequently, raising the resolution by a factor of $1$ through the utilization of Eq. (\ref{eq:derivative_scaling_equation}), we attain a set of homogeneous equations for $T^0_{0l}$:
\begin{eqnarray}
T^0_{ss,0l} &=& 4 \sum_{p,r} H_{0p}H_{lr} \int s'_{p}(x) s'_{r}(x)dx, \nonumber\\
\label{eq:homogeneous_equation_T_k_ss}
&=&  4 \sum_{p,r} h_p h_{r-2l} T^0_{ss,0(r-p)}.
\end{eqnarray}

We proceed to establish an inhomogeneous equation, corresponding to step five, for the variable $D_{ss,0q}$, which, when combined with the homogeneous set of equations (\ref{eq:homogeneous_equation_T_k_ss}), enables the unique determination of $D_{ss,0q}$. 
To formulate this set of equations, we require the decomposition of $1$ (the partition of unity property of the scaling functions), $x$, and $x^2$ in the scaling function basis. The expansion of $1$ and $x$ are already established in Eq. (\ref{eq:expansion_of_x}) and Eq. (\ref{eq:partition_of_unity_2}),
\begin{eqnarray}
&1= \sum_n n s'_n(x),&\\
&x= \sum_n ns_n(x)+\frac{1}{\sqrt{2}}\sum_l lh_l.&
\end{eqnarray}
Now, we will derive the expansion of $x^2$. $x^2$ can be expressed pointwise in the scaling function basis in the following manner:
\begin{eqnarray}
x^2= \sum_n c_n s_n(x).
\end{eqnarray}
where,
\begin{eqnarray}
c_n = \int x^2 s_n(x) dx,
\end{eqnarray}
and can be evaluated as follows,
\begin{eqnarray}
c_n &=& \int x^2 s_n(x)dx \nonumber \\
&=& \int (n+x)^2 s(x) dx \nonumber \\
&=& \int (n^2+2 x+ x^2) s(x) dx\nonumber \\
&=& \left( n^2 +2 \expval{x}_s + \expval{x^2}_s\right),
\end{eqnarray}
$\expval{x}_s$ and $\expval{x^2}_S$ are called the moment of the scaling and can be evaluated as the procedure described in Sec. \ref{subsec:moment_of_scaling_and_wavelet_function}. Therefore, we can write $x^2$ in the scaling function basis as follows,
\begin{eqnarray}
\label{eq:expansion_of_x_2}
x^2= \sum_n \left(n^2+2n\expval{x}_s+\expval{x^2}_s\right)s_n(x).
\end{eqnarray}
Differentiating, Eq. (\ref{eq:expansion_of_x_2}) and utilizing the partition of unity property Eq. (\ref{eq:partition_of_unity_2}) we get
\begin{eqnarray}
\label{eq:expansion_of_2x_in_wavelet_basis}
2x= \sum_n (n^2+2n\expval{x}) s'_n(x)= \sum_n n^2 s'_n (x)+ 2\expval{x}.
\end{eqnarray}
Multiplying both side of Eq. (\ref{eq:expansion_of_2x_in_wavelet_basis}) with $s'(x)$ and integrating, we arrive at:
\begin{eqnarray}
\int 2x s'(x)dx &=& \sum_n \left(n^2 \int s'(x)s'_n(x)dx+2\expval{x}\int s'(x)dx\right)\nonumber\\
\implies -2 &=& \sum_n n^2 \int s'(x)s'_n(x)dx\nonumber\\
\label{eq:inhomogeneous_equation_T_k_ss}
\implies -2 &=& \sum_n n^2 T^0_{ss,0n},
\end{eqnarray}
the inhomogeneous equation for variables $T^0_{ss,0l}$.

The concluding step involves solving the set of homogeneous and inhomogeneous equations to determine the values of the coefficients.The system of linear equations presented in Eq. (\ref{eq:homogeneous_equation_T_k_ss}) and (\ref{eq:inhomogeneous_equation_T_k_ss}) can be accurately solved using built-in functions like "$Solve[]"$ in Mathematica or corresponding tools in other programming languages such as Matlab or Python. The resulting values for $T^0_{ss;0l}$ are rational numbers and were initially computed in reference \cite{doi:10.1137/0729097}. Refer to Table \ref{tab:the_ke_overlap_integrals} for details.
\begin{table}[ht]
\begin{center}
\setlength{\tabcolsep}{1.5pc}
\catcode`?=\active \def?{\kern\digitwidth}
\caption{The values of overlap integrals of product of derivative of scaling functions.}
\label{tab:the_ke_overlap_integrals}
\vspace{3mm}
\begin{tabular}{c c}
\specialrule{.15em}{.0em}{.15em}
\hline
Integrals & Values \\
\hline
$D_{ss,0(-4)}$ & $-3/560$ \\

$D_{ss,0(-3)}$ & $-4/35$ \\

$D_{ss,0(-2)}$ & $92/105$ \\

$D_{ss,0(-1)}$ & $-356/105$ \\

$D_{ss,00}$ & $295/56$ \\

$D_{ss,01}$ & $-356/105$\\

$D_{ss,02}$ & $92/105$\\

$D_{ss,03}$ & $-4/35$\\

$D_{ss,04}$ & $-3/560$\\
\hline
\specialrule{.15em}{.15em}{.0em}
\end{tabular}
\end{center}
\end{table}
\bibliographystyle{unsrt}
\section{The potential energy term}
\label{appen:the_potential_energy_term_1}

\subsection{The Dirac delta function potential}
\label{appen:The_Dirac_delta_function_potential}
The overlap integration involving the product of delta function and two scaling function, Eq. (\ref{eq:overlap_integral_delta_function_and_scaling_function_1}) and Eq. (\ref{eq:overlap_integral_delta_function_scaling_function_2}), can be expressed in terms of $0$th resolution scaling functions using the property of scaling operator $\hat{D}$ and changing the variable of the integration:
\begin{eqnarray}
V^k_{\delta ss,mn}&=&\int \delta(x)s^k_m(x)s^k_n(x),\nonumber\\
&=& 2^k V_{\delta ss,mn},
\end{eqnarray}
where
\begin{eqnarray}
V_{\delta ss,mn}\int \delta(x)s_m(x)s_n(x)dx.
\end{eqnarray}
Now, from the unitary nature of the operator $\hat{D}$ we can express $V_{\delta ss,mn}$ in terms of $V^{-1}_{\delta ss,mn}$ as follows:
\begin{eqnarray}
V^{-1}_{\delta ss,mn}&=& \int \delta(x) s^{-1}_m(x)s^{-1}_n(x)dx,\nonumber\\
&=& \frac{1}{2} \int \delta(x) s_m(x)s_n(x)dx,\nonumber\\
\implies V_{\delta ss,mn}&=& 2 \times V^{-1}_{\delta ss,mn},
\end{eqnarray}
and using Eq. (\ref{eq:scaling_equation_3}) we can find out the set of homogeneous equation for the integral $V_{\delta ss,mn}$,
\begin{eqnarray}
\label{eq:homogeneous_equation_delta_function_and_scaling_function}
V_{\delta ss,mn}=2\times \sum_{p,q} H_{mp}H_{nq}V_{\delta ss,pq}.
\end{eqnarray}
To get the inhomogeneous equation we start with the definition of delta function and the partition of unity property of scaling function Eq. (\ref{eq:partition_of_unity_1}).
\begin{eqnarray}
\int \delta(x)dx&=&1,\nonumber\\
\sum_{m,n} \int \delta(x)s_m(x)s_n(x)dx&=& 1,\nonumber\\
\label{eq:inhomogeneous_equation_delta_function_and_scaling_function}
\sum_{m,n} V_{\delta ss,mn}&=& 1.
\end{eqnarray}
Here, the delta function is centered at the origin. So, the delta function and two scaling functions will overlap for $-4\leq m,n \leq -1$. For all other values of $m$ and $n$ here will be no overlap among them, so the value of the integration will be $0$. We can solve the set of equations (\ref{eq:homogeneous_equation_delta_function_and_scaling_function}) and (\ref{eq:inhomogeneous_equation_delta_function_and_scaling_function}) to get all the possible values of integrals. See Table \ref{tab:overlap_integral_values_delta_function_scaling_function}.
\begin{table}[ht]
\begin{center}
\setlength{\tabcolsep}{1.5pc}
\catcode`?=\active \def?{\kern\digitwidth}
\caption{The values of overlap integrals of product of derivative of scaling functions.}
\label{tab:overlap_integral_values_delta_function_scaling_function}
\vspace{3mm}
\begin{tabular}{c c}
\specialrule{.15em}{.0em}{.15em}
\hline
Integrals & Values \\
\hline
$V_{\delta ss,(-4)(-4)}$ & $0.0000179297$ \\

$V_{\delta ss,(-4)(-3)}$ & $0.000403396$ \\

$V_{\delta ss,(-4)(-2)}$ & $-0.00163377$ \\

$V_{\delta ss,(-4)(-1)}$ & $0.00544679$ \\

$V_{\delta ss,(-3)(-3)}$ & $0.00907591$ \\

$V_{\delta ss,(-3)(-2)}$ & $-0.0367577$\\

$V_{\delta ss,(-3)(-1)}$ & $0.122546$\\

$V_{\delta ss,(-2)(-2)}$ & $0.14887$\\

$V_{\delta ss,(-2)(-1)}$ & $-0.496316$\\

$V_{\delta ss,(-1)(-1)}$ & $1.65466$\\
\hline
\specialrule{.15em}{.15em}{.0em}
\end{tabular}
\end{center}
\end{table}
\subsection{The Triangular potential}
\label{appen:The_triangular_potential}
The potential energy term of the matrix element for the triangular potential is given by,
\begin{eqnarray}
V^k_{\Delta ss,mn}&=& \int_{-\infty}^{\infty}|x|s^k_m(x)s^k_n(x)dx\nonumber\\
&=& -\int_{-\infty}^{0}xs^k_m(x)s^k_n(x)dx + \int_{0}^{\infty}xs^k_m(x)s^k_n(x)dx\nonumber\\
V^k_{\Delta ss,mn} &=& -I^{k-}_{xss,mn}+I^{k+}_{xss,mn},
\end{eqnarray}
where
\begin{eqnarray}
\label{eq:partial_moment_x_scaling_scaling_function_negative}
I^{k-}_{xss,mn}=\int_{-\infty}^{0}xs^k_m(x)s^k_n(x)dx,
\end{eqnarray}
and
\begin{eqnarray}
\label{eq:partial_moment_x_and_scaling_scaling_function_positive}
I^{k+}_{xss,mn}=\int_{0}^{\infty}xs^k_m(x)s^k_n(x)dx.
\end{eqnarray}
The integrals for $-4\leq m,n\leq -1$ are the 'partial integral’ because the end or starting points of these integrals are inside the support of the scaling function. The
evaluation of these integrations has been described later in this section.

To evaluate integrals, $I^{k+}_{xss,mn}$ for $m,n\geq -1$ and $I^{k-}_{xss,mn}$ for $m,n\leq -4$ we need to expand $x$ in the scaling function of resolution $k$ as follows,
\begin{eqnarray}
\label{eq:expansion_of_x_in_wavelet_basis}
x=\sum_n c_n s^k_n(x).
\end{eqnarray}
$c_n$ can be calculated as follows:
\begin{eqnarray}
c_n&=& \int_{-\infty}^{\infty} x s^k_n(x)dx\nonumber\\
&=& \int_{-\infty}^{\infty} x s^k(x-n)dx\nonumber\\
&=& \int_{-\infty}^{\infty} (x+n) s^k(x)dx\nonumber\\
&=& \int_{-\infty}^{\infty} x s^k(x)dx+n\times \frac{1}{2^{\frac{k}{2}}}\nonumber\\
\label{eq:coefficient_c_n}
&=& \expval{x}^k_s+n\times \frac{1}{2^{\frac{k}{2}}},
\end{eqnarray}
where,
\begin{eqnarray}
\label{eq:moment_of_the_scaling_function_4}
\expval{x}^k_s=\int x s^k(x)dx,
\end{eqnarray}
can be evaluated using the method described in Sec. \ref{subsec:moment_of_scaling_and_wavelet_function}. Putting, $m=1$ in Eq. (\ref{eq:moment_of_scaling_function_2}), we get the following relation,
\begin{eqnarray}
\expval{x}_s=\frac{1}{\sqrt{2}}\sum_{l=0}^{2K-1}lh_l.
\end{eqnarray}
From, Eq. (\ref{eq:expansion_of_x_in_wavelet_basis}), Eq. (\ref{eq:coefficient_c_n}), and Eq. (\ref{eq:moment_of_the_scaling_function_4}) we get the expansion of $x$ in wavelet basis,
\begin{eqnarray}
\label{eq:expansion_of_x_in_wavelet_basis_1}
x=\frac{1}{2^{\frac{k}{2}}}\left(\sum_p ps^k_p(x)+\frac{1}{\sqrt{2}}\sum_{l=1}^{2K-1}lh_l \right).
\end{eqnarray}
By utilizing Eq. (\ref{eq:expansion_of_x_in_wavelet_basis_1}) and Eq. (\ref{eq:partial_moment_x_and_scaling_scaling_function_positive}), and leveraging of the orthonormality property of the scaling function, we arrive at the following equation:
\begin{eqnarray}
I^{k+}_{xss,mn}&=&\frac{1}{2^{\frac{k}{2}}}\left(\sum_p p \int_0^{\infty}s^k_p(x)s^k_m(x)s^k_n(x)dx+\expval{x}_s\int_0^\infty s^k_m(x)s^k_n(x)dx \right)\nonumber\\
&=& \frac{1}{\frac{k}{2}}\left(\sum_p p I^{k+}_{sss,pmn}+\expval{x}_s \delta_{mn} \right),
\end{eqnarray}
where,
\begin{eqnarray}
I^{k+}_{sss,pmn}&=& \int_0^{\infty} s^k_p(x)s^k_m(x)s^k_n(x)dx.
\end{eqnarray}
Similarly, employing the same procedure leads to a similar equation for Eq. (\ref{eq:partial_moment_x_scaling_scaling_function_negative}),
\begin{eqnarray}
I^{k-}_{xss,mn} = \frac{1}{\frac{k}{2}}\left(\sum_p p I^{k-}_{sss,pmn}+\expval{x}_s \delta_{mn} \right),
\end{eqnarray}
where,
\begin{eqnarray}
I^{k-}_{sss,pmn}&=& \int^0_{-\infty} s^k_p(x)s^k_m(x)s^k_n(x)dx.
\end{eqnarray}
Due to the compact support of the scaling function, $I^{k+}_{sss,pmn}$ and $I^{k-}_{sss,pmn}$ are the 'full moment' for $m,n\leq -4$ and $m,n\geq -1$. So, we can replace them with the integrals of the following form,
\begin{eqnarray}
I^k_{sss,pmn}=\int_{-\infty}^{\infty} s^k_p(x)s^k_m(x)s^k_n(x) dx.
\end{eqnarray}
From the property of translation invariance, we write the following equation,
\begin{eqnarray}
I^k_{sss,pmn}&=& \int_{-\infty}^{\infty} s^k_0(x)s^k_{m-p}(x)s^k_{n-p}(x) dx\nonumber\\
&=& I^k_{sss,0(m-p)(n-p)}=I^k_{sss,0rq},
\end{eqnarray}
such that,
\begin{eqnarray}
r=(m-p), \quad\quad \text{and}\quad\quad q=(n-p).
\end{eqnarray}
We first decrease the resolution by one unit from the property of dilation operator and change the variable to get the set of homogeneous equation for the variables $I^k_{sss,0rq}$ as follows,
\begin{eqnarray}
I^{k-1}_{sss,0rq} &=& \int D^{-1} s^k(x) D^{-1} s^k_r(x) D^{-1} s^k_q(x)\nonumber\\
&=& \int \frac{1}{\sqrt{2}} s^k\left(\frac{x}{2}\right)\frac{1}{\sqrt{2}} s^k_r\left(\frac{x}{2}\right)\frac{1}{\sqrt{2}} s^k_1\left(\frac{x}{2}\right)dx\nonumber\\
\implies I^{k}_{sss,0rq}&=& \sqrt{2} I^{k-1}_{sss,0rq}.
\end{eqnarray}
From Eq. (\ref{eq:scaling_equation_3}), we get the following homogeneous equation for $I^k_{sss,0rq}$,
\begin{eqnarray}
\label{eq:homogeneous_equation_I_sss_0rq}
I^k_{sss,0rq}=\sqrt{2} \sum_{p,u,v} H_{0p}H_{r,u+p}H_{q,p+v} I^k_{sss,0uv}.
\end{eqnarray}
Here, the variables $p$, $u$, and $v$ range from $-4$ to $4$.

From the normalization condition of scaling function and the partition of unity Eq. (), we will get the following inhomogeneous equations for the variable $I^k_{sss,0rq}$,
\begin{eqnarray}
\int s^k(x)s^k(x)dx &=& 1\nonumber\\
\implies \sum_p \int s^k_p(x)s^k(x)s^k(x)dx &=& 1 \quad \text{(using the partition of unity)}\nonumber\\
\label{eq:inhomogeneous_equation_I_sss_0rq}
\implies \sum_p I^k_{sss,p00} &=& 1.
\end{eqnarray} 
We can solve Eq. (\ref{eq:homogeneous_equation_I_sss_0rq}) and Eq. (\ref{eq:inhomogeneous_equation_I_sss_0rq}) to get the values of $I^k_{sss,0rq}$ which are given in Table \ref{tab:the_triangular_potential_overlap_integrals}.
\begin{table*}[ht]
\begin{center}
\setlength{\tabcolsep}{1.5pc}
\catcode`?=\active \def?{\kern\digitwidth}
\caption{The values of overlap integrals of product of three scaling functions.}
\label{tab:the_triangular_potential_overlap_integrals}
\vspace{1mm}
\begin{tabular}{c c c c}
\specialrule{.15em}{.0em}{.15em}
\hline
Integrals & Values & Integrals & Values\\
\hline
$I^{k}_{sss,0(-4)(-4)}$ & $1.16064\times 10^{-7}$ & $I^{k}_{sss,0(-1)(2)}$ & $-0.000544457$\\

$I^{k}_{sss,0(-4)(-3)}$ & $9.7888\times 10^{-7}$ & $I^{k}_{sss,0(-1)(3)}$ & $9.7888\times 10^{-7}$\\

$I^{k}_{sss,0(-4)(-2)}$ & $-2.81154\times 10^{-6}$ & $I^{k}_{sss,0(0)(0)}$ & $0.910448$ \\

$I^{k}_{sss,0(-4)(-1)}$ & $6.18441\times 10^{-6}$ & $I^{k}_{sss,0(0)(1)}$ & $0.146924$\\

$I^{k}_{sss,0(-4)(0)}$ & $-4.46781\times 10^{-6}$ & $I^{k}_{sss,0(0)(2)}$ & $0.00702793$\\

$I^{k}_{sss,0(-3)(-3)}$ & $0.000202592$ & $I^{k}_{sss,0(0)(3)}$ & $0.000202592$\\

$I^{k}_{sss,0(-3)(-2)}$ & $-0.000544457$ & $I^{k}_{sss,0(0)(4)}$ & $1.16064\times 10^{-7}$\\

$I^{k}_{sss,0(-3)(-1)}$ & $0.00115963$ & $I^{k}_{sss,0(1)(1)}$ & $-0.0866059$\\

$I^{k}_{sss,0(-3)(0)}$ & $-0.000824925$ & $I^{k}_{sss,0(1)(2)}$ & $-0.0304701$\\

$I^{k}_{sss,0(-3)(1)}$ & $6.18441\times 10^{-6}$ & $I^{k}_{sss,0(1)(3)}$ & $-0.000544457$\\

$I^{k}_{sss,0(-2)(-2)}$ & $0.00702793$ & $I^{k}_{sss,0(1)(4)}$ & $9.7888\times 10^{-7}$\\

$I^{k}_{sss,0(-2)(-1)}$ & $-0.0304701$ & $I^{k}_{sss,0(2)(2)}$ & $0.0228326$\\

$I^{k}_{sss,0(-2)(0)}$ & $0.0228326$ & $I^{k}_{sss,0(2)(3)}$ & $0.00115963$\\

$I^{k}_{sss,0(-2)(1)}$ & $0.00115963$ & $I^{k}_{sss,0(2)(4)}$ & $-2.81154\times 10^{-6}$\\

$I^{k}_{sss,0(-2)(2)}$ & $-2.81154\times 10^{-6}$ & $I^{k}_{sss,0(3)(3)}$ & $-0.000824925$\\

$I^{k}_{sss,0(-1)(-1)}$ & $0.146924$ & $I^{k}_{sss,0(3)(4)}$ & $6.18441\times 10^{-6}$\\

$I^{k}_{sss,0(-1)(0)}$ & $-0.0866059$ & $I^{k}_{sss,0(4)(4)}$ & $-4.46781\times 10^{-6}$\\

$I^{k}_{sss,0(-1)(1)}$ & $-0.0304701$ \\
\hline
\specialrule{.15em}{.15em}{.0em}
\end{tabular}
\end{center}
\end{table*}

Now we will calculate the partial moment,
\begin{eqnarray}
PI^{k+}_{xss,mn} &=& \int_0^{\infty} x s^k_m(x)s^k_n(x) dx,\\
PI^{k-}_{xss,mn} &=& \int_{-\infty}^{0} x s^k_m(x)s^k_n(x) dx,
\end{eqnarray}
for,
\begin{eqnarray}
-4\leq m,n \leq -1.
\end{eqnarray}
First, we decrease the resolution of $PI^{k+}_{xss,mn}$ by using the property of dilation operator by one unit as follows,
\begin{eqnarray}
PI^{-1,k-1+}_{xss,mn} &=& \int_0^{\infty} D^{-1}x D^{-1}s^k_m(x) D^{-1} s^k_n(x)dx\nonumber\\
&=& \int \frac{1}{\sqrt{2}}\frac{x}{2}\frac{1}{\sqrt{2}}s^k_m\left(\frac{x}{2}\right)\frac{1}{\sqrt{2}}s^k_n\left(\frac{x}{2}\right)dx\nonumber\\
\implies PI^{k+}_{xss,mn} &=& \sqrt{2} PI^{-1,k-1+}_{xss,mn}.
\end{eqnarray}
Now, changing the variable, we get the following relation
\begin{eqnarray}
PI^{k+}_{xss,mn}=\frac{1}{2}\int_0^{\infty} xs^{k-1}_m(x)s^{k-1}_n(x).
\end{eqnarray}
Now, from Eq. (\ref{eq:scaling_equation_3}), we get the following homogeneous equation,
\begin{eqnarray}
PI^{k+}_{xss,mn} &=& \frac{1}{2}\sum_{p,q} H_{mp} H_{nq} \int_0^{\infty} x s^k_p(x)s^k_q(x)dx\nonumber\\
&=& \frac{1}{2} \sum_{p,q} H_{mp} H_{nq} PI^{k+}_{xss,pq},
\end{eqnarray}
where,
\begin{eqnarray}
H_{mp}=h_{p-2m}\quad \text{and} \quad H_{nq}=h_{q-2n}.
\end{eqnarray}
Now changing the index
\begin{eqnarray}
p-2m=r \quad \text{and} \quad q-2n=s,
\end{eqnarray}
we will get the following form of the equation,
\begin{eqnarray}
\label{eq:partial_moment_positive_triangular_potential}
PI^{k+}_{xss,mn}=\frac{1}{2}\sum_{r,s}h_r h_s PI^{k+}_{xss,(r+2m)(s+2n)}.
\end{eqnarray}
We get the same equation for $PI^{k-1}_{xss,mn}$,
\begin{eqnarray}
\label{eq:partial_moment_negative_triangular_potential}
PI^{k-}_{xss,mn}=\frac{1}{2}\sum_{r,s}h_r h_s PI^{k-}_{xss,(r+2m)(s+2n)},
\end{eqnarray}
for $m\geq -1$ and $n\geq -1$:
\begin{eqnarray}
PI^{k+}_{xss,mn}=I^{k+}_{xss,mn},
\end{eqnarray}
and for $m\leq -4$ and $n\leq -4$:
\begin{eqnarray}
PI^{k-}_{xss,mn}=I^{k-}_{xss,mn}.
\end{eqnarray}
Equation (\ref{eq:partial_moment_positive_triangular_potential}) and Eq. (\ref{eq:partial_moment_negative_triangular_potential}) are the linear system of equations which relates the known integral $I^{k+}_{xss,mn}$ and $I^{k-}_{xss,mn}$ with the unknown integrals  $PI^{k+}_{xss,mn}$ and $PI^{k-}_{xss,mn}$ respectively. Those system of equations can be solved to get the values of partial moments. See Table. \ref{tab:the_partial_moments}
\begin{table*}[ht]
\begin{center}
\setlength{\tabcolsep}{1.5pc}
\catcode`?=\active \def?{\kern\digitwidth}
\caption{The values of partial moments for product of $x$ and two scaling functions.}
\label{tab:the_partial_moments}
\vspace{1mm}
\begin{tabular}{c c c c}
\specialrule{.15em}{.0em}{.15em}
\hline
Integrals & Values & Integrals & Values\\
\hline
$PI^{k+}_{ss,(-4)(-4)}$ & $1.11124\times 10^{-7}$ & $PI^{k-}_{ss,(-4)(-4)}$ & $2.97758$\\

$PI^{k+}_{ss,(-4)(-3)}$ & $1.12319\times 10^{-6}$ & $PI^{k-}_{ss,(-4)(-3)}$ & $0.121044$\\

$PI^{k+}_{ss,(-4)(-2)}$ & $-3.56546\times 10^{-6}$ & $PI^{k-}_{ss,(-4)(-2)}$ & $-0.0192164$ \\

$PI^{k+}_{ss,0(-4)(-1)}$ & $9.01137\times 10^{-6}$ & $PI^{k-}_{ss,(-4)(-1)}$ & $0.000685231$\\

$PI^{k+}_{ss,0(-3)(-3)}$ & $0.000183898$ & $PI^{k-}_{ss,(-3)(-3)}$ & $1.97776$\\

$PI^{k+}_{ss,0(-3)(-2)}$ & $-0.000530252$ & $PI^{k-}_{ss,(-3)(-2)}$ & $0.120513$\\

$PI^{k+}_{ss,0(-3)(-1)}$ & $0.00125024$ & $PI^{k-}_{ss,(-3)(-1)}$ & $-0.0179626$\\

$PI^{k+}_{ss,0(-2)(-2)}$ & $0.00655595$ & $PI^{k-}_{ss,(-2)(-2)}$ & $0.984134$\\

$PI^{k+}_{ss,0(-2)(-1)}$ & $-0.0275912$ & $PI^{k-}_{ss,(-2)(-1)}$ & $0.0934521$\\

$PI^{k+}_{ss,0(-1)(-1)}$ & $0.140356$ & $PI^{k-}_{ss,(-1)(-1)}$ & $0.117934$\\
\hline
\specialrule{.15em}{.15em}{.0em}
\end{tabular}
\end{center}
\end{table*}
\subsection{The potential of the form $x^{2p}$}
\label{appen:The potential of the form x_2n}
The potential energy term for the potential term of the form, $x^{2p}$, in a wavelet basis using scaling-wavelet function, is expressed as:
\begin{eqnarray}
\begin{split}
\label{eq:potential_energy_terms_harmonic_oscillator_2}
V^k_{x^{2p}ss,mn}&=&\int x^{2p}s^k_{m}(x)s^k_n(x)\\
V^{k,l}_{x^{2p}sw,mn}&=&\int x^{2p}s^k_{m}(x)w^l_n(x),\quad l\geq k\\
V^{q,k}_{x^{2p}ws,mn}&=&\int x^{2p}w^q_{m}(x)s^k_n(x), \quad q\geq k\\
V^{q,l}_{x^{2p}ww,mn}&=&\int x^{2p}w^q_{m}(x)w^l_n(x), \quad l\geq q\quad \text{or}\quad q\geq l.
\end{split}
\end{eqnarray}
We can represent $x^{2p}$ in the a wavelet basis using scaling and wavelet function as follows:
\begin{eqnarray}
\label{eq:expansion_of_x_2p_using_scaling_wavelet_basis}
x^{2p}=\sum_{n=-\infty}^{\infty} c^{s,k}_n s^k_n(x)+\sum_{n=-\infty}^{\infty}\sum_{l\geq k} d^{w,l}_n w^{l}_n(x),
\end{eqnarray} 
where,
\begin{eqnarray}
c^{s,k}_n=\int x^{2p} s^k_n(x)dx = \expval{x^{2p}}^k_{s,n}, \quad \text{and} \quad d^{w,l}_n=\int x^{2p} w^l_n(x)dx = \expval{x^{2p}}^k_{w,n}.
\end{eqnarray}
In this context, $c^{s,k}_n$ and $d^{w,l}_n$ denote the moments of the scaling and wavelet functions, respectively, and can be calculated using the procedure outlined in Sec. \ref{subsec:moment_of_scaling_and_wavelet_function}.

Substituting expansion of $x^{2p}$, Eq. (\ref{eq:expansion_of_x_2p_using_scaling_wavelet_basis}), into Eq. (\ref{eq:potential_energy_terms_harmonic_oscillator_2}), we obtain the following expressions for the potential energy term:
\begin{eqnarray}
\begin{split}
V^k_{x^{2p}ss,mn}&=& \sum_{r} c^{s,k}_r I^{k,k,k}_{sss,rmn}+\sum_r \sum_{l \geq k}d^{w,l}_n I^{l,k,k}_{wss,rmn},\\
V^{k,l}_{x^{2p}sw,mn}&=& \sum_{r} c^{s,k}_r I^{k,k,l}_{ssw,rmn}+\sum_r \sum_{q \geq k}d^{w,q}_r I^{q,k,l}_{wsw,rmn},\\
V^{q,k}_{x^{2p}ws,mn}&=& \sum_{r} c^{s,k}_r I^{k,q,k}_{sws,rmn}+\sum_r \sum_{l \geq k}d^{w,l}_r I^{l,q,k}_{wws,rmn},\\
V^{q,l}_{x^{2p}ww,mn}&=& \sum_{r} c^{s,k}_r I^{k,q,l}_{sww,rmn}+\sum_r \sum_{t \geq k}d^{w,t}_r I^{t,q,l}_{www,rmn}.
\end{split}
\end{eqnarray}
Here, $I^{l,k,k}_{wss,rmn}$, $I^{k,k,l}_{ssw,rmn}$, and $I^{k,q,k}_{sws,rmn}$ are equivalent. Similarly $I^{q,k,l}_{wsw,rmn}$, $I^{l,q,k}_{wws,rmn}$, and $I^{k,q,l}_{sww,rmn}$ are also identical. Therefore, the evaluation of the following integrals is adequate:
\begin{eqnarray}
I^{k,k,k}_{sss,rmn}&=&\int s^k_r(x)s^k_m(x)s^k_n(x)dx,\\
I^{k,k,l}_{ssw,rmn}&=&\int s^k_r(x)s^k_m(x)w^l_n(x)dx,\quad l \geq k\\
I^{k,q,l}_{sww,rmn}&=&\int s^k_r(x)w^q_m(x)w^l_n(x)dx,\quad l\geq q\geq k\\
I^{t,q,l}_{sww,rmn}&=&\int w^t_r(x)w^q_m(x)w^l_n(x)dx,\quad l\geq q\geq t
\end{eqnarray}
$I^{k,k,l}_{ssw,rmn}$, $I^{k,q,l}_{sww,rmn}$, and $I^{t,q,l}_{sww,rmn}$ can be represented as the linear sum of $I^{k,k,k}_{sss,rmn}$ using Eq. (\ref{eq:scaling_equation_3}) and Eq. (\ref{eq:wavelet_equation_3}) as follows,
\begin{eqnarray}
I^{k,k,l}_{ssw,rmn}&=& \sum_{p,q} G_{np} H^{l-k-1}_{pq} I^{k,k,k}_{sss,rmn},\\
I^{k,q,l}_{sww,rmn}&=& \sum_{p,u,s,t} G_{mp} G_{ns} H^{q-k-1}_{pu} H^{l-k-1}_{st} I^{k,k,k}_{sss,rut},\\
I^{t,q,l}_{www,rmn}&=& \sum_{p,u,v,i,j} G_{rp} G_{mu} G_{ni} H^{q-t}_{uv} H^{l-t}_{ij} I^{t-1,t-1,t-1}_{sss,pvj}.
\end{eqnarray}
The computation of $I^{k,k,k}_{sss,rmn}$ is given in the Appendix \ref{appen:The_triangular_potential}. Substituting $p=1$ in Eq. (\ref{eq:expansion_of_x_2p_using_scaling_wavelet_basis}), we obtain the potential energy term for the harmonic oscillator.
\subsection{Overlap integral involving the product of four scaling functions}
\label{appen:Overlap_integral_involving_the_product_of_four_scaling_functions}
The general matrix element for the $\Phi^4(x)$, interaction term can be found out from the following matrix elements:
\begin{eqnarray}
\Gamma^k_{ssss,n_1,n_2,n_3,n_4}=\int s^k_{n_1}(x)s^k_{n_2}(x)s^k_{n_3}(x)s^k_{n_4}(x)dx.
\end{eqnarray}
By changing the variable of the integration the integral will take the following form,
\begin{eqnarray}
\Gamma^k_{ssss,n_1,n_2,n_3,n_4}=2^k \Gamma^0_{ssss,n_1,n_2,n_3,n_4},
\end{eqnarray}
where,
\begin{eqnarray}
\label{eq:overlap_integral_Gamma_ssss}
\Gamma^0_{ssss,n_1,n_2,n_3,n_4}=\int s_{n_1}(x)s_{n_2}(x)s_{n_3}(x)s_{n_4}(x)dx.
\end{eqnarray}
Using translational invariance, we can rewrite Eq. (\ref{eq:overlap_integral_Gamma_ssss}) as,
\begin{eqnarray}
\Gamma_{ssss,0,l_1,l_2,l_3}&=&\int s_{0}(x) s_{l_1}(x) s_{l_2}(x)s_{l_3}(x) dx,\nonumber\\
&& \quad l_1=(n_2-n_1),\quad l_2=(n_3-n_1),\text{and}\quad l_3=(n_4-n_1).
\end{eqnarray}
For $K=3$, the scaling function and their derivatives have support on $[0,5]$. This means $\Gamma_{ssss,0,l_1,l_2,l_3}$ will be non-zero only for the values of $l_1$, $l_2$, and $l_3$ lying between $-4$ to $4$. So, there will be $9^3$, which is equal to $729$, nonzero values of the integral. These integrals are related to each other through a set of equations which can be derived by the procedure given in Appendix \ref{appen:the_procedure_for_determining_the_overlap_integrals}.

By leveraging the property of the operator $\hat{D}$, we can scale down the resolution functions by a factor of $-1$. This allows us to establish the relationship between $\Gamma^{-1}_{ssss,0,l_1,l_2,l_3}$ and $\Gamma_{ssss,0,l_1,l_2,l_3}$ as expressed below:
\begin{eqnarray}
\Gamma^{-1}_{ssss,0,l_1,l_2,l_3}&=&\int \hat{D}^{-1}s_{0}(x)\hat{D}^{-1}s_{l_1}(x)\hat{D}^{-1}s_{l_2}(x)\hat{D}^{-1}s_{l_3}(x)dx\nonumber\\
&=& \int 2^{-\frac{1}{2}}s_0(\frac{x}{2})2^{-\frac{1}{2}}s_{l_1}(\frac{x}{2})2^{-\frac{1}{2}}s_{l_2}(\frac{x}{2})2^{-\frac{1}{2}}s_{l_3}(\frac{x}{2})dx\nonumber\\
&=& \frac{1}{2^2}\int s_0(\frac{x}{2})s_{l_1}(\frac{x}{2})s_{l_2}(\frac{x}{2})s_{l_3}(\frac{x}{2})dx \nonumber\\
&=& \frac{1}{2}\int s_0(x)s_{l_1}(x)s_{l_2}(x)s_{l_3}(x)dx\nonumber\\
&=& \frac{1}{2}\Gamma_{ssss,0,l_1,l_2,l_3}\nonumber\\
\label{eq:relation_between_gamma_0_and_gamma_-1}
\Gamma_{ssss,0,l_1,l_2,l_3}&=&2 \Gamma^{-1}_{ssss,0,l_1,l_2,l_3}.
\end{eqnarray}
Now, we can scale up the resolution of $\Gamma^{-1}_{ssss,0,l_1,l_2,l_3}$ by using Eq. (\ref{eq:scaling_equation_2}) and arrive at the following set of homogeneous equation:
\begin{eqnarray}
\Gamma_{ssss,0,l_1,l_2,l_3}&=& 2 \int s^{-1}_0(x)s^{-1}_{l_1}(x)s^{-1}_{l_2}(x)s^{-1}_{l_3}(x)dx\nonumber\\
&=& 2 \sum_{p_0,p_1,p_2,p_3} H_{0p_0}H_{l_1 p_1}H_{l_2 p_2}H_{l_3 p_3}\int s_{p_0}(x)s_{p_1}(x)s_{p_2}(x)s_{p_3}(x)dx\quad\nonumber\\
&=& 2 \sum_{p_0,p_1,p_2,p_3} h_{p_0}h_{p_1-2l_1}h_{p_2-2l_2}h_{p_3-2l_3}\Gamma_{ssss,0,(p_1-p_0),(p_2-p_0),(p_3-p_0)}.
\end{eqnarray}
Upon changing the variables $p_0$, $p_1-2l_1$, $p_2-2l_2$, and $p_3-2l_3$ to $q_0$, $q_1$, $q_2$, and $q_3$, respectively, the resulting equation is as follows:
\begin{eqnarray}
\label{eq:homogeneous_equations_gamma}
\Gamma_{ssss,0,l_1,l_2,l_3}=2 \sum_{q_0,q_1,q_2,q_3} h_{q_0} h_{q_1} h_{q_2} h_{q_3} \Gamma_{ssss,0,(2l_1+q_1-q_0),(2l_2+q_2-q_0),(2l_3+q_3-q_0)}.
\end{eqnarray}
We proceed to derive a set of inhomogeneous equations for the variable $\Gamma_{ssss,0,l_1,l_2,l_3}$. This set, when combined with the homogeneous set of equations (\ref{eq:homogeneous_equations_gamma}), enables the unique determination of $\Gamma_{ssss,0,l_1,l_2,l_3}$. We begin with the partition of unity equation, Eq. (\ref{eq:partition_of_unity_1}) and the normalization condition of scaling functions, Eq. (\ref{eq:normalization_condition_sk_m_sk_n}).
\begin{eqnarray}
\int s_{0}(x)s_0(x)dx &=& 1\nonumber\\
\implies \sum_{q_2,q_3}\int s_0(x)s_0(x)s_{q_2}(x)s_{q_3}(x)dx &=& 1\nonumber\\
\implies \sum_{q_2,q_3} \Gamma_{ssss,0,0,q_2,q_3} &=& 1. 
\end{eqnarray} 
This is the inhomogeneous equation for the variable $\Gamma_{ssss,0,l_1,l_2,l_3}$.

\chapter{Green's Function}
\label{appen:greens_function}
In this section of the appendix, I will discuss the derivation of Eq. (\ref{eq:expression_of_greens_function}) and the integral expression of the Green's function corresponding to a specified Hamiltonian such that the Green's function for the free Hamiltonian is provided.

The equation for the Green's function in a representation-independent form is expressed for the Hamiltonian, $H$, as follows:
\begin{eqnarray}
\label{eq:greens_equation_representation_independent_form}
(E-H)G=\mathds{1}.
\end{eqnarray}
Where, the Hamiltonian $H$ is decomposed into the sum of its free component, $H_0$, and the interacting component, $V$,
\begin{eqnarray}
\label{eq:hamiltonian_h_0_and_v}
H=H_0+V.
\end{eqnarray}
The Green's function equation for the free Hamiltonian is given by,
\begin{eqnarray}
\left(E-H_0\right)G_0 &=& \mathds{1}\nonumber\\
G_0 &=& (E-H_0)^{-1},
\end{eqnarray}
the expression for $G_0$ is only valid, given that the inversion of the operator $(E-H_0)$ exists.

From Eq. (\ref{eq:hamiltonian_h_0_and_v}), Eq. (\ref{eq:greens_equation_representation_independent_form}) can be rewritten as,
\begin{eqnarray}
\left(E-H_0-V\right)G &=& \mathds{1}\nonumber \\
G &=& \left(E-H_0-V\right)^{-1}\nonumber\\
&=& \left[(E-H_0)(\mathds{1}-V(E-H_0)^{-1})\right]^{-1}\nonumber\\
&=& (\mathds{1}-V(E-H_0)^{-1})^{-1}(E-H_0)^{-1}\nonumber\\
&=& (\mathds{1}-VG_0)^{-1}G_0 \nonumber\\
&=& (\mathds{1}+VG_0+VG_0VG_0+...)G_0\nonumber\\
&=& G_0+VG_0(\mathds{1}+VG_0+VG_0VG_0+...)G_0\nonumber\\
\label{eq:expression_for_G_G_0_is_given}
&=& G_0+VG_0G.
\end{eqnarray}
The position basis representation of Eq. (\ref{eq:expression_for_G_G_0_is_given}) yields the following expression:
\begin{eqnarray}
\bra{\textbf{x}}G\ket{\textbf{y}} &=& \bra{\textbf{x}}G_0\ket{\textbf{y}} + \int \int d^3\textbf{z}d^3\textbf{z}^{\prime}\bra{\textbf{x}}G_0\ket{\textbf{z}}\bra{\textbf{z}}V\ket{\textbf{z}^{\prime}}\bra{\textbf{z}^{\prime}}G\ket{\textbf{y}}\nonumber\\
G(E;\textbf{x},\textbf{y})&=& G_0(E;\textbf{x},\textbf{y})+\int d^3\textbf{z}\int d^3\textbf{z}^{\prime}G(E;\textbf{x},\textbf{z})V(\textbf{z}^{\prime})\delta^3(\textbf{z}-\textbf{z}^{\prime})G(E;\textbf{z}^{\prime},\textbf{y})\nonumber\\
&=&G_0(E;\textbf{x},\textbf{y})+\int d^3\textbf{z} G_0(E;\textbf{x},\textbf{z})V(\textbf{z})G(E;\textbf{z},\textbf{y}).
\end{eqnarray}
\bibliography{References}
\addcontentsline{toc}{chapter}{Bibliography}

\newpage
\addcontentsline{toc}{chapter}{\textbf{List of publications}}
\chapter*{List of publications}
\begin{flushleft}
~~~~~\textbf{\textit{Published in International Journals}}
\begin{enumerate}
\item \textbf{Renormalization in a wavelet basis}\\
Mrinmoy Basak and Raghunath Ratabole\\
  \href{https://link.aps.org/doi/10.1103/PhysRevD.107.036015}{PhysRevD.107.036015}.	
\item \textbf{Hamiltonian Flow Equations in Daubechies Wavelet Basis}\\
Mrinmoy Basak and Raghunath Ratabole\\
  \href{https://doi.org/10.48550/arXiv.2501.13618}{Accepted in PRD}.	
\end{enumerate}

~~~~~\textbf{\textit{Conference Publications}}
\begin{enumerate}
\item \textbf{Renormalization in a wavelet basis}\\
Mrinmoy Basak\\
  \href{https://doi.org/10.1016/j.nuclphysbps.2023.11.011}{Nuclear and Particle Physics Proceedings, Vol 343, pages 120-124}.	
\end{enumerate}

\end{flushleft}

\newpage
\addcontentsline{toc}{chapter}{\textbf{Brief Bio-data of the candidate}}
\chapter*{Brief Bio-data of the candidate}
\noindent
Mrinmoy Basak joined the Department of Physics, BITS-Pilani, KK Birla Goa Campus as a Ph.D. student in January 2019. He holds the Bachelor’s degree in Physics (2013) from West Bengal State University, Barasat. He also holds the masters degree in physics from IIt Guwahati (2016). He qualified the NET (July 2019) and the Gate (2018) exam.
\vspace{1mm}
\large
\begin{center}
\textbf{\textit{Conferences and Workshop}}
\end{center}

\normalsize
\begin{itemize}
\item Attended the SERB School during Ph.D. in 2019.
\item Give an oral presentation of my work “Renormalization in a wavelet basis” in the international seminar on QCD at Montpellier, France in 2023. One conference publication resulted from that seminar. 
\item Give an oral presentation on “Exploring wavelets: Quantum mechanics, Renormalization, and Similarity Renormalization group in Daubechies wavelet basis” at a seminar in “BITS-Pilani KK Birla Goa Campus, August 2023.”
\end{itemize}

\newpage
\addcontentsline{toc}{chapter}{\textbf{Brief Bio-data of the supervisor}}
\chapter*{Brief Bio-data of the supervisor}

\large
\begin{flushleft}
\textbf{Personal Information:}
\vspace{2mm}

\normalsize
\textbf{Name}\tab: \hspace{5mm}Raghunath Anand Ratabole

\textbf{Affiliation}\tab: \hspace{5mm}BITS Pilani, K K Birla Goa Campus

\textbf{Designation}\tab: \hspace{5mm}Professor

\textbf{Nationality}\tab: \hspace{5mm}Indian

\textbf{Phone}\tab: \hspace{5mm}0832-2580417

\textbf{Email}\tab: \hspace{5mm}ratabole@goa.bits-pilani.ac.in

\large
\textbf{Educational \& Professional Experience:}
\vspace{2mm}

\normalsize
\textbf{Professor}\tab: \hspace{5mm}2023-current, BITS Pilani, K K Birla Goa Campus

\textbf{Associate Professor}\tab: \hspace{5mm}2013-2023, BITS Pilani, K K Birla Goa Campus

\textbf{Assistant Professor}\tab: \hspace{5mm}2005–2013, BITS Pilani, K K Birla Goa Campus

\textbf{Post-doc}\tab: \hspace{5mm}2003-2005, Institute of Mathematical Sciences, Taramani, Chennai

\textbf{Ph.D.}\tab: \hspace{5mm}1996 – 2003, Indian Institute of Science, Bangalore

\textbf{M.Sc.}\tab: \hspace{5mm}1994-1996, Indian Institute of Technology, Bombay, Powai, Mumbai

\textbf{B.Sc.}\tab: \hspace{5mm}1991-1994, KET’s V. G. Vaze College of Arts, Commerce and Science, Mulund, Mumbai
\end{flushleft}

\large
\noindent
\textbf{Teaching Experience:}
\vspace{2mm}

\normalsize

\noindent
Raghunath Ratabole has taught various courses at the undergraduate and postgraduate levels in the Physics department of the Goa Campus of BITS Pilani. These include Mechanics, Oscillations \& Waves, Classical Mechanics, Electromagnetic Theory I \& II, Methods of Mathematical Physics, Atomic \& Molecular Physics, Quantum Optics, Atoms and Photons, Quantum Theory \& Applications and Introduction to Quantum Field Theory. He has also taught laboratory courses on Mechanics, Oscillations \& Waves, Electromagnetics \& Optics, Modern Physics and Advanced Physics.

\large
\noindent
\textbf{Research Interests:}
\vspace{2mm}

\normalsize

\noindent
His research interests are in the area of High Energy Physics and Quantum Field Theory. He is currently interested in studying non-perturbative wavelet-based methods for analysis of Quantum Field Theories.

\large
\noindent
\textbf{Academic Leadership:}
\vspace{2mm}

\normalsize

\noindent
Raghunath Ratabole serves as the Associate Dean of Work Integrated Learning Programmes (WILP) from 2010 to 2020. He was involved in developing and implementing Digital Learning initiatives for WILP. He currently serves as the nucleus member of the Teaching \& Learning Center.
\end{document}